%% file: texte.tex
\documentclass[12pt,a4paper]{book}

\usepackage{amsmath} 
\usepackage{amscd}
\usepackage{amsfonts}
\usepackage{amsthm,latexsym,multibox,amssymb}
\usepackage{amsbsy}
\usepackage{pb-diagram}   
\usepackage{psfrag}

\usepackage{courier}
\usepackage{graphicx}

\usepackage{cite}
\usepackage{fancyhdr}
\usepackage[
left=1.25in,right=1.25in,
top=1.40in,
bottom=1.40in
]
{geometry} 
\usepackage{setspace}                    
\usepackage{epic}
\usepackage{lscape,array}
\usepackage{xy}
\xyoption{all}

\input amssym.def
\input amssym.tex

\input epsf

\input{preamble}          

\begin{document}

  \frontmatter

\input{titlepage}       

\setcounter{page}{0}

\input{dedicace}

\setcounter{tocdepth}{1}
\setcounter{page}{0}

\tableofcontents
\markboth{Contents}{Contents}

\mainmatter

\input{intro}

\setcounter{section}{0}
\setcounter{equation}{0}
\setcounter{footnote}{0}

\input{higherspins}

\setcounter{section}{0}
\setcounter{equation}{0}
\setcounter{footnote}{0}

\input{dualitetest}             

\setcounter{section}{0}
\setcounter{equation}{0}
\setcounter{footnote}{0}

\input{quantif}

\setcounter{section}{0}
\setcounter{equation}{0}
\setcounter{footnote}{0} 

\input{brst1101}

\input{deformation}

\setcounter{section}{0}
\setcounter{equation}{0}
\setcounter{footnote}{0}

\input{2colldif}              

\setcounter{section}{0}
\setcounter{equation}{0}
\setcounter{footnote}{0}

\input{spin3}

\input{conclusion}

\appendix

\input{young}

\input{chapline}

\input{Zino}

\input{spin2apenprim}

\input{Schouten}

\end{document}

%% file: preamble.tex
\newcommand{\ft}{\footnotesize}
\def\ie{{\it i.e. }}

\def\a{\alpha}
\def\b{\beta}
\def\g{\gamma}
\def\c{\gamma}
\def\d{\delta}
\def\e{\eta}
\def\h{\eta}
\def\eps{\epsilon}
\def\l{\lambda}

\def\m{\mu}
\def\n{\nu}
\def\r{\rho}
\def\o{\omega}
\def\s{\sigma}
\def\S{\Sigma}
\def\t{\tau}
 
\def\x{\xi}

\def\ve{\varepsilon}
\def\veps{\varepsilon}

\def\pa{\partial}

\def\ll{\left} 
\def\rr{\right}

\def\6{\partial}

\def\ca{{\cal A}}
\def\cb{{\cal B}}

\def\cg{{\cal G}}

\def\cl{{\cal L}}

\def\co{{\cal O}}
\def\cp{{\cal P}}

\def\cs{{\cal S}}

\def\cw{{\cal W}}

\def\cy{{\cal Y}}

\def\B{\begin{equation}}
\def\E{\end{equation}}
\def\be{\begin{equation}}
\def\ee{\end{equation}}
\def\bqn{\begin{eqnarray}}
\def\eqn{\end{eqnarray}}
\def\beq{\begin{eqnarray}}
\def\eeq{\end{eqnarray}}
\def\ben{\begin{equation}}
\def\een{\end{equation}}
\def\bena{\begin{eqnarray}}
\def\eena{\end{eqnarray}}

\newcommand{\bea}{\begin{eqnarray}}
\newcommand{\eea}{\end{eqnarray}}
\newcommand{\beasn}{\begin{sneqnarray}}
\newcommand{\eeasn}{\end{sneqnarray}}

\def\nn{\nonumber} 
\newcommand{\nnn}{\nonumber \\}

\newtheorem{theorem}{Theorem}[chapter]
\newtheorem{lemma}{Lemma}[chapter]

\newtheorem{prop}{Proposition}
\newtheorem{proposition}{Proposition}[chapter]

\newcommand{\bref}[1]{(\ref{#1})}

\def\dif{{\rm d}}
\def\restric#1#2{{\left. #1 \right|_{#2}}}


%% file: titlepage.tex
\begin{titlepage}

\thispagestyle{empty}

\begin{centering}

{\Large UNIVERSIT\'E LIBRE DE BRUXELLES\\
Facult\'e des Sciences\\
Service de Physique Th\'eorique et Math\'ematique}

\vspace{4.5cm}  


  \textbf{\Huge{ Higher Spin Gauge Field Theories\\}
       \vspace{1cm}
    \LARGE{Aspects of dualities and interactions}}
  
  \vspace{4.5cm}
  
{\Large Dissertation pr\'esent\'ee en vue de l'obtention du titre de Docteur en Sciences }
  

\vfill

{\large {\bf Sandrine Cnockaert}}

Aspirant F.N.R.S.
\vspace*{2.cm}

{\large Ann\'ee acad\'emique 2005-2006}

\end{centering}

\newpage

\end{titlepage}


%% file: dedicace.tex
\newpage

\thispagestyle{empty}

$\mbox{}$

\newpage

\thispagestyle{empty}

\vspace*{4cm}

\begin{flushright}
\`A mes parents et Nicolas
\end{flushright}

\newpage

\thispagestyle{empty}

$\mbox{}$

\newpage

\thispagestyle{empty}

\vspace*{4cm}

\begin{flushleft}
{\huge {\bf Remerciements}}
\end{flushleft}
\vspace*{2cm}

Je remercie tout d'abord mon promoteur, Marc Henneaux, qui m'a permis de pr\'eparer cette th\`ese sous sa direction. Il y a largement contribu\'e, que ce soit par des collaborations directes, ou par ses id\'ees lumineuses qui ont \'eclair\'e ma voie lorsque je me trouvais dans une impasse sur d'autres projets. Je lui suis reconnaissante pour son soutien tout au long de cette th\`ese.

\vspace{.2cm}

Je remercie chaleureusement Nicolas Boulanger et Xavier Bekaert, qui ont particip\'e \`a une partie importante des travaux de mon doctorat. J'ai beaucoup appr\'eci\'e nos collaborations r\'ep\'et\'ees, qui furent parfois compliqu\'ees par la distance g\'eographi-que nous s\'eparant. 
Je remercie \'egalement Carlo Iazeolla, pour ses discussions stimulantes lors de la r\'edaction des comptes rendus du premier workshop Solvay, ainsi que Serge Leclercq.

\vspace{.2cm}

Un \'el\'ement-clef du bon d\'eroulement de ma th\`ese a \'et\'e l'environnement convivial et stimulant qu'offre le Service de Physique Th\'eorique et Math\'ematique. Les interlocuteurs ne manquent pas pour des discussions tant scientifiques que sur des sujets divers. Ils sont francophones, flamands ou d'ailleurs, on les rencontre aux s\'eminaires de l'ULB, de la VUB ou de la KUL, aux journ\'ees du PAI, aux conf\'erences, aux \'ecoles d'\'et\'e ou d'hiver, ou simplement dans la salle caf\'e, \`a la cantine ou dans les couloirs. Je remercie tous ensemble ceux qui se sentent concern\'es par la description ci-dessus, les nommer s\'epar\'ement me semble impossible.

\vspace{.2cm}

Je remercie pour leur pr\'esence et leur soutien  
mes compagnons doctorants: tout d'abord Sophie de Buyl, avec laquelle j'ai partag\'e \'enorm\'ement d'exp\'eriences (\'etudes, patron de th\`ese, appartement, \'ecoles d'\'et\'e et d'hiver, ...), ensuite Claire No\"el et Laura Lopez Honorez, et tous les autres (dans un ordre al\'eatoire et pardonnez-moi pour ceux que j'oublie..): St\'ephane Detournay, Yannick Kerckx, Olivier Debliquy, Geoffrey Comp\`ere, Nazim Bouatta, Daniel Persson, Elizabete Rodriguez, Paola Aliani, Benoit Roland, Pierre Capel, G\'erald Goldstein, Marc Theeten, Stijn Nevens, Alex Wijns, etc. 

\vspace{.2cm}

Je remercie Pierre Marage, pour sa compr\'ehension et son soutien.

\vspace{.2cm}

Je remercie Glenn Barnich, Olivier Debliquy et Bernard Knaepen pour leur aide dans les probl\`emes informatiques.

\vspace{.2cm}

Je remercie Marc Henneaux, Christiane Schomblond, Sophie de Buyl et Nicolas Borghini pour avoir relu cette th\`ese.

\vspace{.2cm}

Je remercie Glenn Barnich et Geoffrey Comp\`ere qui m'ont aid\'ee \`a mettre cette th\`ese sur arXiv. 

\vspace{.3cm}

Finalement, je remercie mes amis, ma famille et Nicolas de croire en moi.

\thispagestyle{empty}


%% file: intro.tex
\chapter*{Introduction}
\addcontentsline{toc}{chapter}{Introduction}
\markboth{Introduction}{Introduction}

Intriguing open questions of gauge field theory lie in the range of higher-spin gauge fields. These fields arise naturally in the classification of particles propagating in flat space-time. Indeed, as was shown by Bargmann and Wigner around the forties \cite{Wigner:1939cj,Bargmann:1948ck}, group theory imposes that such particles should correspond to irreducible representations of the Poincar\'e group\footnote{For a pedagogical review on the irreducible representations of the Poincar\'e group, in four and higher dimensions, we suggest the thesis by Nicolas Boulanger \cite{thesenico}.}. In four space-time dimensions, these are completely characterized by a mass and a representation of the little group. In the massless case, to which we restrict in this thesis, these representations are labeled by the 
``spin'', a positive integer or half-integer without further restriction.\footnote{Actually, there also exist  ``continuous spin'' massless representations which have an infinite number of components. They are not considered here.} 

\vspace{.2cm}

For some time, the main problem involving higher spins under investigation was the construction of free Lagrangians for fields of increasing spin 
\cite{PaFi,Rarita:1941mf,Fronsdal:1978rb,Fang:1978wz,deWit:1979pe}, sometimes with the help of auxilliary fields. This task was more or less completed by the end of the seventies.
In the eighties, a new approach to higher spins was developped by 
Fradkin and Vasiliev \cite{Vasiliev:1980as,Fradkin:1987ks}, based on a generalization of the vielbeins and spin connections of Mac Dowell and Mansouri \cite{MacDowell:1977jt}. The aim of this approach, appealing by its geometrical structure, was to be able to couple gravity described by spin-2 fields to higher-spin fields.
At the same time, a promising theory for a unified description of the fundamental forces and particles, string theory, prompted a revived interest in higher spin fields. The fundamental objects of this theory are one-dimensional objects that move in space-time and vibrate like the strings of a violin. It was noticed that the spectrum of the vibration modes of the strings includes an infinite number of fields of arbitrary increasing spin.

\vspace{.2cm}

With the advent of string theory, 
one was also confronted with the fact that some theories require the space-time to have more than four dimensions. Indeed, string theories can  be consistently quantized perturbatively only in 10 or 26 dimensions. This observation triggered investigations in a new domain of higher-spin fields.
The exciting fact is that new kinds of fields are allowed in field theories that live in those higher-dimensional space-times. Indeed, more general representations of the Poincar\'e group exist when the space-time dimension $n$ is larger than four. Spin is no longer sufficient to characterize the new representations, therefore it is replaced by a Young diagram in the classification. The word ``spin'' is still used in the higher-dimensional context, where it now denotes the length of the first row of the Young diagrams for bosons, and this length plus one half for fermions.
The usual completely symmetric spin-$s$ field that appears in four dimensions then corresponds  to the simplest Young diagrams of spin $s\,$, \ie  a one-row diagram with $s$ boxes. The new fields include antisymmetric $p$-form fields (which correspond to one-column Young diagrams), and mixed-symmetry fields, the indices of which are neither completely symmetric, nor completely antisymmetric. The latter fields are also called ``exotic''.   
\vspace{.4cm}

In the last two decades, two aspects of higher-spin gauge theories have been mainly studied:
duality and interactions.
We will consider both in this thesis, focussing on massless fields of integer spin $s$.
\vspace{.2cm}

\noindent {\bf (i) Duality}

The first question addressed in this thesis is whether different higher-spin fields are related by dualities. In other words, is it possibler that fields corresponding to different irreducible representations be actually describing the same physical object?  Dualities that relate the components of a same field are considered as well.
These dualities are also important because they often relate theories that are in different coupling regimes, {\it e.g.} a strongly coupled and a weakly coupled theory. 

These issues have already been the focus
of a great interest 
\cite{Nieto:1999pn,Casini:2001gv,CasUrr1,
Boulanger:2003vs,Hull:2001iu,Bekaert:2002dt,deMedeiros:2002ge,All,Francia:2002aa,Francia:2002pt,Francia:2005bu,Bouatta:2004kk,Bekaert:2003az,deMedeiros:2003dc,Henneaux:2004jw,Julia:2005ze}. Dualities were found that relate different representations of the same spin.
In most of these works however, duality is studied at the level of the
equations of motion only (notable exceptions being
Ref.\cite{Nieto:1999pn,Casini:2001gv,CasUrr1}, which deal with the spin-2 case in four
space-time dimensions). One can wonder whether there exists a stronger form of
duality, valid for all spins and in all space-time dimensions, 
which would relate the corresponding actions. This is indeed the case: in specific dimensions, the free theory for completely symmetric spin-$s$ fields is dual at the level of the action to the free theory of some mixed-symmetry fields \cite{Boulanger:2003vs}. The proof of this statement is presented in this thesis for fields propagating in a flat space-time. It relies on the first-order formulation of the action. The proof can be generalized to Anti-de Sitter space-time ($AdS$) \cite{Matveev:2004ac}, and probably also to mixed-symmetry gauge fields, provided one constructs their first-order action. 

Other dualities of field theories are symmetries ``within'' a same theory. String theory exhibits many such dualities. The earliest example of such a duality though is the electric-magnetic duality of electromagnetism. The almost symmetric role of the electric and magnetic fields led Maxwell to complete the symmetry by introducing the ``displacement current''. In this way, he wrote down the correct  equations of electromagnetism.  In the absence of sources, these equations are invariant under duality transformations mixing the electric and the magnetic fields. However, because no isolated magnetic charges have been observed in Nature,
the usual equations  are not invariant in the presence of sources. 
It is nevertheless possible to construct a theory symmetric that is under duality in the presence of sources, by assuming the existence of magnetic monopoles. This was done by Dirac in Ref.\cite{Dirac:1931kp,Dirac:1948um}. In these papers, Dirac
also showed that the existence of magnetic monopoles has a dramatic consequence. Indeed, the presence of a single magnetic monopole implies the quantization of the electric charges. If a magnetic monopole could be found, this would provide a very elegant explanation of why the electric charges of the elementary particles are related by integer factors. Indeed, within the Standard Model, no reason explains why the charges of the ``up'' and ``down'' quarks, $u$ and $d$, are related to the charge of the electron by the simple ratios $Q_u \, :\,  Q_d \, :\,  Q_e\,  =\, 2 \, :\, -1\, :\, -3\; $ (and similarly for the other families of elementary particles). 

Later, the idea of electric-magnetic duality was
 analysed in the context of non-Abelian gauge theories in
\cite{'tHooft:1974qc
,Montonen:1977sn
}, and more
recently it has been generalized to extended objects and $p$-form gauge fields
in \cite{Teitelboim:1985ya
}.
The charge quantization condition becomes more exotic in the latter case.
For example, it is
antisymmetric for $p$-dyons of even spatial dimension $p$, and symmetric
for odd $p$ \cite{Schwinger:1968rq
}: $e\bar{g}\pm g\bar{e}=2\pi n \hbar $, where $(e,g)$ and $(\bar{e},\bar{g})$ are the electric and magnetic charges of two dyons and $n$ is an integer. 
Another feature is that, since in dimensions higher than four duality can relate different kinds of fields, the quantization condition then involves the charges of different fields, like the electric charge of a vector field and the magnetic charge of a $(n-3)$-form.

Finally, magnetic sources and the electric-magnetic duality can be implemented in free higher-spin gauge field theories \cite{Bunster:2006rt}, as we show in this thesis for $n=4$.
The quantization condition now involves the four-momenta of the sources. Thus, for instance for spin-2, the quantized quantity is the product of the energy-momentum four-momenta of the sources, and not the product of the ``electric'' and ``magnetic'' masses.   
A limitation of this generalization is however that, because only the linear theory is considered, the sources are strictly external and their trajectories in space-time are not affected by the backreaction from the higher-spin fields.
These results were obtained for completely symmetric gauge fields, but we expect that the same implementation can also be used in higher dimensions to determine the coupling of magnetic sources to mixed-symmetry fields, and to relate their charges by a quantization condition.

\vspace{.4cm}

\noindent {\bf (ii) Interactions}

The second part of the thesis is related to the following question.
Why do the fields that we see in Nature all have spins lower or equal to two?
A possible answer could be that there is no consistent interacting theory in flat space-time for fields of spin higher than two. There is actually a general belief that this is indeed the case, unless the spectrum of the theory contains an infinite set of higher-spin fields. This is for example what happens in string theory:
an infinite number of higher-spin gauge fields appear in the tower of massive states of this theory, where they even play an important role in the quantum behavior. 

Let us first explain more precisely the present status.
The theory describing the free motion of massless fields of arbitrary spin is by now well established. Several elegant formulations are known, for the completely symmetric fields \cite{Fronsdal:1978rb,deWit:1979pe, Curtright:1980yk,Francia:2002aa,Francia:2002pt,Francia:2005bu, Bouatta:2004kk}
as well as for the mixed-symmetry fields 
\cite{Curtright:1980yk,Aulakh:1986cb,Siegel:1986zi,Labastida:1986gy,Labastida:1987kw,Brink:2000ag,Burdik:2001hj,Dubois-Violette:2001jk,Bekaert:2003az,deMedeiros:2003dc,Zinoviev:2003ix,Alkalaev:2003qv,Alkalaev:2003hc}.
However, the problem of constructing consistent interactions among higher-spin gauge fields is not completely solved.
The first attempts to tackle this problem were reported in 
Ref.\cite{deWit:1979pe,Bengtsson:1983pd,Bengtsson:1986kh,Fradkin:1991iy,Berends:1984wp,Bengtsson:1983bp,Berends:1984rq,Bengtsson:1986bz,Bengtsson:1985iw,Berends:1985xx,Bengtsson:1987jt,Damour:1987fp}, among which some progress was achieved.  These
results describe consistent interactions at first order in a
deformation parameter $g$ and involve more than two derivatives.
 In the light-cone
gauge,  first-order three-point couplings between completely
symmetric\footnote{Light-cone cubic vertices involving 
mixed-symmetry gauge fields were computed in dimensions $n=5,6$
\cite{Metsaev}.} gauge fields with arbitrary spins $s>2$ were
constructed in \cite{Bengtsson:1983pd,Bengtsson:1986kh,Fradkin:1991iy}.
For the spin-$3$ case, a first-order cubic vertex was obtained in a covariant
form by Berends, Burgers and van Dam \cite{Berends:1984wp}. However,
no-go results soon demonstrated the impossibility of extending
these interactions to the next orders in powers of $g$ for the
 spin-$3$ case
\cite{Bengtsson:1983bp,Berends:1984rq,Bengtsson:1986bz}. On the
other hand, the first explicit attempts to introduce interactions
between higher-spin gauge fields and gravity encountered severe
problems \cite{diff}.

Very early, the idea was proposed that a consistent interacting higher-spin
gauge theory could exist, provided the theory contains fields of every possible  spin \cite{Fronsdal:1978rb}. In order to overcome the
gravitational coupling problem, it was also suggested to perturb
around a curved background, like for example $AdS_n$. In such a
case, the cosmological constant $\Lambda$ can be used to cancel
the positive mass dimensions appearing with the increasingly many
derivatives of the vertices. Interesting results have indeed been obtained in
those directions: consistent nonlinear equations of motion have been found (see  \cite{Vasiliev:1990en,
Vasiliev:2004qz,Sagnotti:2005} and references therein), the lowest orders of the interacting action have also been computed 
\cite{Fradkin:1987ks}, but the complete action principle is still missing. 
Infinite towers of higher-spin fields are also studied in the context of the tensionless limit of string theory \cite{Bonelli:2003kh}, where the massive modes become massless.

To tackle the problem of interactions involving a limited number of fields, a new method 
\cite{Barnich:1993vg,Henneaux:1997bm} has been developed in the last decade. It
allows  
for an exhaustive treatment of the consistent local interaction problem while, 
in the aforementioned works
\cite{Bengtsson:1983pd,Berends:1984wp,Bengtsson:1983bp,Bengtsson:1985iw,Berends:1984rq,Berends:1985xx,Bengtsson:1986bz,Bengtsson:1986kh,Bengtsson:1987jt,Fradkin:1991iy},
classes of deformation candidates were rejected {\textit{ab
initio}} from the analysis for the sake of simplicity. For example,
spin-3 cubic vertices containing more than $3$ derivatives were not
considered in the otherwise very general analysis of
\cite{Berends:1984wp}. This ansatz was too restrictive since another
cubic spin-3 vertex with five derivatives exists in dimensions higher than four (it is written explicitly in Section \ref{azerodeux}). 
 In the approach of \cite{Barnich:1993vg}, the standard Noether method
(used for instance in \cite{Berends:1984rq}) is reformulated in the
BRST field-antifield framework 
\cite{Batalin:1981jr,Batalin:1983wj,Batalin:1984jr}, and consistent couplings define deformations of
the solution of the master equation. 
Let us mention that some efforts are still pursued in the light-cone formalism \cite{Metsaev:2005ar}.

The BRST  formulation has been
used recently in different contexts 
\cite{Barnich:1993pa,Henneaux:1997ha,Boulanger:2000rq,Bekaert:2002uh,Bizdadea:2003ht,Boulanger:2004rx,Bekaert:2004dz,Bekaert:2005jf,Boulanger:2005br}, two of which are presented in this thesis: interactions among exotic spin-2 fields \cite{Bekaert:2002uh,Bizdadea:2003ht,Boulanger:2004rx,Bekaert:2004dz} and interactions among symmetric spin-3 fields \cite{Bekaert:2005jf,Boulanger:2005br}.
It is found that no non-Abelian interaction can be built for exotic spin-2 fields. There is thus no analogue to Einstein's gravity for these fields. Nevertheless, some examples of consistent interactions that do not deform the gauge transformations can be written.
For spin-3 fields, non-Abelian first-order vertices exist. On top of the two above-mentioned vertices (the vertex of Berends, Burgers and van Dam and the five-derivative vertex), two extra parity-violating vertices are found, which live in three and five space-time dimensions respectively. However, two of those vertices are obstructed at second order in the coupling constant and further work is needed to check whether the two remaining vertices can be extended to all orders.
It would also be interestiong to determine whether some of these vertices might be related to the nonlinear equations of Vasiliev \cite{Vasiliev:1990en,
Vasiliev:2004qz,Sagnotti:2005}.


\vspace{2cm}
\section*{Overview of the thesis}


This thesis is organized as follows.

In {\bf Chapter  1}, we give a review of the free theory of massless bosonic higher-spin gauge fields \cite{Fronsdal:1978rb}. The concepts presented include gauge invariance, the equations of motion, the action, as well s aconserved charges and the coupling of external electric sources. 

In {\bf Chapter  2},  we introduce 
the first-order reformulation of higher-spin gauge field theories, which has been developped by Vasiliev \cite{Vasiliev:1980as}. In this framework, we prove the duality, at the level of the action, of the free theory of completely symmetric spin-$s$ fields with the free theory of some mixed-symmetry spin-$s$ fields, in specific dimensions \cite{Boulanger:2003vs}.

In four space-time dimensions, the duality procedure of Chapter 2 relates the free theory of a completely symmetric spin-$s$ field with itself. Moreover, the duality interchanges the ``electric'' and ``magnetic'' components of the field. We use this result in {\bf Chapter  3} to couple external magnetic sources to higher-spin fields. Furthermore, we show that the ``electric'' and ``magnetic'' conserved charges are required to satisfy a quantization relation \cite{Bunster:2006rt}. The latter involves the ``electric'' and ``magnetic'' couplings, as well as the four-momenta of the sources. It is a generalization of the Dirac quantization condition for electromagnetism, which constrains the product of electric and magnetic charges.

\vspace{.2cm}

We then turn to the problem of consistent interactions.
In {\bf Chapter  4}, we introduce the framework in which we will work, the BRST field-antifield formalism developped by Batalin and Vilkovisky 
\cite{Batalin:1981jr,Batalin:1983wj,Batalin:1984jr}. We first analyse the general structure of gauge field theories. Then we show how this structure is encoded in the field-antifield formalism. In particular, the consistency of the gauge structure is contained in the {\it master equation}. Finally, we address the problem of constructing consistent local interactions. This is done by deforming the master equation, as was proposed in \cite{Barnich:1993vg,Henneaux:1997bm}.

The theoretical recipes of Chapter 4 are applied to specific examples in the next two chapters.
In {\bf Chapter  5}, we study the self-interactions of exotic spin-two fields \cite{Bekaert:2002uh,Bizdadea:2003ht,Boulanger:2004rx,Bekaert:2004dz}. The symmetries of the indices of these fields are described by Young tableaux made of two columns of arbitrary length $p$ and $q$ (with $p\geq q$). We require $p>1$ to exclude the well-studied usual symmetric spin-two field, the graviton. After computing several cohomology groups, we prove a no-go theorem on interactions with a non-Abelian gauge algebra. We also constrain the interactions that deform the gauge transformations without deforming the algebra.

In {\bf Chapter 6}, we perform the same analysis for completely symmetric spin-three fields \cite{Bekaert:2005jf,Boulanger:2005br}. The computation of some cohomology groups is complicated with respect to the spin-2 case by the additional condition of vanishing trace on the gauge parameter.
At first order in the deformation parameter, we find four consistent deformations of the free Lagrangian and gauge transformations, among which the vertex found by Berends, Burgers and van Dam. The latter deformation and another one  are shown to be obstructed at second order by the requirement that the algebra should close.

After  brief {\bf Conclusions}, some appendices follow. 
An introduction to Young tableaux is given in {\bf Appendix A}.
In {\bf Appendix B}, we  present a generalization of Chapline-Manton interactions that involves exotic spin-two fields or spin-$s$ fields.
{\bf Appendix C} is devoted to the first-order formulation of the free theory for exotic spin-two fields.
The lengthy proof of a theorem stated in Chapter 5 is given in {\bf Appendix D}, as well as technicalities involving Schouten identities, which are needed in Chapter 6.


%% file: higherspins.tex
\chapter{Free higher-spin gauge fields} \label{appendixA}

In this section we review the free theory of bosonic higher-spin gauge
fields. A wider recent review on this topic can be found in \cite{Bouatta:2004kk}.

\section{Spin-$s$ field and gauge invariance}

A massless bosonic spin-$s$ field can be
described by a gauge potential which is a totally symmetric tensor
$h_{\m_1 \m_2 \cdots \m_s}$ subject to the ``double-tracelessness
condition'' \cite{Fronsdal:1978rb}, $$ h_{\m_1 \m_2 \cdots \m_s} =
h_{(\m_1 \m_2 \cdots \m_s)}, \; \; \; h_{\m_1 \m_2 \m_3 \m_4 \cdots
\m_s} \eta^{\m_1 \m_2} \eta^{\m_3 \m_4} = 0 \,.$$  The gauge
transformation reads \be h_{\m_1 \m_2 \cdots \m_s} \rightarrow
h_{\m_1 \m_2 \cdots \m_s} +
\partial_{(\m_1}\xi_{\m_2 \cdots \m_s)} \,,\label{gauges}\ee where the gauge
parameter $\xi_{\m_2 \cdots \m_s}$ is totally symmetric and traceless, 
 $$ \xi_{\m_2 \m_3
\cdots \m_s} \eta^{\m_2 \m_3}= 0 \,.$$  The trace condition on the
gauge parameter appears for spins $\geq 3$, while the double
tracelessness condition on the field appears for spins $\geq 4$.

{}From the field $h_{\m_1 \m_2 \cdots \m_s}$, one can construct a
curvature $R_{\m_1 \n_1 \m_2 \n_2 \cdots \m_s \n_s}$ that contains
$s$ derivatives of the field and that is gauge invariant under the
transformations (\ref{gauges}) even if the gauge parameter is not
traceless, \be R_{\m_1 \n_1 \m_2 \n_2 \cdots \m_s \n_s} =-2\
h_{[\m_1[ \m_2 \cdots [\m_s,\n_s]\cdots \n_2]\n_1]}
\,,\label{DefR}\ee where one antisymmetrizes over $\m_k$ and $\n_k$
for each $k$. This is the analog of the Riemann tensor of the spin-2
case.  The curvature $R_{\m_1 \n_1 \m_2 \n_2 \cdots \m_s \n_s}$ has
the symmetry characterized by the Young tableau \be
\begin{picture}(85,15)(0,2)
\multiframe(0,11)(13.5,0){1}(10.5,10.5){\tiny{$\m_1$}}
\multiframe(0,0)(13.5,0){1}(10.5,10.5){\tiny{$\n_1$}}
\multiframe(11,11)(13.5,0){1}(10.5,10.5){\tiny{$\m_2$}}
\multiframe(11,0)(13.5,0){1}(10.5,10.5){\tiny{$\n_2$}}
\multiframe(22,11)(13.5,0){1}(19.5,10.5){$\cdots$}
\multiframe(22,0)(13.5,0){1}(19.5,10.5){$\cdots$}
\multiframe(42,11)(13.5,0){1}(10.5,10.5){\tiny{$\m_s$}}
\multiframe(42,0)(13.5,0){1}(10.5,10.5){\tiny{$\n_s$}}
\end{picture}
\label{Young}\ee 
\ie  it is symmetric for the exchange of pairs of indices $\m_i \n_i$ and antisymmetrization over any three indices yields zero.
The curvature also fulfills the Bianchi identity \be
\partial_{[\a} R_{\m_1 \n_1] \m_2 \n_2 \cdots \m_s \n_s} = 0\,.
\label{Bianchis}\ee

Conversely, given a tensor $R_{\m_1 \n_1 \m_2 \n_2 \cdots \m_s
\n_s}$ with the Young tableau symmetry (\ref{Young}) and fulfilling
the Bianchi identity (\ref{Bianchis}), there exists a ``potential''
$h_{\m_1 \m_2 \cdots \m_s}$ such that Eq.(\ref{DefR}) holds.  This
potential is determined up to a gauge transformation (\ref{gauges})
where the gauge parameter $\xi_{\m_2 \cdots \m_s}$ is unconstrained 
(\ie its trace can be non-vanishing)
\cite{Olver}.

\section{Equations of motion} 
The trace conditions on the gauge parameter for spins
$\geq 3$ are necessary in order to construct second-order invariants
-- and thus, in particular, gauge invariant second-order equations
of motion. One can show that the Fronsdal tensor $$ F_{\m_1 \m_2
\cdots \m_s} =\Box h_{\m_1 \m_2 \cdots \m_s} -s
\pa_{(\m_1}\pa^{\r} h_{\m_2 \cdots \m_s)\r} + \frac{s(s-1)}{2}
\pa_{(\m_1\m_2}h_{\m_3 \cdots \m_s)\r}^{\hspace{1.2cm} \r} \,, $$
which contains only second derivatives of the potential, transforms
under a gauge transformation (\ref{gauges}) into the trace of the
gauge parameter
$$ F_{\m_1 \m_2 \cdots \m_s} \rightarrow F_{\m_1 \m_2 \cdots \m_s} +
\frac{(s-1)(s-2)}{2} \pa_{(\m_1 \m_2\m_3}\xi_{\m_4\cdots
\m_s)\r}^{\hspace{1.2cm} \r}\,,$$ and is thus gauge invariant when
the gauge parameter is requested to be traceless.  The Fronsdal
tensor is related to the curvature by the relation \be R_{ \m_1 \n_1
\m_2 \n_2 \cdots \m_s\n_s} \eta^{\n_1\n_2}=-\frac{1}{2}F_{\m_1\m_2
[\m_3 [\cdots [\m_s,\n_s] \cdots ]\n_3]}\,.\label{FR}\ee

The equations of motion that follow from
a variational principle 
are \be G_{\m_1 \m_2 \cdots \m_s} = 0 \,,\ee where the ``Einstein''
tensor is defined as \be G_{\m_1 \m_2 \cdots \m_s} = F_{\m_1 \m_2
\cdots \m_s} -\frac{s(s-1)}{4} \eta_{(\m_1\m_2}F_{\m_3\cdots
\m_s)\r}^{\hspace{1.2cm} \r}\,.\label{eomsp}\ee 
These equations are derived from the Fronsdal  action 
\beq \cs [h_{\m_1 \cdots \m_s}(x)]=\int d^4x \  {\cal L}  \,,\label{freeactions} \eeq where 
\beq {\cal
L}
\!\!\!&=&\!\!\! - \textstyle{\frac{1}{2}}\, \pa_{\l}h_{\m_1 \cdots \m_s}
\pa^{\l}h^{\m_1 \cdots \m_s}
+ \textstyle{\frac{s}{2}}\, \pa^{\l}h_{\l \m_2 \cdots \m_s}
\pa_{\r}h^{\r \m_2 \cdots \m_s}
+ \textstyle{\frac{s(s-1)}{2}}\, \pa^{\l\r}h_{\l\r\m_3 \cdots \m_s}h_{\a}^{\ \a \m_3 \cdots \m_s}
\nnn
& &+ \textstyle{\frac{s(s-1)}{4}}\, 
\pa_{\l}h^{\a}_{\ \a\m_3 \cdots \m_s}\pa^{\l}h_{\b}^{\ \ \b \m_3 \cdots \m_s} 
+\textstyle{ \frac{s(s-1)(s-2)}{8}}\, \pa^{\l}h^{\a}_{\ \a\l\m_4 \cdots \m_s}
\pa_{\r}h_{\b}^{\ \b \r\m_4 \cdots \m_s}
\,.\nonumber \eeq
Indeed, one can check that $\frac{\d {\cal L}}{\d
h^{\g_1\cdots\g_s}}=G_{\g_1\cdots\g_s}\,.$
Furthermore, these equations of motion  obviously imply
\be R_{ \m_1 \n_1 \m_2 \n_2 \cdots \m_s\n_s}
\eta^{\n_1\n_2}=0\label{eoms}\,,\ee and the inverse implication is
true as well \cite{Bekaert:2003az}. Indeed, Eq.(\ref{eoms}) implies
that the Fronsdal tensor has the form $ F_{\m_1 \m_2 \cdots \m_s}
=\pa_{(\m_1 \m_2\m_3}\Sigma_{\m_4\cdots \m_s)}$, which can be made
to vanish by a gauge transformation with an unconstrained gauge
parameter (see \cite{Francia:2002pt} for a discussion of the
subtleties associated with the double tracelessness of the spin-$s$
field $h_{\m_1 \cdots \m_s}$). The interest of the equations
(\ref{eoms}) derived from the Einstein equations is that they
contain the same number of derivatives as the curvature. Thus, they
are useful to exhibit duality, which rotates the equations of motion
and the cyclic identities on the curvature.

\section{Fixing the gauge }

Let us check that  when the gauge is completely fixed the right degrees of freedom remain.

If the theory at hand describes a completely symmetric massless spin-$s$ field, then there should be a completely fixed gauge in which the field is transverse to a timelike direction $u^\a$ and traceless.  We prove in this section  that this is indeed the case.
We first give the gauge conditions, then we check that they can be obtained by gauge transformations and that they completely fix the gauge. 

The appropriate gauge conditions are
\bqn
&(i) &\ H_{\m_1 \ldots \m_{s-1}}\equiv  s \, \pa^{\a}h_{\a \m_1 \ldots \m_{s-1}}- \frac{s(s-1)}{2}
\pa_{(\m_1}h_{\m_2 \ldots \m_{s-1})\,\a}^{\hspace{46pt}\a}=0 \;, \nn\\
&(ii)&\ h_{\m_1 \ldots \m_{s-2}\,\a}^{\hspace{46pt}\a}=0 \;,\nn
\eqn
and  (iii) the vanishing of the components with at least one ``minus'' index and the other indices transverse.

The gauge variation of $H_{\m_1 \ldots \m_{s-1}}$ is 
$
\d H_{\m_1 \ldots \m_{s-1}}=\Box \ \xi_{\m_1 \ldots \m_{s-1}}\,.
$
The gauge in which the condition (i) is satisfied can thus be attained by performing a gauge transformation such that 
\be
\xi_{\m_1 \ldots \m_{s-1}}= - \frac{1}{\Box}H_{\m_1 \ldots \m_{s-1}}\;. \nn 
\ee
In this gauge, there is a residual gauge invariance. Indeed, gauge transformations with parameters satisfying $\Box \ \xi_{\m_1 \ldots \m_{s-1}}=0\,$ are still allowed, as they do not modify condition (i). The solution of this equation is 
$$\xi_{\m_1 \ldots \m_{s-1}}=s\,\int d^n k\ Re[\, -i\, c_{\m_1 \ldots \m_{s-1}}(k)\, exp(i k_\a x^\a)]\;, $$ where $k_\a k^\a=0\,$ and $c_{\m_1 \ldots \m_{s-1}}(k)$ is an arbitrary function of $k_\a$. 

We now perform a Fourrier expansion of the field and all the gauge
conditions.  So, e.g.  $h_{\m_1 \ldots \m_s} = \int d^n k\ Re [\
\widehat{h}_{\m_1 \ldots \m_s} \, exp(i k_\a x^\a)]\,$.  Quite
generally, we can consider each Fourrier component separately, which
we will do in the sequel.

  Without loss of generality, we can choose $k^\a=(k^+, 0\ldots 0)\, $
  and 

  \noindent $u^\a=(1,0\ldots 0)\,$.  We first use the residual
  invariance to cancel the traces of the field (gauge condition
  (ii)). Their gauge transformation is \bqn \d h_{\m_1 \ldots
    \m_{s-2}\a}{}^{\a}&=&\d \Big( Re [\ \widehat{h}_{\m_1 \ldots
    \m_{s-2}\a}{}^{\a}\, exp(i k_\b x^\b)]\Big)\nnn &=&
  \textstyle{\frac{2}{s}}\,\pa^{\a} \xi_{\a\m_1 \ldots \m_{s-2}}=Re [-
  2\, k^+ c_{+\m_1 \ldots \m_{s-2}}\, exp(i k_\b x^\b)]\,,\nn \eqn so
  by a gauge transformation with $c_{+\m_1 \ldots
    \m_{s-2}}=\frac{1}{2k^+}\widehat{h}_{\m_1 \ldots
    \m_{s-2}\a}^{\hspace{40pt}\a}$ one can make the traces of the
  field vanish.  The tracelessness condition of the gauge parameter,
  $\xi^{\n}_{~\n\m_3 \ldots \m_{s-1}}=0$ implies that $2\eta^{+-}c_{+-
    \m_3 \ldots \m_{s-1}}+c^{i}_{~i\m_3 \ldots \m_{s-1}}=0$ , which
  means that all the transverse traces of $c$ are fixed by the above
  gauge transformation. Indeed, further gauge transformations with
  non-vanishing transverse traces would spoil the gauge condition
  (ii).
 
The gauge condition (i) now reads $$\pa^{\a}h_{\a \m_1 \ldots \m_{s-1}}= Re [\, i k^+ \widehat{h}_{+ \m_1 \ldots \m_{s-1}} \, exp(i k_\b x^\b)] =0\,.$$ Thus, when (i) and (ii) are satisfied,   all the field components with  at least one ``plus''  and all the traces of the field vanish. To reach the transverse traceless gauge, the residual gauge invariance must be used to cancel the components with at least one ``minus''.
The latter components $h_{- m_1 \ldots m_{s-1}}\,$, where $m\in \{-,i\}$, are not all independent because of the tracelessness of the field. Indeed, it implies that their transverse traces are given by $$h^{~~i}_{- ~i m_1 \ldots m_{s-3}}=-2\eta^{+-}h_{- +- m_1 \ldots m_{s-3}}\,.$$ It is thus enough to cancel the transverse-traceless part of $h_{- m_1 \ldots m_{s-1}}$ .
The gauge transformation of $h_{- m_1 \ldots m_{s-1}}$  reads
\bqn
\d h_{- m_1 \ldots m_{s-1}}&=& Re[\ \d \widehat{h}_{- m_1 \ldots m_{s-1}}\, exp(i k_\b x^\b)] \nnn
&=&   \pa_{(-} \xi_{m_1 \ldots m_{s-1})} = Re[\ k_- c_{m_1 \ldots m_{s-1}} \, exp(i k_\b x^\b)] 
\,.
\nn
\eqn
By the choice of a gauge transformation with $c_{m_1 \ldots m_{s-1}}$ being the transverse-traceless part of $-\frac{1}{k_-} \widehat{h}_{- m_1 \ldots m_{s-1}}$ we attain the desired goal. As we have now used all components of $c_{\m_1 \ldots \m_{s-1}}$, the gauge is completely fixed. QED.

\vspace{.4cm}

It is interesting to study the form of the Lagrangian as one fixes the gauge. Upon gauge fixing, the Fronsdal Lagrangian becomes the gauge fixed Lagrangian
$$\cl^{GF}= \frac{1}{2}\,(h^{GF}_{i_1\ldots i_s}\Box \,h^{GF\,i_1\ldots i_s} -\textstyle{\frac{s(s-1)}{4}}\, h'{}^{\,GF}_{i_1\ldots i_{s-2}}\Box \,h'{}^{GF\,i_1\ldots i_{s-2}})\;.$$
It is obvious that by a mere redefinition of the form $\tilde{h}=h+\eta\, h'$ one gets the action $$\cl^{GF}= \frac{1}{2}\,\tilde{h}^{GF}_{i_1\ldots i_s}\Box \,\tilde{h}^{GF\,i_1\ldots i_s}\;,$$ which yields the Klein-Gordon equations of motion for $\tilde{h}^{GF}_{i_1\ldots i_s}\,.$ (Remember that the double trace of the field vanishes.)

From another point of view, by relaxing the gauge fixing conditions one can generate the Fronsdal Lagrangian from the Klein-Gordon equations of motion. 
To prove this, let us consider the completely fixed gauge. Since the equations of motion for the physical degrees of freedom are the Klein-Gordon equations, $ \Box \, h^{GF}_{i_1\ldots i_s}=0\,,$
the Lagrangian must be $$\cl^{GF}= a\,(h^{GF}_{i_1\ldots i_s}\Box \,h^{GF\,i_1\ldots i_s} + b\, h'{}^{\,GF}_{i_1\ldots i_{s-2}}\Box \,h'{}^{GF\,i_1\ldots i_{s-2}})\;,$$
where $a$ and $b$ are some a priori arbitrary constants. The constant $a$ is actually just an overal factor, which we take equal to $\frac{1}{2}\,.$

Relaxing the gauge conditions (ii) and (iii) does not change the structure of the Lagrangian, it basically widens the range of values that the indices can take. One has 
$$\cl= \frac{1}{2}\,(h^{GF}_{\m_1\ldots \m_s}\Box \,h^{GF\,\m_1\ldots \m_s} + b\, h'{}^{\,GF}_{\m_1\ldots \m_{s-2}}\Box \,h'{}^{GF\,\m_1\ldots \m_{s-2}})\;,$$
where $h^{GF}_{\m_1\ldots \m_s}$ satisfies the gauge condition (i).
To reach the gauge (i) from the covariant theory, one had to perform a gauge transformation 
\be 
h^{GF}_{\m_1\ldots \m_s}=h_{\m_1\ldots \m_s}+\pa_{(\m_1}\xi_{\m_2\ldots \m_s)}\,
\label{gaugefix}\ee
with parameter $$\xi_{\m_1 \ldots \m_{s-1}}= - \frac{1}{\Box}H_{\m_1 \ldots \m_{s-1}}=- \frac{1}{\Box} \Big(s \, \pa^{\a}h_{\a \m_1 \ldots \m_{s-1}}- \textstyle{\frac{s(s-1)}{2}}
\pa_{(\m_1}h_{\m_2 \ldots \m_{s-1})\,\a}{}^{\a}\Big)
\;. $$
We now ``reverse'' this gauge transformation by inserting the expression \bref{gaugefix} for $h^{GF}_{\m_1\ldots \m_s}$ into the above Lagrangian, substituting for $\xi_{\m_1 \ldots \m_{s-1}}$ its expression in terms of the field $h_{\m_1\ldots \m_s}\,.$
Because the guage transformation is not local, non-local terms appear in the Lagrangian. To cancel them, one must impose that $b=-\frac{s(s-1)}{4}\,.$
It turns out that the obtained Lagrangian now exactly matches the Fronsdal Lagrangian \bref{freeactions}.
\footnote{
This procedure to generate the Lagrangian can be generalized to fields with mixed symmetry. New features for the latter are the reducibility of the gauge transformations and the presence of several gauge parameters (if one considers irreducible parameters).

Let us sketch how to proceed in the simplest case, for a $2$-form $A_{\m\n}\,$.
The gauge fixed Lagrangian is $\cl^{GF}=A^{GF}_{\m\n}\Box A^{GF\,\m\n}\,.$
 The gauge transformation reads 
$\d_\xi A_{\m\n} =\pa_\n \xi_\m-\pa_\m \xi_\n\,,$ and is reducible, \ie 
$\d_\xi A_{\m\n} =0$ for parameters $\xi_\m = \pa_\m \l\;.$ To fix the reducibility, one can ask that only  gauge parameters that satisfy the Lorentz condition $\pa^\n\xi_\n=0\,$ be allowed.
The equivalent of the condition (i) is the Lorentz condition 
$H_\n\equiv \pa^\m A_{\m\n}=0\;.$ Since 
$\d_\xi H_\n=\pa_\n\pa^\r \xi_\r -\Box \xi_\n= -\Box \xi_\n\,,$ the gauge transformation to be ``undone'' in $\cl^{GF}$ is $A^{GF}_{\m\n}=A_{\m\n} +\d_\xi A_{\m\n}$ where $\xi_\n= \frac{1}{\Box}\pa^\m A_{\m\n}\,.$ As expected, the resulting Lagrangian is the usual one.


}

\section{Dual curvature} The dual of the curvature tensor is
defined by
$$S_{\m_1 \n_1 \m_2 \n_2 \cdots \m_s\n_s} =-\frac{1}{2}
\ \ve_{\m_1\n_1 \r \s}R^{\r\s}_{~~\m_2 \n_2 \cdots \m_s\n_s} \,,$$ and,
as a consequence of the equations of motion (\ref{eoms}), of the
symmetry of the curvature  and of the Bianchi identity
(\ref{Bianchis}), it has the same symmetry as the curvature and
fulfills the equations $ S_{ \m_1 \n_1 \m_2 \n_2 \cdots \m_s\n_s}
\eta^{\n_1\n_2}=0 $, $
\partial_{[\a} S_{\m_1 \n_1] \m_2 \n_2 \cdots \m_s\n_s}=0$.

\section{Conserved charges}

Non-vanishing conserved charges can be associated with the gauge
transformations (\ref{gauges}) that tend to Killing tensors at
infinity (``improper gauge transformations''). They can be computed
from the Hamiltonian constraints \cite{Regge:1974zd} or equivalently from the knowledge of
their associated conserved antisymmetric tensors $k^{[\a\b]}_\xi\,.$
These generalize the
electromagnetic $F_{\mu\nu}$ and have been computed in \cite{Barnich:2005bn}. Their divergence vanishes in
the absence of sources. The corresponding charge is given by $
Q_\xi=\frac{1}{2} \int_S \star k_\xi\,_{[\a\b]}\, dx^\alpha \wedge
dx^\beta$, where the integral is taken at constant time, over the
$2$-sphere at infinity. The tensors $k^{[\a\b]}_\xi $ read \beq
k^{[\a\b]}_\xi&=&\pa^\a h^{\b\m_1 \cdots \m_{s-1}}\xi_{\m_1 \cdots
\m_{s-1}}+
\frac{(s-1)}{2}\pa^{\b}h_\r^{~\r\m_1\cdots \m_{s-2}}\xi^\a_{~\m_1 \cdots \m_{s-2}} \nonumber \\
&&+(s-1)\pa_\r h^{\r\a\m_1 \cdots \m_{s-2}}\xi^\b_{~\m_1 \cdots
\m_{s-2}}
-\frac{(s-1)^2}{2} \pa^{(\a}h_\r^{~\m_1\cdots \m_{s-2})\r}\xi^\b_{~\m_1 \cdots \m_{s-2}}\nonumber \\
&&- (\a \leftrightarrow \b) +\cdots\,, \nonumber \eeq where the
dots stand for terms involving derivatives of the gauge
parameters.

Of particular interest are the charges corresponding to  gauge
transformations that are ``asymptotic translations'', \ie
$\xi^{\m_1 \cdots \m_{s-1}}\rightarrow_{r\rightarrow \infty}
\epsilon^{\m_1 \cdots \m_{s-1}}$ for some traceless constant tensor
$\epsilon^{\m_1 \cdots \m_{s-1}}$ . For these transformations, the
charges become, using Stokes' theorem and the explicit expression
for $ k^{[\a\b]}_\xi$,
$$ Q_\epsilon =\epsilon_{\m_1 \cdots \m_{s-1}}\int_V
G^{0\m_1 \cdots \m_{s-1}}d^3x .$$ As these charges are conserved for
any traceless $\epsilon_{\m_1 \cdots \m_{s-1}}$, the quantities
$P^{\m_1 \cdots \m_{s-1}}$ defined as the traceless parts  of $
\int_V G^{0\m_1 \cdots \m_{s-1}}d^3x$ are conserved as well. In the
spin-2 case, $P^\m$ is the energy-momentum 4-vector.

\section{Electric sources}

In the presence of only electric sources, a new term is added to the action (\ref{freeactions}),
 $$
S[h_{\m_1 \cdots \m_s}(x), t^{\m_1 \cdots \m_s}]=\int d^4x \  (\ {\cal L}+\ t^{\m_1 \cdots \m_s} h_{\m_1 \cdots \m_s}\ )\,.
$$
The tensor $t^{\m_1 \cdots \m_s}  $ is called the electric ``energy-momentum'' tensor. It is conserved and thus divergence-free,  $\pa_{\m_1} t^{\m_1 \cdots \m_s}=0  \, $. Since the spin-$s$ field $h_{\m_1 \cdots \m_s}$ is double-traceless, it couples only to the double-traceless part of $t_{\m_1 \cdots \m_s} $, which we denote by $T_{\m_1 \cdots \m_s} \, $.

The  equations of motion then
read: \be G_{\m_1 \m_2 \cdots \m_s} + \  T_{\m_1 \m_2
\cdots \m_s} =0\,,\label{Einsts}\ee or equivalently \be R_{ \m_1
\n_1 \m_2 \n_2 \cdots \m_s\n_s} \eta^{\n_1\n_2}=\frac{1}{2} \
\bar{T}_{\m_1\m_2 [\m_3 [\cdots [\m_s,\n_s] \cdots ]\n_3]}\ee where
$\bar{T}_{\m_1\m_2 \cdots \m_s}=T_{\m_1\m_2 \cdots \m_s}-
\frac{s}{4} \eta_{(\m_1\m_2}T^{\prime}_{\m_3\cdots \m_s)}\,$
and primes denote traces,
$T^{\prime}_{\m_3\cdots \m_s}=T_{\m_1\cdots \m_s}\eta^{\m_1\m_2}\,.$
The curvature tensor has the Young symmetry (\ref{Young}) and
fulfills the Bianchi identity (\ref{Bianchis}), as in the case
without sources.

On the other hand, while the trace of the dual curvature tensor
still vanishes, the latter has no longer the Young symmetry
(\ref{Young}) and its Bianchi identity gets modified as well. The
new symmetry is described by the Young tableau
 \be
\begin{picture}(85,15)(0,2)
\multiframe(0,11)(13.5,0){1}(10.5,10.5){\tiny{$\m_1$}}
\multiframe(0,0)(13.5,0){1}(10.5,10.5){\tiny{$\n_1$}}
\multiframe(31,11)(13.5,0){1}(10.5,10.5){\tiny{$\m_2$}}
\multiframe(31,0)(13.5,0){1}(10.5,10.5){\tiny{$\n_2$}}
\multiframe(42,11)(13.5,0){1}(19.5,10.5){$\cdots$}
\multiframe(42,0)(13.5,0){1}(19.5,10.5){$\cdots$}
\multiframe(62,11)(13.5,0){1}(10.5,10.5){\tiny{$\m_s$}}
\multiframe(62,0)(13.5,0){1}(10.5,10.5){\tiny{$\n_s$}}
\put(15,7.5){$\otimes$}
\end{picture}\,, \label{Young1}
\ee as the dual curvature now satisfies $$ S_{ [\m_1 \n_1 \m_2] \n_2
\cdots \m_s\n_s}=-\frac{1}{6}\ \ve_{\m_1 \n_1\m_2
\r}\bar{T}^{\r}_{~\n_2  [\m_3 [\cdots [\m_s,\n_s] \cdots ]\n_3]}\,,$$
while the Bianchi identity becomes $$\partial_{[\a} S_{\m_1 \n_1]
\m_2 \n_2 \cdots \m_s\n_s} =\frac{1}{3}\
\ve_{\a\m_1\n_1\r}\bar{T}^{\r}_{~[\m_2  [\m_3 [\cdots [\m_s,\n_s]
\cdots ]\n_3] \n_2]}\;.$$


%% file: dualitetest.tex
\chapter{Spin-$s$ duality}
%
%

In this section, we prove that some free theories for higher-spin gauge fields are connected by a form of duality that goes beyond equivalence at the level of the equations of motion, because it relates their corresponding actions.
A familiar example
in which duality goes beyond mere on-shell equivalence is given by
the set of a free $p$-form gauge field and a free $(n-p-2)$-form
gauge field in $n$ space-time dimensions. The easiest way to
establish the equivalence of the two theories in that case is to
start from a first-order ``mother'' action involving simultaneously
the $p$-form gauge field $A_{\mu_1 \cdots \mu_p}$ and the field 
strength $H_{\mu_1 \cdots \mu_{n-p-1}}$ of the $(n-p-2)$-form
$B_{\mu_1 \cdots \mu_{n-p-2}}$ treated as independent variables
\begin{equation}
S[A,H] \sim \int dA \wedge H - \frac{1}{2}H \wedge \! ^*H
\label{action0}
\end{equation} 
The field $H$ is an auxiliary field that can be eliminated through
its own equation of motion, which reads $H = \! ^* dA$. Inserting
this relation in the action (\ref{action0}) yields the familiar
second-order Maxwell action $\sim \int dA \wedge \! ^* dA$ for
$A$. Conversely, one may view $A$ as a Lagrange multiplier for the
constraint $dH = 0$, which implies $H = dB$. Solving for the
constraint inside (\ref{action0}) yields the familiar second-order
action $\sim \int dB \wedge \! ^* dB$ for $B$.

Following Fradkin and Tseytlin \cite{Fradkin:1984ai}, we shall
reserve the terminology ``dual theories'' for theories that can be
related through a ``parent action'', referring to ``pseudo-duality''
for situations when there is only on-shell equivalence. The parent
action may not be unique. In the above example, there is another,
``father'' action in which the roles of $A$ and $B$ are
interchanged ($B$ and $F$ are the independent variables, with $S 
\sim \int dB \wedge F - \frac{1}{2} F \wedge \! ^*F$ and $F = dA$
on-shell). That the action of dual theories can be related through
the above transformations is important for establishing
equivalence of the (local) ultraviolet quantum properties of the
theories, since these transformations can formally be implemented 
in the path integral \cite{Fradkin:1984ai}.

Recently, dual formulations of massless spin-2 fields have
attracted interest in connection with their possible role in
uncovering the hidden symmetries of gravitational theories
\cite{CJLP,OP1998,West,H-L1,Damour,Henry-Labordere:2002xh,EHTW}. In these
formulations, the massless spin-2 field is described by a tensor 
gauge field with mixed Young symmetry type. The corresponding
Young diagram has two columns, one with $n-3$ boxes and the other
with one box. The action and gauge symmetries of these dual
gravitational formulations have been given in the free case by
Curtright \cite{Curtright:1980yk}. The connection with the more
familiar Pauli-Fierz formulation \cite{PaFi} was however not clear and
direct attempts to prove equivalence met problems with trace 
conditions on some fields. The difficulty that makes the spin-$1$
treatment not straightforwardly generalizable is that the
higher-spin ($s \geq 2$) gauge Lagrangians are not expressed in
terms of strictly gauge-invariant objects, so that gauge
invariance is a more subtle guide. One of the results of this chapter
 is the explicit proof that the Curtright action and the
Pauli-Fierz action both come from the same parent action and are 
thus dual in the Fradkin-Tseytlin sense. The analysis is carried
out in any number of space-time dimensions and has the useful
property, in the self-dual dimension four, that both the original
and the dual formulations are described by the same Pauli-Fierz
Lagrangian and variables. 

We then extend the analysis to higher-spin gauge fields described
by completely symmetric tensors. The Lagrangians for these
theories, leading to physical second-order equations, have been
given long ago in \cite{Fronsdal:1978rb} and are reviewed in 
Section \ref{appendixA}. We show that the
spin-$s$ theory described  by a totally
symmetric tensor with $s$ indices and subject to the 
double-tracelessness condition is dual to a theory with a field
of mixed symmetry type $[n-3, 1, 1, \cdots, 1]$ (one column with
$n-3$ boxes, $s-1$ columns with one box; cf Appendix \ref{young}), for which we give
explicitly the Lagrangian and gauge symmetries. This field is also
subject to the double tracelessness condition on any pair of pairs 
of indices. A crucial tool in the analysis is given by the
first-order reformulation of the Fronsdal action due to Vasiliev
\cite{Vasiliev:1980as}, which is in fact our starting point. We
find again that in the self-dual dimension four, the original
description and the dual description are the same.

\section{Spin-2 duality} 
\label{spin2duality}
\setcounter{equation}{0}
\setcounter{theorem}{0}
\setcounter{lemma}{0}

\subsection{Parent actions}
%
We consider the first-order action \cite{West}
\be \cs [e_{a b}, Y^{ab|}_{~~~
c }] = -2 \int d^nx \left[Y^{a b|c} \pa_{[a}e_{b]c} -
\frac{1}{2}Y_{ab|c}Y^{ac|b} +\frac{1}{2(n-2)}Y_{ab|}^{~~b}
Y^{ac|}_{~~~c} \right] \label{actioneY} \ee where $e_{ab}$ has
both symmetric and antisymmetric parts and where $ Y^{ab|}_{~~~c}
= - Y^{ba|}_{~~~c}$ is a once-covariant, twice-contravariant mixed
tensor. Neither $e$ nor $Y$ transform in irreducible 
representations of the general linear group since $e_{ab}$ has no
definite symmetry while $Y^{ab|}_{~~~c}$ is subject to no trace
condition. Latin indices run from $0$ to $n-1$ and are lowered or
raised with the flat metric, taken to be of ``mostly plus''
signature $(-,+,\cdots ,+)$. The space-time dimension $n$ is $\geq
3$. The factor $2$ in front of (\ref{actioneY}) is inserted to
follow the conventions of \cite{Vasiliev:1980as}.

The action (\ref{actioneY}) differs from the standard first-order
action for linearized gravity, in which the vielbein $e_{ab}$ and the spin
connection $\omega_{ab \vert c}$ are treated as independent variables, by a
mere change of variables $\omega_{ab \vert c} \rightarrow Y^{ab|}_{~~~c}$
such that the coefficient of the antisymmetrized derivative of the
vielbein in the action is just $Y^{ab|}_{~~~c}\,$, up to the inessential 
factor of $-2$. This change of
variables reads 
$$ Y_{ab \vert c} = \omega_{c|a | b} +\h_{ac} \omega^{i}_{~|b| i}-\h_{bc} 
\omega^{i}_{~|a | i}; \;\;
\omega_{a|b | c} = Y_{bc \vert a}+ \frac{2}{n-2}\h_{a[b} 
Y_{c]d \vert }^{~~~d} .$$
It was
considered (for full gravity) previously in \cite{West}.

By examining the equations of motion for $Y^{ab|}_{~~~c}$, one 
sees that $Y^{ab|}_{~~~c}$ is an auxiliary field that can be
eliminated from the action. The resulting action is \be {\cs
[e_{ab}]= 4 \int d^nx \left[C_{ca |}^{~~~a}C^{cb|}_{~~~b} -
\frac{1}{2}C_{ab|c}C^{ac|b}- \frac{1}{4}C_{ab|c}C^{ab|c}\right] }
\label{actione} \ee where $C_{ab|c} = \pa_{[a}e_{b]c}$. This
action depends only on the symmetric part of $e_{ab}$ (the
Lagrangian depends on the antisymmetric part of $e_{ab}$ only 
through a total derivative) and is a rewriting of the linearized
Einstein action of general relativity (Pauli-Fierz action).

{}From another point of view, $e_{ab}$ can be considered in the
action (\ref{actioneY}) as a Lagrange multiplier for the
constraint $\pa_a Y^{ab|}_{~~~c}$ = 0. This constraint can be
solved explicitely in terms of a new field $Y^{abe|}_{~~~~c} =
Y^{[abe]|}_{~~~~~c}$, as $Y^{ab|}_{~~~c} = \pa_e Y^{abe|}_{~~~~c}$. 
The action then becomes \be \cs [Y^{abe|}_{~~~~c} ] = 2 \int d^nx
\left[\frac{1}{2}Y_{ab|c}Y^{ac|b} -\frac{1}{2(n-2)}Y_{ab|}^{~~b}
Y^{ac|}_{~~~c} \right] \label{actionY} \ee where $Y^{ab|c}$ must
now be viewed as the dependent field $Y^{ab|c} = \pa_e Y^{abe|c}$.
The field $Y^{abe|}_{~~~~c}$ can be decomposed into irreducible
components: $Y^{abe|}_{~~~~~c} = X^{abe|}_{~~~~c} +
\d_c^{[a}Z^{be]}$, with $X^{abc|}_{~~~~c}=0$, $X^{abe|}_{~~~~c} = 
X^{[abe]|}_{~~~~~c}$ and $Z^{be} = Z^{[be]}$. A direct but
somewhat cumbersome computation shows that the resulting action
depends only on the irreducible component $X^{abe|}_{~~~~c} $,
\ie it is invariant under arbitrary shifts of $Z^{ab}$ (which
appears in the Lagrangian only through a total derivative). One
can then introduce in $n \geq 4$ dimensions the field $T_{a_1
\cdots \,a_{n-3}|c}= \frac{1}{3!} \ve_{a_1 \cdots 
\,a_{n-3}efg}X^{efg|}_{~~~~c}$ with $T_{[a_1 \cdots \,a_{n-3}
|c]}= 0$ because of the trace condition on $X^{efg|}_{~~~~c}$, and
rewrite the action in terms of this field\footnote{For $n=3$, the
field $X^{efg|}_{~~~~c}$ is identically zero and the dual
Lagrangian is thus ${\cal L} = 0$. The duality transformation
relates the topological Pauli-Fierz Lagrangian to the topological
Lagrangian ${\cal L} = 0$. We shall assume $n> 3$ from now on.}. 
Explicitly, one finds the action given in
\cite{Curtright:1980yk,Aulakh:1986cb}: 
\bqn 
\cs[T_{a_1 \cdots \,a_{n-3}|c} ]
\!\!\!&=&\!\!\! \frac{-1}{(n-3)!} \int d^n x \Big[ \pa^e T^{b_1 ...
b_{n-3}|a}\pa_e T_{b_1 ... b_{n-3}|a} 
-\pa_e T^{b_1 ... b_{n-3}|e
}\pa^f T_{b_1 ... b_{n-3}|f }\nnn&&
- (n-3)[ - 3 \pa_e T^{e b_2 ... b_{n-3}|a}\pa^f 
T_{f b_2 ... b_{n-3}|a}\nnn
&&\quad \quad \quad\quad-2 T_{g}^{~ b_2 ... b_{n-3}|g }\pa^{e f }
T_{e b_2 ... b_{n-3}|f } 
-\pa^e T_{g}^{~ b_2 ... b_{n-3}|g }\pa_e
T^{f}_{~ b_2 ... b_{n-3}|f }\nnn
&& \quad \quad\quad\quad +(n-4)\pa_ e T_{g}^{~ e b_3 ... b_{n-3}|g }\pa^h
T^{f}_{~~ h b_3 ... b_{n-3}|f }] \Big]\,.
\label{action0T}
\eqn 
By construction, this dual action is equivalent to the initial
Pauli-Fierz action
for linearized general relativity. We shall compare it in the next 
subsections to
the Pauli-Fierz ($n=4$) and Curtright ($n=5$) actions.

One can notice that the equivalence between the actions 
(\ref{actione}) and (\ref{actionY}) can also be proved using the
following parent action: 
\bqn \cs [C_{ab|c},Y_{abc|d} ] = 4\! \int\!
d^nx \Big[ -
\frac{1}{2}C_{ab|c}\pa_dY^{dab|c}+C_{ca|}^{~~~a}C^{cb|}_{~~~b} \qquad
\nnn
-\frac{1}{2}C_{ab|c}C^{ac|b}
- \frac{1}{4}C_{ab|c}C^{ab|c}\Big],
\label{actionCY}\eqn 
where $C_{ab|c}=C_{[ab]|c}$ and $Y_{abc|d}
=Y_{[abc]|d} $. The field $ Y_{abc|d}$ is then a Lagrange
multiplier for the constraint $\pa_{[a} C_{bc]|d}$ = 0, this
constraint implies $C_{ab|c}=\pa_{[a}e_{b]c}$ and, eliminating it, 
one finds that the action (\ref{actionCY}) becomes the action
(\ref{actione}). On the other hand, $C_{ab|c}$ is an auxiliary
field and can be eliminated from the action (\ref{actionCY}) using
its equation of motion, the resulting action is then the action
(\ref{actionY}).

\subsection{Gauge symmetries}
The gauge invariances of the action (\ref{actione}) are known:
$\d e_{ab} = \pa_a\x_b + \pa_b\x_a +\o_{ab}$, where $\o_{ab}=\o_{[ab]}.$
These transformations can be extended to the auxiliary fields 
(as it is always the case \cite{aux}) leading to the gauge invariances of
the parent action (\ref{actioneY}): \bqn
\d_\xi e_{ab} &=& \pa_a\x_b + \pa_b\x_a, \label{gtdiff1}\\
\d_\xi  Y^{ab|}_{~~~~d}&=& - \, 6 \, \pa_c \, \pa^{[a} \x^b \d_d^{c]}
\label{gtdiff2}
\eqn
and
\bqn 
\d_\o e_{ab}& = &\o_{ab},\\
\d_\o  Y^{ab|}_{~~~~d}&=& 3 \,\pa_c \, \o^{[ab}\d_d^{c]}. \eqn
Similarly, the corresponding invariances for the
other parent action (\ref{actionCY}) are:
\bqn
\d_\xi  C_{ab|c} &=& \pa_c \pa_{[a}\x_{b]},\\
\d_\xi  Y^{abc|}_{~~~~~d}&=& -\, 6 \,\pa^{[a} \x^b \d_d^{c]} 
\eqn
and
\bqn
\d_\o  C_{ab|c}& =& \pa_{[a}\o_{b]c},\\
\d_\o  Y^{abc|}_{~~~~d}&=& 3 \, \o^{[ab}\d_d^{c]} . \eqn
These transformations affect only the irreducible
component $Z^{be}$ of $Y^{abe|}_{~~~~c}$.
[Note that one can redefine the gauge parameter $\omega_{ab}$ 
in such a way that $\d e_{ab} = \pa_a\x_b + \omega_{ab}$. In that case,
(\ref{gtdiff1}) and (\ref{gtdiff2}) become simply
$\d_\xi  e_{ab} = \pa_a\x_b$, $\d_\xi  Y^{ab|}_{~~~~d} = 0$.]

Given $Y^{ab|}_{~~~c}$, the equation $Y^{ab|}_{~~~c} = \pa_e Y^{abe|}_{~~~~c}$ 
does not
entirely determine $Y^{abe|}_{~~~~c}$. Indeed $Y^{ab|}_{~~~c}$ is
invariant under the transformation \be \d Y^{abe|}_{~~~~c}= \pa_f
\,( \phi^{abef|}_{~~~~~c}) \label{invceY}\ee of $Y^{abe|}_{~~~~c}$, with
$\phi^{abef|}_{~~~~~c} = \phi^{[abef]|}_{~~~~~~c}$. As the action
(\ref{actionY}) depends on $Y^{abe|}_{~~~~~c}$ only through
$Y^{ab|}_{~~~~c}$, it is also invariant under the gauge
transformations (\ref{invceY}) of the field $Y^{abe|c}$. 
In addition, it is invariant under arbitrary shifts of the irreducible
component $Z^{ab}$,
$$
\d_\o Y^{abc|}_{~~~~d}= 3 \, \o^{[ab}\d_d^{c]} .$$
The gauge invariances of the action (\ref{action0T}) involving only
$X^{abe|}_{~~~~c}$ (or,

\noindent equivalently, 
$T_{a_1 \cdots \,a_{n-3}|c}$) are simply
(\ref{invceY}) projected on the irreducible component
$X^{abe|}_{~~~~c}$ (or $T_{a_1 \cdots \,a_{n-3}|c} $).

It is of interest to note that it is the same $\omega$-symmetry that
removes the antisymmetric component of the tetrad in the action
(\ref{actione}) (yielding the Pauli-Fierz action for $e_{(ab)}$)
and the trace $Z^{ab}$ of the field $Y^{abe|}_{~~~~~c}$ (yielding 
the action (\ref{action0T}) for $T_{a_1 \cdots \,a_{n-3}|c} $ (or
$X^{abe|}_{~~~~c}$)). Because it is the same invariance that
is at play, one cannot
eliminate simultaneously both $e_{[ab]}$ and the trace of $Y^{ab|}_{~~~c}$
in the parent actions, even though these fields can
each be eliminated individually in their corresponding ``children'' actions 
(see \cite{West:2002jj} in this context).

\subsection{n=4: ``Pauli-Fierz is dual to Pauli-Fierz''}
In $n=4$ space-time dimensions,
the tensor $T_{a_1 \cdots \,a_{n-3}|c} $ has just two indices and
is symmetric, $T_{ab} = T_{ba}$. A direct computation shows that the action
(\ref{action0T}) then becomes
\be 
\cs[T_{ab}] =\int d^4 x \, [\pa^a T^{bc}\pa_a T_{bc}-
2 \pa_a T^{ab}\pa^c T_{cb}-2 T_a^{~a} \pa^{bc}T_{bc}
-\pa_a T_b^{~b} \pa^a T_c^{~c}]
\ee
which is the Pauli-Fierz action for the symmetric massless
tensor $T_{ab}$. At the same time, the gauge parameters
$\phi^{abef|}_{~~~~~c}$ can be written as $\phi^{abef|}_{~~~~~c} =
\ve^{abef} \gamma_c$ and the gauge transformations reduce to
$\delta T_{ab} \sim \partial_a \gamma_b + \partial_b \gamma_a$, as they
should. 
Our dualization procedure possesses thus the distinct feature, in four space-time
dimensions, of mapping the Pauli-Fierz action on itself.
Note that the electric (respectively, the magnetic) part of the (linearized) Weyl
tensor of the original Pauli-Fierz field $h_{ab} \equiv e_{(ab)}$ is
equal to the magnetic (respectively, minus the electric)
part of the (linearized) Weyl tensor of
the dual Pauli-Fierz $T_{ab}$,
as expected for duality \cite{DesNep,Hull:2001iu}. More precisely,  the curvatures $R^{ab\vert ce}(h)=2\pa^{[a}h^{b][c,e]}$ and $R^{ab\vert ce}(T)=2\pa^{[a}T^{b][c,e]}$ are related on-shell by the simple expression $K^{ab\vert ce}(h)\propto \ve^{abgh}K_{gh}{}^{\vert ce}(T)\,.$

An alternative, interesting, dualization procedure has been discussed in 
\cite{CasUrr1}.
In that procedure, the dual theory is described by a different action,
which has an additional antisymmetric field, denoted $\omega_{ab}$. This field 
does nontrivially enter the Lagrangian through its divergence
$\partial^a \omega_{ab}\,$.\footnote{In
the Lagrangian (27) of \cite{CasUrr1},
one can actually dualize the 
field $\omega_{ab}$
to a scalar $\Phi$ (\ie (i) replace $\partial^a \omega_{ab}$ by a vector 
$k_b$ in the
action; (ii) force $k_b = \partial^a \omega_{ab}$ through a Lagrange multiplier
term $\Phi \partial^a k_a$ where $\Phi $ is the Lagrange multiplier; and (iii)
eliminate the auxiliary field $k_a$ through its equations of motion).
A redefinition of the symmetric field $\tilde{h}_{ab}$ of \cite{CasUrr1} 
by a term $\sim \eta_{ab} \Phi$ enables one to absorb the scalar $\Phi$,
yielding the Pauli-Fierz action for the redefined symmetric field.}

\subsection{n=5: ``Pauli-Fierz is dual to Curtright''}
In $n=5$ space-time dimensions, the dual field is $T_{ab|c}= \frac{1}{3!}
\ve_{abefg}X^{efg|}_{~~~~c}$ , and has the
symmetries $T_{ab|c }= T_{[ab]|c} $ and $T_{[ab|c]}= 0.$ The action found 
by substituting this field into (\ref{actionY}) reads 
\bqn
&\cs[T_{ab|c }] = \frac{1}{2} \int d^5 x &[\pa^a T^{bc|d}\pa_a T_{bc|d} 
-2 \pa_a T^{ab|c}\pa^d T_{db|c}
-\pa_a T^{bc|a}\pa^d T_{bc|d} \nonumber \\
&&-4 T_a^{~b|a}\pa^{cd}T_{cb|d}-2 \pa_a T_b^{~c|b}\pa^a T^d_{~~c|d}
+2 \pa_a T_b^{~a|b}\pa^c T^d_{~~c|d}]\nn
\eqn
It is the action given by Curtright in
\cite{Curtright:1980yk} for such an ``exotic'' field. 

The gauge symmetries also match, as can be seen by redefining the
gauge parameters as $\psi_{gc}=
-\frac{1}{4!}\ve_{abefg}\phi^{abef|}_{~~~~~c}$. The gauge
transformations become 
\be 
\d T_{ab|c }= -2\pa_{[a}S_{b]c}- \frac{1}{3}
[\pa_aA_{bc}+\pa_bA_{ca}-2\pa_cA_{ab}], 
\ee 
where $\psi_{ab}=S_{ab}+A_{ab}, S_{ab}=S_{ba}, A_{ab}=-A_{ba}$. 
These are exactly the gauge transformations of \cite{Curtright:1980yk}.

It was known from \cite{Hull:2001iu} that the equations of motion for a
Pauli-Fierz field were equivalent to the equations of motion for a
Curtright field, \ie that the two theories were ``pseudo-dual''.
We have established here that they are, in fact, dual. The
duality transformation considered here contains the duality
transformation on the curvatures considered in \cite{Hull:2001iu}. 
Indeed, when the equations of motion hold, one has $R_{\m \n \a
\b}[h] \propto \varepsilon_{\m \n \r \s \tau} R^{\r \s
\tau}_{\hspace{.5cm} \a \b}[T]$ where $R_{\m \n \a \b}[h]$
(respectively $R_{\r \s \tau \a \b}[T]$) is the linearized
curvature of $h_{ab}\equiv e_{(ab)}$ (respectively, $T_{ab |c}$).

%
\section{Vasiliev description of higher-spin fields}
\setcounter{equation}{0}
\setcounter{theorem}{0}
\setcounter{lemma}{0}
\label{vasidescr}

In the discussion of duality for spin-two gauge fields, a crucial
role is played by the first-order action (\ref{actioneY}), in
which both the (linearized) vielbein and the (linearized)
spin connection (or, rather, a linear combination of it) are
treated as independent variables. This first-order action is
indeed one of the possible parent actions. In order to extend the
analysis to higher-spin massless gauge fields, we need a similar
description of higher-spin theories. Such a first-order 
description has been given in \cite{Vasiliev:1980as}. In this
section, we briefly review this formulation, alternative to the
more familiar second-order approach of \cite{Fronsdal:1978rb} 
(see Section \ref{appendixA} for the latter). We
assume $s >1$ and $n>3$.

\subsection{Generalized vielbein and spin connection}
The set of bosonic fields introduced in \cite{Vasiliev:1980as}
consists of a generalized vielbein 

\noindent $e_{\m\vert a_1\ldots a_{s-1}}$ 
and a generalized spin connection $\o_{\m\vert b\vert a_1\ldots
a_{s-1}}$. The vielbein is completely symmetric and traceless in
its last $s-1$ indices. The spin connection is not only completely
symmetric and traceless in its last $s-1$ indices but also
traceless between its second index and one of its last $s-1$ indices.
Moreover, complete symmetrization in all its indices but the first
gives zero. Thus, one has
\bqn 
&e_{\m\vert a_1\ldots 
a_{s-1}}=e_{\m\vert(a_1\ldots a_{s-1})}\,,~~ e_{\m\vert~b\ldots
a_{s-1}}^{~\;\, b}=0\,,&
\nonumber \\
&\o_{\m\vert b\vert a_1\ldots a_{s-1}}=\o_{\m\vert b\vert (a_1\ldots a_{s-1})}
\,,~~ \o_{\m\vert (b\vert a_1\ldots a_{s-1})}=0\,,&
\nonumber \\
&\o_{\m\vert b\vert ~ c\ldots a_{s-1}}^{~~~\,c}=0\,,~~ \o_{\m\vert
~\vert b\ldots a_{s-1}}^{~~b}=0\,.& \label{algcond1} 
\eqn 
The 
first index of both the vielbein and the spin connection may be
seen as a space-time form-index, while all the others are regarded
as internal indices. As we work at the linearized level, no
distinction will be made between both kinds of indices and they
will both be labelled either by Greek or by Latin letters, running
over $0,1,\cdots,n-1$.

The action was originally written in \cite{Vasiliev:1980as} in
four dimensions as \be \cs^s[e,\o]=\int d^4x \, \ve^{\m\n\r\s} \,
\ve_{abc\s} \, \o_{\r\vert}^{~\; b\vert a i_1\ldots i_{s-2}}\Big(
\pa_{\m}e_{\n\vert i_1\ldots i_{s-2}}^{\hspace*{36pt}c}
-1/2\o_{\m\vert\n\vert i_1\ldots i_{s-2}}^{\hspace*{43pt}c}\Big).
\label{Vasaction} \ee 
By expanding out the product of the two
$\ve$-symbols, one can rewrite it in a form valid in any 
number of space-time dimensions, 
\bqn 
\cs^s[e,\omega]=- 2 \int
d^nx\Big[ (B_{a_1[\n\vert\m]a_2\ldots a_{s-1}}
-\frac{1}{2(s-1)}B_{\n\m\vert a_1\ldots a_{s-1}})
K^{\m\n\vert a_1\ldots a_{s-1}}
\nonumber \\
\hspace{3.6cm} +(2B^{\r}_{~\m\vert a_2\ldots a_{s-1}\r}+
(s-2)B^{\r}_{~a_2\vert a_3 \ldots a_{s-1}\m\r})
K^{\m\n\vert a_2\ldots a_{s-1}}_{\hspace*{45pt}\n}\Big]\quad\quad
\label{VasaB} \eqn where \be B_{\m b\vert a_1\ldots a_{s-1}}\equiv 
2 \o_{[\m \vert b]\vert a_1 \ldots a_{s-1}} \label{Bdefinition}
\ee and where \be K^{\m\n\vert a_1\ldots a_{s-1}} =
\pa^{[\m}e^{\n]\vert a_1\ldots a_{s-1}}
-\frac{1}{4}B^{\m\n\vert a_1\ldots a_{s-1}}. \ee
The field $B_{\m b\vert a_1\ldots a_{s-1}}$ is antisymmetric in 
the first two indices, symmetric in the last $s-1$ internal
indices and traceless in the internal indices, \be B_{\m b\vert
a_1\ldots a_{s-1}} = B_{[\m b]\vert a_1\ldots a_{s-1}}, \; B_{\m
b\vert a_1\ldots a_{s-1}} = B_{\m b\vert (a_1\ldots a_{s-1})}, \;
B_{\m b\vert a_1\ldots a_{s-2}}^{\hspace{40pt} a_{s-2}} = 0\,, \ee
but it is otherwise arbitrary : given $B$ subject to these
conditions, one can always find an $\o$ such that
(\ref{Bdefinition}) holds \cite{Vasiliev:1980as}.

The invariances of the action (\ref{Vasaction}) are \cite{Vasiliev:1980as}
\bqn
\d e_{\m\vert a_1\ldots a_{s-1}} &=& \pa_{\m}\xi_{a_1\ldots a_{s-1}} 
+\a_{\m\vert a_1\ldots a_{s-1}}\,,
\label{Vasinva1}\\
\d \o_{\m\vert b\vert a_1\ldots a_{s-1}}&=&
\pa_{\m}\a_{b\vert a_1\ldots a_{s-1}}+\S_{\m\vert b\vert a_1\ldots a_{s-1}}\,,
\label{Vasinva2}
\eqn
where the parameters $\a_{\m\vert a_1\ldots a_{s-1}}$ and
$\S_{\m\vert b\vert a_1\ldots a_{s-1}}$ 
possess
the following algebraic properties
\bqn
&\a_{\n |(a_1 ... a_{s-1})}=\a_{\n |a_1 ... a_{s-1}}\,,\;
\a_{(\n |a_1 ... a_{s-1})}=0\,,\;
\a^{\n}_{~|\n a_2 ... a_{s-1}}=0 \,,\;
\a_{\n |a_1 ... a_{s-3}b}^{~~~~~~~~~~~~b}=0, \nonumber \\
&\S_{\m\vert b\vert a_1\ldots a_{s-1}}=\S_{(\m\vert b)\vert 
a_1\ldots a_{s-1}}=\S_{\m\vert b\vert (a_1\ldots a_{s-1})}\,,~~~
\S_{\m\vert (b\vert a_1\ldots a_{s-1})}=0\,,\nonumber \\
&\S^b_{~\vert b\vert a_1\ldots a_{s-1}}=0\,,~~~ \S^b_{~\vert
c\vert b a_2\ldots a_{s-1}}=0\,,~~~ \S_{\m\vert b\vert a_1\ldots
a_{s-3}c}^{\hspace*{53pt}c}=0\,. 
\label{algcond2} 
\eqn 
Moreover, the
parameter $\xi$ is traceless and completely symmetric.

The invariance under the transformation with the parameter $\xi$ can easily be checked in the action (\ref{Vasaction}). Indeed, the latter involves the vielbein only through its  antisymmetrized derivative  $\pa_{[\m} e_{\n]\vert a_1 \ldots a_{s-1}} \,$, which is invariant under the given transformation. 

 The parameter $\a$ generalizes the Lorentz parameter for gravitation
in the vielbein formalism. To show that the action is invariant under the transformation related to it, one must notice that the term bilinear in $\o$ is symmetric under the exchange of the $\o$'s:
\be \ve^{\m\n\r\s} \,
\ve_{abc\s} \, \o^{1}_{\r\vert}{}^{ b\vert a i_1\ldots i_{s-2}}
\o^{2}_{\m\vert\n\vert i_1\ldots i_{s-2}}{}^{c}
= \ve^{\m\n\r\s} \,
\ve_{abc\s} \, \o^{2}_{\r\vert}{}^{ b\vert a i_1\ldots i_{s-2}}
\o^{1}_{\m\vert\n\vert i_1\ldots i_{s-2}}{}^{c}\,.
\label{symomega}\ee
A way to prove this property is to expand the product of $\ve$-symbols and compare both sides of the equation. Schematically, the variation of the action (\ref{Vasaction}) then reads
\bqn
\d_\a \cs^s &=& \int d^4x\, [ \d_\a \o\, (\pa e - \frac{1}{2} \o)+\o\, \d_\a (\pa  e - \frac{1}{2} \o) ] \nnn
&=&\int d^4x\, [ \d_\a \o \pa e +\o \,\d_\a (\pa e -  \o) ] 
=\int d^4x\, [ -\pa (\d_\a \o) \, e +\o \,\d_\a (\pa  e -  \o) ] \,.
\nn
\eqn
We used \bref{symomega} for the second equality, and, for the last equality, 
we supposed that there is no border term. The first term vanishes because 
the explicit derivative is antisymmetrized with the derivative in 
$\d_\a \o$. The second term vanishes because the variation of $\pa  e$ is exactly the variation of $ \o\,$.

To understand the invariance involving the parameter $\Sigma$, let us decompose the fields $\o$, $B$ and $\Sigma$ into their traceless irreducible components. One has (see Appendix \ref{young})
\bqn
\begin{picture}(305,15)(0,2)
\multiframe(70,5.5)(13.5,0){1}(10.5,10.5){\tiny{$\m$}}
\multiframe(101,11)(13.5,0){1}(10.5,10.5){\tiny{$a_1$}}
\multiframe(101,0)(13.5,0){1}(10.5,10.5){\tiny{$b$}}
\multiframe(112,11)(13.5,0){1}(19.5,10.5){$\cdots$}
\multiframe(132,11)(13.5,0){1}(19.5,10.5){\tiny{$a_{s-1}$}}
\put(0,7.5){$\o_{\m\vert b\vert a_1\ldots a_{s-1}} \sim$}
\put(88,7.5){$\otimes$}
\end{picture}
\nn
\\
\begin{picture}(250,19)(0,2)
\put(0,7.5){$=$}
\multiframe(15,0)(13.5,0){1}(6.5,6.5){}
\multiframe(15,7)(13.5,0){1}(6.5,6.5){}
\multiframe(15,14)(13.5,0){1}(6.5,6.5){}
\multiframe(22,14)(13.5,0){1}(19.5,6.5){$\cdots$}
\multiframe(42,14)(13.5,0){1}(6.5,6.5){}
\put(50,16){\tiny{s-1}}
\put(57,7.5){$\oplus$}
\multiframe(70,3)(13.5,0){1}(6.5,6.5){}
\multiframe(70,10)(13.5,0){1}(6.5,6.5){}
\multiframe(77,3)(13.5,0){1}(6.5,6.5){}
\multiframe(77,10)(13.5,0){1}(6.5,6.5){}
\multiframe(84,10)(13.5,0){1}(19.5,6.5){$\cdots$}
\multiframe(104,10)(13.5,0){1}(6.5,6.5){}
\put(112,12){\tiny{s-1}}
\put(122,7.5){$\oplus$}
\multiframe(135,3)(13.5,0){1}(6.5,6.5){}
\multiframe(135,10)(13.5,0){1}(6.5,6.5){}
\multiframe(142,10)(13.5,0){1}(19.5,6.5){$\cdots$}
\multiframe(162,10)(13.5,0){1}(6.5,6.5){}
\put(170,12){\tiny{s}}
\put(181,7.5){$\oplus$}
\multiframe(195,3)(13.5,0){1}(6.5,6.5){}
\multiframe(195,10)(13.5,0){1}(6.5,6.5){}
\multiframe(202,10)(13.5,0){1}(19.5,6.5){$\cdots$}
\multiframe(222,10)(13.5,0){1}(6.5,6.5){}
\put(230,12){\tiny{s-2}}
\put(240,7.5){$\oplus$}
\multiframe(255,7)(13.5,0){1}(6.5,6.5){}
\multiframe(262,7)(13.5,0){1}(19.5,6.5){$\cdots$}
\multiframe(282,7)(13.5,0){1}(6.5,6.5){}
\put(290,9){\tiny{s-1}}
\put(305,7.5){,}
\end{picture}
\nn \\
\begin{picture}(305,25)(0,2)
\multiframe(70,0)(13.5,0){1}(10.5,10.5){\tiny{$\n$}}
\multiframe(70,11)(13.5,0){1}(10.5,10.5){\tiny{$\m$}}
\multiframe(101,7.5)(13.5,0){1}(10.5,10.5){\tiny{$a_1$}}
\multiframe(112,7.5)(13.5,0){1}(19.5,10.5){$\cdots$}
\multiframe(132,7.5)(13.5,0){1}(19.5,10.5){\tiny{$a_{s-1}$}}
\put(0,7.5){$B_{\m\n\vert a_1\ldots a_{s-1}} \sim$}
\put(88,7.5){$\otimes$}
\end{picture}
\nn
\\
\begin{picture}(250,19)(0,2)
\put(0,7.5){$=$}
\multiframe(15,0)(13.5,0){1}(6.5,6.5){}
\multiframe(15,7)(13.5,0){1}(6.5,6.5){}
\multiframe(15,14)(13.5,0){1}(6.5,6.5){}
\multiframe(22,14)(13.5,0){1}(19.5,6.5){$\cdots$}
\multiframe(42,14)(13.5,0){1}(6.5,6.5){}
\put(50,16){\tiny{s-1}}
\put(57,7.5){$\oplus$}
\multiframe(75,3)(13.5,0){1}(6.5,6.5){}
\multiframe(75,10)(13.5,0){1}(6.5,6.5){}
\multiframe(82,10)(13.5,0){1}(19.5,6.5){$\cdots$}
\multiframe(102,10)(13.5,0){1}(6.5,6.5){}
\put(110,12){\tiny{s}}
\put(121,7.5){$\oplus$}
\multiframe(135,3)(13.5,0){1}(6.5,6.5){}
\multiframe(135,10)(13.5,0){1}(6.5,6.5){}
\multiframe(142,10)(13.5,0){1}(19.5,6.5){$\cdots$}
\multiframe(162,10)(13.5,0){1}(6.5,6.5){}
\put(170,12){\tiny{s-2}}
\put(180,7.5){$\oplus$}
\multiframe(195,7)(13.5,0){1}(6.5,6.5){}
\multiframe(202,7)(13.5,0){1}(19.5,6.5){$\cdots$}
\multiframe(222,7)(13.5,0){1}(6.5,6.5){}
\put(230,9){\tiny{s-1}}
\put(245,7.5){,}
\end{picture}
\nn \\
\begin{picture}(305,25)(0,2)
\put(0,7.5){$\Sigma_{\m\vert b\vert a_1\ldots a_{s-1}} \sim$}
\multiframe(70,0)(13.5,0){1}(10.5,10.5){\tiny{$\m$}}
\multiframe(70,11)(13.5,0){1}(10.5,10.5){\tiny{$a_1$}}
\multiframe(81,0)(13.5,0){1}(10.5,10.5){\tiny{$b$}}
\multiframe(81,11)(13.5,0){1}(10.5,10.5){\tiny{$a_2$}}
\multiframe(92,11)(13.5,0){1}(19.5,10.5){$\cdots$}
\multiframe(112,11)(13.5,0){1}(19.5,10.5){\tiny{$a_{s-1}$}}
\put(147,7.5){.}
\end{picture}
\nn
\eqn
The field $B$ is defined as a projection of $\o$. The decomposition into irreducible components shows that $B$ contains all the irreducible components of $\o$, except the one that has the symmetry of $\Sigma$, which we call $\o_\Sigma\,$. Conversely, 
all components of $\o$ except the latter can be expressed in terms of $B\,$.
Since the action  \bref{Vasaction} can be written in terms of only $B$ as \bref{VasaB}, it is thus invariant under any shift of the component $\o_\Sigma\,$. since this is exactly how the transformation with parameter $\Sigma$ acts,  the action  \bref{Vasaction} is invariant under these transformations.

In the Vasiliev formulation, the fields and gauge parameters are
subject to the tracelessness conditions contained in (\ref{algcond1})
and (\ref{algcond2}). It would be of interest to investigate
whether these conditions can be dispensed with as in
\cite{Francia:2002aa,Francia:2002pt}. 

\subsection{Equivalence with the standard second-order formulation}
Since the action (\ref{VasaB}) depends on $\o$ only through $B$, extremizing it with respect
to $\o$ is equivalent to extremizing it with respect to $B$. Thus, we
can view $\cs^s[e,\o]$ as $\cs^s[e,B]$.
In the action $\cs^s[e,B]$, the 
field $B^{\m\n\vert a_1\ldots a_{s-1}}$ is an auxiliary field.
Indeed, the field equations for $B^{\m\n\vert a_1\ldots a_{s-1}}$
enable one to express $B$ in terms of the vielbein and its derivatives as,
\be
B^{\m\n\vert a_1\ldots a_{s-1}}=2 \pa^{[\m}e^{\n]\vert a_1\ldots a_{s-1}}
\label{sol0forB}
\ee
(the field $\o$ is thus fixed up to the pure gauge component related to $\S$.) 
When substituted into (\ref{VasaB}), (\ref{sol0forB}) gives an action 
$S^s[e,B(e)]$
invariant under (\ref{Vasinva1}).

The field $e_{\m\vert a_1\ldots a_{s-1}}$ can be represented by
\bqn e_{\m\vert a_1\ldots a_{s-1}} &=& h_{\m a_1\ldots a_{s-1}}
+\frac{(s-1)(s-2)}{2s}[\h_{\m (a_1}h^{\prime}_{a_2\ldots a_{s-1})}
-\h_{(a_1 a_2} h^{\prime}_{\m a_3\ldots a_{s-1})}]
\nonumber \\ 
&+&\b_{\m\vert a_1\ldots a_{s-1}}\,, \eqn where $h_{\m a_1\ldots
a_{s-1}}$ is completely symmetric, $h^{\prime}_{a_2\ldots a_{s-1}}=
h^{\m}_{\hspace{5pt} \m \ldots a_{s-1}}$ is its trace, and the
component $\b_{\m\vert a_1\ldots a_{s-1}}$ possesses the
symmetries of the parameter $\a$ in (\ref{Vasinva1}) and thus
disappears from $S^s[e,\o(e)]$. Of course, the double trace $h^{\m 
\n }_{\hspace{10pt} \m \n \ldots a_{s-1}}$ of $h_{\m a_1\ldots
a_{s-1}}$ vanishes. The action $S^s[e(h)]$ is nothing but the one
given in \cite{Fronsdal:1978rb}  for a completely symmetric and
double-traceless bosonic spin-$s$ gauge field $h_{\m a_1\ldots
a_{s-1}}$, \ie the action (\ref{freeactions}).

In the spin-2 case, the Vasiliev fields are $e_{\mu |a}$ and 
$\o_{\nu |b |a}$ with $\omega_{\nu |b |a} = - \o_{\nu |a |b}$. The
$\S$-gauge invariance is absent since the conditions $\S_{\n| b
|a} = - \S_{\n |a |b}$, $\S_{b |c |a} = \S_{c |b |a}$ imply
$\S_{\n |a| b} = 0$. The gauge transformations read \be \d e_{\n
|a} = \pa_ \n \xi_a + \a_{\n |a}, \; \; \d \o_{\n |b |a} = \pa_\n
\a_{b |a} \ee with $\a_{\n |a} = - \a_{a |\n}$. The relation 
between $\o$ and $B$ is invertible and the action (\ref{VasaB}) is
explicitly given by \be S^2[e,B]=\!- 2\!\int\!\! d^nx\Big[
(B_{a[\n\vert\m]}
-\frac{1}{2}B_{\n\m\vert a})
(\pa^{[\m}e^{\n]\vert a}
-\frac{1}{4}B^{\m\n\vert a})
+ 2B^{\r}_{~\m\vert \r} 
(\pa^{[\m}e^{\n]\vert }_{\hspace*{10pt}\n}
-\frac{1}{4}B^{\m\n\vert }_{\hspace*{13pt}\n})\Big]
\ee
Up to the front factor $-2$, the coefficient $Y_{\m \n |a}$ of the antisymmetrized derivative
$\pa^{[\m}e^{\n]\vert a}$ of the vielbein is given in terms
of $B$ by 
\be
Y_{\m \n |a} = B_{a[\m|\n]} -\frac{1}{2}B_{\m \n |a} -
2 \h_{a[\m}B_{\n] b|}^{\hspace*{13pt}b}.
\ee
This relation can be inverted to yield $B$ in terms of $Y$,
\be
B_{\m \n |a} = 2 Y_{a [\m| \n]} - 
\frac{2}{n-2}\h_{a[\m}Y_{\n] b|}^{\hspace*{13pt}b}.
\ee 
Re-expressing the action in terms of
$e_{\mu a}$ and $Y_{\m \n a}$ gives the action (\ref{actioneY})
considered previously.
%
\section{Spin-3 duality} 
\setcounter{equation}{0}
\setcounter{theorem}{0}
\setcounter{lemma}{0}
%
Before dealing with duality in the general spin-$s$ case,
we treat in detail the spin-3 case. 

\subsection{Arbitrary dimension $\geq 4$}
Following the spin-2 procedure, we first rewrite the action (\ref{VasaB})
in terms of $e_{\n|\r\s}$ and the coefficient $Y_{\m\n|\r\s}$
of the antisymmetrized derivatives of $e_{\n|\r\s}$ in the action.
In terms of $\o_{\m|\n|\r\s}$, this field is given by
\bqn
Y_{\m\n|\r\s}=2 [\o_{\r|[\n|\m]\s}+\o_{\s|[\n|\m]\r}- 
2\o^{\l}_{~~|[\l|\m](\r}\h_{\s)\n}
+ 2\o^{\l}_{~~|[\l|\n](\r}\h_{\s)\m}]\nn
\eqn
or, equivalently,
\be
Y_{\m\n|a_1 a_{2}}=B_{a_1 \m|\n a_2}- \frac {1}{4} 
B_{\m\n|a_1 a_{2}}+ 2 \h_{\m a_1} B^{\l}_{~\n|\l a_2 }
+ \h_{\m a_1} B^{\l}_{~a_2|\l \n }
\label{YB3}
\ee
where antisymmetrization in $\m$, $\n$ and symmetrization in
$a_1$, $a_2$ is understood.
The field $Y_{\m\n|\r\s} $ fulfills the algebraic relations 
$Y_{\m\n|\r\s} =Y_{[\m\n]|\r\s} =Y_{\m\n|(\r\s)} $ and
$Y_{\m\n|\b}^{~~~~\b}=0$. 

One can invert (\ref{YB3}) to express the field $B_{\m\n|\r\s}$ in
terms of $Y_{\m\n|\r\s}$. One gets 
$$ B_{\m\n|\r\s} =
\frac{4}{3} \Big[Y _{\m\n|\r\s}+2 [Y _{\r[\m|\n]\s}+Y
_{\s[\m|\n]\r}]+
\textstyle{\frac{2}{n-1}}[-2\h_{\r\s}Y_{\l[\m|\n]}^{\hspace{20pt}\l}
+Y^{\l}_{\r|\l[\n}\h_{\m]\s}+Y^{\l}_{\s|\l[\n}\h_{\m]\r}]\Big]
$$ 
When inserted into the action, this yields \bqn 
\cs(e_{\m|\n\r},Y_{\m\n|\r\s} )&=& -2 \int d^nx \left\{ \right.
Y_{\m\n|\r\s} \pa^{\m} e^{\n|\r\s}
\nonumber \\
&&+ {\frac{4}{3}}[{\frac{1}{4} }Y^{\m\n|\r\s} Y_{\m\n|\r\s}
-Y^{\m\n|\r\s} Y_{\r\n|\m\s}
+{\frac{1}{n-1}} Y^{\r\m|\n}_{~~~~~\r} Y_{\l\n|\m}^{~~~~\l}] \left. 
\right\}\,. \nn\eqn

The generalized vielbein $e_{\n|\r\s}$ may again be viewed as a
Lagrange multiplier since it occurs linearly. Its equations of
motion force the constraints \be \pa^{\m}Y_{\m\n|\r\s}=0 
\label{constraintspin3} \ee The solution of these equations is
$Y_{\m\n|\r\s}=\pa^{\l}Y_{\l\m\n|\r\s}$ where
$Y_{\l\m\n|\r\s}=Y_{[\l\m\n]|\r\s}=Y_{\l\m\n|(\r\s)}$ and
$Y_{\l\m\n|\r}^{~~~~~\r}=0$. The action then becomes \bqn
\cs(Y_{\l\m\n|\r\s})= \frac{8}{3} \int d^nx \,[-\frac{1}{4}
Y^{\m\n|\r\s} Y_{\m\n|\r\s} +Y^{\m\n|\r\s} Y_{\r\n|\m\s}
-\frac{1}{n-1} Y^{\r\m|\n}_{~~~~~\r} Y_{\l\m|\n}^{~~~~\l}] \,,\nn 
\eqn
where $Y_{\m\n|\r\s}$ must now be viewed as the dependent field
$Y_{\m\n|\r\s}=\pa^{\l}Y_{\l\m\n|\r\s}$ .

One now decomposes the field $Y_{\l\m\n|\r\s}$ into irreducible
components, \be Y^{\l\n\m|}_{~~~~~\r\s} = X^{\l\n\m|}_{~~~~~\r\s} 
+ \d_{(\r}^{[\l}Z^{\m\n]}_{~~~\s)} \label{decomp0} \ee with
$X^{\l\n\m|}_{~~~~~\r\m}=0$, $X^{\l\n\m|}_{~~~~~\r\s}=
X^{[\l\n\m]|}_{~~~~~\r\s}$,
$X^{\l\n\m|}_{~~~~~\r\s}=X^{\l\n\m|}_{~~~~~(\r\s)}$ and $Z^{\m\n}
_{~~~\s}= Z^{[\m\n]}_{~~~\s}$. Since $Z^{\m\n} _{~~~\s}$ is
defined by Eq.(\ref{decomp0}) only up to the addition of a term like
$\delta^{[\m}_{\s} k^{\n]}$ with $k^\n$ arbitrary, one may assume 
$Z^{\m\n}_{~~~\n} = 0$. 

The new feature compared to spin $2$
is that the field $Z^{\m\n} _{~~~\s}$ is no longer entirely pure
gauge. However, the component of $Z^{\m\n} _{~~~\s}$ that is
not pure gauge is entirely determined by
$X^{\l\n\m|}_{~~~~~\r\s}$.
Indeed, the tracelessness condition $Y^{\l\n\m|}_{~~~~~\r\s}
\eta^{\r \s} = 0$ implies \be Z^{[\l \m \vert \n]} = - 
X^{\l\n\m|}_{~~~~~\r\s} \eta^{\r \s} \label{ZenfonctiondeX} \ee
One can further decompose $Z_{\l \m \vert \n} = \Phi_{\l \m \n} +
\frac{4}{3} \Psi_{[\l \vert \m] \n} $ with $\Phi_{\l \m \n} =
\Phi_{[\l \m \n]}= Z_{[\l \m \vert \n]}$ and $\Psi_{\l \vert \m
\n} = \Psi_{\l \vert (\m \n)} = Z_{\l (\m \vert \n)}$. In
addition, $\Psi_{(\l \vert \m \n)} = Z_{(\l \m \vert \n)}= 0$ and
$\Psi_{\l \vert \m \n} \eta^{\m \n} = Z_{\l \m \vert \n} \eta^{\m 
\n} = 0$. {} Furthermore, the $\alpha$-gauge symmetry reads $ \d
Z_{\l \m \vert \n} = \a_{[\l \vert \m] \n}$ i.e, $\d \Phi_{\l \m
\n} = 0$ and $\d \Psi_{\l \vert \m \n} = \frac{3}{4} \a_{\l \vert
\m \n}$. Thus, the $\Psi$-component of $Z$ can be gauged away
while its $\Phi$-component is fixed by $X$. The only remaining
field in the action is $X^{\l\n\m|}_{~~~~~\r\s}$, as in the
spin-2 case.

Also as in the spin-2 case, there is a redundancy in the solution of
the constraint (\ref{constraintspin3})
for $Y_{\n\a|\b\g}$, leading to the gauge symmetry
(in addition to the $\alpha$-gauge symmetry)
\be
\d Y^{\l\m\n|}_{~~~~~a_1 a_{2}}~=\pa_\r
\psi^{\r\l\m\n|}_{~~~~~~~a_1 a_{2}} 
\label{invaraspin3}
\ee
where $\psi^{\r\l\m\n|}_{~~~~~~~a_1 a_{2}}$ is antisymmetric in
$\r$, $\l$, $\m$, $\n$ and symmetric in $a_1$, $a_2$ and is traceless
on $a_1$, $a_2$, \ie $\psi^{\r\l\m\n|}_{~~~~~~~a_1 a_{2}} \eta^{a_1 a_2} = 0$.
This gives, for $X$,
\be \d X^{\l\m\n|}_{~~~~~a_1 a_{2}}~=\pa_\r
\big( \psi^{\r\l\m\n|}_{~~~~~~~a_1 a_{2}} + \frac{6}{n-1} \d^{[\l}_{(a_1} 
\psi^{\m\n]\r\s|}_{~~~~~~~a_2) \s} \big) \ee

\subsection{$n=5$ and $n=4$}
One can then trade the
field $X$ for a field $T$ obtained by 
dualizing on the indices $\l$, $\m$, $\n$ with the $\ve$-symbol.
We shall carry out the computations
only in the case $n=5$ and
$n=4$, since the case of general dimensions will be covered
below for general spins.
Dualising in $n=5$ gives
$X^{\l\n\m|}_{~~~~~\r\s} = \frac{1}{2}\ve^{\l\n\m\a\b}T_{\a\b|\r\s} $
and the action becomes:
\bqn 
\cs(T_{\m\n|\r\s})&\!\!=& \!\!\frac{2}{3} \int d^5x
[-\pa_{\l}T_{\m\n|\r\s}\pa^{\l}T^{\m\n|\r\s}
+2 \pa^{\l}T_{\l\n|\r\s}\pa_{\m}T^{\m\n|\r\s}+2 \pa^{\r}T_{\m\n|\r\s}\pa_{\l}
T^{\m\n|\l\s}
\nonumber \\
&&\quad+ 8 T_{\m\n|\r\s}\pa^{\m\r}T_{\l}^{~\n|\l\s}+2T_{\m\n|\r\s} 
\pa^{\r\s}T^{\m\n|\l}_{~~~~~\l}+4 \pa_{\r}T_{\l}^{~~\n|\l\s}\pa^{\r}
T^{\m}_{~~\n|\m\s}
\nonumber \\
&&\quad-4 \pa_{\n}T_{\l}^{~~\n|\l\s}\pa^{\r}
T^{\m}_{~~\r|\m\s}+4 \pa_{\s}T_{\l}^{~~\n|\l\s}\pa^{\r}T_{\r\n|\m}^{~~~~~\m}
+\pa_{\l}T^{\m\n|\r}_{~~~~~\r}\pa^{\l}T_{\m\n|\s}^{~~~~~\s}] \nn
\eqn
with $T_{\m\n|\r\s}=T_{\m\n|(\r\s)}=T_{[\m\n]|\r\s}$ and $T_{[\m\n|\r]\s}=0$.
The gauge symmetries of the $T$ field following from
(\ref{invaraspin3}) are
\be
\d T_{\m\n|\r\s} =-\pa_{[\m}\varphi_{\n]|\s\r} +
 \frac{3}{4}[\pa_{[\m}\varphi_{\n|\s]\r}+\pa_{[\m}\varphi_{\n|\r]\s}] \,,
\ee
where the gauge parameter $\varphi_{\a|\r\s} \sim
\ve_{\a \l\m\n\t} \psi^{\l\m\n\t|}_{\hspace*{23pt}\r\s}$ is such that 
$\varphi_{\a|\r\s} =\varphi_{\a|(\r\s)} $ and 
$\varphi_{\a|\r}^{\hspace*{13pt}\r}=0$.
The parameter $
\varphi_{\a|\r\s} $ can be decomposed into irreducible components:
$\varphi_{\a|\r\s}=\chi_{\a\r\s}+\phi_{\a(\r|\s)}$ where 
$\chi_{\a\r\s}=\varphi_{(\a|\r\s)}$
and $\phi_{\a\r|\s}=\frac{3}{4} \varphi_{[\a|\r]\s}$ . 
The gauge transformation then reads
\be
\d T_{\m\n|\r\s} =\pa_{[\m}\chi_{\n]\r\s}
+\frac{1}{8}[-2\pa_{[\m}\phi_{\n]\r|\s}+3 \phi_{\m\n|(\s,\r)}]\,, 
\ee
and the new gauge parameters are constrained by the condition
$\chi_{\a|\r}^{\hspace*{13pt}\r}+\phi_{\a|\r}^{\hspace*{13pt}\r}=0$.

These are the action and gauge symmetries for the field $T_{\m\n|\r\s}$ dual to
$e_{(\m\n\r)}$ in $n=5$ ; they coincide with the ones given in 
\cite{Chung:1987mv,Labastida:1987kw,Burdik:2001hj,Bekaert:2003az}.

In four space-time dimensions, dualization reads 
$T_{\a\r\s } = \ve_{\l \m \n \a} X^{\l\m\n|}_{~~~~~\r\s}$.
The field $T_{\a\r\s}$ is totally symmetric because of
$X^{\l\n\m|}_{~~~~~\r\m}=0$.
The action reads 
\bqn
\cs (T_{\m\n\r}) = -\frac{4}{3}\int d^4 x \Big[ \pa_{\l}T_{\m\n\r}
\pa^{\l}T^{\m\n\r}-3\pa^{\m}T_{\m\n\r}\pa_{\l}T^{\l\n\r}-6T_{\l}^{~~\l\m} 
\pa^{\n\r}T_{\m\n\r} \nonumber \\
-3 \pa_{\l}T_{\m}^{~~\m\n}\pa_{\l}T_{\r}^{~~\r\n}-\frac{3}{2}\pa_{\l}
T^{\l\m}_{~~~\m}\pa_{\n}T^{\n\r}_{~~~\r}\Big]
\eqn
The gauge parameter
$\psi^{\r\l\m\n|}_{~~~~~~~a_1 a_{2}}$ can be rewritten as
$\psi^{\r\l\m\n|}_{~~~~~~~a_1 a_{2}} = (-1/2) 
\ve^{\r\l\m\n} k_{a_1 a_{2}}$ where $k_{a_1 a_{2}}$ is
symmetric and traceless. The gauge transformations are, in terms
of $T$, $\d T_{\r\s \a} = \pa_\r k_{\s \a} + \pa_\s k_{\a \r} +
\pa_\a k_{\r \s}$. The dualization procedure yields back the
Fronsdal action and gauge symmetries \cite{Fronsdal:1978rb}. Note
also that the gauge-invariant curvatures of the original field
$h_{\m\n\r}\equiv e_{(\m \n \r)}$ and of $T_{\m\n\r}$, which now involve three
derivatives \cite{deWit:1979pe,Damour:1987vm}, are again related on-shell 
by an $\ve$-transformation 
$R_{\a\b\m\n\r\s}[h]\propto\ve_{\a\b\a'\b'}$ 
$R^{\a'\b'}_{\hspace{13pt}\m\n\r\s}[T]$, as they should.
%
\section{Spin-$s$  duality}
\setcounter{equation}{0}
\setcounter{theorem}{0}
\setcounter{lemma}{0} 
%
The method for dualizing the spin-$s$  theory follows exactly the same pattern
as for spins two and three:
\begin{itemize}
\item First, one rewrites the action in terms of $e$ and $Y$ (coefficient 
of the antisymmetrized derivatives of the generalized vielbein in the action);
\item Second, one observes that $e$ is a Lagrange multiplier for a differential
constraint on $Y$, which can be solved explicitly in terms of a new field
 with one more index;
\item Third, one decomposes this new field into irreducible components;
only one component (denoted $X$) remains in the action; using the
$\ve$-symbol, this component can be replaced by the ``dual field''
$T$.
\item Fourth, one derives the gauge invariances of the dual theory 
from the redundancy in the description of the solution of the constraint
in step 2.
\end{itemize}
We now implement these steps explicitly.

\subsection{Trading $B$ for $Y$}
The coefficient of $\pa^{[\n} e^{\m]|a_1 ... a_{s-1}}$ in the
action (\ref{VasaB}) is given by \bqn Y_{\m\n|a_1 ... a_{s-1}}&=&B_{a_1 \m|\n 
a_2 ... a_{s-1}}- \frac {1}{2(s-1)} B_{\m\n|a_1 ... a_{s-1}}+ 2
\h_{\m a_1} B^{\l}_{~\n|\l a_2 ... a_{s-1}}
\nonumber \\
&+& (s-2) \h_{\m a_1} B^{\l}_{~a_2|\l \n a_3 ... a_{s-1}}\,,
\label{YYBB} \eqn where the r.h.s. of this expression must be
antisymmetrized in $\m,\n$ and symmetrized in the indices $a_i$. 
The field $Y_{\m\n|a_1 ... a_{s-1}}$ is antisymmetric in $\m$ and
$\n$, totally symmetric in its internal indices $a_i$ and
traceless on its internal indices. One can invert Eq.(\ref{YYBB}) to
express $B_{\m\n|a_1 ... a_{s-1}}$ in terms of $Y_{\m\n|a_1 ...
a_{s-1}}$. To that end, one first computes the trace of
$Y_{\m\n|a_1 ... a_{s-1}}$. One gets \bqn && Y^\l_{~~\m |\l a_2
\cdots a_{s-1}} = \frac{n + s -4}{2(s-1)} \left(2 B^\l_{~~\m |\l
a_2 \cdots a_{s-1}} + (s-2) B^\l_{~~(a_2 | a_3 \cdots a_{s-1}) \l
\m } \right)\nn \\ 
&& \Leftrightarrow \; B^\l_{~~\m |\l a_2 \cdots a_{s-1}} =
\frac{2(s-1)^2}{s(n+s-4)} \left( Y^\l_{~~\m |\l a_2 \cdots
a_{s-1}} - \left(\frac{s-2}{s-1}\right) Y^\l_{~~(a_2 | a_3 \cdots
a_{s-1}) \l \m } \right) \nn\eqn Using this expression, one can then
easily solve Eq.(\ref{YYBB}) for $B_{\m\n|a_1 ... a_{s-1}}$, \bqn 
B_{\m\n|a_1 ... a_{s-1}} &=& 2 \frac{(s-1)}{s}\Big[ (s-2)
Y_{\m\n|a_1\ldots a_{s-1}} - 2 ( s-1) Y_{\m a_1|\n a_2 ...
a_{s-1}}
\nonumber \\
&+&2 \frac {(s-1)}{(n+s-4)}[(s-2)
\h_{a_1 a_2} Y_{\m\r|\n a_3 ... a_{s-1}}^{~~~~~~~~~~~~\r}
\nonumber \\
&-& (s-2)\h_{a_1 \m} Y_{a_2\r|\n a_3 ...
a_{s-1}}^{~~~~~~~~~~~~~\r}+(s-3) \h_{a_1 \m} Y_{\n\r| a_2 ... 
a_{s-1}}^{~~~~~~~~~~~\r} ] \Big] \label{BBYY} \eqn where the
r.h.s. must again be antisymmetrized in $\m,\n$ and symmetrized in
the indices $a_i$. We have checked Eq.(\ref{BBYY}) using FORM
(symbolic manipulation program \cite{form}).

The action (\ref{VasaB}) now reads \bqn \cs^s &=& - 2 \int d^n x 
\Big[ Y_{\m\n|a_1 ... a_{s-1}} \pa^{[\n} e^{\m]|a_1 ... a_{s-1}} +
\textstyle{\frac{(s-1)^2}{s} }\Big[ -Y_{\m\n|a_1 ... a_{s-1}}Y^{\m a_1|\n a_2
... a_{s-1}}
\nonumber \\
&+&\textstyle{\frac {(s-2)}{2(s-1)}}Y_{\m\n|a_1 ... a_{s-1}}Y^{\m\n|a_1 ... a_{s-1}} \
+ \textstyle{\frac{1}{(n+s-4)}} [ (s-3) Y_{\m\n |a_1 ... a_{s-2}}^{\hspace{45pt}\m}
Y^{\n\r | a_1 ... a_{s-2}}_{\hspace*{48pt}\r} 
\nonumber \\
&-&(s-2)Y_{\m\n |a_1 ... a_{s-2}}^{\hspace{45pt}\m}
Y^{a_1\r | \n a_2 ... a_{s-2}}_{\hspace{56pt}\r}]\Big]\Big]\,.\label{actidual}
\eqn
It is invariant under the transformations (\ref{Vasinva1}) and (\ref{Vasinva2})
\bqn
\d e_{\n |a_1 ... a_{s-1}}&=& \pa_{\m}\xi_{a_1\ldots a_{s-1}} +
\a_{\n |a_1 ... a_{s-1}}
\nonumber \\ 
\d Y^{\m\n|} _{~~~~a_1 ... a_{s-1}}&=& 3 \pa_{\l} \d^{[\l}_{(a_1}
\a^{\m|\n]}_{~~~~a_2 ... a_{s-1})}\nn
\eqn
Remember that $ \a_{\n |a_1 ... a_{s-1}}$ satisfies the relations
\be 
\a_{(\n |a_1 ... a_{s-1})}=0\,,~~~
\a^{\n}_{~|\n a_2 ... a_{s-1}}=0 \,,~~~
\a_{\n |a_1 ... a_{s-3}b}^{~~~~~~~~~~~~b}=0 \,.\label{rela}
\ee
while $\xi_{a_1\ldots a_{s-1}}$ is completely symmetric and traceless.

\subsection{Eliminating the constraint}
The field equations for $e^{\m|a_1 ... a_{s-1}} $  are
constraints for the field $Y$,
\be \pa^{\n}Y_{\n\m|a_1 ... a_{s-1}}=0 \,,
\label{constraintforY}\ee
which imply 
\be Y_{\m\n|a_1 ... a_{s-1}}=\pa^{\l}Y_{\l\m\n|a_1 ... a_{s-1}}\,,
\ee where $Y_{\l\m\n|a_1 ... a_{s-1}}=Y_{[\l\m\n]|a_1 ... a_{s-1}} 
=Y_{\l\m\n|(a_1 ... a_{s-1})}$ and $Y^{\l\m\n|a}_{~~~~~~~a a_3 ...
a_{s-1}}=0$ . If one substitutes the solution of the constraints
inside the action, one gets \bqn &&\cs(Y_{\l\m\n|a_1 ... a_{s-1}})
=- 2 \textstyle{\frac{(s-1)^2}{s}} \int d^n x \Big[-Y_{\m\n|a_1 ...
a_{s-1}}Y^{\m a_1|\n a_2 ... a_{s-1}}
\nonumber \\
&&+\textstyle{\frac {(s-2)}{2(s-1)}}Y_{\m\n|a_1 ... a_{s-1}}Y^{\m\n|a_1 ... a_{s-1}}
+ \textstyle{\frac{1}{(n+s-4)}} [ (s-3) Y_{\m\n |a_1 ... a_{s-2}}^{\hspace{45pt}\m}
Y^{\n\r | a_1 ... a_{s-2}}_{\hspace{48pt}\r} 
\nonumber \\
&&-(s-2)Y_{\m\n |a_1 ... a_{s-2}}^{\hspace{45pt}\m}
Y^{a_1\r | \n a_2 ... a_{s-2}}_{\hspace{58pt}\r}]\Big]\,,
\eqn
where $Y_{\m\n|a_1 ... a_{s-1}} \equiv \pa^{\l}Y_{\l\m\n|a_1 ... a_{s-1}} $. 
This action
is invariant under the transformations
\be 
\d Y^{\l\m\n|}_{~~~~~a_1 ... a_{s-1}}~=3 \, \d^{[\l}_{(a_1}
\a^{\m|\n]}_{~~~~a_2 ... a_{s-1})}\,,\label{invara}
\ee
where $\a_{\n |a_1 ... a_{s-1}}$ satisfies the relations (\ref{rela}),
as well as under the transformations
\be
\d Y^{\l\m\n|}_{~~~~~a_1 ... a_{s-1}}~=\pa_\r
\psi^{\r\l\m\n|}_{~~~~~~~a_1 ... a_{s-1}}\,.\label{invarpsi} 
\ee
that follow from the redundancy of the parametrization
of the solution of the constraints
(\ref{constraintforY}).
The gauge parameter $\psi^{\r\l\m\n|}_{~~~~~~~a_1 ... a_{s-1}}$ is subject
to the algebraic conditions
$
\psi^{\r\l\m\n|}_{~~~~~~~a_1 ... a_{s-1}} = 
\psi^{[\r\l\m\n]|}_{~~~~~~~a_1 ... a_{s-1}}=
\psi^{\r\l\m\n|}_{~~~~~~~(a_1 ... a_{s-1})} $ and $$
\psi^{\r\l\m\n|}_{~~~~~~~a_1 a_2 ... a_{s-1}} \eta^{a_1 a_2} = 0\,.$$

\subsection{Decomposing $Y_{\l\m\n|a_1 ... a_{s-1}}$ -- Dual action}
The field $Y_{\l\m\n|a_1 ... a_{s-1}}$ can be decomposed into the following
irreducible components
\be Y^{\l\m\n|}_{~~~~~a_1 ... a_{s-1}}= X^{\l\m\n|}_{~~~~~a_1 ... a_{s-1}}+
\d^{[\l}_{(a_1} Z^{\m\n ]|}_{~~~~a_2 ... a_{s-1})} \label{decompy}\ee
where $X^{\l\m\n|}_{~~~~~ \l a_2 ... a_{s-1}}=0$ ,
$Z^{\m\n |}_{~~~~ \m a_3 ... a_{s-1}}=0$.
The condition $Y^{\l\m\n| a}_{~~~~~~~ a a_3 ... a_{s-1}}=0$ implies
\bqn 
Z^{\m\n |a}_{~~~~~ a a_4 ... a_{s-1}}&=&0\,, \label{zz} \\
Z^{[\m\n |\l]}_{~~~~~~~ a_3 ... a_{s-1}}&=& - \frac{(s-1)}{2}
X^{\m\n \l |a}_{~~~~~~~ a a_3 ... a_{s-1}}\,.\label{zx} \eqn The
invariance (\ref{invara}) of the action involves only the field Z
and reads \bqn \d X^{\l\m\n|}_{~~~~~a_1 ... a_{s-1}}&=&0 \cr \d
Z_{\m\n |a_1 ... a_{s-2}}&=&\a_{[\m|\n] a_1 ... a_{s-2}} 
\label{inva} \eqn Next, one rewrites $Z_{\m\n |a_1 ... a_{s-2}}$
as \be Z_{\m\n |a_1 ... a_{s-2}} = \frac{3(s-2)}{s} \Phi_{\m\n
(a_1 \vert a_2 ... a_{s-2})} + \frac{2(s-1)}{s} \Psi_{[\m \vert
\n] a_1 ... a_{s-2}} \ee with $\Phi_{\m\n a_1 \vert a_2 ...
a_{s-2}} = Z_{[\m\n |a_1] a_2 ... a_{s-2}}$ and $\Psi_{\m \vert \n
a_1 ... a_{s-2}} = Z_{\m(\n |a_1 ... a_{s-2})}$. So the
irreducible component $\Phi_{\m\n a_1 \vert a_2 ... a_{s-2}}$ of
$Z$ can be expressed in terms of $X$ by the relation (\ref{zx}), 
while the other component $\Psi_{\m \vert \n a_1 ... a_{s-2}}$ is
pure gauge by virtue of the gauge symmetry (\ref{inva}), which
does not affect $\Phi_{\m\n a_1 \vert a_2 ... a_{s-2}}$ and reads
$\d \Psi_{\m \vert \n a_1 ... a_{s-2}} = (1/2) \a_{\m \vert \n a_1
... a_{s-2}}$ (note that $\Psi_{\m \vert \n a_1 ... a_{s-2}}$ is
subject to the same algebraic identities (\ref{rela}) as $\a_{\m
\vert \n a_1 ... a_{s-2}}$). As a result, the only independent
field appearing in $\cs(Y^{\l\m\n|}_{~~~~~a_1 ... a_{s-1}}) $ is
$X^{\l\m\n|}_{~~~~~a_1 ... a_{s-1}}$. 

Performing the change of variables 
\be 
X^{\l\m\n|}_{~~~~~a_2 ...
a_s} = \frac{1}{(n-3)!} \ve^{\l\m\n b_1 ... b_{n-3}} T_{b_1
... b_{n-3}|a_2 ... a_s}\,, 
\label{decompx} \ee 
the action for this field reads 
\bqn 
\cs &=& - \frac{2(s-1)}{s(n-3)!} \int d^n
x \Big[ \pa^e T^{b_1 ... b_{n-3}|a_2 ... a_s}\pa_e T_{b_1 ... 
b_{n-3}|a_2 ... a_s}
\nonumber \\
&&- (n-3)\pa_e T^{e b_2 ... b_{n-3}|a_2 ... a_s}\pa^f
T_{f b_2 ... b_{n-3}|a_2 ... a_s} \nonumber \\
&&+(s-1)[ -\pa_e T^{b_1 ... b_{n-3}|e a_3 ... a_s}\pa^f
T_{b_1 ... b_{n-3}|f a_3 ... a_s} \nonumber \\
&&-2(n-3) T_{g}^{~ b_2 ... b_{n-3}|g a_3 ... a_s}\pa^{e f }
T_{e b_2 ... b_{n-3}|f a_3 ... a_s} \nonumber \\
&&-(s-2)T^{b_1 ... b_{n-3}|c ~~ a_4 ... a_s}_{~~~~~~~~~~~c}\pa^{e f } 
T_{b_1 ... b_{n-3}|e f a_4 ... a_s} \nonumber \\
&& -(n-3)\pa^e T_{g}^{~ b_2 ... b_{n-3}|g a_3 ... a_s}\pa_e
T^{f}_{~ b_2 ... b_{n-3}|f a_3 ... a_s} \nonumber \\
&&-\frac{1}{2}(s-2) \pa^e
T^{ b_1 ... b_{n-3}|c ~~ a_4 ... a_s}_{~~~~~~~~~~~c} \pa_e
T_{ b_1 ... b_{n-3}|d ~~ a_4 ... a_s}^{~~~~~~~~~~~d} 
\nonumber \\
&&+(n-3)(n-4)\pa_ e T_{g}^{~ e b_3 ... b_{n-3}|g a_3 ... a_s}\pa^h
T^{f}_{~~ h b_3 ... b_{n-3}|f a_3 ... a_s} \nonumber \\
&&- (s-2)(n-3) \pa_ e T_{g}^{~ b_2 ... b_{n-3}|g e a_4 ... a_s} \pa^f
T_{f b_2 ... b_{n-3}|c ~~ a_4 ... a_s}^{~~~~~~~~~~~~c} \nonumber \\
&&+\frac{1}{4}(s-2)(n-3) \pa_e
T^{e b_2 ... b_{n-3}|c ~~ a_4 ... a_s}_{~~~~~~~~~~~~c} \pa^f
T_{f b_2 ... b_{n-3}|d ~~ a_4 ... a_s}^{~~~~~~~~~~~~d}
\nonumber \\ 
&&-\frac{1}{4}(s-2)(s-3) \pa_e
T^{ b_1 ... b_{n-3}|c ~~ e a_5 ... a_s}_{~~~~~~~~~~~c} \pa^f
T_{ b_1 ... b_{n-3}|d ~~ f a_5 ... a_s}^{~~~~~~~~~~~d}]\Big]\,. 
\label{actionduale}
\eqn
The field $T_{b_1... b_{n-3}|a_2 ... a_s} $
fulfills the following algebraic properties,
\begin{eqnarray} 
&& T_{b_1 ... b_{n-3} |a_2 ... a_s} =
T_{[b_1 ... b_{n-3}]|a_2 ... a_s} \,,\nn \\
&& T_{b_1 ... b_{n-3} |a_2 ... a_s} =
T_{b_1 ... b_{n-3} |(a_2 ... a_s)} \,,\nn\\
&&T_{[b_1 ... b_{n-3} |a_2] ... a_s} = 0\,,\nn \\
&& T_{b_1 ... b_{n-3} |a_2 a_3 a_4 a_5... a_s}
\eta^{a_2 a_3} \eta^{a_4 a_5} = 0 \,,\nn\\ 
&& T_{b_1 ... b_{n-3} |a_2 a_3 a_4... a_s} \eta^{b_1 a_2}
\eta^{a_3 a_4}= 0\,,\nn
\end{eqnarray}
the last two relations coming from Eqs.(\ref{zx}) and (\ref{zz}).

Conversely, given a tensor $T_{b_1 ... b_{n-3} |a_2 ... a_s} $ fulfilling the
above algebraic conditions, one may first reconstruct
$X^{\l\m\n|}_{~~~~~a_2 ... a_s}$ such that $X^{\l\m\n|}_{~~~~~a_2 ... a_s} 
= X^{[\l\m\n]|}_{~~~~~a_2 ... a_s}$, $X^{\l\m\n|}_{~~~~~a_2 ... a_s}
= X^{\l\m\n|}_{~~~~~(a_2 ... a_s)}$
and $X^{\l\m\n|}_{~~~~~\n a_3 ... a_s} = 0$. One then gets the
$\Phi$-component of $Z^{\m\n |}_{~~~~a_2 ... a_{s-1}}$
through Eq.(\ref{zx}) and finds that it
is traceless thanks to the double tracelessness conditions on
$T_{b_1 ... b_{n-3} |a_2 ... a_s} $. 

The equations of motion for the action (\ref{actionduale}) are
\be
G_{b_1 ... b_{n-3} |a_2 ... a_s}=0\,, \ee
where
\bqn
G_{b_1 ... b_{n-3} |a_2 ... a_s}=F_{b_1 ... b_{n-3} |a_2 ... a_s}
-\frac{(s-1)}{4}\Big[2(n-3) \h_{b_1 a_2} 
F^c_{~b_2 ... b_{n-3} |c a_3 ... a_s}\nonumber \\
+(s-2)\h_{ a_2 a_3}
F^{\hspace{42pt}c}_{b_1 ... b_{n-3} |c ~ ~a_4 ... a_s} \Big], 
\nonumber 
\eqn
and
\bqn
F_{b_1 ... b_{n-3} |a_2 ... a_s}\!\! &= &\pa_c \pa^c 
T_{b_1 ... b_{n-3} |a_2 ... a_s} \nonumber \\
&-&\!\!(n-3)\pa_{b_1}\pa^c T_{c b_2 ... b_{n-3} |a_2 ... a_s}
-(s-1)\pa_{a_2}\pa^c T_{b_1 ... b_{n-3} |ca_3 ... a_s} \nonumber \\
&+&\!\!(s-1) \Big[(n-3)\pa_{a_2 b_1}T^{c}_{~b_2 ... b_{n-3} |c a_3 ... a_s}+ 
\textstyle{\frac{(s-2)}{2}}\pa_{a_2 a_3}
T^{\hspace{42pt}c}_{b_1 ... b_{n-3} |c~~a_4 ... a_s}\Big]\,,\nonumber
\eqn
and where the r.h.s. of both expressions has to be antisymmetrized in 
$b_1 ... b_{n-3}$
and symmetrized in $a_2 ... a_s$.

\subsection{Gauge symmetries of the dual theory}
As a consequence of (\ref{invarpsi}), (\ref{decompy}) and
(\ref{decompx}), the dual action is invariant under the gauge 
transformations: $$ \d T_{b_1 ... b_{n-3}|a_2 ... a_s} =
\pa_{[b_1}\phi_{b_2 ... b_{n-3}]|a_2 ... a_s} +
\frac{(s-1)(n-2)}{(n+s-4)}\pa_f \phi_{c_1 ... c_{n-4}|g a_3 ...
a_s} \d^{[fg c_1 ... c_{n-4}]}_{[a_2 b_1 ... b_{n-3}]}\,,
$$
 where the r.h.s. must be symmetrized in the
indices $a_i$ and where the gauge parameter 
$\phi_{b_1 ...
b_{n-4}|a_2 ... a_s} \sim \ve_{ b_1 ... b_{n-4} \r \l \m \n}
\psi^{\r\l\m\n \vert }_{~~~~~~a_2 ... a_{s}}$ is such that 
$$\phi_{b_1 ... b_{n-4}|a_2 ... a_s} = \phi_{[b_1 ... b_{n-4}]|a_2
... a_s}=\phi_{b_1 ... b_{n-4}|(a_2 ... a_s)}\,,$$ and
$\phi^{~~~~~~~~~~a}_{b_1 ... b_{n-4}|~~ a a_4 ... a_s} =0$.

This completes the dualization procedure and provides the dual
description, in terms of the field $T_{b_1 ... b_{n-3}|a_2 ... 
a_s} $, of the spin-$s$  theory in $n$ space-time dimensions. Note that
in four dimensions, the field $T_{b_1|a_2 ... a_s}$ has $s$
indices, is totally symmetric and is subject to the double
tracelessness condition. In that case, one gets back  the original
Fronsdal action, equations of motion and gauge symmetries.

\section{Comments on interactions} 
\setcounter{equation}{0}
\setcounter{theorem}{0}
\setcounter{lemma}{0}

We have investigated so far duality only at the level of the free theories.
It is well known that duality becomes far more tricky in the presence
of interactions. The point is that consistent, local interactions for 
one of the children theories may not be local for the other. For instance,
in the case of $p$-form gauge theories,
Chern-Simons terms are in that class since they
involve ``bare'' potentials. An exception where the same interaction is
local on both sides is given by the Freedman-Townsend model
\cite{Free-Town} in four dimensions, where duality relates
a scalar theory (namely, a nonlinear 
$\sigma$-model) to an interacting $2$-form theory.

It is interesting to analyse the difficulties at the level of the
parent action. We consider the definite case of spin $2$. The
second-order action $\cs[e_{ab}]$ (Eq.(\ref{actione})) can of
course be consistently deformed, leading to the Einstein action.
One can extend this deformation to the action (\ref{actioneY})
where the auxiliary fields are included (see e.g. \cite{West}). In 
fact, auxiliary fields are never obstructions since they do not
contribute to the local BRST cohomology \cite{aux,Bekaert:2002uh}. The
problem is that one cannot go any more to the other single-field
theory action $\cs[Y]$. The interacting parent action has only
one child. The reason why one cannot get rid of the vielbein
field $e_{a \m}$ is that it is no longer a Lagrange multiplier. 
The equations of motion for $e_{a \m}$ are not constraints on $Y$. 
Rather, they mix both $e$
and $Y$. One is thus prevented from ``going down'' to $\cs[Y]$ 
(the possibility of doing so is in fact prevented by the
no-go theorem of \cite{Bekaert:2002uh}). At
the same time, the other parent action corresponding to
(\ref{actionCY}) does not exist once interactions are switched
on. By contrast, in the Freedman-Townsend model, the Lagrange 
multiplier remains a Lagrange multiplier.


%% file: quantif.tex
\chapter{Spin-$s$ electric-magnetic duality}

Since duality can be defined for higher spins, and since conserved
external electric-type sources can easily be coupled to them,
 one might wonder whether magnetic sources
can be considered as well. This chapter solves positively this question
for all spins at the linearized level and provides additional
insight in the full nonlinear theory for spin 2.

We show that conserved external sources of both types can be coupled
to any given higher (integer) spin field within the context of the
linear theory.  The presence of magnetic sources requires the
introduction of Dirac strings, as in the spin-1 case. To preserve
manifest covariance, the location of the string must be left
arbitrary and is, in fact, classically unobservable.  The
requirement that the Dirac string is unobservable
quantum-mechanically forces a quantization condition of the form \be
\frac{1  }{2 \pi \hbar} \, Q_{\g_1 \cdots \g_{s-1}}(v)P^{\g_1 \cdots
\g_{s-1}}(u) \in \Bbb{Z} \,.\ee  Here, the symmetric tensor $P^{\g_1
\cdots \g_{s-1}}(u)$ is the conserved electric charge associated
with the asymptotic symmetries of the spin-$s$ field, while $Q_{\g_1
\cdots \g_{s-1}}(v)$ is the corresponding ``topological'' magnetic
charge. For $s = 1$, the asymptotic symmetries are internal symmetries
and, actually, just constant phase transformations. The conserved
charge $P$ is the electric charge $q$ while $Q$ is the magnetic
charge $g$, yielding the familiar Dirac quantization condition for
the product of electric and magnetic charges.  {}For $s = 2$ the
conserved charges have a space-time index and the quantization
condition reads (after rescaling the conserved quantities so that they
have dimensions of mass) \be \frac{4 G P_\g Q^\g}{\hbar} \in \Bbb{Z}
\, . \label{QC1}\ee The quantity $P_\g$ is the ``electric''
4-momentum associated with constant linearized diffeomorphisms
(translations), while $Q_\g$ is the corresponding magnetic
four-\\
\noindent momentum.  {}For a point particle source, $P_\g = M u_\g$ where
$M$ is the ``electric'' mass and $u_\g$ the 4-velocity of the
electric source. Similarly, $Q_\g = N v_\g$ where $N$ is the
``magnetic'' mass and $v_\g$ the 4-velocity of the magnetic source.

All this is just a generalization of the familiar spin-1 case,
although the explicit introduction of the Dirac string is more
intricate for higher spins because the gauge invariance is then more
delicate to control.  Indeed, there is no gauge invariant object
that involves first derivatives of the fields only ($s > 1$). Hence,
the Lagrangian is not strictly gauge invariant, contrary to what
happens for electromagnetism, but is gauge invariant only up to a
total derivative.

A serious limitation of the linear theory for $s>1$ is that the
sources must move on straight lines.  This follows from the strict
conservation laws implied by the field equations, which are much
more stringent for $s>1$ than they are for $s = 1$. Thus the sources
must be treated as externally given and cannot be freely varied in
the variational principle. One cannot study the backreaction of the
spin-$s$ field on the sources without introducing self-interactions.
This problem occurs already for the spin-2 case and has nothing to
do with the introduction of magnetic sources.

We do not investigate the backreaction problem for general spins
$s>2$ since the nonlinear theory is still a subject of investigation
 even in the absence
of sources. We discuss briefly the spin-2 case, for which
the nonlinear theory is given by the Einstein theory of gravity.
The remarkable Taub-NUT solution \cite{Newman:1963yy}, which
represents the vacuum exterior field of a gravitational dyon,
indicates that Einstein's theory can support both electric and
magnetic masses. 

This chapter is organized as follows. In Section \ref{lingrav}, we consider in
detail the linearized spin-2 case with point particle electric and magnetic
sources. We introduce Dirac strings and derive the quantization
condition. We then extend the formalism to higher spins (Section
\ref{higherspins0}), again with point particle sources. In Section
\ref{sect4} we comment on the extension of magnetic sources and the
quantization condition to the nonlinear context.

\hfill

\section{Linearized gravity with electric and 
magnetic masses}
\setcounter{equation}{0}\label{lingrav}

\subsection{Electric and magnetic sources}

The equations of motion for linearized gravity coupled to both
electric and magnetic sources are naturally written in terms of
the linearized Riemann tensor $R_{\a \b \l \m}$, hereafter just
called ``Riemann tensor'' for simplicity. This is the physical,
gauge-invariant object analogous to the field strength $F_{\m
\n}$ of electromagnetism.  How to introduce the ``potential'',
\ie the symmetric spin-2 field $h_{\m \n} = h_{\n \m}$ will be
discussed below.  The dual to the Riemann tensor is defined as $$
S_{\a \b \l \m} = - \frac{1}{2} \ve_{\a \b \g \d} R^{\g
\d}_{\;\;\;\; \l \m}.$$

We denote the ``electric'' energy-momentum tensor by $T^{\m \n}$ and
the ``magnetic'' energy-momentum tensor by $\Theta^{\m \n}$. These
are both symmetric and conserved, $T^{\m \n} = T^{\n \m}$,
$\Theta^{\m \n} = \Theta^{\n \m}$, $ T^{\m \n}_{\; \; \; \; , \, \n}
= 0$, $ \Theta^{\m \n} _{\; \; \; \; , \, \n} = 0 $. It is also
useful to define $ \bar{T}^{\m \n} = T^{\m \n} - \frac{1}{2} \,
\eta^{\m \n}\, T $, $ \bar{\Theta}^{\m \n} = \Theta^{\m \n} -
\frac{1}{2} \, \eta^{\m \n}\, \Theta $ where $T$ and $\Theta$ are
the traces. We assume that $T^{\m \n}$ and $\Theta^{\m \n}$ have the
units of an energy density.  We set $c=1$ but keep $G$.

The form of the equations in the presence of both types of sources
is fixed by: (i) requiring duality invariance with respect to the
$SO(2)$-rotations of the curvatures and the sources
\cite{Henneaux:2004jw},  \beq R'_{\a \b \l \m}  &=& \cos \a \, R_{\a
\b \l \m} + \sin \a \, S_{\a \b \l \m} , \;\; \; \; S'_{\a \b \l \m}
= - \sin \a \, R_{\a \b \l \m} + \cos \a \, S_{\a \b \l \m},
\nonumber\\ T'_{\a \b} &=& \cos \a \, T_{\a \b} + \sin \a \,
\Theta_{\a \b} , \; \; \; \;\; \; \; \; \; \; \; \; \; \; \; \;
\Theta'_{\a \b} = - \sin \a \, T_{\a \b} + \cos \a \, \Theta_{\a
\b}, \nonumber \eeq and, (ii) using the known form of the equations
in the presence of electric masses only.  One finds explicitly the
following:
\begin{itemize}
\item The Riemann tensor is antisymmetric in the first two indices
and the last two indices, but in general is not symmetric for the
exchange of the pairs, \ie $R_{\a \b \l \m} = - R_{\b \a \l
\m}$, $R_{\a \b \l \m} = - R_{\a \b \m \l}$ with $R_{\a \b \l \m}
\not= R_{\l \m \a \b }$ (in the presence of magnetic sources).
\item In the presence of
magnetic sources the cyclic identity is \footnote{In terms of the
Riemann tensor, this ``identity'' is a nontrivial equation and not
an identity. It becomes an identity only after the Riemann tensor is
expressed in terms of the spin-2 field $h_{\m \n}$ introduced below.
We shall nevertheless loosely refer to this equation as the
(generalized) cyclic identity. A similar remark holds for the
Bianchi identity below.} \be R_{\a \b \l \m} + R_{\b \l \a \m} +
R_{\l \a \b \m} = 8 \pi G\, \epsilon_{\a \b \l \n} \,
\bar{\Theta}^{\n}_{\; \; \m} \label{cyclic}\,.\ee This enables one
to relate $R_{\a \b \l \m}$ to $R_{\l \m \a \b }$ through \be R_{\a
\b \g \d} - R_{\g \d \a \b} = 4 \pi G \left( \ve_{\a \b \g \l}
\bar{\Theta}^\l_{\; \; \d} - \ve_{\a \b \d \l} \bar{\Theta}^\l_{\; \;
\g} + \ve_{\b \g \d \l} \bar{\Theta}^\l_{\; \; \a} - \ve_{\a \g \d \l}
\bar{\Theta}^\l_{\; \; \b} \right). \label{symm}\ee It follows that
the Ricci tensor is symmetric, $ R_{\l \m} = R_{\m \l} $.  The
Einstein tensor $G_{\l \m} = R_{\l \m} - (1/2) \eta_{\l \m} R $  is
then also symmetric.
\item The Bianchi identity is \be
\partial_\epsilon R_{\a \b \g \d} +
\partial_\a R_{\b \epsilon \g \d} + \partial_\b R_{\epsilon\a \g \d} = 8 \pi
G \, \ve_{\epsilon\a \b \r} ( \partial_\g \bar{\Theta}^\r_{\; \; \d} -
\partial_\d \bar{\Theta}^\r_{\; \; \g} )\,. \label{BI}\ee  Although
there is now a right-hand side in the Bianchi identity, the
contracted Bianchi identities are easily verified to be unaffected
and still read \be G^{\l \m}_{\; \; \; \; , \, \m} = 0 \,.\ee \item
The Einstein equations are \be G^{\l \m} = 8 \pi G \, T^{\l \m}\,,
\label{Einstein}\ee or equivalently, $R^{\l \m} = 8 \pi G \,
\bar{T}^{\l \m}$, and force exact conservation of the sources
because of the contracted Bianchi identity, as in the absence of
magnetic mass.
\end{itemize}

The equations are completely symmetric under duality.  Indeed, one
easily checks that one gets the same equations for the dual
curvature $S_{\a \b \l \m}$ with the roles of the electric and
magnetic energy-momentum tensors exchanged.  In the course of the
verification of this property, the equation $$
\partial^\m R_{\m \r \g \d} = 8 \pi G \, \left(
\partial_\g \bar{T}_{\r \d} -
\partial_\d \bar{T}_{\r \g} \right),$$ which
follows from Eqs.(\ref{symm}), (\ref{BI}) and the conservation of
$\Theta^{\m \n}$ is useful. {}Furthermore, in the absence of
magnetic sources, one recovers the equations of the standard
linearized Einstein theory since the cyclic and Bianchi identities
have no source term in their right-hand sides.

The formalism can be extended to include a cosmological constant
$\Lambda$.  The relevant curvature is then the MacDowell-Mansouri
curvature \cite{MacDowell:1977jt} linearized around (anti) de Sitter space
\cite{Julia:2005ze}. In terms of this tensor, the equations
(\ref{cyclic}), (\ref{BI}) and (\ref{Einstein}) take the same
form, with ordinary derivatives replaced by covariant derivatives
with respect to the (anti) de Sitter background.

\subsection{Decomposition of the Riemann tensor - Spin-2 field}

We exhibit a variational principle from which the equations of
motion follow. To that end, we first need to indicate how to
introduce the spin-2 field $h_{\mu\nu}$.

Because there are right-hand sides in the cyclic and Bianchi
identities, the Riemann tensor is not directly derived from a
potential $h_{\m \n}$. To introduce $h_{\m \n}$, we split $R_{\l \m
\a \b}$ into a part that obeys the cyclic and Bianchi identities and
a part that is fixed by the magnetic energy-momentum tensor.  Let
$\Phi^{\a \b}_{\; \; \; \; \g}$ be such that \be \partial_\a \,
\Phi^{\a \b}_{\; \; \; \; \g} = 16 \pi G \, \Theta^\b_{\; \; \g} ,
\; \; \; \Phi^{\a \b}_{\; \; \; \; \g} = - \, \Phi^{\b \a}_{\; \; \;
\; \g} \label{phi}\,.\ee We shall construct $\Phi^{\a \b}_{\; \; \;
\; \g}$ in terms of $\Theta^\b_{\; \; \g}$ and Dirac strings below.
We set \be R_{\l \m \a \b} = r_{\l \m \a \b} + \frac{1}{4}\, \ve_{\l
\m \r \s} \left(\partial_\a \bar{\Phi}^{\r \s}_{\; \; \; \; \b} -
\partial_\b \bar{\Phi}^{\r \s}_{\; \; \; \; \a} \right)\,, \label{split0}
\ee with \be \bar{\Phi}^{\r \s}_{\; \; \; \; \a} = \Phi^{\r \s}_{\;
\; \; \; \a} + \frac{1}{2} \left( \d^\r _{\; \a} \Phi^\s - \d^\s
_{\; \a} \Phi^\r \right) , \; \; \; \Phi^\r \equiv \Phi^{\r \s}_{\;
\; \; \; \s} \nonumber \,.\ee Using $ \partial_\a \, \bar{\Phi}^{\a
\b}_{\; \; \; \; \g} = 16 \pi G \, \bar{\Theta}^\b_{\; \; \g} -
\partial_\g \bar{\Phi}^\b $, $ \bar{\Phi}^\b = -
\frac{1}{2} \Phi^\b $, one easily verifies that the cyclic and
Bianchi identities take the standard form when written in terms of
$r_{\a \b \l \m}$, namely, \be r_{\a \b \l \m} + r_{\b \l \a \m} +
r_{\l \a \b \m} = 0 , \; \; \; \; \;
\partial_\epsilon r_{\a \b \g \d} +
\partial_\a r_{\b \epsilon \g \d} + \partial_\b r_{\epsilon \a \g \d} = 0
.\nonumber \ee   Hence, there exists a symmetric tensor $h_{\m
\n}$ such that $ r_{\a \b \l \m} = - 2
\partial_{[\b} h_{\a][\l , \, \m ]} $.

If one sets $ y^{\l \m}_{\; \; \; \; \g} = \ve^{\l \m \r \s}
\partial _\r h_{\s \g} = - y^{\m \l}_{\; \; \; \; \g}, $ one
may rewrite the curvature as \be R_{\l \m \a \b} =  \frac{1}{4}\,
\ve_{\l \m \r \s} \left(\partial_\a \bar{Y}^{\r \s}_{\; \; \; \; \b}
- \partial_\b \bar{Y}^{\r \s}_{\; \; \; \; \a} \right) \,,
\label{curvature}\ee with \be \label{ybar} Y^{\r \s}_{\; \; \; \;
\b} = y^{\r \s}_{\; \; \; \; \b} + \Phi^{\r \s}_{\; \; \; \; \b} = -
Y^{\s \r}_{\; \; \; \; \b}, \; \; \; \;  \bar{Y}^{\r \s}_{\; \; \;
\; \a} = Y^{\r \s}_{\; \; \; \; \a} + \frac{1}{2} \left( \d^\r _{\;
\a} Y^\s - \d^\s _{\; \a} Y^\r \right) , \; \; \; Y^\r \equiv Y^{\r
\s}_{\; \; \; \; \s} \,,\ee (note that $\bar{y}^{\r \s}_{\; \; \; \;
\a} = y^{\r \s}_{\; \; \; \; \a}$ and that $\partial_\r y^{\r
\s}_{\; \; \; \; \a} = 0$).

\subsection{Dirac string}
We consider point particle sources.  The particles must be forced to
follow straight lines because of the conservation equations $T^{\m
\n}_{\; \; \; , \n} = 0$ and $\Theta^{\m \n}_{\; \; \; , \n} = 0$.
If $u^\m$ is the 4-velocity of the electric source and $v^\m$ the
4-velocity of the magnetic source, one has \be T^\m_{\; \; \n} = M
u_\n \int d\l \delta^{(4)}(x-z(\l)) \dot{z}^\m , \; \; \; \; \;
\Theta^\m_{\; \; \n} = N v_\n \int d\l \delta^{(4)}(x-\bar{z}(\l))
\dot{\bar{z}}^\m \,,\ee where $z^\m(\l)$ and $\bar{z}^\m(\l))$ are
the worldlines of the electric and magnetic sources respectively,
e.g. $u^\m = dz^\m /ds$. Performing the integral, one finds
$$ T^{\m\n} = \frac{u^\m u^\n}{u^0} \delta^{(3)}(\vec{x} -
\vec{z}(x^0)), \; \; \; \; \; \Theta^{\m\n} = \frac{v^\m v^\n}{v^0}
\delta^{(3)}(\vec{x} - \vec{\bar{z}}(x^0))\,.$$

The tensor $\Phi^{\a \b}_{\; \; \; \; \g}$ introduced in Eq.(\ref{phi})
can be constructed \`a la Dirac \cite{Dirac:1948um}, by attaching a
Dirac string $y^\m(\l, \s)$ to the magnetic source, $y^\m(\l,
0)=\bar{z}^\m(\l)$. (The parameter $\s$ varies from $0$ to $\infty\,$.) One has explicitly \be \Phi^{\a \b}_{\; \; \; \;
\g} = 16 \pi G \, N v_\g \int d \l d \s (y'^\a \dot{y}^\b -
\dot{y}^\a y'^\b) \delta^{(4)}(x-y(\l, \s)) \,,\label{exprforphi}\ee
where  $$ \dot{y}^\a = \frac{\partial y^\a}{\partial \l}, \; \; \;
\; y'^\a = \frac{\partial y^\a}{\partial \s}\, .
$$ One verifies exactly as for electromagnetism that the
divergence of $\Phi^{\a \b}_{\; \; \; \; \g}$ is equal to the
magnetic energy-momentum tensor (up to the factor $16 \pi G$). What
plays the role of the magnetic charge $g$ in electromagnetism is now
the conserved product $N v_\m$ of the magnetic mass of the source by
its $4$-velocity.  This is the magnetic $4$-momentum.

\subsection{Variational principle}

\subsubsection{Action}

When the curvature is expressed in terms of $h_{\m \n}$ as in
Eq.(\ref{curvature}), the expressions (\ref{cyclic}) and (\ref{BI}) are
identically fulfilled and the relations (\ref{Einstein}) become
equations of motion for $h_{\m \n}$.  These equations can be derived
from a variational principle which we now describe.

The action that yields (\ref{Einstein}) is \be S[h_{\m
\n}(x),y^\m(\l, \s)] = \frac{1}{16 \pi G} \int \frac{1}{4}\left(
\bar{Y}_{\a \b \g} \bar{Y}^{\a \g \b} - \bar{Y}_\a
\bar{Y}^\a \right) d^4 x + \frac{1}{2} \int h_{\m \n} T^{\m \n} d^4
x \,.\label{actionq} \ee One varies the fields $h_{\m \n}$ and the
coordinates $y^\m$ of the string (with the condition that it remains
attached to the magnetic source), but not the trajectories of the
sources, which are fixed because of the conservation laws
$\partial_\m T^{\m \n}=0$ and $\partial_\m \Theta^{\m \n}=0$. This
is a well known limitation of the linearized theory, present already
in the pure electric case.  To treat the sources as dynamical, one
needs to go beyond the linear theory.

If there is no magnetic source, the first term in the action
reduces to $$ S^{PF} = \frac{1}{16 \pi G} \int\,d^4x\, \frac{1}{4}\left( -
\partial_\l h_{\a \b} \partial^\l h^{\a \b} + 2 \partial_\l h^{\l \a}
\partial^\m h_{\m \a} -2 \partial^\l h \partial_\m h^{\m \l}
+ \partial_\l h \partial^\l h \right), $$ which is the Pauli-Fierz
action. Its variation with respect to $h_{\a \b}$ gives
$-\frac{1}{16 \pi G}$ times the linearized Einstein tensor $G^{\a
\b}$.  It is straightforward to verify that the variation of the
first term in the action with respect to $h_{\a \b}$ still gives
$-\frac{1}{16 \pi G}$ times the linearized Einstein tensor $G^{\a
\b}$ with correct $\Phi^{\m \n}_{\; \; \; \; \a}$ contributions even
in the presence of magnetic sources. So, the equations of motion
that follow from (\ref{actionq}) when one varies the gravitational
field are the Einstein equations (\ref{Einstein}).

Extremization with respect to the string coordinates does not bring
in new conditions provided that the Dirac string does not go through
an electric source (Dirac veto).

The action (\ref{actionq}) was obtained by using the analysis of
source-free linearized gravity in terms of two independent fields
given in Section \ref{spin2duality} \cite{Boulanger:2003vs}, which enables one to go from the electric
to the magnetic formulations and vice-versa, by elimination of
magnetic or electric variables. As one knows how to introduce
electric sources in the electric formulation, through standard
minimal coupling, one can find how these sources appear in the
magnetic formulation by eliminating the electric variables and
keeping the magnetic potentials.  So, one can determine how to
introduce electric poles in the magnetic formulation, or, what is
equivalent, magnetic poles in the electric formulation.

\subsubsection{Gauge invariances}
\paragraph{\it \small Diffeomorphism invariance\\} 

\vspace{.2cm}

The action (\ref{actionq}) is invariant under linearized
diffeomorphisms and under displacements of the Dirac string
(accompanied by appropriate transformations of the spin-2 field).
The easiest way to show this is to observe that the first term in
the action (\ref{actionq}) is invariant if one shifts $Y^{\m \n}_{\;
\; \; \; \a}$ according to \be Y^{\m \n}_{\; \; \; \; \a}
\rightarrow Y^{\m \n}_{\; \; \; \; \a} + \delta^\m_\a
\partial_\r z^{\n \r} - \delta^\n_\a \partial_\r z^{\m \r} +
\partial_\a z^{\m \n} \,,\label{gauge1}\ee where $z^{\m\n} = - z^{\n
\m}$ is arbitrary. This is most directly verified by noting that
under (\ref{gauge1}), the tensor $\bar{Y}^{\m \n}_{\; \; \; \; \a}$
defined in Eq.(\ref{ybar}) transforms simply as \be \bar{Y}^{\m \n}_{\;
\; \; \; \a} \rightarrow \bar{Y}^{\m \n}_{\; \; \; \; \a}  +
\partial_\a z^{\m \n} \label{gauge2}\ee and this leaves invariant
the first term in (\ref{actionq}) up to a total derivative. Note that
the Riemann tensor (\ref{curvature}) is strictly invariant.  The
transformation (\ref{gauge1}) can be conveniently rewritten as \be
Y^{\m \n}_{\; \; \; \; \a} \rightarrow Y^{\m \n}_{\; \; \; \; \a} +
\ve^{\m \n \r \s}
\partial_\r a_{\s \a} \,, \label{gauge3}\ee where $a_{\s \a} = - a_{
\a \s}$ is given by $ a_{\s \a} = \frac{1}{2} \ve_{\s \a \b
\g} z^{\b \g} $.

A (linearized) diffeomorphism \be h_{\m \n} \rightarrow h_{\m \n}
+ \partial_\m \xi_\n + \partial_\n \xi_\m \ee (with the string
coordinates unaffected) modifies $Y^{\m \n}_{\; \; \; \; \a}$ as
in (\ref{gauge3}) with $ a_{\s \a} =
\partial_\a \xi_\s - \partial_\s \xi_\a $ (note that the term
$\partial_\s \xi_\a$ in $a_{\s \a}$ does not contribute because
$\partial_{[\r}\partial_{\s]} \xi_\a = 0$).  Hence, the first term
in the action (\ref{actionq}) is invariant under diffeomorphisms. The
minimal coupling term is also invariant because the energy-momentum
tensor is conserved. It follows that the complete action
(\ref{actionq}) is invariant under diffeomorphisms. 


\paragraph
{\it \small Displacements of the Dirac string\\} 

An arbitrary displacement of the Dirac string, \be y^\a (\tau, \s)
\rightarrow y^\a (\tau, \s) + \delta y^\a (\tau, \s)
\label{displ1}\ee also modifies $Y^{\m \n}_{\; \; \; \; \a}$ as in
(\ref{gauge3}) provided one transforms simultaneously the spin-2
field $h_{\m \n}$ appropriately. Indeed, under the displacement
(\ref{displ1}) of the Dirac string, the quantity $\Phi^{\m \n}_{\;
\; \; \; \a}$ changes as $ \Phi^{\m \n}_{\; \; \; \; \a} \rightarrow
\Phi^{\m \n}_{\; \; \; \; \a} + k^{\m \n}_{\; \; \; \; \a}$ where
$k^{\m \n}_{\; \; \; \; \a}$ can be computed from $\delta y^\a
(\tau, \s)$ through (\ref{exprforphi}) and has support on the old
and new string locations.  Its explicit expression will not be
needed. What will be needed is that it fulfills \be
\partial_\m k^{\m \n}_{\; \; \; \; \a} = 0 \,,\label{eqfork}\ee because
the magnetic energy-momentum tensor is not modified by a
displacement of the Dirac string. The field $Y^{\m \n}_{\; \; \; \;
\a}$ changes then as \be Y^{\m \n}_{\; \; \; \; \a} \rightarrow
Y^{\m \n}_{\; \; \; \; \a} + \ve^{\m \n \r \s} \partial_\r
\delta h_{\s \a} + k^{\m \n}_{\; \; \; \; \a} \label{eqforY}\ee
where $\delta h_{\s \a}$ is the sought for variation of $h_{\s
\a}$. By using Eq.(\ref{eqfork}), one may rewrite the last term in
Eq.(\ref{eqforY}) as $\partial_\r t^{\m \n \r}_{\; \; \; \; \; \; \a}$
for some $ t^{\m \n \r}_{\;  \; \; \; \; \; \a} =  t^{[\m \n
\r]}_{\; \; \; \; \; \; \; \a}$.  Again, we shall not need an
explicit expression for $t^{\m \n \r}_{\;  \; \; \; \; \; \a}$, but
only the fact that because $k^{\m \n}_{\; \; \; \; \a}$ has support
on the string locations, which do not go through the electric
sources (Dirac veto), one may choose $t^{\m \n \r}_{\; \; \; \; \;
\; \a}$ to vanish on the electric sources as well.  In fact, one may
take $t^{\m \n \r}_{\; \; \; \; \; \; \a}$ to be non-vanishing only
on a membrane supported by the string. Decomposing $t^{\m \n \r}_{\;
\; \; \; \; \; \a}$ as $ t^{\m \n \r}_{\;  \; \; \; \; \; \a} =
\ve^{\m \n \r \s}
 \left( s_{\s \a} + a_{\s \a} \right)$,
$s_{\s \a} = s_{(\s \a)}$, $a_{\s \a} = a_{[\s \a]}$ and taking
$h_{\s \a}$ to transform as $ h_{\s \a} \rightarrow h_{\s \a} -
s_{\s \a} $ one sees from Eq.(\ref{eqforY}) that the variation of
$Y^{\m \n}_{\; \; \; \; \a}$ takes indeed the form (\ref{gauge3}).
The first term in the action is thus invariant. The minimal
coupling term is also invariant because the support of the
variation of the spin-2 field does not contain the electric
worldlines.

One can also observe that the variation $\d r_{\a \b \r \s}$
vanishes outside the original and displaced string locations. This
implies $\d h_{\a \b} = \partial_\a \xi _\b +
\partial_\b \xi_\a$ except on the location of both strings, where
$\xi_\a$ induces a delta function contribution on the string
(``singular gauge transformation''). The explicit expressions will
not be given here. \vspace{.3cm}

\paragraph{\it  Identities\\} 

The identities which follow from the invariance (\ref{gauge1}), or
(\ref{gauge3}), of the first term $$ {\cal L} = \frac{1}{64 \pi
G}\left( \bar{Y}_{\a \b \g} \bar{Y}^{\a \g \b} -
\bar{Y}_\a \bar{Y}^\a \right) $$ in the action may be written as \be
\partial_\r \left(\frac{\partial {\cal L}}{\partial Y^{\a \b}_{\;
\; \; \; \g}}\right) \ve^{\a \b \r \s} =
\partial_\r \left(\frac{\partial {\cal L}}{\partial Y^{\a \b}_{\;
\; \; \; \s}}\right) \ve^{\a \b \r \g}\,.\ee They imply that
\beq - \frac{1}{16 \pi G} G^{\a \b} = \frac{\delta {\cal L}}{\delta
h_{\a \b}} &=& -\partial_\r \left(\frac{\partial {\cal L}}{\partial
Y^{\m \n}_{\; \; \; \; \b}}\right) \ve^{\m \n \r \a} \\ &=& -
\partial_\r \left(\frac{\partial {\cal L}}{\partial Y^{\m \n}_{\;
\; \; \; \a}}\right) \ve^{\m \n \r \b} \,,\eeq from which the
contracted Bianchi identities are easily seen to indeed hold.

The expression (\ref{curvature}) of the Riemann tensor in terms of
$\bar{Y}^{\m \n}_{\; \; \; \; \a}$ makes it clear that it is
invariant under (\ref{gauge2}) and thus, invariant under both
diffeomorphisms and displacements of the Dirac string.

\subsection{Quantization condition} Because of the gauge
invariances just described, the Dirac string is classically
unobservable.  In the Hamiltonian formalism, this translates itself
into the existence of first-class constraints expressing the momenta
conjugate to the string coordinates in terms of the remaining
variables. Demanding that the string remains unobservable in the
quantum theory imposes a quantization condition on the charges,
which we now derive. The argument follows closely that of Dirac in
the electromagnetic case \cite{Dirac:1948um}.

Working for simplicity in the gauge $y^0 = \l$ (which eliminates
$y^0$ as an independent variable), one finds the constraints \be
\label{just} \pi_m = - 32 \pi G \, N y'^n \, v_\g \frac{\partial
{\cal L}}{\partial Y^{mn}_{\; \; \; \; \g}} \,. \ee  The right-hand
side of Eq.(\ref{just}) generates the change of the gravitational field
that accompanies the displacement of the Dirac string. 
It is obtained as the coefficient of the variation of $\dot{y}_m$ 
in the action.

In the quantum theory, the wave functional $\psi$ must therefore
fulfill $$ \frac{\hbar}{i} \frac{\delta \psi}{\delta y^m(\s)} = - 32
\pi G \, N y'^n \, v_\g \frac{\partial {\cal L}}{\partial Y^{mn}_{\;
\; \; \; \g}} \psi \,.$$  We integrate this equation as in
\cite{Dirac:1948um}, along a path in the configuration space of the
string that encloses an electric source.  One finds that the
variation of the phase of the wave functional is given by \be \Delta
\Psi = - \frac{16 \pi G \, N \, v_\g}{\hbar} \int \frac{\partial
{\cal L}}{\partial Y^{mn}_{\; \; \; \; \g}} \left(\dot{y}^m y'^n -
\dot{y}^n y'^m \right) d \s d \l \,,\ee where the integral is taken
on the two-dimensional surface enclosing the electric source. Using
the Gauss theorem, this can be converted to a volume integral, $$
\Delta \Psi = - \frac{16 \pi G \, N \, v_\g}{\hbar} \int d^3 x \,
\ve^{mnp}\,
\partial_p \left(\frac{\partial {\cal L}}{\partial Y^{mn}_{\; \; \;
\; \g}} \right) \,.$$ Because $ \ve^{mnp}
\partial_p \left(\frac{\partial {\cal L}}{\partial Y^{mn}_{\; \;
\; \; \g}} \right) = \frac{\delta {\cal L}}{\delta h_{0 \g}} $, the
variation of the phase becomes, upon use of the constraint (initial
value) Einstein equations $G^{0 \g} = 8 \pi G T^{0 \g}$, $$ \Delta
\Psi = \frac{8 \pi G \, N \, v_\g}{\hbar} \int d^3 x\ T^{0 \g} =
\frac{8 \pi G \, N \, M \, v_\g u^\g}{\hbar}.$$ For the wave
functional to be single-valued, this should be a multiple of $2
\pi$.  This yields the quantization condition \be \frac{4 N M G v_\g
u^\g}{\hbar} = n , \; \; \; n \in \Bbb{Z} \,.\ee Introducing the
conserved charges $P^\g$, $Q^\g$ associated with the spin-2 theory
(electric and magnetic 4-momentum), this can be rewritten as \be
\frac{4  G P_\g Q^\g}{\hbar}  \in \Bbb{Z} \,.\ee It is to be
stressed that the quantization condition is not a condition on the
electric and magnetic masses, but rather, on the electric and
magnetic 4-momenta. In the rest frame of the magnetic source, the
quantization condition becomes \be \frac{4  G E N}{\hbar} \in
\Bbb{Z} \,,\ee where $E$ is the (electric) energy of the electric
mass. Thus, it is the energy which is quantized, not the mass.

We have taken above a pure electric source and a pure magnetic pole.
We could have taken dyons, one with charges $(P^\g, Q^\g)$, the
other with charges $(\bar{P}^\g, \bar{Q}^\g)$. Then the quantization
condition reads \be \frac{4 G \left(P_\g \bar{Q}^\g- \bar{P}_\g
Q^\g\right)}{\hbar} \equiv \frac{4 G
 \ve_{ab}Q^{a}_\g \bar{Q}^{b\g}}{\hbar}\in \Bbb{Z} \,,\ee
since the sources are pointlike (0-dyons). Here $Q^{a}_\g \equiv
(P_\g, Q_\g)$, $a,b = 1,2$ and $\ve_{ab}$ is the
$SO(2)$-invariant Levi-Civita tensor in the 2-dimensional space of
the charges.

\subsection{One-particle solutions}
\subsubsection{Electric mass} We consider a point
particle electric mass at rest at the origin of the coordinate
system. The only non-vanishing component of its electric energy
momentum tensor is $ T^{00}(x^0, \vec{x}) = M \delta^{(3)}(\vec{x})$
while $\Theta^{\m \n}$ vanishes. There is no Dirac string since
there is no magnetic mass. The metric generated by this source is
static. The linearized Einstein equations are well known to imply in
that case the linearized Schwarzschild solution, namely in polar
coordinates $$ h_{00} = \frac{2 GM}{r} = h_{rr}, \; \; \;
\hbox{other components vanish}\,,
$$ or in Cartesian coordinates $$ h_{00} = \frac{2 GM}{r}, \; \;
\; h_{ij} = \frac{2 GM}{r^3}x_i x_j, \; \; \; \hbox{other
components vanish.} $$ Indeed, one then finds \beq R_{0s0b} &=& M
\left(-\frac{3x_sx_b}{r^5} + \frac{\d_{sb}}{r^3} +
\frac{4 \pi}{3}\, \d_{sb} \, \d (\vec{x}) \right) \nonumber \\
R_{0sab} &=& 0  \; = \;
R_{ab0s}\,, \nonumber \\
R_{pqab} &=& \left(\d_{pa} \d_{qb} - \d_{pb} \d_{qa} \right)\left(
\frac{2M}{r^3} + \frac{8 \pi}{3} \d(\vec{x}) \right) \,,\nonumber
\\ && - 3 M \left( \d_{pa} \frac{x_b x_q}{r^5} - \d_{qa}
\frac{x_b x_p}{r^5} - \d_{pb} \frac{x_a x_q}{r^5} + \d_{qb}
\frac{x_a x_p}{r^5}\right) \,,\nonumber \eeq and thus $ R_{00} = 4
\pi G \, M \delta^{3}(\vec{x})$, $R_{ab} = 4 \pi G \, M \, \d_{ab}
\, \delta^{3}(\vec{x})$. The solution can be translated and boosted
to obtain a moving source at an arbitrary location.

\subsubsection{Magnetic mass} We now consider the dual solution,
that is, a point magnetic mass sitting at the origin. We have
$\Theta^{00}(x^0, \vec{x}) = N \delta^{(3)}(\vec{x})$ as the only
non-vanishing component of the magnetic energy-momentum tensor.
{}Furthermore, $T^{\m \n} = 0$. The solution is linearized Taub-NUT
\cite{Newman:1963yy}, with only magnetic mass, \ie, in polar
coordinates, $$ h_{0\varphi} = -2 N (1- \cos \theta), \; \; \;
\hbox{other components vanish.} $$ With this choice of
$h_{0\varphi}$ the string must be taken along the negative $z$-axis
in order to cancel the singularity at $\theta=\pi$. The tensor
$\Phi^{\a \b}_{\; \; \; \; \l}$ is given by 

\noindent $ \Phi^{0z}_{\; \; \; \;
0} = - 16 \pi N \theta(-z) \d(x) \d(y) $ (other components vanish).

One then finds the only non-vanishing components (in Cartesian
coordinates)
$$ \bar{Y}'^{0s}_{\; \; \; \; 0} = - 2 N \frac{x^s}{r^3}, \; \; \;
\; \bar{Y}'^{rs}_{\; \; \; \; c} = 2 N \left( \d^r_c \,
\frac{x^s}{r^3} - \d^s_c \, \frac{x^r}{r^3} \right) \,.$$ Here,
$\bar{Y}'^{\a \b}_{\; \; \; \; \g}$ differs from $\bar{Y}^{\a
\b}_{\; \; \; \; \g}$ by a gauge transformation (\ref{gauge2}) with
$z^{lm} = \ve^{lmp} h_{0p}$, $z^{0m} = 0$, and hence gives the
same curvature. Dealing with $\bar{Y}'^{\a \b}_{\; \; \; \; \g}$
rather than $\bar{Y}^{\a \b}_{\; \; \; \; \g}$ simplifies the
computations. It follows that the curvature is given by \beq
R_{0s0b} &=& 0, \; \; \; \; R_{lmab} = 0, \nonumber
\\R_{lm0b} &=& N \ve_{lms} \left(\frac{3 x_b x_s}{r^5} -
\frac{\d_{bs}}{r^3} - \frac{4 \pi}{3} \, \d_{bs}\, \d(\vec{x})
\right) , \nonumber \\ R_{0mab} &=& 2N
\ve_{abm}\left(\frac{1}{r^3} + \frac{4 \pi}{3} \d(\vec{x})
\right) - 3N \left( \ve_{mak} \frac{x_b x_k}{r^5} -
\ve_{mbk} \frac{x_a x_k}{r^5}\right)\,, \nonumber \eeq which
satisfies the equations of motion, $ R_{\a \b} = 0\,,$ and $$ R_{0ijk} +
R_{ij0k} + R_{j0ik} = 4 \pi N \ve_{ijk} \d(\vec{x}) = - 8 \pi
\ve_{0ij\l} \bar{\Theta}^\l_{\; k} \;.$$

{}Finally, one easily checks that the linearized Riemann tensor of
linearized Taub-NUT is indeed dual to the linearized Riemann tensor
of linearized Schwarschild.  In that respect, the reason that it was
more convenient to work with $\bar{Y}'^{\a \b}_{\; \; \; \; \g}$
instead of $\bar{Y}^{\a \b}_{\; \; \; \; \g}$ above is that it is
$\bar{Y}'^{\a \b}_{mag \; \g}$ that is dual to $\bar{Y}^{\a
\b}_{Schw\; \g}$. While the curvatures are dual, the original
quantities $\bar{Y}^{\a \b}_{\; \; \; \; \g}$ are dual up to a gauge
transformation (\ref{gauge2}).

\section{Magnetic sources for bosonic higher spins}
\setcounter{equation}{0} \label{higherspins0}

We now indicate how to couple magnetic sources to spins greater than
two. The procedure parallels what we have just done for spin 2 but
the formulas are somewhat cumbersome because of the extra indices on
the fields and the extra trace conditions to be taken into account.  The
formalism describing higher spin fields in the absence of magnetic
sources has been recalled in Section \ref{appendixA}.

The spin-$s$ curvature $R_{ \m_1 \n_1 \m_2 \n_2 \cdots \m_s\n_s}$ is
the gauge invariant object in terms of which we shall first write
the equations of the theory.  Its index symmetry is described by
the Young tableau \be
\begin{picture}(85,15)(0,2)
\multiframe(0,11)(13.5,0){1}(10.5,10.5){\tiny{$\m_1$}}
\multiframe(0,0)(13.5,0){1}(10.5,10.5){\tiny{$\n_1$}}
\multiframe(31,11)(13.5,0){1}(10.5,10.5){\tiny{$\m_2$}}
\multiframe(31,0)(13.5,0){1}(10.5,10.5){\tiny{$\n_2$}}
\multiframe(42,11)(13.5,0){1}(19.5,10.5){$\cdots$}
\multiframe(42,0)(13.5,0){1}(19.5,10.5){$\cdots$}
\multiframe(62,11)(13.5,0){1}(10.5,10.5){\tiny{$\m_s$}}
\multiframe(62,0)(13.5,0){1}(10.5,10.5){\tiny{$\n_s$}}
\put(15,7.5){$\otimes$}
\end{picture}\,, \label{Young00}
\ee \ie, \be R_{ \m_1 \n_1 \cdots \m_i \n_i \cdots \m_s\n_s} = -
R_{ \m_1 \n_1 \cdots \n_i \m_i \cdots \m_s\n_s}, \; \; \; i = 1,
\cdots, s \ee and \be R_{ \m_1 \n_1 \cdots [\m_i \n_i
\m_{i+1}]\cdots \m_s\n_s} = 0 , \; \; \; i = 2, \cdots, s-1 \, .
\label{cycyclic}\ee Its dual, defined through
$$S_{\m_1 \n_1 \m_2 \n_2 \cdots \m_s\n_s} =-\frac{1}{2} \ve_{\m_1\n_1
\r \s}R^{\r\s}_{~~\m_2 \n_2 \cdots \m_s\n_s} \, ,$$ has the same
symmetry structure. Note that, just as in the spin-2 case, this does
not define an irreducible representation of the linear group.  But,
also as in the spin-2 case, we shall find that only the irreducible
part described by \be
\begin{picture}(85,15)(0,2)
\multiframe(0,11)(13.5,0){1}(10.5,10.5){\tiny{$\m_1$}}
\multiframe(0,0)(13.5,0){1}(10.5,10.5){\tiny{$\n_1$}}
\multiframe(11,11)(13.5,0){1}(10.5,10.5){\tiny{$\m_2$}}
\multiframe(11,0)(13.5,0){1}(10.5,10.5){\tiny{$\n_2$}}
\multiframe(22,11)(13.5,0){1}(19.5,10.5){$\cdots$}
\multiframe(22,0)(13.5,0){1}(19.5,10.5){$\cdots$}
\multiframe(42,11)(13.5,0){1}(10.5,10.5){\tiny{$\m_s$}}
\multiframe(42,0)(13.5,0){1}(10.5,10.5){\tiny{$\n_s$}}
\end{picture}\, \label{Young000} \ee (\ie, fulfilling also
Eq.(\ref{cycyclic}) for $i=1$) corresponds to the independent degrees
of freedom (the rest being determined by the sources).

The electric and magnetic  energy-momentum tensors will be denoted
by 

\noindent $t_{\m_1 \m_2 \cdots \m_s} $ and $\theta_{\m_1 \m_2 \cdots
\m_s}$. They are conserved, \ie divergence-free, $$ \pa_\m
t^{\m\n_1 \cdots \n_{s-1}} =0\;,\; \pa_\m \theta^{\m \n_1 \cdots
\n_{s-1}} =0\;.$$ Their double traceless parts are written $T_{\m_1
\m_2 \cdots \m_s} $ and $\Theta_{\m_1 \m_2 \cdots \m_s} $, and are
the tensors that actually couple to the spin-$s$ field.

\subsection{Electric and magnetic sources}

The equations in the presence of both electric and magnetic sources
are determined again by the requirements: (i) that they reduce to
the known equations with electric sources only when the magnetic
sources are absent, and (ii) that they be invariant under the
duality transformations that rotate the spin-$s$ curvature and its
dual, as well as the electric and magnetic sources.

Defining $\bar{\Theta}_{\m_1\m_2 \cdots \m_s}=\Theta_{\m_1\m_2
\cdots \m_s}- \frac{s}{4}
\eta_{(\m_1\m_2}\Theta^{\prime}_{\m_3\cdots \m_s)}\,,$ and 
$\bar{T}_{\m_1\m_2 \cdots \m_s}$ similarly,
 one finds
the following set of equations for the curvature: \beq R_{ \m_1
\n_1 \m_2 \n_2 \cdots \m_s\n_s} \eta^{\n_1\n_2}&=&\frac{1}{2}\
\bar{T}_{\m_1\m_2 [\m_3 [\cdots [\m_s,\n_s] \cdots ]\n_3]}\,, \label{CC1}\\
R_{ [\m_1 \n_1 \m_2] \n_2 \cdots \m_s\n_s}&=&\frac{1}{6} \
\ve_{\m_1 \n_1\m_2
\r}\bar{\Theta}^{\r}_{~\n_2  [\m_3 [\cdots [\m_s,\n_s] \cdots ]\n_3]}\,,\label{CC2}\\
\partial_{[\a} R_{\m_1 \n_1] \m_2 \n_2 \cdots \m_s\n_s} &=&-\frac{1}{3}\
\ve_{\a\m_1\n_1\r}\bar{\Theta}^{\r}_{~[\m_2  [\m_3 [\cdots
[\m_s,\n_s] \cdots ]\n_3] \n_2]}\,.\label{CC3}\eeq The first
equation is the analog of the Einstein equation (\ref{Einstein}),
the second is the analog of the modified cyclic identity
(\ref{cyclic}), while the third is the analog of the modified
Bianchi identity (\ref{BI}). It follows from these equations that
the dual curvature obeys similar equations, \beq S_{ \m_1 \n_1 \m_2
\n_2 \cdots \m_s\n_s} \eta^{\n_1\n_2}&=&\frac{1}{2}\
\bar{\Theta}_{\m_1\m_2 [\m_3 [\cdots [\m_s,\n_s] \cdots ]\n_3]} \,,\\
S_{ [\m_1 \n_1 \m_2] \n_2 \cdots \m_s\n_s}&=&-\frac{1}{6} \
\ve_{\m_1 \n_1\m_2 \r}\bar{T}^{\r}_{~\n_2  [\m_3 [\cdots [\m_s,\n_s] \cdots ]\n_3]}\,,\\
\partial_{[\a} S_{\m_1 \n_1] \m_2 \n_2 \cdots \m_s\n_s}& =&\frac{1}{3}\
\ve_{\a\m_1\n_1\r}\bar{T}^{\r}_{~[\m_2  [\m_3 [\cdots [\m_s,\n_s]
\cdots ]\n_3] \n_2]}\label{CC6}\, ,\eeq exhibiting manifest
duality symmetry.

\subsection{Decomposition of the curvature tensor}

As in the spin-2 case, the curvature tensor can be expressed in
terms of a completely symmetric potential $h_{\m_1 \cdots \m_s}$
and of a tensor $\Phi^{\r\s}_{~~~\m_1 \cdots \m_{s-1}}$ fixed by
the magnetic energy-momentum tensor, so that the cyclic and
Bianchi identities do indeed become true identities.

Let $\Phi^{\r\s}_{~~~\m_1 \cdots \m_{s-1}}$ be such that \be\pa_\r
\Phi^{\r\s}_{~~~\m_1 \cdots \m_{s-1}}=\theta^\s_{~\m_1 \cdots
\m_{s-1}}\,, \label{phis}\ee and let $\hat{\Phi}^{\r\s}_{~~~\m_1
\cdots \m_{s-1}}$ be the part of $\Phi^{\r\s}_{~~~\m_1 \cdots
\m_{s-1}}$ that is traceless in the indices $\m_1 \cdots \m_{s-1}$.
For computations, it is useful to note that $$ \pa_\r
\hat{\Phi}^{\r\s}_{~~~\m_1 \cdots \m_{s-1}}=\Theta^\s_{~\m_1 \cdots
\m_{s-1}}-\frac{(s-2)}{4} \eta_{(\m_1\m_2}\Theta'^{\s}_{~\m_3\cdots
\m_{s-1})}\,, $$ where primes denote traces. The expression of the tensor $\Phi^{\r\s}_{~~~\m_1
\cdots \m_{s-1}}$ in terms of the Dirac string is given below. The
appropriate expression of the curvature tensor in terms of the spin-$s$
 field and the Dirac string contribution is \be R_{ \m_1 \n_1 \m_2
\n_2 \cdots \m_s\n_s} =-\frac{1}{2}\ \ve_{\m_1\n_1 \r\s}
\bar{Y}^{\r\s}_{~~~[\m_2 [\m_3 [\cdots [\m_s,\n_s] \cdots ]\n_3]
\n_2]}\,, \ee where \beq\bar{Y}^{\r\s}_{~~~~\m_1 \cdots
\m_{s-1}}&=&Y^{\r\s}_{~~~~\m_1 \cdots
\m_{s-1}}+\frac{2(s-1)}{s}\d^{[\r}_{(\m_1}Y^{\s]\t}_{~~~~\m_2
\cdots \m_{s-1})\t}\,,\label{YYY1}\\
Y^{\r\s}_{~~~~\m_1 \cdots \m_{s-1}}&=&\pa_\t
X^{\r\s\t}_{~~~\m_1 \cdots \m_{s-1}}+
\hat{\Phi}^{\r\s}_{~~~\m_1 \cdots \m_{s-1}}\,,\nn
\\
X^{\r\s\t}_{~~~~\m_1 \cdots \m_{s-1}}&=&\ve^{\r\s\t\l}h_{\l\m_1
\cdots \m_{s-1}} -\frac{3(s-1)(s-2)}{2s}\eta_{\a
(\m_1}\d_{\m_2}^{[\r}\ve^{\s\t]\a\b}h_{\m_3\cdots
\m_{s-1})\b}^{\prime}\,.\nn
 \eeq The split of
$Y^{\r\s}_{~~~~\m_1 \cdots \m_{s-1}}$ into an $X$-part and a
$\Phi$-part defines a split of the Riemann tensor analogous to the
split (\ref{split0}) introduced for spin 2.  The Dirac string
contribution ($\Phi$-term) removes the magnetic terms violating the
standard cyclic and Bianchi identities, leaving one with a tensor
$r_{ \m_1 \n_1 \m_2 \n_2 \cdots \m_s\n_s}$ that fulfills $$r_{ [\m_1
\n_1 \m_2] \n_2 \cdots \m_s\n_s}=0, \;\;\;
\partial_{[\a} r_{\m_1 \n_1] \m_2 \n_2 \cdots \m_s\n_s} =0 \,,$$ and
thus derives from a symmetric potential (the spin-$s$ field $h_{\m_1
\cdots \m_s}$) as \be r_{\m_1 \n_1 \m_2 \n_2 \cdots \m_s \n_s} =-2\
h_{[\m_1[ \m_2 \cdots [\m_s,\n_s]\cdots \n_2]\n_1]} \label{DefR0}\ee
(see Section \ref{appendixA}).  The $X$-term in the curvature is a
rewriting of (\ref{DefR0}) that is convenient for the subsequent
analysis. The potential $h_{\m_1 \cdots \m_s}$ is determined from
the curvature up to a gauge transformation with unconstrained trace.
The fact that only $\hat{\Phi}^{\r\s}_{~~~\m_1 \cdots \m_{s-1}}$
appears in the curvature and not $\Phi^{\r\s}_{~~~\m_1 \cdots
\m_{s-1}}$ is a hint that only the double traceless part
$\Theta_{\m_1 \cdots \m_{s}}$ of the magnetic energy-momentum tensor
plays a physical role.

\subsection{Equations of motion for the spin-$s$ field}
In terms of the potential, the remaining equation (\ref{CC1}) is of
order $s$.  In the sourceless case, one replaces it by the second-order
equation written first by Fronsdal \cite{Fronsdal:1978rb}. This can
be done also in the presence of both electric and magnetic sources
by following the procedure described in
\cite{Francia:2002pt,Bekaert:2003az}. The crucial observation is
that the curvature is related as in Eq.(\ref{FR}), namely, \be R_{ \m_1
\n_1 \m_2 \n_2 \cdots \m_s\n_s}
\eta^{\n_1\n_2}=-\frac{1}{2}F_{\m_1\m_2 [\m_3 [\cdots [\m_s,\n_s]
\cdots ]\n_3]}\,,\label{FR0}\ee to the generalized Fronsdal tensor
given by \be F_{\g_1\cdots\g_s}=-\frac{1}{2}
\ve_{\g_1\m\n\l}\Big(\pa^\l \bar{Y}^{\m\n}_{~~~\g_2\cdots\g_s}
-(s-1)\
\pa_{(\g_2}\bar{Y}^{\m\n\l}_{~~~~~\g_3\cdots\g_s)}\Big)\,,\label{fronsgen}
\ee so that Eq.(\ref{CC1}) is equivalent to $ F_{\m_1\m_2 [\m_3 [\cdots
[\m_s,\n_s] \cdots ]\n_3]} +\bar{T}_{\m_1\m_2 [\m_3 [\cdots
[\m_s,\n_s] \cdots ]\n_3]} = 0 $.  This implies $F_{\m_1\m_2 \m_3
\cdots \m_s}+\bar{T}_{\m_1\m_2 \m_3 \cdots \m_s} = \pa_{(\m_1
\m_2\m_3}\Lambda_{\m_4\cdots \m_s)}$ for some $\Lambda_{\m_4\cdots
\m_s}$ \cite{Olver,DVH,Dubois-Violette:2001jk}.  By making a gauge transformation on the spin-$s$
field, one can set the right-hand side of this relation equal to
zero (see Section \ref{appendixA}), obtaining the field equation
\be \label{poppast}F_{\m_1\m_2 \m_3 \cdots \m_s}+\bar{T}_{\m_1\m_2
\m_3 \cdots \m_s} =0\,,\ee which fixes the trace of the gauge
parameter. When $s = 3$ this is the end of the story.

For s $\geq 4$ additional restrictions are necessary, namely, we
shall demand that the gauge transformation that brings the field
equation to the form (\ref{poppast}) eliminates at the same time the
double trace of the field $h_{\m_1 \cdots \m_s}$ (see
\cite{Francia:2002pt} for a discussion).

In terms of the generalized Einstein tensor defined as in
(\ref{eomsp}), \ie \be G_{\m_1 \m_2 \cdots \m_s} =F_{\m_1 \m_2
\cdots \m_s} -\frac{s(s-1)}{4} \eta_{(\m_1\m_2}F_{\m_3\cdots
\m_s)\r}^{\hspace{1.2cm} \r} ,\label{EinsteinS0} \ee the equations
become \be G_{\m_1\m_2 \m_3 \cdots \m_s}+T_{\m_1\m_2 \m_3 \cdots
\m_s} =0.\label{EinHSEq0}\ee

We shall thus adopt (\ref{EinHSEq0}), with the Einstein tensor,
Fronsdal tensor and $Y$-tensor defined as in (\ref{EinsteinS0}),
(\ref{fronsgen}) and (\ref{YYY1}), respectively,  as the equations
of motion for a double traceless spin-$s$ field $h_{\m_1 \cdots
\m_s}$. These equations imply Eqs.(\ref{CC1}) through (\ref{CC6}) and
define the theory in the presence of both electric and magnetic
sources. It is these equations that we shall derive from a
variational principle. 

\subsection{Point particles sources - Dirac string}

For point sources, the tensors that couple to the spin-$s$ field read
$$ t^{\m\n_1 \cdots \n_{s-1}} = M u^{\n_1} \cdots u^{\n_{s-1}}\int
d\l \delta^{(4)}(x-z(\l)) \dot{z}^\m =M \frac{u^\m u^{\n_1} \cdots
u^{\n_{s-1}}}{u^0} \delta^{(3)}(\vec{x} - \vec{z}(x^0))$$ and $$
 \theta^{\m \n_1 \cdots \n_{s-1}} =
N v^{\n_1} \cdots v^{\n_{s-1}} \int d\l \delta^{(4)}(x-\bar{z}(\l))
\dot{\bar{z}}^\m \\
=N \frac{v^\m v^{\n_1} \cdots v^{\n_{s-1}}}{v^0}
\delta^{(3)}(\vec{x} - \vec{\bar{z}}(x^0))\,.$$ One can check that
they are indeed conserved.

A tensor $\Phi^{\a \b}_{\; \; \; \; \g_1 \cdots \g_{s-1}}$ that
satisfies Eq.(\ref{phis}) can again be constructed by attaching a
Dirac string $y^\m(\l, \s)$ to the magnetic source, $y^\m(\l,
0)=\bar{z}^\m(\l)$. One has $$ \Phi^{\a \b}_{\; \; \; \; \g_1
\cdots \g_{s-1}} =  N v_{\g_1} \cdots v_{\g_{s-1}} \int d \l d \s
(y'^\a \dot{y}^\b  - \dot{y}^\a y'^\b) \delta^{(4)}(x-y(\l, \s))
\,.$$

One can compute explicitly the conserved charges associated with
asymptotic symmetries for electric point sources (see Section
\ref{appendixA}). Using the equations of motion, they read $$
P^{\m_1 \cdots \m_{s-1}}=M f^{\m_1 \cdots \m_{s-1}}(u)\,,$$ where
$f^{\m_1 \cdots \m_{s-1}}(u)$ is the traceless part of $u^{\m_1}
\cdots u^{\m_{s-1}}$. It reads
$$ f^{\m_1 \cdots \m_{s-1}}(u)= \sum_l \a_l \ \eta^{(\m_1\m_2}
\cdots \eta^{\m_{2l-1} \m_{2l}}  u^{\m_{2l+1}} \cdots
u^{\m_{s-1})}\vert u\vert^{2l} \,,$$ where the sum goes over
$l=0,1,\cdots$ such that $2l\leq s-1 $, $\a_0=1$ and
$\a_{l+1}=-\frac{(s-1-2l)(s-2-2l)}{4(l+1)(s-1-l)}\ \a_l\,.$ 

The dual magnetic charges
$$ Q^{\m_1 \cdots \m_{s-1}}=N f^{\m_1 \cdots \m_{s-1}}(v)\,$$ are
also conserved.

\subsection{Variational Principle}

\subsubsection{ Action}

The second-order equations of motion $
G_{\g_1\cdots\g_s}+\ T_{\g_1\cdots\g_s}=0$ equivalent to
Eq.(\ref{CC1}), are the Euler-Lagrange derivatives with respect to
$h^{\g_1\cdots\g_s}$ of the action \beq S[h_{\m_1 \cdots \m_s}(x),
y^\m(\l,\s)]=\int d^4x \ ( {\cal L} +\ h_{\m_1 \cdots
\m_s} t^{\m_1 \cdots \m_s} )\,,\label{actions} \eeq where \beq {\cal
L}=-\frac{(s-1)}{2}Y_{\m\n \a_1 \cdots \a_{s-1}}
\Big[-Y^{\m\a_1 \n\a_2 \cdots \a_{s-1}}+\frac{(s-2)}{2(s-1)}
Y^{\m\n \a_1 \cdots \a_{s-1}}\hspace{2cm}\nonumber \\
+\frac{(s-3)}{s}\eta^{\m\a_1}Y^{\n\r \a_2 \cdots
\a_{s-1}}_{\hspace{1.6cm}\r}-\frac{(s-2)}{s}\eta^{\m\a_1}Y^{\a_2\r
\n\a_3 \cdots \a_{s-1}}_{\hspace{2cm}\r}\Big]\,.\nonumber \eeq
One can check that this action reduces to the usual action (\ref{freeactions}) in the absence of sources.
As in the spin-2 case,
the trajectories of the electric and magnetic sources are kept
fixed, \ie, the sources are not dynamical. The magnetic coupling in
the action was obtained by introducing the familiar minimal electric
coupling in the ``parent action'' (\ref{actidual}) of the preceding chapter, which contains
two potentials, and determining what it becomes in the dual
formulation.

\subsubsection{Gauge invariances}

We now verify that the action (\ref{actions}) is invariant under the
gauge symmetries (\ref{gauges}) of the spin-$s$ field as well as under
displacements of the Dirac string (accompanied by an appropriate
redefinition of $h_{\m_1 \cdots \m_s}$). 

To that end, we first observe that the first term in the action
(\ref{actions}) is invariant under the following shifts of $Y^{\m
\n}_{\; \; \; \; ~\a_1\cdots \a_{s-1}}$ : \be \d Y^{\m
\n}_{\; \; \; \;~ \a_1\cdots \a_{s-1}} =
\partial_\r\delta^\m_{(\a_1} z^{\n \r}_{ \; \; \; \; \a_2\cdots
\a_{s-1})}- \partial_\r\delta^\n_{(\a_1} z^{\m \r}_{ \; \; \; \;
\a_2\cdots \a_{s-1})}+
\partial_{(\a_1} z^{\m \n}_{ \; \; \; \; \a_2\cdots \a_{s-1})} \,,\label{gauge1s}\ee
where $z^{\m\n}_{ \; \; \; \; \a_1\cdots \a_{s-2}}  = z^{[\m\n]}_{
\; \; \; \; \a_1\cdots \a_{s-2}} =z^{\m\n}_{ \; \; \; \;
(\a_1\cdots \a_{s-2})} $  is an arbitrary traceless tensor that
satisfies $\eta^{\a_1[\l}z^{\m\n]}_{ \; \; \; \; \;\a_1\cdots
\a_{s-2}} =0$  when $s>2$. Under this transformation, $\bar{Y}^{\m
\n}_{\; \; \; \; ~\a_1\cdots \a_{s-1}}$ transforms as $ \d
\bar{Y}^{\m \n}_{~\; \; \; \; \a_1\cdots
\a_{s-1}}=\partial_{(\a_1} z^{\m \n}_{ \; \; \; \; \a_2\cdots
\a_{s-1})} $,  which makes it obvious that the curvature and the
Fronsdal tensor are invariant under (\ref{gauge1s}).

The transformation (\ref{gauge1s}) can be conveniently written \be
\d Y^{\m \n}_{~\; \; \; \; \a_1\cdots \a_{s-1}} =\ve^{\m \n \r
\s} \partial_\r a_{\s \a_1 \a_2 \cdots \a_{s-1}}
\,,\label{gauge3s} \ee where $a_{\s \a_1 \a_2\cdots \a_{s-1}}
= - a_{ \a_1 \s  \a_2\cdots \a_{s-1}}=a_{\s \a_1 (
\a_2\cdots \a_{s-1})} $ is given by \be a_{\s \a_1 \a_2\cdots
\a_{s-1}} = \frac{1}{2} \ve_{\s \b \g\a_1 } z^{\b \g}_{ ~~~
\a_2\cdots \a_{s-1} }\,,\ee is traceless and satisfies $a_{[\s
\a_1 \a_2] \a_3\cdots \a_{s-1}} =0$ when $s>2$.

\paragraph{\it Standard spin-$s$ gauge invariance\\}

Direct computation shows that the gauge transformation
(\ref{gauges}) of the spin-$s$ field acts on $Y^{\m \n}_{\; \; \;
\; ~\a_1\cdots \a_{s-1}} $ as the transformation (\ref{gauge3s})
with \beq a_{\r \s ( \a_1\cdots \a_{s-2})}
&=&-2\frac{(s-1)}{s}
\pa_{[\r} \xi_{\s ]\a_1\cdots \a_{s-2}} \nonumber \\
&&+\frac{(s-1)(s-2)}{s^2}\Big[\eta_{\r(\a_1}\pa^{\l}\xi_{\a_2\cdots
\a_{s-2})\l\s}-\eta_{\s (\a_1}\pa^{\l}\xi_{\a_2\cdots
\a_{s-2})\l\r}\Big]\nonumber \,.\eeq It follows from this fact and
the conservation of the energy-momentum tensor that the action
(\ref{actions}) is invariant under the standard gauge transformation
(\ref{gauges}) of the spin-$s$ field.

\paragraph{\it Displacements of the Dirac string\\}

The displacements of the Dirac string change $\Phi^{\m \n}_{\; \; \;
\; \a_1\cdots \a_{s-1}} $ as 

\noindent $ \d \Phi^{\m \n}_{\; \;\; \;
\a_1\cdots \a_{s-1}} = k^{\m \n}_{\; \;\; \; \a_1\cdots \a_{s-1}}\,, $
where $\pa_\m k^{\m \n}_{\; \;\; \; \a_1\cdots \a_{s-1}} =0\,.$ The
latter equation implies that $ k^{\m \n}_{\; \;\; \; \a_1\cdots
\a_{s-1}} =\pa_\l K^{\m \n\l}_{\; \;\; \; ~~\a_1\cdots \a_{s-1}} $,
where $K^{\m \n\l}_{\; \;\; \; ~\a_1\cdots \a_{s-1}}=K^{[\m
\n\l]}_{\; \;\; \; ~~\a_1\cdots \a_{s-1}}$. Let $\hat{K}^{\m
\n\l}_{\; \;\; \; ~\a_1\cdots \a_{s-1}}$ be the part of $K^{\m
\n\l}_{\; \;\; \; ~\a_1\cdots \a_{s-1}}$ that is traceless in
$\a_1\cdots \a_{s-1}$; it can be  decomposed as  $$ \hat{K}^{\m
\n\l}_{\; \;\; \; ~~\a_1\cdots \a_{s-1}} =x^{\m \n\l}_{\; \;\; \;
~~\a_1\cdots \a_{s-1}} +\d^{[\l}_{(\a_1} y^{\m\n]}_{~~\a_2\cdots
\a_{s-1})}\,,$$ where $x^{\m \n\l}_{\; \;\; \; ~~\a_1\cdots
\a_{s-1}}$ and $y^{\m\n}_{~~\a_2\cdots \a_{s-1}}$ satisfy \beq x^{\m
\n\l}_{\; \;\; \; ~~\a_1\cdots \a_{s-1}}=x^{[\m \n\l]}_{\; \;\; \;
~~\a_1\cdots \a_{s-1}}=x^{\m \n\l}_{\; \;\; \; ~~(\a_1\cdots
\a_{s-1})}\,,\
 x^{\m \n\l}_{\; \;\; \; ~~\a_1\cdots \a_{s-1}}\d_\l^{\a_1}=0 \,, \nonumber \\
y^{\m\n}_{~~\a_2\cdots \a_{s-1}}=y^{[\m\n]}_{~~\a_2\cdots
\a_{s-1}}= y^{\m\n}_{~~(\a_2\cdots \a_{s-1})}\,, \
y^{\m\n}_{~~\a_2\cdots
\a_{s-1}}\d_\n^{\a_2}=0\,, \nonumber \\
\eta^{\a_1[\l}y^{\m\n]}_{~~~\a_1 \cdots
\a_{s-1}}=-\textstyle{\frac{(s-1)}{2}}\ x^{\m \n\l}_{\; \;\; \; ~~\a_1\cdots
\a_{s-1}}\eta^{\a_1\a_2}\,, y^{\m\n}_{~~~\a_1 \cdots
\a_{s-1}}\eta^{\a_1\a_2}=0\,. \nonumber \eeq For the action to be
invariant under displacements of the string, the variation of
$\Phi^{\m \n}_{\; \; \; \; \a_1\cdots \a_{s-1}} $ has to be
supplemented with an appropriate transformation of $h_{\a_1\cdots
\a_{s}} $. This transformation reads $ \d h_{\a_1\cdots
\a_{s}}=\frac{1}{6} \ \ve_{\m\n\l(\a_1}\ x^{\m\n\l}_{~~~~\a_2\cdots
\a_{s})}\,.$ Indeed, when one performs both variations, $Y^{\m
\n}_{\; \; \; \; ~\a_1\cdots \a_{s-1}} $ transforms as in
(\ref{gauge1s}), so the first term in the action is invariant.
Furthermore, the electric coupling term is invariant as well
because the support of the variation of the spin-$s$ field does not
contain the electric worldlines.

\paragraph{\it Identities\\}

The identities that follow from the invariance (\ref{gauge1s}) --
or (\ref{gauge3s}) -- of the first term  ${\cal L}$  in the action
may be written conveniently in terms of $$ A^{\s \g_1\cdots
\g_{s-1}}=\ve^{\s\m\n\l }
\partial_\l \left(\frac{\partial {\cal L}}{\partial Y^{\m \n}_{\;
\; \; \; ~\g_1\cdots \g_{s-1}}}\right)\,, $$ and its trace $A^{
\prime\g_2\cdots \g_{s-1}}=A^{\s \g_1\cdots
\g_{s-1}}\eta_{\s\g_1}$ . They read $$ 0=A^{\s \g_1\cdots
\g_{s-1}}-A^{(\g_1 \g_2\cdots
\g_{s-1})\s}-\frac{s-2}{s}\Big(\eta^{\s(\g_1}A^{ \prime\g_2\cdots
\g_{s-1})}-\eta^{(\g_1\g_2}A^{ \prime\g_3\cdots \g_{s-1})\s}\Big)\,.
$$ Using these identities, one checks the following 
relation, \beq G^{\g_1 \cdots \g_{s}}=\frac{\delta {\cal L}}{\delta
h_{ \g_1 \cdots \g_{s}}} &=&A^{(\g_1\g_2 \cdots
\g_{s})}+\frac{(s-1)(s-2)}{2s}
\eta^{(\g_1\g_2}A^{\prime \g_3\cdots \g_{s})}\nnn
&=&A^{\g_1\g_2 \cdots
\g_{s}}+\frac{(s-1)(s-2)}{2s}\eta^{(\g_2\g_3}A^{\prime \g_4\cdots
\g_{s})\g_1}\,, \label{useful} \eeq which will be used in the
following section.

\subsection{Quantization condition}

As for spin 2, the unobservability of the Dirac string in the
quantum theory leads to a quantization condition.  The computation
proceeds exactly as in the spin-2 case.

In the gauge $y^0 = \l$, the unobservability constraints read $$ \pi_m = -2 N y'^n
\, f_{\g_1 \cdots \g_{s-1}}(v)\frac{\partial {\cal L}}{\partial
Y^{mn}_{\; \; \;\; \; ~\g_1 \cdots \g_{s-1}}} \,.$$ In the
quantum theory, the wave functional $\psi$ must thus fulfill $$
\frac{\hbar}{i} \frac{\delta \psi}{\delta y^m(\s)} = -2N y'^n \,
f_{\g_1 \cdots \g_{s-1}}(v)\frac{\partial {\cal L}}{\partial
Y^{mn}_{\; \; \; \; \; ~\g_1 \cdots \g_{s-1}}} \psi \,.$$
Integrating this equation along a path that encloses an electric
source, one finds the following variation of the phase of the wave
functional  $$ \Delta \Psi = - \frac{ N }{\hbar} \, f_{\g_1 \cdots
\g_{s-1}}(v)\int \frac{\partial {\cal L}}{\partial Y^{mn}_{\;
\; \; \; \; ~\g_1 \cdots \g_{s-1}}} \left(\dot{y}^m y'^n - \dot{y}^n
y'^m \right) d \s d \l \,,$$ where the integral is taken on the
two-dimensional surface enclosing the electric source. Using the
Gauss theorem, this can be converted into a volume integral, $$
\Delta \Psi =  -\frac{N  }{\hbar} \, f_{\g_1 \cdots \g_{s-1}}(v)\int
d^3 x\ \ve^{mnp} \partial_p \left(\frac{\partial {\cal L}}{\partial
Y^{mn}_{\; \; \; \;\;~ \g_1 \cdots \g_{s-1}}} \right)\,. $$
Using the relations (\ref{useful}), one checks that $$ \ve^{mnp}
\partial_p \left(\frac{\partial {\cal L}}{\partial Y^{mn}_{\;~\; \;
\; \; \g_1 \cdots \g_{s-1}}} \right) = \frac{\delta {\cal L}}{\delta
h_{0 \g_1 \cdots \g_{s-1}}} +\cdots\,,$$ where the dots stand for
terms of the form $\eta^{(\g_1\g_2}X^{\g_3 \cdots \g_{s-1})}$. Upon
use of the Einstein equations $G^{0 \g_1 \cdots \g_{s-1}} =
-T^{0 \g_1 \cdots \g_{s-1}}$, the variation of the phase
becomes,  $$ \Delta \Psi = \frac{N }{\hbar} \, f_{\g_1 \cdots
\g_{s-1}}(v) \int d^3 x\ T^{0 \g_1 \cdots \g_{s-1}} = \frac{ MN
}{\hbar} \, f_{\g_1 \cdots \g_{s-1}}(v)f^{\g_1 \cdots \g_{s-1}}(u)\,.
$$  For the wave functional to be single-valued, this should be a
multiple of $2 \pi$.  This yields the quantization condition \be
\frac{MN  }{2 \pi \hbar} \, f_{\g_1 \cdots \g_{s-1}}(v)f^{\g_1
\cdots \g_{s-1}}(u)= n , \; \; \; n \in \Bbb{Z} \,.\ee Introducing
the conserved charges $P^{\g_1 \cdots \g_{s-1}}$, $Q^{\g_1 \cdots
\g_{s-1}}$, this can be rewritten as \be \frac{1  }{2 \pi \hbar} \,
Q_{\g_1 \cdots \g_{s-1}}(v)P^{\g_1 \cdots \g_{s-1}}(u) \in \Bbb{Z}
\,.\ee

\section{Beyond the linear theory for spin two}
\label{sect4}
We have seen that magnetic sources can be introduced for linearized
gravity and linearized higher-spin theories, and that an appropriate
generalization of the Dirac quantization condition on the sources
must hold. However in the linear theory the treatment is already
unsatisfactory since the sources must be external.  In the full 
nonlinear theory even the introduction of external sources is not
possible.  For spin 2 the difficulty stems from the fact that the
source must be covariantly conserved and for spins $\geq 2$ the formulation
of the nonlinear theory is still incomplete.

Nevertheless, we shall comment on the issue of duality in the spin-2
case, for which the nonlinear theory in the absence of sources is
the vacuum Einstein theory of gravitation. This is the ``electric''
formulation.  Electric sources are coupled through their standard
energy-momentum momentum tensor.  We do not know whether magnetic
sources should appear as independent fundamental degrees of freedom
(the complete action with these degrees of freedom included is
unknown and would presumably be non-local, as the results of \cite{Deser:2005sz} suggest) or whether they should
appear as solitons somewhat like in Yang-Mills theory
\cite{Montonen:1977sn}.

Whatever the answer, there are indications that duality invariance
and quantization conditions are valid beyond the flat space, linear
regime studied above. One indication is given by dimensional
reduction of the full Einstein theory, which reveals the existence
of ``hidden symmetries'' that include duality \cite{Geroch}. Another
indication that nonlinear gravity enjoys duality invariance is
given by the existence of the Taub-NUT solution
\cite{Newman:1963yy}, which is an exact solution of the vacuum
Einstein theory describing a gravitational dyon. 
The Taub-NUT metric is given by $$ ds^2 = - V(r)[
dt + 2N(k-\cos\theta) \, d\phi ]^2 + V(r)^{-1} dr^2 +
(r^2+N^2)(d\theta^2 + \sin^2 \theta\, d\phi^2) \,, $$ with $$ V(r)
= 1 - \frac{2(N^2+Mr) }{(r^2 + N^2)} = \frac{r^2 - 2Mr -N^2}{r^2 +
N^2} \,, $$ where $N$ and $M$ are the magnetic and electric masses
as follows from the asymptotic analysis of the metric and our
discussion of the linear theory. A pure magnetic mass has $M=0$. 
 The quantization
condition on the energy of a particle moving in the Taub-NUT
geometry is a well known result which has been discussed by many
authors \cite{Zeeetal} and which can be viewed as a consequence of
the existence of closed timelike lines \cite{Misner1}. 
For further discussions on this topic, see \cite{Bunster:2006rt}.


%% file: brst1101.tex
\chapter{Field-Antifield Formalism}

The purpose of this chapter is to provide an introduction to the field-antifield formalism for gauge field theories, as well as to the construction of consistent interactions for these fields. 
An excellent review on the field-antifield formalism, also called BRST, antibracket or Batalin-Vilkovisky formalism, is  \cite{Gomis:1994he}. The content of the first sections is based on this reference, which we refer to for further details.
The problem of finding consistent interactions in the BRST formalism has been developped in \cite{Barnich:1993vg,Henneaux:1997bm}. As we will show, it is related to the consistent deformations of the BRST master equation.

In this chapter, we first review the structure of general gauge field theories in Section \ref{s:ssgt}. 
Then we introduce the ghosts and the antifields, as well as relevant mathematical tools in Section \ref{s:faf}.
Finally, we present the deformation technique in Section \ref{cons}.

\section{Structure of Gauge Theories}
\label{s:ssgt}

\hspace{\parindent}
The most familiar example of a gauge theory
is the one associated
with a non-Abelian Yang-Mills theory
\cite{Yang:1954ek},
namely a compact Lie group.
The gauge structure is then determined by
the structure constants of the corresponding Lie algebra,
which satisfy the Jacobi identity.

In more general theories, 
the transformation rules can involve
field-dependent structure functions.
The determination of the gauge algebra 
(called ``soft algebra''\cite{Sohnius:1982rs}) is then more
complicated than in the Yang-Mills case.
The Jacobi identity must be appropriately generalized
\cite{Batalin:1981jr,dewitt84a}.
Furthermore, new structure tensors\footnote{Throughout the section, we will call  the objects that characterize the gauge structure ``tensors'', which they are not strictly speaking. The reason we use this terminology is because they have indices that behave like covariant and contravariant indices under linear transformations of the fields, gauge parameters, etc.}
 may appear which then
 need to obey new consistency identities.
In other types of theories,
the generators of the gauge transformations
are not independent.
This occurs when there is ``a gauge invariance''
for gauge transformations.
One says that the system is {\it reducible}.
Yet another complication occurs
when the commutator of two gauge transformations
produces a term that vanishes only on-shell,
\ie when the equations of motion are used.

In this section we discuss the above-mentioned complications
for a generic gauge theory.
The main issues are to find the relevant
gauge-structure tensors and
the equations that
they need to satisfy.

\subsection{Gauge Transformations}
\label{ss:gt}

\hspace{\parindent}
This subsection introduces the notions
of a gauge theory and a gauge transformation.
It also defines notations.

Consider a dynamical system  governed by
a classical action $\cs_0 [ \phi ]$,
which depends on $N$ different fields $\phi^i(x)$, $i=1,\cdots,N$.
The index $i$ can label the space-time indices $\mu$, $\nu$
of tensor fields,
the spinor indices of fermionic fields,
and/or an index distinguishing different types of generic fields.

The action is invariant
under a set of $m_0$ ($m_0\leq N$)
nontrivial gauge transformations,
which, when written in infinitesimal form, read
\be
\delta\phi^i (x) =
\left( R^i_{\alpha} (\phi)\varepsilon^{\alpha} \right) (x) \quad ,
\quad {\rm where}
\ \alpha=1,\, 2,  \, \ldots  \ m_0 \quad .
\label{trans gauge}
\ee
Here, $\varepsilon^{\alpha} (x)$
are infinitesimal gauge parameters,
that is, arbitrary functions of the space-time variable $x$,
and $R^i_{\alpha}$ are
the generators of gauge transformations.
These generators are operators
that act on the gauge parameters.
In kernel form,
$\left( R^i_{\alpha} (\phi)\varepsilon^{\alpha} \right) (x)$
can be represented as
$\int {\dif y \ R_{\alpha }^i\left( {x,y} \right)
\varepsilon ^{\alpha }\left( y \right)}$ ,
where $$R_{\alpha }^i\left( {x,y} \right)=
 r^{i}_\alpha \delta(x-y) + r^{i\mu}_\alpha 
\delta,_\mu(x-y) + \cdots + r^{i\mu_1 \dots \mu_s}_\alpha 
\delta,_{\mu_1 \dots \mu_s}(x-y) 
\,$$ 
and $r^{i}_{\a} $, $r^{i\mu}_\a$, $\ldots \ $ are functions of $x$ and $\phi (x)$.

One often adopts the  compact notation
\cite{dewitt64a} where the appearance of a discrete index
also indicates the presence of a space-time variable. Summation over 
a discrete index then also implies integration over the space-time variable.
With this convention, the transformation laws become
\be
   \delta\phi^i=R^i_{\alpha} \varepsilon^{\alpha} =
   \sum\limits_{\alpha }
   \int {\dif y \ R_{\alpha }^i\left( { x,y } \right)
   \varepsilon ^{\alpha }\left( y \right)}\quad .
\label{transformation rule}
\ee

Let $\cs_{0,i} \left( { \phi , x} \right) $
denote the Euler-Lagrange variation of the action
with respect to $\phi^i (x)$:
\be
   \cs_{0,i}\left( { \phi , x} \right) \equiv
  \frac{\d^R \cs_0 [ \phi ] }{\d \phi^i (x) } \quad ,
\label{def s0i}
\ee
where the supscript $R$
indicates that the derivative is
to be taken from the right.

The statement that the action is invariant
under the gauge transformation in Eq.\bref{trans gauge} means that
the Noether identities
\be
  \int {\dif x} \sum\limits_{i=1}^N 
 \cs_{0,i}\, R^i_{\alpha}\left( { x,y } \right)
  \left( { x} \right) = 0
\label{ident noether long form}
\ee
hold, or equivalently, in compact notation,
\be
  \cs_{0,i}  R^i_{\alpha} =0 \quad .
\label{ident noether}
\ee
Eq.\bref{ident noether long form} (or Eq.\bref{ident noether}) is derived by varying $\cs_0$
with respect to the right variations
of the $\phi^i$ given by Eq.\bref{trans gauge}.
It sometimes vanishes because the integrand
is a total derivative.
We assume that surface terms can be dropped in such integrals --
this is indeed the case when
Eq.\bref{ident noether long form} is applied
to gauge parameters that fall off sufficiently fast
at spatial and temporal infinity.

Notice that the gauge generators are not unique, one can
take linear combinations of them to form a new
set and the gauge-structure tensors will depend on this choice.
Another source of non-uniqueness is the presence of trivial 
gauge transformations defined by
\be
   \delta_\mu\phi^i = \cs_{0,j}\mu^{ji}
\ ,\quad \quad
   \mu^{ji} = - (-1)^{\epsilon_i \epsilon_j}\mu^{ij} \quad ,
\label{trivials}
\ee
where $\mu^{ji}$ are arbitrary functions and $\epsilon_i$ is the parity of $\phi^i$.
It is easily demonstrated that,
as a consequence
of the symmetry properties of $\mu^{ji}$,
the transformations \bref{trivials} 
leave the action invariant. These transformations are of no physical
interest and lead to no conserved currents.
However, in studying the structure of the gauge transformations,
it is necessary to take them into consideration.
Indeed, in general
the commutator of two nontrivial gauge transformations
can produce trivial gauge transformations.

\subsection{Irreducible and Reducible Gauge Theories}
\label{ss:irgt}

\hspace{\parindent}
To determine the independent degrees of freedom,
it is important to know any relations
among the gauge generators. 
The simplest gauge theories,
for which all gauge transformations
are independent, are called {\it irreducible}.
When dependences exist,
the theory is {\it reducible}.
In reducible gauge theories,
there is a ``gauge invariance for gauge transformations'',
 called ``level-one'' gauge invariance.
If the level-one gauge transformations
are independent,
then the theory is called {\it first-stage reducible}.
This may not happen.
Then, there are ``level-two'' gauge invariances, \ie
gauge invariances for the level-one gauge invariances
and so on.
This leads to the concept
of an {\it $L$-th stage reducible theory}.
In what follows we let $m_s$ denote
the number of gauge generators at the $s$-th stage
regardless of whether they are independent.

Let us define the above concepts with equations.
Assume that all gauge invariances of a theory are known
and that some regularity conditions (see  \cite{Gomis:1994he}) are satisfied.
Then, the most general solution of the
Noether identities \bref{ident noether}
is a gauge transformation, up to terms that vanish
when the equations of motion are satisfied:
\be
  \cs_{0,i}  \, \lambda^i= 0 \Leftrightarrow
  \lambda^i =  R^i_{0\alpha_0}\lambda^{\prime \alpha_0}  +
   \cs_{0,j}\,  T^{ji}
\quad ,
\label{completesa}
\ee
where $T^{ij}$ must satisfy the graded symmetry property
\be
  T^{ij} = -(-1)^{\eps_i \eps_j } T^{ji}
\quad .
\label{symmetry of Tij}
\ee
The $R^i_{0\alpha_0}$ are the gauge generators
in Eq.\bref{trans gauge}, to which we added the subscripts $0$ 
to indicate the level of the gauge transformation.
The second term $\cs_{0,j}T^{ji}$ in Eq.\bref{completesa}
is  a trivial gauge transformation.
The first term $R^i_{0\alpha_0} \lambda^{\prime \alpha_0}$
in Eq.\bref{completesa} is similar
to a nontrivial gauge transformation
of the form of Eq.\bref{trans gauge} with
$\varepsilon^{\alpha_0} = \lambda^{\prime \alpha_0}$.
The key assumption to have Eq.\bref{completesa}
is that the set of functionals $R^i_{0\alpha_0}$
exhausts on-shell the relations among the equations of motion,
namely the Noether identities.
In other words, the gauge generators
form a complete set on-shell.

Let us consider a {\it reducible} theory, \ie  
there are dependences among the gauge
generators.
If $m_0-m_1$ of the generators are independent on-shell,
then there are $m_1$ linear combinations of them that vanish on-shell.
In other words, there exist
$m_1$ functionals $R^{\alpha_0}_{1\alpha_1}$ such that
\bea
    R^i_{0\alpha_0} R^{\alpha_0}_{1\alpha_1}&=&
    \cs_{0,j} V^{ji}_{1\alpha_1} \ ,
    \quad\quad\quad \alpha_1=1,\ldots,m_1 \quad  ,
\label{depen generadors}
\eea
for some $V^{ji}_{1\alpha_1}$
satisfying
$V^{ij}_{1\alpha_1} = -(-1)^{\eps_i \eps_j } V^{ji}_{1\alpha_1}$.
The $R^{\alpha_0}_{1\alpha_1}$
are the on-shell null vectors for
$R^i_{0\alpha_0}$ since
$\restric{R^i_{0\alpha_0} R^{\alpha_0}_{1\alpha_1}}{\Sigma}=0$ ,
where $\Sigma$ is the surface on which the equations of motion hold.
Notice that,
if
$
  \varepsilon^\alpha = R^{\alpha}_{1\alpha_1} \varepsilon^\alpha_1
$
for any $\varepsilon^\alpha_1$,
then $\delta \phi^i$
in Eq.\bref{transformation rule}
is zero on-shell,
so that no gauge transformation is produced.
In Eq.\bref{depen generadors}
it is assumed that the reducibility of the $R^i_{0\alpha_0}$
is completely contained in $R^{\alpha_0}_{1\alpha_1}$, \ie
$R^{\alpha_0}_{1\alpha_1}$ also constitute a complete set
\be
   R^i_{0\alpha_0}\lambda^{\alpha_0}=
    \cs_{0,j}\, M_0^{ji} \Rightarrow
   \lambda^{\alpha_0}=
   R^{\alpha_0}_{1\alpha_1}\lambda^{\prime \alpha_1}
   + \,T_0^{j\alpha_0}
\quad ,
\label{consequence of regularity}
\ee
for some $\lambda^{\prime \alpha_1}$ and some $T_0^{j\alpha_0}$.

If the functionals $R^{\alpha_0}_{1\alpha_1}$
are independent on-shell, 
then the theory is called {\it first-stage reducible}.
If the functionals $R^{\alpha_0}_{1\alpha_1}$
are not all independent on-shell,
relations exist among them
and the theory is second-or-higher-stage reducible.
Then, the on-shell null vectors of $R^{\alpha_0}_{1\alpha_1}$
and higher $R$-type tensors
must be found.

Most generally, a theory is
$L$-{\it th stage reducible}
\cite{Batalin:1984jr}
if there exist functionals
\be
   R^{\alpha_{s-1}}_{s\alpha_s},\quad\quad
   \alpha_s=1,\ldots,m_s \ , \quad \quad s=0,\ldots,L \quad  ,
\label{generadors r}
\ee
such that $R^i_{0\alpha_0}$ satisfies Eq.\bref{ident noether},
\ie $\cs_{0 , i}\, R^i_{0\alpha_0} = 0$,
and such that, at each stage, the $R^{\alpha_{s-1}}_{s\alpha_s}$
constitute a complete set, \ie 
$$
   R^{\alpha_{s-1}}_{s\alpha_s} \lambda^{\alpha_s} =
  \cs_{0,j} \,  M_s^{j\alpha_{s-1}}\Rightarrow
   \lambda^{\alpha_s} = R^{\alpha_s}_{s+1,\alpha_{s+1}}
    \lambda^{\prime \alpha_{s+1}} + \cs_{0,j} \, T_s^{j\alpha_s}\quad ,
$$
$$
    R^{\alpha_{s-2}}_{s-1,\alpha_{s-1}}R^{\alpha_{s-1}}_{s\alpha_s}
    =\cs_{0,i}\, V^{i\alpha_{s-2}}_{s\alpha_s} \ ,
  \quad\quad\quad s=1,\ldots,L \quad ,
$$
where we have defined
$
R^{\alpha_{-1}}_{0\alpha_0}\equiv R^i_{0\alpha_0}$
and
$\alpha_{-1}\equiv i$.
The $R^{\alpha_{s-1}}_{s\alpha_s}$
are the on-shell null vectors for
$R^{\alpha_{s-2}}_{s-1 \alpha_{s-1}}$.

\subsection{The Gauge Structure}
\label{ss:gs}

\hspace{\parindent}
In this section we restrict ourselves to the simplest case
of irreducible  gauge theories. The same developpements 
can be performed for gauge theories with reducibilities, 
but the number of equations and structure tensors 
increases rapidly while the philosophy stays the same.
To avoid cumbersome notation,
we use $R_{\alpha}^i$ for $R_{0 \alpha_0}^i$,
so that the index $\alpha_0$ corresponds to $\alpha$.

The general strategy to obtain the gauge structure
is as follows \cite{deWit:1978cd}.
The first gauge-structure tensors are
the gauge generators themselves,
and the first gauge-structure equations
are the Noether identities
\bref{ident noether}.
One computes commutators, commutators of commutators, etc.,
of gauge transformations.
These are still gauge transformations, so they must also verify the 
Noether identity. Generic solutions are obtained
by exploiting the consequences of the completeness of the set
of gauge transformations.
In this process,
additional gauge-structure tensors appear.
They enter in higher-order
 identity equations like the Jacobi identity, produced by
 the graded symmetrization of commutators of commutators, etc. 
The completeness is again used to solve these equations
and introduces new tensors.
The process is continued until it terminates.

Consider the commutator of two gauge
transformations of the type \bref{trans gauge}.
On one hand, a direct computation leads to
$$
   [\delta_1,\delta_2]\phi^i=
 \left( R^i_{\alpha,j} R^j_\beta -
  (-1)^{\epsilon_\alpha \epsilon_\beta}
 R^i_{\beta,j} R^j_\alpha\right)
   \varepsilon_1^\beta\varepsilon_2^\alpha
\quad ,
$$
where $\epsilon_\a$ is the Grassman parity of $\varepsilon^\a$. (Note that the Grassman parity of $ R^j_\a$ is $\epsilon_j + \epsilon_\a\,$.)
On the other hand, this commutator is also a gauge symmetry of the action. 
So it satisfies the Noether identity. 
Factoring out the gauge parameters
$\varepsilon_1^\beta$ and $\varepsilon_2^\alpha$,
one may write
$$
\cs_{0,i} \,\left(R^i_{\alpha,j}R^j_\beta
 - (-1)^{\epsilon_\alpha \epsilon_\beta}
  R^i_{\beta,j}R^j_\alpha\right) = 0
\quad .
$$
Taking into account the completeness property \bref{completesa},
the above equation implies
the following important relation among the generators
\be
 R^i_{\alpha,j}R^j_\beta -
  (-1)^{\epsilon_\alpha \epsilon_\beta}
 R^i_{\beta,j}R^j_\alpha
   = R^i_\gamma T^\gamma_{\alpha\beta} -
   \cs_{0,j} E^{ji}_{\alpha\beta}
\quad ,
\label{algebra oberta}
\ee
for some gauge-structure tensors
$T^\gamma_{\alpha\beta}$ and
$E^{ji}_{\alpha\beta}$.
This equation defines
$T^\gamma_{\alpha\beta}$ and $E^{ji}_{\alpha\beta}$. 

Restoring the dependence on the gauge parameters
$\varepsilon_1^\beta$ and $\varepsilon_2^\alpha$,
the last two equations imply
\be
   [\delta_1,\delta_2]\phi^i \equiv
    R^i_\gamma T^\gamma_{\alpha\beta}
   \varepsilon_1^\beta\varepsilon_2^\alpha -
   \cs_{0,j} E^{ji}_{\alpha\beta}
   \varepsilon_1^\beta\varepsilon_2^\alpha
\quad ,
\label{com of two gens}
\ee
where $T^\gamma_{\alpha\beta}$ are known
as the ``structure constants''
of the gauge algebra.
The words {\it structure constants} are in quotes
because in general the $T^\gamma_{\alpha\beta}$
depend on the fields of the theory and are not ``constant''.

The possible presence
of the $E^{ji}_{\alpha\beta}$ term is due
to the fact that the commutator of two gauge
transformations may give rise to trivial gauge transformations
\cite{deWit:1978cd,Batalin:1981jr,Batalin:1984ss}.
The gauge algebra
generated by the $R^i_\alpha$ is said to be {\it open}
if $E^{ij}_{\alpha\beta}\neq 0$,
whereas the algebra is said to be {\it closed}
if $E^{ij}_{\alpha\beta}=0$.
Moreover, Eq.\bref{algebra oberta} defines a {\it Lie algebra}
if the algebra is closed, $E^{ij}_{\alpha\beta}=0$, and
 the $T^\gamma_{\alpha\beta}$ do not depend on the fields $\phi^i$.

\vspace{.2cm}

The next step determines
the restrictions imposed by the Jacobi identity.
In general, it leads to
new gauge-structure tensors and equations
\cite{Kallosh:1978de,VanNieuwenhuizen:1981ae,dewitt84a,Batalin:1985qj}.
The identity
$$
   \sum_{\rm cyclic \  over \ 1,\ 2,\ 3}
[\delta_1,[\delta_2,\delta_3]]=0 \quad ,
$$
implies the following relations among
the tensors $R$, $T$ and $E$ :
\be
   \sum_{\rm cyclic \ over \ 1,\ 2,\ 3 }
   \left(R^i_\delta A^\delta_{\alpha\beta\gamma}-
   \cs_{0,j} B^{ji}_{\alpha\beta\gamma}\right)
   \veps^\gamma_1\veps^\beta_2\veps^\alpha_3=0 \quad ,
\label{cons jacobi}
\ee
where we have defined
$$
   3 A^\delta_{\alpha\beta\gamma} \equiv
   \left( T^\delta_{\alpha\beta,k}R^k_\gamma -
   T^\delta_{\alpha\eta} T^\eta_{\beta\gamma} \right) +
\nonumber
$$
\be
  (-1)^{\epsilon_\alpha ( \epsilon_\beta + \epsilon_\gamma )}
   \left( T^\delta_{\beta\gamma,k} R^k_\alpha -
   T^\delta_{\beta\eta} T^\eta_{\gamma\alpha}\right) +
   (-1)^{\epsilon_\gamma (\epsilon_\alpha + \epsilon_\beta )}
   \left( T^\delta_{\gamma\alpha,k}R^k_\beta -
   T^\delta_{\gamma\eta} T^\eta_{\alpha\beta} \right)
\quad ,
\label {Adef}
\ee
and
\bqn
   3B^{ji}_{\alpha\beta\gamma} &\equiv& \left(
   E^{ji}_{\alpha\beta,k} R^k_\gamma -
   E^{ji}_{\alpha\delta} T^\delta_{\beta\gamma} -
   (-1)^{\epsilon_i \epsilon_\alpha }
   R^j_{\alpha,k} E^{ki}_{\beta\gamma} +
   (-1)^{\epsilon_j (\epsilon_i + \epsilon_\alpha )}
   R^i_{\alpha,k} E^{kj}_{\beta\gamma}
\right)
\nonumber\\
  && + (-1)^{\epsilon_\alpha (\epsilon_\beta + \epsilon_\gamma )}
\Big( {\rm r.h.s. \ of \ above \ line \ with \
{{\scriptstyle \alpha \rightarrow \beta} \;,\;
{ {\scriptstyle \beta \rightarrow \gamma \;,\; \gamma \rightarrow \alpha }}
}
} \Big)
\nnn
 &&  + (-1)^{\epsilon_\gamma (\epsilon_\alpha + \epsilon_\beta )}
\Big( {\rm \rm r.h.s. \ of \ first \ line \ with \
{{\scriptstyle \alpha \rightarrow \gamma} \;,\; 
{ {\scriptstyle \beta \rightarrow \alpha \;,\;  \gamma \rightarrow \beta }}
}
} \Big)
\; .
\label{Bdef}
\eqn

For an irreducible theory,
the on-shell independence of the generators 
and their completeness \bref{completesa}
lead to the following solution of Eq.\bref{cons jacobi} :
\be
   A^\delta_{\alpha\beta\gamma} =\cs_{0,j}
  D^{j\delta}_{\alpha\beta\gamma} 
\quad ,
\label{jacobi1}
\ee
where $D^{j\delta}_{\alpha\beta\gamma}$ are new structure functions. (Were the theory  reducible, other new structure tensors could be present in the solution.)
On the other hand, using this solution in the original equation
\bref{cons jacobi}, one obtains the following condition
on the $D^{j\delta}_{\alpha\beta\gamma}$ :
\be
   \sum_{\rm cyclic \ over \ \ve_1,\ \ve_2,\ \ve_3 }
   \cs_{0,j}\left(B^{ji}_{\alpha\beta\gamma} -
   (-1)^{\epsilon_j ( \epsilon_i + \epsilon_\delta ) }
   R^i_\delta D^{j\delta}_{\alpha\beta\gamma}\right)
   \ve_1^\gamma \ve_2^\beta \ve_3^\alpha = 0
\label{cond sup}
\ee
or, equivalently,
$$   \sum_{\rm cyclic \ over \ \ve_1,\ \ve_2,\ \ve_3 }
   \cs_{0,j}\left(B^{ji}_{\alpha\beta\gamma} +
   (-1)^{\epsilon_i \epsilon_\delta }
   R^j_\delta D^{i\delta}_{\alpha\beta\gamma}  -
   (-1)^{\epsilon_j ( \epsilon_i + \epsilon_\delta ) }
   R^i_\delta D^{j\delta}_{\alpha\beta\gamma}\right)
   \ve_1^\gamma \ve_2^\beta \ve_3^\alpha = 0\;,$$
where we have added vanishing terms. 
Again, the completeness of the generators implies that
the general solution of the preceding equation is of the form
\be
   B^{ji}_{\alpha\beta\gamma} +
   (-1)^{\epsilon_i \epsilon_\delta }
   R^j_\delta D^{i\delta}_{\alpha\beta\gamma} -
   (-1)^{\epsilon_j (\epsilon_i + \epsilon_\delta ) }
   R^i_\delta D^{j\delta}_{\alpha\beta\gamma} =
   - \cs_{0,k} M^{kji}_{\alpha\beta\gamma}
\quad .
\label {jacobi2}
\ee
The reason to include the ``trivial'' second terms is to have nice symmetry properties for the indices $i,j$ of
$M^{kji}_{\alpha\beta\gamma}\,$. 

In this way, the Jacobi identity leads to
the existence of two new gauge-structure tensors
$D^{j\delta}_{\alpha\beta\gamma}$ and
$M^{kji}_{\alpha\beta\gamma}$
which, for a generic theory, are
different from zero and must satisfy
Eqs.\bref{jacobi1} and \bref{jacobi2}.

New structure tensors with increasing numbers of indices are 
obtained from the commutators of more and more gauge transformations.
These tensors are called the structure functions of the gauge algebra
and they determine the nature of the set of gauge transformations
of the theory.
In the simplest gauge theories,
such as Yang-Mills,
they vanish.

For reducible theories, the same procedure as above is also applied 
to the reducibility transformations, which produces
more structure functions and more equations to be 
satisfied for consistency.

\vspace{.2cm}

Having in mind the problem of constructing consistent interactions,
it is obvious that this formalism is highly inadequate to investigate
the most general theories, given the number of structure functions and 
equations that they should satisfy.
In the next section, we will see that the BRST formalism
\cite{Batalin:1981jr,Batalin:1983wj,Batalin:1984jr} is far more convenient.
Indeed, the generic 
gauge-structure tensors then correspond to coefficients of
the expansion of a generating functional
in terms of auxiliary fields.
Furthermore, a single simple equation,
 when expanded in terms of auxiliary fields,
generates the entire set of gauge-structure equations.

\hfill

\section{Fields and Antifields}
\label{s:faf}

\hspace{\parindent}
Consider the classical system defined
in Section \ref{s:ssgt},
described by the action $\cs_0 [ \phi^i ]$
and having gauge invariances.
The field-antifield formalism
was developed to achieve the quantization of this theory in a covariant way. However, at the classical level, it can also be used for the classification of consistent deformations of the theory. As we are interested in the latter, we present only the field-antifield formalism
at the classical level.

The ingredients of the field-antifield formalism are the following:
(i) The original configuration space,
consisting of the $\phi^i$,
is enlarged to include additional fields such as
ghost fields, ghosts for ghosts, etc.
One also introduces antifields for these fields.
(ii) On the space of fields and antifields,
one defines an odd symplectic structure $( \ , \ )$
called the antibracket.
(iii) The classical action $\cs_0$ is extended to $W_0$, which
includes ghosts and antifields.
(iv) The {\it classical master equation}
is defined to be $(W_0,\ W_0)=0$ and the solution
starting as $\cs_0$ is determined.

The action $W_0$ is the generating functional
for the structure functions and the master equation generates 
all the equations relating them.
Hence, the field-antifield formalism is a compact
and efficient way of obtaining
the gauge structure derived in Section \ref{s:ssgt}.

\subsection{Fields and Antifields}
\label{ss:fa}

\hspace{\parindent}
For an irreducible theory 
with $m_0$ gauge invariances,
one introduces $m_0$ ghost fields.
Hence, the field set $\Phi^A$ is
$\Phi^A=\left\{\phi^i, \ {\cal C}^{\alpha_0}_0 \right\}$
where $\alpha_0=1,\ldots,m_0$.
If the theory is first-stage reducible,
there are gauge invariances for gauge invariances
and one introduces ghosts for ghosts.
If there are $m_1$ first-level gauge invariances
then, to the above set of fields,
one adds the ghost-for-ghost fields
$ {\cal C}^{\alpha_1}_1 $ where
$\alpha_1=1,\ldots,m_1$.
In general for an $L$-th stage reducible theory,
the total set of fields $\Phi^A$ is
\be
   \Phi^A=\left\{ \; \phi^i, \  {\cal C}^{\alpha_s}_s; \ \
   s=0,\ldots,L; \ \  \alpha_s=1,\ldots,m_s\right\}
\quad .
\label{field set}
\ee
A graduation
called ghost number
is assigned to each of these fields.
The  fields $\phi^i$ are assigned ghost number zero,
whereas ordinary ghosts have ghost number one.
Ghosts for ghosts,
\ie level-one ghosts,
have ghost number two, etc. So a level-$s$ ghost has
ghost number $s+1$.
Similarly, ghosts have statistics opposite
to those of the corresponding gauge parameter,
but ghosts for ghosts have the same statistics
as the gauge parameter, and so on,
with the statistics alternating for higher-level ghosts.
More precisely,
\be
   {\rm gh} \left[ {\cal C}^{\alpha_s}_s \right] = s+1 \ ,\quad .
\label{ghost numbers and statistics}
\ee
Next, one introduces an antifield $\Phi^*_A$
for each field $\Phi^A$.
The antifields do not have
any direct physical meaning.
They can however be interpreted
as source coefficients
for BRST transformations (see e.g. \cite{Gomis:1994he} for more details).

The ghost number  of $\Phi^*_A$ is
\be
   {\rm gh} \left[ \Phi^*_A \right] =
 - {\rm gh} \left[ \Phi^A   \right] - 1 \ , 
\label{antifield ghost numbers and statistics}
\ee
and its statistics is opposite to that of $\Phi^A$.

One also defines the ``antifield number'' $antif$ by $antif=0$ for the fields $\Phi^A$, and $antif=-gh$ for the antifields. Finally the ``pureghost number'' $puregh$ is defined by $puregh=gh$ for the fields (including ghosts) and $puregh=0$ for the antifields.

\subsection{The Antibracket}
\label{ss:a}

\hspace{\parindent}
In the space of fields and antifields, an antibracket
is defined by \cite{zinnjustin75a,Batalin:1981jr}
\be
 (X,Y) \equiv \frac{\d^R X}{\d\Phi^A}\frac{\d^LY}
  {\d\Phi^*_A}-\frac{\d^R X}{\d\Phi^*_A}
  \frac{\d^L Y}{\d\Phi^A}
\quad .
\label{antibracket def}
\ee
Many properties of $( X , Y )$ are
similar to those of a graded version of the Poisson bracket,
with the grading of $X$ and $Y$
being $\epsilon_X+1$ and $\epsilon_Y+1$
instead of $\epsilon_X$ and $\epsilon_Y$.
The antibracket satisfies
$$
 (Y,X) = -(-1)^{(\epsilon_X+1)(\epsilon_Y+1)}(X,Y)
\quad ,
$$
$$
  \left( \left( X, Y \right) , Z \right) +
  (-1)^{( \epsilon_X + 1 )( \epsilon_Y + \epsilon_Z )}
  \left( \left( Y, Z \right) , X \right) +
  (-1)^{( \epsilon_Z + 1 )( \epsilon_X + \epsilon_Y )}
  \left( \left( Z, X \right) , Y \right) = 0
\ ,
$$
$$
 {\rm gh} [(X,Y)] = {\rm gh}[X] + {\rm gh}[Y] + 1
\quad ,
$$
\be
 \epsilon[(X,Y)] =
  \epsilon_X + \epsilon_Y + 1
  \ ( {\rm mod \  } 2 )
\quad .
\label{antibracket properties}
\ee
The first equation says that $( \ , \ )$ is graded antisymmetric.
The second equation shows that $( \ , \ )$ satisfies
a graded Jacobi identity.
The antibracket ``carries'' ghost number one and
has odd statistics.

The antibracket $(X,Y)$ is also a graded derivation
with ordinary statistics for $X$ and $Y$:
$$
  ( X , YZ ) = ( X , Y ) Z + (-1)^{\eps_Y \eps_Z } ( X , Z ) Y
\quad ,
$$
\be
  ( XY , Z ) = X ( Y , Z ) + (-1)^{\eps_X \eps_Y } Y ( X , Z )
\quad .
\label{bracket derivation}
\ee

The antibracket defines an odd symplectic structure
because it can be written as
\be
 (X,Y) = \frac{\partial^R X}{\partial z^a } \zeta^{ab}
         \frac{\partial^L Y}{\partial z^b }
\ , \quad \quad {\rm where } \quad
  \zeta^{ab} \equiv \left(
  \begin{array}{cc}
   0 & \delta^A_B\\
   -\delta^A_B & 0
  \end{array}\right)
\quad ,
\label{sym form of antibracket}
\ee
when one groups the fields and antifields collectively
into $z^a = \{\Phi^A , \Phi^*_A \}$.
The expression for the antibracket
in Eq.\bref{sym form of antibracket}
is sometimes useful in abstract proofs.

One defines {\it canonical transformations} as the transformations that preserve the antibracket.
They mix the fields and antifields as  $\Phi^A \rightarrow \bar \Phi^A$ and
$\Phi^*_A \rightarrow \bar \Phi^*_A$,
where $\bar \Phi^A$ and $\bar \Phi^*_A$ are functions
of the $\Phi$ and $\Phi^*$.
Similarly to the result of Hamiltonian mechanics, the infinitesimal canonical transformations
\cite{Batalin:1981jr}
have the form
\be
  \bar \Phi^A = \Phi^A +
    \varepsilon \left( \Phi^A , F \right)
    + O(\varepsilon^2)
\ , \quad \quad
  \bar \Phi^*_A = \Phi^*_A +
    \varepsilon \left( \Phi^*_A , F \right)
    + O(\varepsilon^2)
\quad ,
\label{canonical transformation}
\ee
where $F$ is an arbitrary function of the fields and antifields,
with ${\rm gh} [F] = - 1 $ and $ \eps (F) = 1 $.

\subsection{Classical Master Equation}
\label{ss:cmebc}

\hspace{\parindent}
Let $W_0[ \Phi, \Phi^* ]$ be an arbitrary functional of the
fields and antifields, with
the dimensions of an action, and
with ghost number zero and even statistics:
$\eps (W_0) = 0$ and ${\rm gh} [W_0] = 0$.
The equation
\be
   (W_0,W_0)  = 2{\frac{\d^R W_0}
   {\d \Phi^A}}{\frac{\d^L W_0}
    {\d \Phi_A^*}} =0
\label{master equation}
\ee
is called the {\it classical master equation}.

One can regard $W_0$ as an action
for the fields and antifields.
The variations of $W_0$ with respect to
$\Phi^A$ and $\Phi_A^*$
are the equations of motion:
\be
  \frac{\d^L W_0}{\d \Phi^A} = 0 \quad ,\quad
  \frac{\d^L W_0}{\d \Phi^*_A} = 0  
\quad .
\label{eqs of motion for S}
\ee

Not every solution of Eq.\bref{master equation} is of interest.
Usually, only solutions for which
the number of independent nontrivial
gauge invariances
is the number of antifields
 are interesting. They are called {\it proper solutions} (for a precise definition, 
see \cite{Gomis:1994he}).
 
\vspace{.2cm}

To make contact with the original theory,
one looks for a proper solution $W_0$ that contains
the original action $\cs_0 [ \phi ]$ as its 
antifield-independent component:
\be
  \restric{W_0 \left[ \Phi,\Phi^* \right] }{\Phi^*=0} =
   \cs_0 \left[ \phi \right]
\quad .
\label{classical bc}
\ee
An additional requirement is
\be
   \restric{\frac{\d^L \d^R W_0}
    {\d {\cal C}^*_{s-1,\alpha_{s-1}}
    \d {\cal C}^{\alpha_s}_s}}{\Phi^*=0} =
    R^{\alpha_{s-1}}_{s\alpha_s}(\phi)
\ ,\quad \quad s=0,\ldots,L
\quad ,
\label{hessian bc}
\ee
where ${\cal C}^*_{s-1,\alpha_{s-1}}$ is the antifield of
${\cal C}^{\alpha_{s-1}}_{s-1}$:
$    {\cal C}^*_{s,\alpha_{s}} \equiv
    \left( {\cal C}^{\alpha_{s}}_{s} \right)^*
\, .
$
For notational convenience, we have defined
$
  {\cal C}^{\alpha_{-1}}_{-1} \equiv \phi^i
 \ , \
  {\cal C}^*_{-1,\alpha_{-1}} \equiv \phi^*_i
\ ,\
 {\rm with \ } \alpha_{-1} = i
\,.
$
Actually, Eq.\bref{hessian bc} does not need to be imposed
as a separate condition.
Although it is not obvious,
the requirement of being proper
 and
the condition \bref{classical bc}
necessarily imply that a solution $W_0$ must
satisfy Eq.\bref{hessian bc}
\cite{Batalin:1985qj}. 
Comments on the unicity of such a solution follow below.

\subsection{The Proper Solution and the Gauge Algebra}
\label{ss:psga}

\hspace{\parindent}
The proper solution $W_0$ is
the generating functional for the structure
functions of the gauge algebra.
Indeed, all relations among the structure functions
are contained in Eq.\bref{master equation},
thereby reproducing the equations
of Section \ref{ss:gs}
and generalizing them
to the generic $L$-th stage reducible theory.
Let us sketch the connection between the proper solution
of the classical master equation, the gauge-structure tensors
and the equations that the latter
must satisfy.

The proper solution $W_0$ can be expanded as a power series
in the ghosts and antifields.
Given the conditions \bref{classical bc}
and \bref{hessian bc},
the expansion necessarily begins as
\be
     W_0 \left[ \Phi,\Phi^* \right] = \cs_0 \left[ \phi \right]
     +\sum_{s=0}^L {\cal C}^*_{s-1,\alpha_{s-1}}
       R^{\alpha_{s-1}}_{s\alpha_s}
      {\cal C}^{\alpha_s}_s +O(C^{*2})
\quad .
\label{beg fsr proper solution}
\ee
For the further terms, let us consider an irreducible theory, for which 
the set of fields is $\phi^i$ and 
${\cal C}^{\alpha_0}_0$ (which we call ${\cal C}^\alpha$).
Most generally, one has
\bqn
  W_0\left[ {\Phi ,\Phi^*} \right]& =&
  \cs_0\left[ \phi  \right] +
  \phi_i^*R_\alpha^i{\cal C}^\alpha +
  {\cal C}_\alpha^* 
   T_{\beta\gamma }^\alpha
   {\cal C}^\gamma
    {\cal C}^\beta
\nonumber \\
&&  +\phi_i^*\phi_j^* 
    E_{\alpha \beta }^{ji}
    {\cal C}^\beta {\cal C}^\alpha 
  +{\cal C}_\delta^*\phi_i^*
     D_{\alpha \beta \gamma }^{i\delta }
    {\cal C}^\gamma {\cal C}^\beta {\cal C}^\alpha 
\nonumber
\\
 && + \phi_i^*\phi_j^*\phi_k^*
    M_{\alpha \beta \gamma }^{kji}
   {\cal C}^\gamma {\cal C}^\beta {\cal C}^\alpha +\ldots
\quad ,\label{fsr proper solution}
\eqn
where,
with the exception of $R_\alpha^i$
which is fixed by (\ref{hessian bc}),
 the tensors 
$T_{\alpha \beta }^\gamma$,
$E_{\alpha \beta }^{ji}$,
etc.\ in Eq.\bref{fsr proper solution}
are {\it a priori} unknown.
However, inserting the above expression for $W_0$ into the classical master equation \bref{master equation}, one finds that the latter
is satisfied if the tensors,
$T_{\alpha \beta }^\gamma$,
$E_{\alpha \beta }^{ji}$, etc.\ in Eq.\bref{fsr proper solution}
are the ones
of Section \ref{ss:gs} (up to some irrelevant signs and numerical factors).
In other words,
Eq.\bref{fsr proper solution}
with the tensors identified
as the ones
of Section \ref{ss:gs}
is a proper solution of the master equation.
The result is similar for gauge theories with reducibilities.

The reason why one equation $(W_0,W_0)=0$ is able to generate
many equations is that the coefficients
of each ghost and antifield term must vanish separately.
Summarizing,
the antibracket formalism
using fields and antifields
allows a simple determination of the relevant
gauge structure tensors.
The proper solution to the classical master equation
provides a compact way of
expressing the relations among the structure tensors.


\vspace{.2cm}

One might wonder whether there always exists a proper solution
to the classical master equation
and whether the proper solution is unique.
Given reasonable conditions,
there always exists a proper solution with the required 
ghost-independent piece. This was proved
in \cite{Voronov:1982cp,Batalin:1985qj}
for the case of an irreducible theory
and in \cite{Fisch:1989rp}
for a general $L$-th stage reducible theory.
Furthermore, as was shown in 
\cite{Voronov:1982cp,Batalin:1985qj,Fisch:1989rp}, 
given the  set of fields \bref{field set},
the proper solution
of the classical master equation is unique
up to canonical transformations.
Indeed, if one has found a proper solution $W_0$
such that $(W_0,W_0)=0$ and  performs
an infinitesimal canonical transformation,
the transformed proper solution 
$W_0^\prime = W_0 + \varepsilon (W_0, F) + O(\varepsilon^2) $
also satisfies the master equation and the condition 
\bref{classical bc} up to field redefinitions (see Section \ref{cons}).
Actually, canonical transformations
correspond to the freedom
of redefining the fields and the gauge generators, which was already mentionned
in the end of Section \ref{ss:gt}.

\subsection{The Classical BRST Symmetry}
\label{ss:brsts}

\hspace{\parindent}
Via the antibracket, the proper solution $W_0$
is the generator  of the so-called BRST symmetry $s$.
Indeed, one defines the  BRST transformation
of a functional $X$ of fields and antifields by
\be
  s X \equiv \left( {W_0,X} \right)
\quad .
\label{def of BRST}
\ee
The transformation rule for fields
and antifields is therefore
\be
  s \Phi^A=-{\frac{\d^R W_0} {\d \Phi_A^*}}
\ , \quad \quad
 s \Phi_A^* =
   {\frac{\d^R W}  {\d \Phi^A}} 
\quad .
\label{BRST trans of fields and antifields}
\ee
The field-antifield action $W_0$ is
 BRST-symmetric
\be
 s   W_0=0
\label{BRST inv of action}
\ee
as a consequence of
$
  \left( {W_0,W_0} \right) = 0
$.

The BRST-operator $s$
is a nilpotent graded derivation:
Given two functionals $X$ and $Y$,
$$
   s \left( {XY} \right) =
  (s X)  Y+
   (-1)^{\epsilon_X}Xs Y
\quad .
\label{der prop of BRST}
$$
and
\be
s^2 X=0
\quad .
\label{nilpotency of BRST}
\ee
The nilpotency follows from two properties
of the antibracket: the graded Jacobi identity
and the graded antisymmetry
(see Eq.\bref{antibracket properties}).

\subsection{Algebraic structure}

The algebraic structure of the field-antifield formalism 
is related to two crucial ingredients of the BRST-differential: 
the Koszul-Tate resolution $\d$, generated by the antifields,
 which implements the equations of motion in (co)homology;
 and the longitudinal exterior derivative $\g$, which implements 
gauge invariance. These operators are the first components in 
the decomposition of the BRST-differential $s$  according to 
the antifield number:
$$s=\d +\g+s_1+ "higher \, order"\,,$$ where $\d$ has $antif= -1$, 
$\g$ has $antif= 0$, $s_1$ has $antif=1$ and $"higher \, order"$ has $antif \geq 1$.
The complete action of the operators $\d$ and $\g$ on the fields and antifields can be found in \cite{Henneaux:1992ig}; let us just mention to illustrate the above statements that 
$$\d \phi^*_i=\frac{\d {\cal L}}{\d \phi^i}\;, \hspace{2cm} 
\g \phi^i=R^i_{\alpha}  {\cal C}^{\alpha}_1\;.$$
Explicit examples of these operators will be given in the chapters \ref{spin2} and \ref{spin3}.

From the nilpotency of $s$, one deduces that the Koszul-Tate resolution is a differential, $\d^2=0\,$. Furthermore, 
\bqn
&\g \d+ \d \g=0&\\
&\g^2=-(\d s_1+ s_1 \d)\,.&
\eqn

\subsection{Definitions and general theorems}
\label{genthm}

In this section, we provide definitions and introduce some further notations. We also state useful general theorems, the proof of which can be found in \cite{Henneaux:1992ig,Barnich:1994db} and references therein. They concern the cohomology groups involving the total derivative $d$ and the Koszul-Tate differential $\d$.

\vspace{.2cm}

One of the key assumption used in the sequel is locality. 
A local function of some set of fields $\phi^i$ is a smooth function
of the fields $\phi^i$ and their
derivatives $\partial\phi^i$, $\partial^2\phi^i$, ... up to some
{\it finite} order, say $k$, in the number of derivatives.
Such a set of variables $\phi^i$, $\partial\phi^i$, ...,
$\partial^k\phi^i$ will be collectively denoted by $[\phi^i]$.
Therefore, a local function of $\phi^i$ is denoted by $f([\phi^i])$. A local
$p$-form $(0\leq p \leq n)$ is a differential $p$-form the components of which are local
functions: 
\bqn \omega =
\frac{1}{p!}\,\omega_{\m_1\ldots\m_p}(x, [\phi^i])\,
dx^{\m_1} \wedge \cdots \wedge dx^{\m_{p}}\,.\nonumber 
\eqn
A local functional is the integral of a local $n$-form.

If $A$
is a local functional
that vanishes for all allowed field configurations, $A=\int a =0$, then,
the $n$-form $a$
is a ``total derivative'', $a=dj$, where $d$ is the space-time exterior
derivative (see e.g. \cite {Henneaux:1992ig}, Chapter 12). That is,
one can ``desintegrate''
equalities involving local functionals but the integrands are determined up
to $d$-exact terms.

\vspace{.2cm}

Let us now recall the definition of a cohomology group.
Consider  operators $\co , \cp$ acting within a space $E\,$, and  let $e,f$ be elements of $E$.
\begin{itemize}
\item
Elements $e$ that are annihilated by $\co$, $\co e=0$ , are called {\it cocycles}, or $\co$-cocycles.
\item
Elements $e$ that are in the image of $\co$, $e=\co f$, are called {\it coboundaries}, or $\co$-coboundaries. They are also said to be $\co$-exact.
\item
The cohomology group of $\co$ in the space $E$, denoted $H(\co, E)$, is the group of equivalence classes of cocycles of $E$, where two elements are equivalent if they differ by a coboundary:
$$ H(\co,E)= \{ e \in E \; \vert \; \co e=0\;,\; e \sim e + \co f \;,\, f\in E
\}\;.$$
When the space $E$ in which the operators act is unambiguous, the reference to $E$ is often dropped: $H(\co,E)$ is written $H(\co)\,$.
\item
If a cohomology group is denoted $H(\co \vert \cp, E)$, then all relations are ``up to $\cp$-exact terms'':
$$ H(\co \vert \cp,E)= \{ e \;\in E\;\vert \; \co e=\cp f\;,\; e \sim e + \co f+ \cp g \;,\;  f,\,g \in E
\}\;.$$
\end{itemize}

\vspace{.2cm}

We now turn to the general theorems. The space in which these cohomology groups are computed is the space of local forms depending on the space-time coordinates, the fields and the antifields.
The supscript $p$ of a cohomology group $H^p_k(\ldots)$ denotes the form-degree, while the subscript $k$ denotes the antifield number.

\begin{theorem}\label{acycl}{\em{\bf (Acyclicity)}:}
The cohomology of the Koszul-Tate differential is trivial in strictly positive antifield number: 
\be H_k(\d)=0\,,\;k>0 \,. \ee
\end{theorem}

\begin{theorem}
\label{d_cohomology}{\em{\bf (Algebraic Poincar\'e lemma)}:}
The cohomology of $d$ in the algebra of local $p$-forms is given by
\begin{eqnarray}
H^0(d)\simeq {\bf R},\nonumber\\
H^k(d)=0\ for\ k\neq 0,\ k\neq n,\nonumber\\
H^n(d)\simeq  space\ of\ equivalence\ classes\ of\ local\ n-forms,
\end{eqnarray}
where two local $n$-forms $\alpha = f dx^0\dots dx^{n-1}$ and
$\alpha^\prime=f^\prime dx^0\dots dx^{n-1}$
are equivalent if and only if $f$ and $f^\prime$ have identical 
Euler-Lagrange derivatives with respect to
all the fields and antifields,
\begin{eqnarray}
\frac{\delta (f-f^\prime)}{\delta\phi^A} =0=\frac{\delta (f-
f^\prime)}{\delta\phi^*_A}\Longleftrightarrow
\alpha \ and\ \alpha^\prime\ are\ equivalent.
\end{eqnarray}
\end{theorem}
\noindent
Note that if one does not allow for explicit coordinate dependencies, then the
groups $H^k(d)$ no longer vanish for $k\neq 0$ and $k\neq n\,$. Indeed, in that case, constant forms are not  $d$-exact; so  $H^k(d)$ is isomorphic to the set of constant $k$-forms.

\begin{theorem}
\label{deltad_cohomology}{\em{\bf :}}
In the algebra of local forms,
\begin{eqnarray}
H_k(\delta|d)=0
\end{eqnarray}
for $k>0$ {\underbar {and}} pureghost number $>0$.
\end{theorem}

\begin{theorem}\label{descent_eq_deltad}{\em{\bf :}}
if $p\geq 1$ and $k>1$, then
\begin{eqnarray}
H^p_k(\delta|d)\simeq H^{p-1}_{k-1}(\delta|d)\label{descent_iso}.
\end{eqnarray}
\end{theorem}

\begin{theorem}\label{gen_sym}{\em{\bf :}}
if $p\geq 1$ and $k\geq 1$ with $(p,k)\neq (1,1)$, then
\begin{eqnarray}
H^p_k(\delta|d)\simeq H^{p-1}_{k-1}(d|\delta)\label{inversion}
\end{eqnarray}
Furthermore,
\begin{eqnarray}
H^1_1(\delta|d)\simeq H^0_0(d|\delta)/{\bf R}.
\end{eqnarray}
\end{theorem}
\noindent If one does not allow for an explicit $x$-dependence in the local
forms, then, (\ref{inversion})
must be replaced by $H^p_1(\delta|d)\simeq H^{p-
1}_{0}(d|\delta)/\{constant\ forms\}$ for $k=1\,$.

\begin{theorem}\label{9.1}{\em{\bf :}}
For a linear gauge theory of reducibility order $r$, one has,
\begin{eqnarray}
H^n_j(\delta|d)=0,\qquad j>r+2
\end{eqnarray}
whenever $j$ is strictly greater than $r+2$ (we set $r=-1$ for a theory
without gauge freedom).
\end{theorem}

\begin{theorem}\label{quadr}{\em{\bf :}}
for linear gauge theories, there is no nontrivial element of
$H^n_2(\delta|d)$ that is purely
quadratic in the antifields $\phi^*_i$ and their derivatives.
That is, if $\mu$ is quadratic in
the antifields $\phi^*_i$ and their derivatives and if
$\delta\mu + d b =0$ then
$\mu =\delta C+d V$.
\end{theorem}

Let us now introduce some definitions and notations related to $H(\g)$, the space of solutions of $\g a
= 0$ modulo trivial coboundaries of the form $\g b$. 
Elements of $H(\g)$ are called ``invariants'' and often denoted by Greek letters. To understand the terminology,
remember that the operator $\g$ implements the gauge invariance in the field-antifield formalism.

Let $\left\{\omega^I\right\}$ be
 a basis of the algebra of polynomials in the ghosts of $H(\g)$.
 Any element of $H(\g)$ can be decomposed in
 this basis, hence for any $\g$-cocycle $\a$
 \be\gamma \a=0 \quad\Leftrightarrow\quad
 \a=\a_I\;
 \omega^I + \g \b
 \label{gammaa1}\ee where the $\a_I$ depend only on (a subset of) the field $\phi$, the antifields and their derivatives. 
If $\a$ has a finite ghost number and a bounded number of derivatives, then the $\a_I$ are polynomials. For this reason, the $\a_I$ are often referred to as {\it invariant polynomials}. 
 An obvious property is that $\a_I\omega^I$ is $\g$-exact if and only if all the
 coefficients $\a_I$ are zero \be \a_I\omega^I=\gamma\b,\quad
 \Leftrightarrow\quad \a_I=0,\quad\mbox{for
 all}\,\,I.\quad\label{gammab1}\ee 

Other useful concepts are the $D$-differential and the $D$-degree. The  differential $D$ acts on the field $\phi$ and on the antifields in the same way as $d$, while its action on the ghosts is determined by the two following conditions: (i)
the operator $D$ coincides with $d$ up to $\g$-exact terms and (ii)  $D\o^J=A^J_{~I}\o^I$ for some matrix $A^J_{~I}$ that involves the $dx^\m$.
A grading is associated with the $D$-differential, the $D$-degree.
The $D$-degree is chosen to be zero for elements that do not involve derivatives of the ghosts. It is defined so that it is increased by one by the action of the $D$-differential on ghosts. Explicit examples of the $D$-differential and the $D$-degree will follow in Chapters \ref{spin2} and \ref{spin3}.


%% file: deformation.tex
\section{Construction of interactions}
\label{cons}

The purpose of this section is to analyse the  problem of
constructing
consistent local interactions among fields with a gauge freedom in the light of
the antibracket formalism. 
This formulation has been used to solve the question of consistent
self-interactions in flat background in several cases:
for vector  gauge fields in
\cite{Barnich:1993pa}, for $p$-forms  in
\cite{Henneaux:1997ha}, for Fierz-Pauli  in
\cite{Boulanger:2000rq}, for  $[p,q]$-fields ($p>1\,$) in
\cite{Bekaert:2002uh,Bizdadea:2003ht,Boulanger:2004rx,Bekaert:2004dz}
and for spin-3 fields in \cite{Bekaert:2005jf,Boulanger:2005br}
. The results for the latter $[p,q]$-fields ($p>1\,$) and spin-3 fields are presented in the chapters \ref{spin2} and \ref{spin3}. 

\vspace{.2cm}

The problem of
constructing
consistent local interactions can be economically reformulated as a
deformation problem, namely that of
{\em{deforming
consistently the master equation}}.  
Consider the
``free'' action
$\cs_0 [\phi^i]$ with ``free'' gauge symmetries
\begin{equation}
\delta_\varepsilon \phi^i = \stackrel{\smash{(0)}}{R}^i_\alpha
\varepsilon^\alpha ,
\end{equation}
\begin{equation}
\stackrel{\smash{(0)}}{R}^i_\alpha
\frac{\delta \cs_0}{\delta\phi^i}=0\ .
\end{equation}
 One wishes to introduce consistent interactions, \ie to modify $\cs_0$
\begin{equation}
\cs_0\longrightarrow \cs = \cs_0 +
g \cs_1 +
g^2 \cs_2 + ...\label{fullaction}
\end{equation}
in such a way that one can consistently deform the original gauge symmetries,
\begin{equation}
\stackrel{\smash{(0)}}{R}^i_\alpha \longrightarrow R^i_\alpha =
\stackrel{\smash{(0)}}{R}^i_\alpha
+ g \stackrel{\smash{(1)}}{R}^i_\alpha +
g^2 \stackrel{\smash{(2)}}{R}^i_\alpha
+ ...\label{fullsymmetries} .
\end{equation}
The deformed gauge transformations
$\delta_\varepsilon \phi^i =
R^i_\alpha \varepsilon^\alpha$ 
are called ``consistent'' if they are
gauge symmetries of the full
action (\ref{fullaction}),
\begin{equation}
(\stackrel{\smash{(0)}}{R}^i_\alpha
+ g \stackrel{\smash{(1)}}{R}^i_\alpha + g^2 \stackrel{\smash{(2)}}
{R}^i_\alpha + ...)
\frac{\delta(\cs_0 + g\cs_1 +
g^2\cs_2 + ...)}{\delta\phi^i}=0 \label{fullnoetheridentities} \quad.
\end{equation}
This implies automatically that the modified gauge transformations close
on-shell
for the interacting action (see \cite{Henneaux:1992ig}, Chapter 3).
If the original gauge transformations are reducible,
one should also demand that
(\ref{fullsymmetries}) remain reducible. 
Indeed, the deformed theory should
possess the same number of (possibly deformed) independent
gauge symmetries, reducibility identities, {\it etc.}, as the system one
started with, so that the number of physical degrees of
freedom is unchanged.

The deformation procedure is perturbative: one tries to construct the interactions order by order in the deformation parameter $g\,$.

A trivial type of consistent interactions is obtained by making field
redefinitions
$\phi^i\longrightarrow\bar{\phi}^i = \phi^i + g F^i + ...$  .
One gets
\begin{equation}
\cs_0 [\phi^i]\longrightarrow \cs[\bar{\phi}^i] \equiv \cs_0[\phi^i[\bar{\phi}^i]]
=\cs_0[\bar{\phi}^i -g F^i + ...] =
\cs_0 [\bar{\phi}^i]- g \,\cs_{0,i}\,F^i
+ ...\label{trivialinteractions}\ .
\end{equation}
Since interactions that can be eliminated by field redefinitions are usually
thought of
as being no interactions, one says that a theory is rigid if the only
consistent deformations are
proportional to $\cs_0$ up to field redefinitions.
In that case, the interactions can be summed as
\begin{equation}
{\cs_0}\longrightarrow \cs = (1 + k_1 g + k_2 g^2 + ...)\,
{\cs_0}
\end{equation}
and simply amount to a change of the coupling constant in front of the
unperturbed action.

The problem of constructing consistent interactions is a complicated one
because one must
simultaneously modify ${\cs_0}$ and $\stackrel{\smash{(0)}}{R}^i_\alpha$
in such a way that
(\ref{fullnoetheridentities}) is valid order by order in $g$.
One can formulate economically the problem in terms of the
solution $W_0$ of the master equation. Indeed, if the interactions can be
consistently
constructed, then the solution $W_0$ of the master equation
for the free theory
can be deformed into the solution $W$ of the master equation for the
interacting theory,
\begin{equation}
W_0 \longrightarrow W = W_0 +g W_1 + g^2 W_2 + ...
\end{equation}
\begin{equation}
(W_0,W_0)=0\longrightarrow
(W,W)=0 \label{fullmasterequation}.
\end{equation}
The master equation $(W,W)=0$ guarantees that the consistency requirements
on $\cs$
and $R^i_\alpha$ are fulfilled.

The master equation for $W$ splits according to the deformation parameter
$g$ as
\begin{eqnarray}
(W_0,W_0)&= 0 \label{deformation1}\\
2(W_0,W_1)&= 0 \label{deformation2}\\
2(W_0,W_2) + (W_1,W_1)&= 0 \label{deformation3}\\
&\vdots\ \ \ .\nonumber
\end{eqnarray}
The first equation is satisfied by assumption, while the second implies that
$W_1$
is a cocycle for the free BRST-differential $s\equiv
(W_0,\cdot)$.

Suppose that $W_1$ is a coboundary, $W_1=(W_0,T)$.
This corresponds to a trivial deformation because ${\cs_0}$
is then modified
as in (\ref{trivialinteractions})
\begin{eqnarray}
{\cs_0}\longrightarrow
\cs_0 +g\ [(W_0,T)]_{\Phi^*=0}&=&
\cs_0
+g\ \Big[\frac{\d^RW_0}{\d \Phi^A}\frac{\d^LT}{\d \Phi^*_A}
-\frac{\d^RW_0}{\d \Phi^*_A}\frac{\d^LT}{\d \Phi^A}\Big]_{\Phi^*=0}\nnn
&=&{\cs_0} + g \ 
\frac{\delta^R \cs_0}{\delta\phi^i}\
\Big[\frac{\delta^L {T}}{\delta\phi^*_i}\Big]_{\Phi^*=0}
\end{eqnarray}
(the other modifications induced by ${T}$ affect the terms with ghosts, \ie the higher-order
structure functions which carry some intrinsic ambiguity \cite {Henneaux:1989jq}). Trivial deformations thus correspond to $s$-exact quantities, \ie  trivial elements of the cohomological
space
$H({s})$ of the undeformed theory in ghost number zero.
Since the deformations must be $s$-cocycles, nontrivial deformations are thus determined by the equivalence classes of 
$H({s})$  in ghost number zero.

The next equation, Eq.(\ref{deformation3}), implies that $W_1$
should be such that
$(W_1,W_1)$ is trivial in
$H({s})$ in ghost number one.

\vspace{.2cm}

We now wish to implement locality in the analysis. The deformation of the gauge transformations, {\it etc.},  must be local functions, as well as the field redefinitions. If this were not the case, the deformation procedure would not provide any constraint (see \cite{Barnich:1993vg,Henneaux:1997bm}).

Let $W_k=\int {\cal{L}}_k$ where
${\cal{L}}_k$ is a local $n$-form, which thus depends on the field variables and only a finite number
of their derivatives. We also denote by $(a,b)$ the antibracket for 
$n$-forms, \ie,
\begin{equation}
(A,B) = \int \, (a,b) \label{localantibracket}
\end{equation}
if $A = \int a$ and $B = \int b$. Because $(A,B)$ is a local functional,
there exists $(a,b)$
such that Eq.(\ref{localantibracket}) holds, but $(a,b)$ is defined only up
to $d$-exact terms.
This ambiguity plays no role in the subsequent developments.
The equations (\ref{deformation2}-\ref{deformation3}) for $W_k$
read
\begin{eqnarray}
2 s{\cal{L}}_1&=&d j_1
\label{localdeformation}\\
s{\cal{L}}_2 +
({\cal{L}}_1,{\cal{L}}_1)
&=&dj_2\\
&\vdots&\nonumber
\end{eqnarray}
in terms of the integrands ${\cal{L}}_k$.
The equation (\ref{localdeformation}) expresses that
${\cal{L}}_1$ should
be BRST-closed modulo $d$ and again, it is easy to see that a BRST-exact
term modulo $d$
corresponds to trivial deformations. Nontrivial local deformations of the
master equation are
thus determined by
$H^{n,0}({s}|d)$, the cohomology of the BRST-differential $s$ modulo the
total derivative $d\,$, in maximal form-degree $n$ and in ghost
number $0\,$.

\subsection{Computation of $H^{n,0}(s \vert\, d)$}
\label{ss:coh}

The purpose of this
section is  to show how to compute $H^{n,0}(s \vert\, d)$. 
Although this cohomology depends on the theory at hand, one can provide a general framework to compute it,  assuming some properties that have to be proved separately for each theory.
They are the following:
\begin{itemize}
\item[(i)]
The BRST-differential decomposition in antifield number reads $s=\g + \d \,,$ \ie all higher-order components vanish. 
The operator $\g$ then satisfies the nilpotency relation  \be  \gamma^2 = 0\,. \ee
\item[(ii)]
If $a$ has strictly positive antifield number (and involves
 possibly the ghosts), the equation $\gamma a + d b = 0$ is
 equivalent, up to trivial redefinitions, to $\gamma a = 0.$ That
 is, if $antif(a)>0\,,$ then
 \begin{equation}
 \gamma a + d b = 0
 \Leftrightarrow 
a=a'+dc\,,\ 
 \gamma a' = 0\,.\label{blip1}
 \end{equation}
\item[(iii)]
At given pureghost number, there is an upper bound on the $D$-degree defined at the end of Section \ref{genthm}.
\end{itemize}
If the above properties are verified by the theory at hand, one can compute $H^{n,0}(s \vert\, d)$ in the following way.

One must find the general solution of
the cocycle condition \be s a^{n,0} + db^{n-1,1} =0, \label{coc1}
\ee where $a^{n,0}$ is a topform of ghost number zero and
$b^{n-1,1}$ a $(n-1)$-form of ghost number one, with the
understanding that two solutions of Eq.(\ref{coc1}) that differ by a
trivial solution should be identified \bqn a^{n,0}\sim a^{n,0} + s
m^{n,-1}  + dn^{n-1,0} \nonumber \eqn as they define the same
interactions up to field redefinitions (\ref{trivialinteractions}). The
cocycles and coboundaries $a,b,m,n,\ldots\,$ are local forms of
the field variables (including ghosts and antifields)

\vspace{.2cm}

Let $a^{n,\,0}$ be a solution of Eq.(\ref{coc1}) with
ghost number zero and form-degree $n$. For convenience, we will
frequently omit to write the upper indices. One can expand
$a$($=a^{n,\,0})$ as $a=a_0+a_1 + \ldots + a_k$  where $a_i$ 
has antifield number $i$. The expansion can be assumed to stop at
some finite value of the antifield number under the sole
hypothesis of locality \cite{Barnich:1994db, Barnich:1994mt} or Chapter 12 of \cite{Henneaux:1992ig}. 
One can also expand $b$ according to the antifield number: $b = b_0 + b_1 + ...+b_j \, $. This  expansion  also stops at some finite antifield number by locality.

Using the decomposition of the BRST-differential as $s=\g + \d$ and separating the components of different antifield number,
the equation $sa+db=0$ is equivalent to \bqn
\d a_1 + \g a_0 + d b_0 &=&0 \,,\nonumber \\
\d a_2 + \g a_1 + d b_1 &=&0 \,,\nonumber \\
&\vdots &\nonumber \\
\d a_k + \g a_{k-1} + d b_{k-1}&=&0 \,,\nonumber \\
\g a_k&=&0 \,.\label{descente1} \eqn 
Without loss of generality, we have assumed that $b_j=0$ for
$j\geq  k$. Indeed, if $j>k$ the last equation is $db_j=0$ and implies $b_j=dc_j$ by the algebraic Poincar\'e lemma (Theorem \ref{d_cohomology}), as $b$ is a $(n-1)$-form. One can thus absorb $b_j$ into a redefinition of $b$. If $j=k$, the last equation is $\g a_k+db_k=0$. Using the property (\ref{blip1}), it can be rewritten as $\g a_k=0$ modulo a field redefinition of $a$: $a\rightarrow a +dc$ for some $c$.

The next step consists in the analysis of the term $a_k$ with highest antifield
number and the determination of whether it can be removed by trivial
redefinitions or not. We here show that
the  terms $a_k$ ($k>1$) may be discarded one after another from the aforementioned descent if the cohomology group  $H^{inv}_k(\delta \vert d)$ vanishes. (The group $H^{inv}_k(\delta \vert d)\equiv H_k(\d \vert d ,H(\g))$ is the space of invariants $a_k$ of antifield number $k$ that are solutions of the equation $\d a_k+db=0$, modulo trivial coboundaries $\d m+dn$ where $m$ and $n$ are invariants.)
This result is independent of any condition on the number of
 derivatives or of Poincar\'e invariance.

The last equation of the descent (\ref{descente1}) implies that $a_k=\a_J\,\o^J$ where $\a_J$ is an invariant
polynomial and $\o^J$ is a polynomial in the  ghosts of $H(\g)$, up to a trivial term $\g c$ that can be removed by the trivial redefinition $a\rightarrow a-sc\,$. 

One now considers the next equation of the descent, $\d a_k + \g a_{k-1} + d b_{k-1}=0 \,.$ Acting with $\g$ on it and using $\g^2=0$, one gets $d\g b_{k-1}=0$, which, by the Poincar\'e lemma and (\ref{blip1}), implies that $b_{k-1}$ is also invariant: $b_{k-1}=\b_J\, \o^J \,.$
Substituting the expressions for $a_k$ and $b_{k-1}$ into the equation yields
$\d(\a_J\,\o^J)+D(\b_J \,\o^J )=\g(\ldots)\,,$ or, using (\ref{gammab1}),
$$\d(\a_J)\,\o^J+D(\b_J \,\o^J )=0\,.$$ 
To analyze this equation, one expands it according to the $D$-degree. The term of degree zero reads $$ \d(\a_{J_0})+d(\b_{J_0}  )=0\,,$$ where $J_i$ labels the $\o^J$ of $D$-degree $i$.
If the cohomology group $H^{inv}_k(\delta \vert d)$ vanishes, then the solution to this equation is
$\a_{J_0}=\d \m_{J_0}+d\n_{J_0}\,,$ where $\m_{J_0}$ and $\n_{J_0}$ are invariants. The $D$-degree zero component of $a_k$, denoted $a_k^0$, then reads 
$$a_k^0= (\d \m_{J_0}+d\n_{J_0})\o^{J_0}\,.$$
This is equal to $s(\m_{J_0}\o^{J_0})+ d(\n_{J_0}\o^{J_0})$ up to terms arising from $d\o^{J_0}$, which can be written as $d\o^{J_0}=  D\o^{J_0}+\g u^{J_0}=A^{J_0}_{J_1} \o^{J_1}+\g u^{J_0}\,.$ The term $\n_{J_0}A^{J_0}_{J_1} \o^{J_1}$ has $D$-degree one and can be removed by redefining $a^1_k$. The term $\n_{J_0}\g u^{J_0}$ differs from $s (\n_{J_0} u^{J_0})$ by a term of lower antifield number ($\sim \d(\n_{J_0})u^{J_0}$), it can thus be removed by a redefinition of $a_{k-1}\,$.

With the same procedure, one can successively remove all the terms with higher $D$-degree, until one has completely redefined away $a_k\,$. 
One might wonder if the number of redefinitions needed is finite, but this is secured by the fact that at given pureghost number there is an upper limit for the $D$-degree. Remember that one should check the latter property for the theory at hand.

We stress  that the crucial ingredient for the removal of $a_k$ is the vanishing of the cohomology group $H^{inv}_k(\delta \vert d)\,.$ 
More precisely, if one looks for Poincar\'e-invariant theories, it is enough that there be no nontrivial elements without explicit $x$-dependence in $H^{inv}_k(\delta \vert d)\,.$ Indeed, the Lagrangian (\ie $a_0$) of a Poincar\'e-invariant theory should not depend explicitely on $x$ and it can be shown \cite{Henneaux:1992ig} that then  the whole cocycle $a=a_0+a_1 +
\ldots + a_k$ satisfying $sa+db=0$ can be chosen $x$-independent  (modulo
trivial redefinitions).

\vspace{.2cm}

The next steps depend too much on the studied theory to be explained here. They are left for the next chapters, in which particular cases are treated.


%% file: 2colldif.tex
\chapter{Interactions for exotic spin-2 fields}
\label{spin2}

In this chapter, we address the problem of switching
on consistent self-interactions in flat background 
among exotic spin-2 tensor gauge
fields, the symmetry of which is characterized by the Young diagram $[p,q]$
with $p>1\,$. We do not consider the case $p=q=1\,$, which corresponds to the usual graviton. The physical degrees of freedom of such theories correspond to a traceless tensor carrying an irreducible representation of $O(n-2)$ associated with the Young diagram $[p,q]$. Therefore, we work in space-time dimension $n\geq p+q+2\,$. Indeed, there are no propagating degrees of freedom when $n<p+q+2\,$.
We use the BRST-cohomological reformulation of the Noether method
for the problem of consistent interactions, which has been developped in Section \ref{cons}. For
an alternative Hamiltonian-based deformation point of view, we suggest the reference
\cite{Bizdadea:2001re}.

The main (no-go) result  \cite{Bekaert:2002uh,Bizdadea:2003ht ,Boulanger:2004rx,Bekaert:2004dz} proved in this chapter  can be stated as follows, spelling out explicitly
the assumptions:

{\textit{In flat space and under the assumptions of locality and
translation invariance, there is no consistent smooth deformation of the
free theory for $[p,q]$-type tensor gauge fields with $p>1$ that
modifies the gauge algebra, which remains Abelian.
Furthermore, for $q>1$, when there is no positive integer $s$ such that
$p+2=(s+1)(q+1)$, there exists no smooth deformation that alters the
gauge transformations either. Finally, if one excludes deformations that
involve more than two derivatives in the Lagrangian and that 
are not Lorentz-invariant, then the only smooth
deformation of the free theory is a cosmological-constant like term for
$p=q\,.$}}

One can reformulate this result in more
physical terms by saying that no analogue of Yang-Mills nor
Einstein theories seems to exist for more exotic fields (at least
not in the range of local perturbative theories).

 Without the extra condition on the derivative order, one can e.g.
 introduce Born-Infeld-like interactions that involve powers of the
 gauge-invariant curvatures $K\,$, but modify neither the gauge algebra nor the gauge
 transformations. 
When involving other fields, nontrivial interactions are also possible. Indeed, one can build interactions that couple $[p,q]$-fields and $p'$-forms generalizing the Chapline-Manton interaction among $p$-forms (see Appendix \ref{chapline}). The latter interactions do not modify the gauge transformations of the spin-2 field but those of the $p'$-form.
 No general systematic analysis has yet been done about
interactions modifying the gauge transformations of the exotic spin-2 field when coupling them with different $[p,q]$-type fields (where ``different'' means e.g. $[p_1,q_1] \neq [p_2,q_2]$),
or with other types of fields.

\vspace*{.3cm}

This chapter is organized as follows. In Section \ref{freetheory}, we
review the free theory of $[p,q]$-type tensor gauge fields. In
Section \ref{BRST}, we construct the BRST spectrum and differentials for the
theory. Sections \ref{cohogamma} to
\ref{Invariantcharacteristiccohomology} are devoted to the proof
of cohomological results. We compute $H(\g)$ in Section
\ref{cohogamma}, an invariant Poincar\'e lemma is proved in
Section \ref{InvariantPoincarelemma}, the  cohomologies $H_k^n(\d
\vert d)$ and $H_k^{n,\, inv}(\d \vert d)$ are computed
respectively in Sections \ref{Characteristiccohomology} and 
\ref{Invariantcharacteristiccohomology}, and partly in the appendix \ref{append}.
 The self-interaction
question is answered in Section \ref{self-interactions}.

\section{Free theory}
\label{freetheory}

As stated above, we consider theories for mixed tensor gauge
fields $\phi_{\m_1 \dots \m_{p} \vert \n_1 \dots \n_q }$ whose
symmetry properties are characterized by two columns of arbitrary
lengths $p$ and $q$, with $p >1$. These gauge fields thus obey  the
conditions (see Appendix \ref{young})
\bqn
&\phi_{\m_1 \dots \m_{p} \vert \n_1 \dots \n_q }
=\phi_{[\m_1 \dots \m_{p}] \vert \n_1 \dots \n_q
}=\phi_{\m_1 \dots \m_{p} \vert [\n_1 \dots \n_q]}\,,&
\nonumber \\
&\phi_{[\m_1 \dots \m_{p} \vert \n_1]\n_2 \dots \n_q }=0\,,&
\nonumber
\eqn
where square brackets denote strength-one complete antisymmetrization.
We consider the second-order free theory. There also exists  a first-order formulation of the theory, which can be found in the appendix \ref{zino}.

\subsection{Lagrangian and gauge invariances}
\label{Lagrangianandgaugeinvariance}

The Lagrangian of the free theory is \bqn \cl =
-\frac{1}{2\,(p+1)!\,q!}\;\delta^{[\r_1 \dots \r_q \m_1 \dots
\m_{p+1}]}_{[\n_1 \dots \n_q \s_1 \dots \s_{p+1}]}\;
\pa^{[\s_1}\phi^{\s_2 \dots \s_{p+1}]\vert}_{\hspace*{1.4cm} \r_1
\dots \r_q}\; \pa^{}_{[\m_1} \phi_{\m_2 \dots
\m_{p+1}]\vert}^{\hspace*{1.4cm}\n_1 \dots \n_q }\,, \nn \eqn
where the generalized Kronecker delta has strength one: $\delta^{\m_1 \ldots \m_n}_{\n_1 \ldots \n_n} \equiv \delta^{[\m_1 }_{[\n_1 } \ldots \delta^{\m_n] }_{\n_n ]}$. This
Lagrangian was obtained for $[2,1]$-fields in
\cite{Curtright:1980yk}, for $[p,1]$-fields in \cite{Aulakh:1986cb}
and, for the general case of $[p,q]$-fields, in  \cite{deMedeiros:2003dc}. 

The quadratic action \bqn \cs_0 [\phi]=\int d^nx\,{\cal
L}(\partial\phi) \label{action} \eqn is invariant under gauge
transformations with gauge parameters $\a^{(1,0)}$ and
$\a^{(0,1)}$ that have respective symmetries $[p-1,q]$ and
$[p,q-1]\,$. In the same manner as for $p$-forms, these gauge
transformations are \textit{reducible}, their order of
reducibility growing with $p$. We identify the gauge field
$\phi\,$ with $\a^{(0,0)}$, the zeroth order parameter of
reducibility. The gauge transformations and their reducibilities
 are\footnote{We
introduce the short notation $\m_{[p]}\equiv [\m_1 \dots \m_{p}]\,$.
A comma stands for a derivative: $\a_{,\n}\equiv \pa_{\n}\a$.}
\bqn \label{gaugetransfo} \d \a^{(i,j)}_{\m_{[p-i]} \vert \n_{[q-j]}}&=
&\pa_{[\m_1}\a^{(i+1,j)}_{\m_2 \dots \m_{p-i}]
\vert\, \n_{[q-j]}}\\
 &&+\, b_{i,j}  \left( \a^{(i,j+1)}_{\m_{[p-i]} \vert\, [\n_{[q-j-1]},\n_{q-j}]}
+ a_{i,j}\,\a^{(i,j+1)}_{\n_{[q-j]}[\m_{q-j+1} \dots
\m_{p-i}\vert\, \m_{[q-j-1]},\m_{q-j}]} \right)
 \nonumber \eqn where $i=0, ..., p-q$ and $j=0, ...,
q\,$. The coefficients $a_{i,j}$ and $b_{i,j}$ are given by \bqn
a_{i,j}=\frac{(p-i)!}{(p-i-q+j+1)! \, (q-j)!}\,,\quad
b_{i,j}=(-)^i \, \frac{(p-q+j+2)}{(p-i-q+j+2)}\,.\nonumber \eqn To
the above formulae, we must add the convention that, for all
$j\,$,  $\a^{(p-q+1,j)}=0=\a^{(j,q+1)}\,$. The symmetry properties
of the parameters $\a^{(i,j)}$ are those of Young diagrams with
two columns of lengths $p-i$ and $q-j\,$: 
\begin{eqnarray}
&\a^{(i,j)}_{\m_1 \dots \m_{p-i} \vert \n_1 \dots \n_{q-j}}=\a^{(i,j)}_{[\m_1 \dots \m_{p-i} ] \vert \n_1 \dots \n_{q-j}}=\a^{(i,j)}_{\m_1 \dots \m_{p-i} \vert [\n_1 \dots \n_{q-j}]}\,,&
\nonumber \\
&\a^{(i,j)}_{[\m_1 \dots \m_{p-i} \vert \m_{p-i+1} ] \n_2 \dots \n_{q-j}}=0\,.&
\end{eqnarray}
More details on the
reducibility parameters $\a^{(i,j)}_{\m_1 \dots \m_{p-i}\vert\,
\n_1 \dots \n_{q-j}}$ will be given in Section
\ref{BRSTghostsofghosts}.

The fundamental gauge-invariant object is the field strength or curvature
$K\,$, which is the $[p+1,q+1]$-tensor defined as the double curl of the
gauge field: \bqn K_{\m_1 \dots \m_{p+1} \vert\, \n_1 \dots
\n_{q+1} }\equiv\pa_{[\m_1}\phi_{\m_2 \dots \m_{p+1} ]\,\vert\, [\n_1
\dots \n_q \,,\,\n_{q+1} ]}\,.\nonumber \eqn By definition, it
satisfies the Bianchi (BII) identities \be
\partial_{[\m_1} K_{\m_2 \dots \m_{p+2}] \vert\, \n_1 \dots \n_{q+1} }=0\,,
\quad K_{\m_1 \dots \m_{p+1} \,\vert\, [\n_1 \dots \n_{q+1},\n_{q+2}]}=0\,.
\label{Bianchi}
\ee
 Its vanishing implies that $\phi_{\m_1 \dots \m_p \vert \n_1 \dots \n_q}$ is pure gauge \cite{Bekaert:2002dt}.

 The most general gauge-invariant object depends on the
 field $\phi_{\m_1 \dots \m_p \vert \n_1 \dots \n_q}$ and its derivatives only through the
 curvature $K$ and its derivatives.  

\subsection{Equations of motion}
\label{equationsofmotion}

The equations of motion are expressed in terms of the field
strength: \bqn G^{\m_1 \dots \m_p \vert\,}_{ \hspace*{1cm}\n_1
\dots \n_q} \equiv \frac{\d {\cal L}}{\d\phi_{\m_1 \dots \m_p
\vert\,}^{\hspace*{1cm} \n_1 \dots \n_q}}=
\frac{1}{(p+1)!q!}\;\delta^{[\r_1 \dots \r_{q+1} \m_1 \dots
\m_{p}]}_{[\n_1 \dots \n_q \s_1 \dots \s_{p+1}]}\; K^{\s_1\dots
\s_{p+1}\vert\,}_{\hspace*{1.3cm}\r_1 \dots \r_{q+1}}\approx 0 \,,
\nonumber \eqn where a weak equality ``$\approx$'' means ``equal
on the surface of the solutions of the equations of motion''.
This is a generalization of the vacuum Einstein equations, linearized
around the flat background. Taking successive traces of the
equations of motion, one can show that they are equivalent to the
tracelessness of the field strength \be \eta^{\s_1\r_1}K_{\s_1
\dots \s_{p+1}\vert\, \r_1 \dots \r_{q+1}}\approx 0\,.
\label{Ricci} \ee This equation generalizes the vanishing of the
Ricci tensor (in the vacuum), and is nontrivial only when $p+q+2
\leq n$. Together with the ``Ricci equation'' (\ref{Ricci}),
the Bianchi identities (\ref{Bianchi}) imply \cite{Hull:2001iu}
\be
\partial^{\s_1}K_{\s_1 \dots \s_{p+1}\vert\, \r_1 \dots \r_{q+1}}\approx 0
\approx \partial^{\r_1}K_{\s_1 \dots \s_{p+1}\vert\, \r_1\dots \r_{q+1}}\,.
\label{divergence}
\ee
The gauge invariance of the action is equivalent to the
divergenceless of the tensor $G^{\m_{[p]} \vert \n_{[q]}}$, that is, the latter
satisfies the Noether identities
\be
\partial^{\s_1}G_{\s_1 \dots \s_{p+1}\vert\, \r_1 \dots
\r_{q+1}}=0=\partial^{\r_1}G_{\s_1 \dots \s_{p+1}\vert\, \r_1
\dots \r_{q+1}}\,. \label{Noether} \ee These identities are a
direct consequence of the Bianchi ones (\ref{Bianchi}). The
Noether identities (\ref{Noether}) ensure that the equations of
motion can be written as \bqn 0\approx G^{\m_1 \dots \m_p \vert\,
\n_1 \dots \n_q} =\pa_{\a}H^{\a\m_1 \dots \m_p \vert\, \n_1 \dots
\n_q}\,, \nonumber \eqn where  \bqn H^{\a\m_1 \dots \m_p \vert\,}_{ \hspace*{1.2cm}\n_1 \dots \n_q}
=\frac{1}{(p+1)!q!} \;\delta^{[\r_1 \dots \r_q \a \m_1 \dots
\m_{p}]}_{[\n_1 \dots \n_q \b \s_1 \dots \s_p]}\;
\partial^{[\beta}\phi^{\s_1 \dots \s_p]\vert\,}_{\hspace*{1.1cm}\r_1
\dots \r_q} \,. \nonumber \eqn The symmetries of the tensor
$H$ correspond to the Young diagram $[p+1,q]\,$. This property will be useful in the
computation of the local BRST cohomology.

\section{BRST construction}
\label{BRST}

In this section, we apply the rules of Section \ref{s:faf} to build the field-antifield formulation of the theory of free $[p,q]$-fields. We introduce the new fields and antifields in Section \ref{BRSTspectrum}, and the BRST transformation in Section \ref{BRSTdifferential}.

\subsection{BRST spectrum}\label{BRSTghostsofghosts}
\label{BRSTspectrum}

According to the general
rules of the field-antifield formalism,
 we associate with each
gauge parameter $\a^{(i,j)}$ a ghost, and then with any field
(including ghosts) a corresponding antifield  of opposite Grassmann
parity. More precisely, the spectrum of fields (including ghosts)
and antifields is given by
\begin{itemize}
\item {\underline{the fields}}: $A^{(i,j)}_{\m_{[p-i]}
\vert\,\n_{[q-j]}}\,$, where $A^{(0,0)}$ is identified with
$\phi\,$; \item {\underline{the antifields}}:
$A^{*(i,j)\;\m_{[p-i]}\vert\, \n_{[q-j]}}\,$,
\end{itemize}
where  $i=0, ..., p-q$ and $j=0,..., q\,$. The symmetry properties
of the fields $A^{(i,j)}_{\m_{[p-i]} \vert\,\n_{[q-j]}}$ and
antifields $A^{*(i,j)\;\m_{[p-i]}\vert\, \n_{[q-j]}}$ are those of
Young diagrams with two columns of lengths $p-i$ and $q-j\,$. With
each field and antifield are associated a pureghost number and an
antifield  number. The pureghost number is given by
$i+j$ for the fields $A^{(i,j)}$ and $0$ for the antifields, while
the antifield number is $0$ for the fields and $i+j+1$ for the
antifields $A^{*(i,j)}\,$. The Grassmann parity is given by the
pureghost number, resp. the antifield number, modulo $2\,$ for fields and antifields. All
this is summarized in Table \ref{table1}.

\begin{figure}
\centerline{
\includegraphics[scale=0.8]{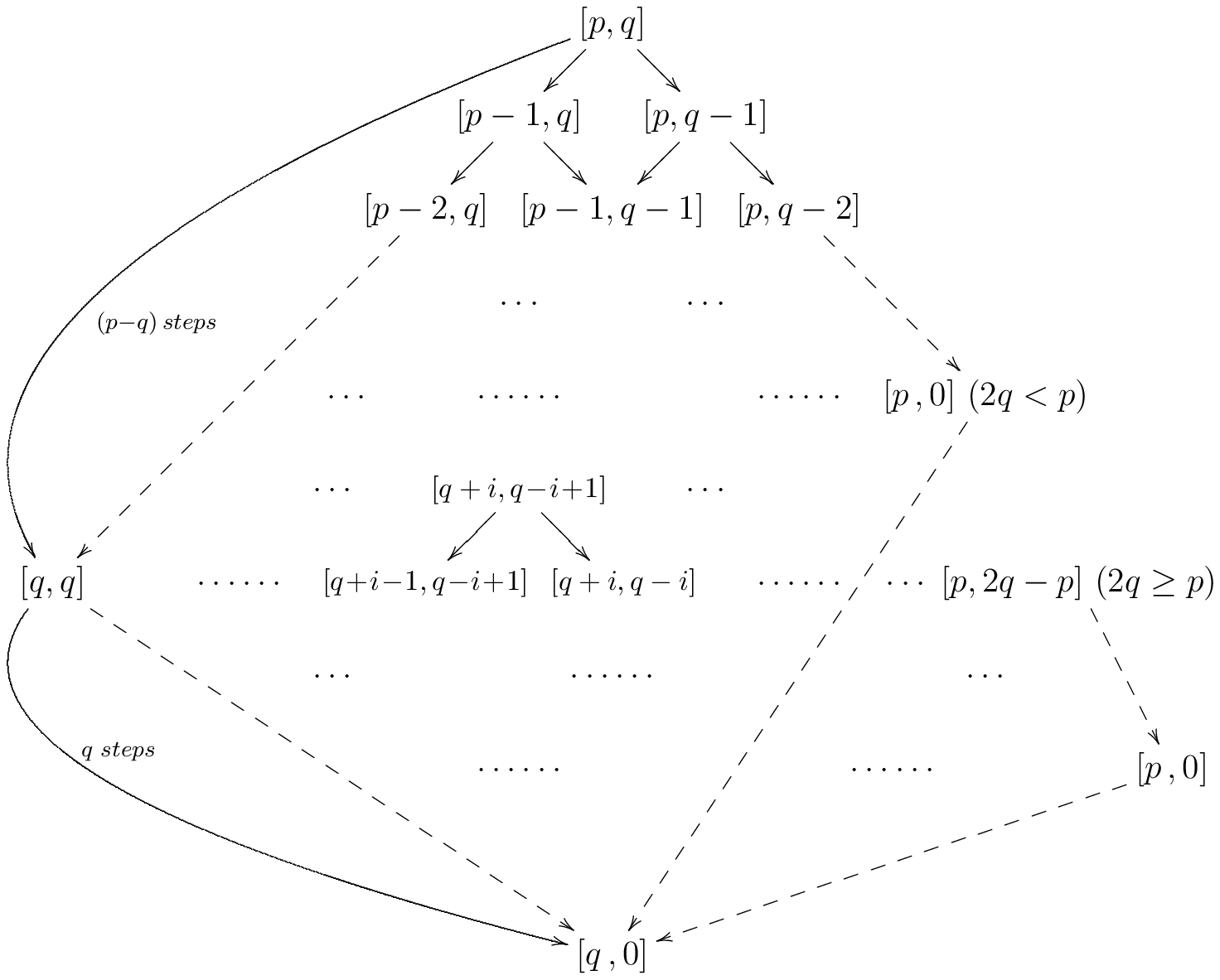}
}
\caption{\textit{Antifield-zero BRST spectrum of $[p,q]-$type gauge field. \label{figure1}}}
\end{figure}

\begin{table}[!ht]
  \centering
\begin{tabular}{|c|c|c|c|c|}\hline
  & Young & $pureghost$ & $antifield $ & Parity \\\hline
  $A^{(i,j)}$ & $[p-i,q-j]$ & $i+j$ & $0$ & $i+j$ \\\hline
  $A^{*(i,j)}$ & $[p-i,q-j]$ & $0$ & $i+j+1$ & $i+j+1$ \\ \hline
\end{tabular}
\caption{\it Symmetry, pureghost number, antifield number
and
 parity of the (anti)fields.\label{table1}}
\end{table}
One can visualize the whole BRST spectrum in vanishing antifield
number as well as the procedure that gives all the ghosts
starting from $\phi_{\m_{[p]}\,\vert\, \n_{[q]}}$ on Figure \ref{figure1}, where the pureghost number increases from top down, by
one unit at each line. The fields are represented by the Young diagram corresponding to their symmetry.


At the top of Figure \ref{figure1} lies the gauge field
$\phi_{\m_{[p]}\,\vert\, \n_{[q]}}$ with pureghost number zero. At
the level below, one finds the pureghost number one gauge
parameters $A^{(1,0)}_{\m_{[p-1]}\vert\, \n_{[q]}}$ and
$A^{(0,1)}_{\m_{[p]}\vert\, \n_{[q-1]}}\,$ whose respective
symmetries are obtained by removing a box in the first (resp.
second) column of the Young diagram $[p,q]$ corresponding to the
gauge field $\phi_{\m_{[p]}\,\vert\, \n_{[q]}}\,$. 


\begin{center}
\begin{picture}(150,80)(0,-60)
\multiframe(0,10)(10.5,0){1}(10,10){\ft$ 1$}
\multiframe(0,-20)(10.5,0){1}(10,29.5){$ $}
\multiframe(10.5,10)(10.5,0){1}(10,10){\ft$1 $}
\multiframe(10.5,-9.5)(10.5,0){1}(10,19){$ $}
\multiframe(10.5,-20)(10.5,0){1}(10,10){\ft$q$}
\multiframe(0,-39.5)(10.5,0){1}(10,19){$ $}
\multiframe(0,-50)(10.5,0){1}(10,10){\ft$p$}
\put(30,-10){~~~~$\longrightarrow$}
\multiframe(80,10)(10.5,0){1}(10,10){\ft$ 1$}
\multiframe(80,-20)(10.5,0){1}(10,29.5){$ $}
\multiframe(90.5,10)(10.5,0){1}(10,10){\ft$1 $}
\multiframe(90.5,-9.5)(10.5,0){1}(10,19){$ $}
\multiframe(90.5,-20)(10.5,0){1}(10,10){\ft$q$}
\multiframe(80,-39.5)(10.5,0){1}(10,19){$ $}
\put(120,-10){{$\oplus$}}
\multiframe(150,10)(10.5,0){1}(10,10){\ft$ 1$}
\multiframe(150,-20)(10.5,0){1}(10,29.5){$ $}
\multiframe(160.5,10)(10.5,0){1}(10,10){\ft$1 $}
\multiframe(160.5,-9.5)(10.5,0){1}(10,19){$ $}
\multiframe(150,-39.5)(10.5,0){1}(10,19){$ $}
\multiframe(150,-50)(10.5,0){1}(10,10){\ft$p$}
\put(180,-10){.}\put(0,-65){$\phi_{[p,q]}$}
\put(70,-65){$A^{(1,0)}_{[p-1,q]}$}
\put(160,-65){$A^{(0,1)}_{[p,q-1]}$}
\end{picture}
\end{center}

The rules that
give the $(i+1)$-th generation ghosts from the $i$-th generation
ones can be found in \cite{Labastida:1986gy,Bekaert:2002dt}.
In short, the Young diagrams of the ghosts
are obtained by removing boxes from the Young diagrams of the ghosts
 with lower pureghost number, with the rule that one is not allowed to remove  two boxes from the same row.

Some ghosts that play a particular role arise at pureghost level $p-q$, $q$ and $p$. They correspond to the edges of the figure.

In pureghost number $p-q$, the set of ghosts contains
$A^{(p-q,0)}_{\m_{[q]}\n_{[q]}}$ $\sim [q,q]\,$. The Young diagram
corresponding to the latter ghost is obtained by removing $p-q$
boxes from the first column of $[p,q]$. Removing any box from this diagram yields $[q,q-1]$.

At the pureghost
level $q$, one finds the $p$-form ghost $A^{(0,q)}_{\m_{[p]}}\,$
$\sim~[p\,,0]\,$, obtained from the field by removing all the boxes of the second column of $[p,q]$ in order to
empty it completely. For this ghost there is also only one way to remove a box.
 
The procedure  terminates at pureghost number
$p$ with the $q$-form ghost\\ $A^{(p-q,q)}_{\m_{[q]}}\sim [q,0]\,$.
There are no ghosts $A_{\m_{[r]}\vert \n_{[s]}}$
with $r,s<q\,$, since it would mean that two boxes from a same row
would have been removed from $[p,q]$.

\vspace{.2cm}

The antifield sector has exactly the same structure as the ghost sector of
Figure \ref{figure1}, where each ghost $A^{(i,j)}$ is replaced by its antifield $A^{*(i,j)}$.

\subsection{BRST-differential}
\label{BRSTdifferential}

The BRST-differential $s$ of  the free theory (\ref{action}), (\ref{gaugetransfo}) is generated by the functional
\bqn
W_0 = \cs_0 [\phi] \;+ \int d^nx\; \!\!\!&\Big[& \sum_{i=0}^{p-q} \sum_{j=0}^{q}
(-)^{i+j} \,
A^{*(i,j)\;\m_1 \dots \m_{p-i} \vert\, \n_1 \dots \n_{q-j}} \nonumber \\
&\,&\!\!\times(\pa^{}_{[\m_1}A^{(i+1,j)}_{\m_2 \dots \m_{p-i}] \vert\,
\n_1 \dots \n_{q-j}} - b_{i+1,j}\,A^{(i,j+1)}_{\m_1 \dots \m_{p-i}
\vert\, [\n_1 \dots \n_{q-j-1},\n_{q-j}]})\Big]\;, \nonumber \eqn
with the convention that $ A^{(p-q+1,j)}= A^{(i,q+1)}=
A^{*(-1,j)}= A^{*(i,-1)}=0\,$. More precisely, $W_0$ is the
generator of the BRST-differential $s$ of the free theory through
\bqn s A = (W_0, A)\,, \nonumber \eqn where the antibracket
$(~,~)$ is defined by Eq.\bref{antibracket def}. The functional $W_0$ is a
solution of the \emph{master equation}
$$(W_0,W_0)=0\,.$$
The BRST-differential
$s$ decomposes into $s=\g + \d \,$. The first piece $\g\,$, the differential along the gauge orbits,
increases the pureghost number by one unit, whereas the Koszul-Tate differential $\d$
decreases the antifield number by one unit.
These gradings are related to the ghost number by $$gh=pureghost - antifield \,.$$
The action of $\g$ and $\d$ on the fields and antifields is zero, except in the following cases:
\bqn
\g A^{(i,j)}_{\m_{[p-i]} \vert\, \n_{[q-j]}}
&= &\pa_{[\m_1}A^{(i+1,j)}_{\m_2 \dots \m_{p-i}]
\vert\,
\n_{[q-j]}}\nonumber \\
 &+&\, b_{i,j}  \left( A^{(i,j+1)}_{ \m_{[p-i]} \vert\,
 [\n_{[q-j-1]},\n_{q-j}]} +
 a_{i,j}A^{(i,j+1)}_{\n_{[q-j]}[\m_{q-j+1} \dots \m_{p-i}\vert\,
\m_{[q-j-1]},\m_{q-j}]}\right)\nonumber\\
\d  A^{*(0,0)\; \m_{[p]} \vert\,  \n_{[q]} } &=& G^{\m_{[p]}
\vert\,\n_{[q]}}\nonumber \\
\d  A^{*(i,j)\;\m_{[p-i]} \vert\,  \n_{[q-j]} }
&=&
(-)^{i+j}\Big( \pa_{\s}A^{*(i-1,j)\;\s\m_{[p-i]} \vert\,
\n_{[q-j]}}\nnn
&&\hspace{2cm} -\frac{1}{p-i+1}\,\pa_{\s}A^{*(i-1,j)\; \n_1\m_{[p-i]}
\vert\, \s \n_2 \dots \n_{q-j}} \Big)
\nonumber \\
&&+(-)^{i+j+1}b_{i+1,j-1}\pa_{\s} A^{*(i,j-1)\;\m_{[p-i]} \vert\,\n_{[q-j]}\s}\,,
\nonumber
\eqn
where the last equation holds only for $(i,j)$ different from $(0,0)$.

One can check that  \be \delta^2 = 0\,, \; \delta \gamma +
 \gamma \delta = 0\,, \; \gamma^2 = 0\,. \ee

For later computations, it is useful to define a unique antifield
for each antifield number: \bqn C^{* \; \m_1 \dots \m_q \vert\,
\n_1 \dots \n_j}_{p+1-j}=
\sum_{k=0}^{j}\epsilon_{k,j}A^{*(p-q-j+k,q-k)\;\m_1 \dots \m_q
[\n_{k+1}\dots \n_j \vert\, \n_1 \dots \n_k]} \nonumber \eqn for
$0 \leq j \leq p\,$, and, in antifield number zero, the
following specific combination of single derivatives of the field
\bqn C^{* \; \m_1 \dots \m_q \vert\, \n_1 \dots \n_{p+1}}_{0}=
\epsilon_{q,p+1}H^{\m_1 \dots \m_q [\n_{q+1}\dots \n_{p+1} \vert\,
\n_1 \dots \n_q]} \,, \nonumber \eqn where $\epsilon_{k,j}$
vanishes for $k>q$  and for $j-k>p-q\,$, and is given in the other
cases by: \bqn \epsilon_{k,j}=(-)^{pk+j(k+p+q)+
\frac{k(k+1)}{2}}\frac{(^k_{p+1})\, (^k_j)}{(^k_q)} \nonumber \eqn
where $(_n^m)$ are the binomial coefficients ($n\geq m$). Some properties of
the new variables $C^{*}_{k}$ are summarized in Table
\ref{table2}.

\begin{table}[!ht]
  \centering
\begin{tabular}{|c|c|c|c|c|}\hline
  & Young diagram & $pureghost$ & $antifield$ & Parity \\\hline
  $C^{*}_{k}$ & $[q] \otimes [p+1-k]- [p+1] \otimes [q-k] $ & $0$ & $k$ & $k $\\ \hline
\end{tabular}
\caption{\it Young diagram, pureghost number, antifield
number and parity of the antifields $C^{*}_{k}$.}\label{table2}
\end{table}
\noindent The symmetry properties of $C^{*}_{k}$ are denoted by
\bqn [q] \otimes [p+1-k]\,-\,[p+1] \otimes [q-k] \nonumber \eqn
which means that this field  has the symmetry properties corresponding
to the tensor product of a column $[q]$ by a column $[p+1-k]$ from
which one should substract (when $k\leq q$) all the Young
diagrams appearing in the tensor product $[p+1] \otimes [q-k]$.
\vspace*{.2cm}

The antifields $C^{*\; \m_{[q]} \vert\, \nu_{[p+1-k]}}_k$ have
been defined in such a way that they obey the following relations: \bqn \d
C^{* \; \m_1 \dots \m_q \vert\, \n_1 \dots \n_j}_{p+1-j} &=&
\pa_{\s} C^{* \;\m_1 \dots \m_q \vert\, \vert\, \n_1 \dots \n_j
\s}_{p-j}\quad
{\rm{for}} \quad 0 \leq j \leq p \,,\nonumber \\
\d C^{* \; \m_1 \dots \m_q \vert\, \n_1 \dots \n_{p+1}}_{0}
&=&0\,. \label{beau} \eqn
 We further define the
inhomogeneous form \bqn \tilde{H}^{\m_1 \dots \m_q
}\equiv\sum_{j=0}^{p+1}C^{* \, n-j\; \m_1 \dots \m_q  }_{p+1-j}\;,
\nonumber \eqn where  \bqn C^{* \, n-j\; \m_1 \dots \m_q
}_{p+1-j}\equiv (-)^{jp+\frac{j(j+1)}{2}}\frac{1}{j!(n-j)!} \,C^{*
\; \m_1 \dots \m_q \vert\,\n_1 \dots \n_j}_{p+1-j}\epsilon_{\n_1
\ldots\n_n}dx^{\n_{j+1}}\ldots dx^{\n_n}\,. \nonumber \eqn 
Then, as a
consequence of Eqs.(\ref{beau}), any polynomial $P(\tilde{H})$ in
$\tilde{H}^{\m_1 \dots \m_q }$  satisfies \bqn
(\d+d)P(\tilde{H})=0\,. \label{htilde} \eqn

The polynomial $\tilde{H}$ is not invariant under gauge transformations. It is therefore useful to introduce another polynomial, $\tilde{\cal H}\,$, with an explicit $x$-dependence, that {\it is} invariant. $\tilde{\cal H}\,$ is defined by $$\tilde{\cal H}_{\m_{[q]}}\equiv \sum_{j=1}^{p+1}C^{* \, n-p-1+j}_{j\, \m_{[q]}}+\tilde{a} \,\epsilon_{[\m_{[q]}\s_{[p+1]} \t_{[n-p-q-1]}]} K^{q+1\,\s_{[p+1]}}x^{\t_1}dx^{\t_2} \ldots dx^{\t_{n-p-q-1}}\,,$$
where $\tilde{a} =(-)^{\frac{p(p-1)+q(q-1)}{2}}\frac{1}{q!q!(p+q+1)!(p+1-q)!(n-p-q-1)!}$.  One can check that $\tilde{\cal H}=\tilde{H}+dm^{n-p-2}_0$ for some $m^{n-p-2}_0 $. This fact has the consequence that polynomials in $\tilde{\cal H}$ also satisfy $(\d+d)P(\tilde{\cal H})=0$.

\section{Cohomology of $\g$}
\label{cohogamma}

We hereafter give the content of $H(\g)$, \ie  the space of solutions of $\g a
= 0$ modulo trivial coboundaries of the form $\g b$. Subsequently, we
explain the procedure that we followed in order to obtain that
result.

\begin{theorem}\label{Hgamma} The cohomology of $\g$ is isomorphic to the space of
functions depending on
\begin{itemize}
  \item the antifields and their derivatives $[A^{*(i,j)}]\,$,
  \item the curvature and its derivatives $[K]\,$,
  \item the $p\,$-th generation ghost $A^{(p-q,q)}$ and
  \item the curl $D^0_{\m_1 \ldots \m_{p+1}} \equiv (-)^q\pa^{}_{[\m_1}A^{(0,q)}_{\m_2 \ldots
\m_{p+1} ]}$ of the $q\,$-th generation ghost $A^{(0,q)}$.
\end{itemize}
\begin{eqnarray}
    H(\g)\simeq \left\{ f\left([A^{*(i,j)}],[K],A^{(p-q,q)},D^0_{\m_1 \ldots \m_{p+1}}
     \right)\right\}\nn\,.
\end{eqnarray}
\end{theorem}

\vspace*{.2cm} \noindent{ \bfseries{Proof :}} The antifields and
all their derivatives are annihilated by $\g\,$. Since they carry
no pureghost degree by definition, they cannot be equal to the
$\g\,$-variation of any quantity. Hence, they obviously belong
to the cohomology of $\g\,$.

To compute the $\g\,$-cohomology in the sector of the field, the
ghosts and all their derivatives, we split the variables into
three sets of {\it independent} variables obeying respectively $\g
u^{\ell} = v^{\ell}\,$, $\g v^{\ell} = 0\,$ and $\g w^i=0\,$. The
variables $u^{\ell}$ and $v^{\ell}$ form so-called ``contractible
pairs'' and the cohomology of $\g$ is therefore generated by the
variables $w^i\,$ (see e.g. \cite{Henneaux:1992ig}, Theorem 8.2).

We decompose the spaces spanned by the derivatives
$\pa_{\m_1\ldots\m_k}A^{(i,j)}\,$, $k\geq 0\,$, $0\leq i
\leq p-q\,$, $0\leq j\leq q\,$, into irreps of
$GL(n,\mathbb{R})\,$ and use the structure of the reducibility
conditions (see Figures  2. and 3.) in order to group the
variables into contractible pairs.

\hspace*{2cm}
\xymatrix @!=.3cm {
 A^{(i,j-1)}\ar@{~}[dr]_{d^{\{2\}}} & & A^{(i-1,j)}\ar@{~}[dl]^{d^{\{1\}}} \\
  & A^{(i,j)}\ar@{~}[dr]_{d^{\{2\}}}&  \\
  & & \\}
  \hspace*{3cm}
\xymatrix @!=.3cm {
 & & \\
 & A^{(i,j)}\ar@{~}[dl]_{d^{\{1\}}}\ar@{~}[dr]^{d^{\{2\}}}\ar@{~}[ul]_{d_{\{2\}}}& \\
  A^{(i+1,j)} & & A^{(i,j+1)} \\}

\vspace*{.1cm}
\hspace*{3.5cm}{\footnotesize{Figure 2}}\hspace*{5.6cm}{\footnotesize{Figure 3}}
\vspace*{.2cm}

\noindent We use the differential operators $d^{\{i\}}\,$,
$i=1,2,...$ (see \cite{Bekaert:2002dt} for a general definition)
that act, for instance on Young-symmetry type tensor fields
$T_{[2,1]}$, as follows:

\vspace*{.5cm}\hspace*{1.5cm}
\begin{picture}(38,45)(-10,-30)
\put(-30,0){$T~~\sim~~$}
\multiframe(10,3)(10.5,0){2}(10,10){}{}
\multiframe(10,-7.5)(10.5,0){1}(10,10) {}
\put(35,0){$\longrightarrow$}\put(35,-10){\ft{$d^{\{1\}}$}}
\multiframe(60,3)(10.5,0){2}(10,10){}{}
\multiframe(60,-7.5)(10.5,0){1}(10,10){}
\multiframe(60,-18)(10.5,0){1}(10,10){\ft$\pa$}
\put(90,0){,}
\multiframe(110,3)(10.5,0){2}(10,10){}{}
\multiframe(110,-7.5)(10.5,0){1}(10,10) {}
\put(135,0){$\longrightarrow$}\put(135,-10){\ft{$d^{\{2\}}$}}
\multiframe(160,3)(10.5,0){2}(10,10){}{}
\multiframe(160,-7.5)(10.5,0){2}(10,10){}{\ft$\pa$}
\put(190,0){,}
\multiframe(210,3)(10.5,0){2}(10,10){}{}
\multiframe(210,-7.5)(10.5,0){1}(10,10) {}
\put(235,0){$\longrightarrow$}\put(235,-10){\ft{$d^{\{3\}}$}}
\multiframe(260,3)(10.5,0){3}(10,10){}{}{\ft$\pa$}
\multiframe(260,-7.5)(10.5,0){1}(10,10){}
\put(300,0){,}
\put(320,0){$etc.$}
\end{picture}

For fixed $i$ and $j$ the set of ghosts $A^{(i,j)}$ and all
their derivatives decompose into four types of independent
variables: \bqn [A^{(i,j)}]\quad \longleftrightarrow \quad \co
A^{(i,j+1)}\,,\,\co d^{\{1\}}A^{(i,j+1)}\,,\,\co
d^{\{2\}}A^{(i,j+1)}\,,\,\co d^{\{1\}}d^{\{2\}}A^{(i,j+1)} \nonumber \eqn
where $\co$ denotes any operator of the type $\prod_{m \geq
3}d^{\{m\}}\,$  or the identity.

Different cases arise depending on the position of the field
$A^{(i,j)}$ in Figure 1. We have to consider fields that sit in
the interior, on a border or at a corner of the diagram.

\begin{itemize}
\item[$\bullet$]\underline{Interior}

\noindent In this case, all the ghosts $A^{(i,j)}$ and their
derivatives form $u^{\ell}$ or $v^{\ell}$ variables. The general relations
involving $\g$ to have in mind are (for any $k,l$, provided the $A$'s are nonvanishing):
\begin{eqnarray}
\g  A^{(k,l)}&\propto&\big[d^{\{1\}}A^{(k+1,l)}+d^{\{2\}}A^{(k,l+1)}\big]\,,\nonumber\\
\g \big[d^{\{1\}}A^{(k+1,l)}+d^{\{2\}}A^{(k,l+1)}\big]&=&0\,,\nonumber\\
\g \big[d^{\{1\}}A^{(k+1,l)}-d^{\{2\}}A^{(k,l+1)}\big] &\propto& d^{\{1\}}d^{\{2\}}A^{(k+1,l+1)}\,,\nonumber\\
\g \big[d^{\{1\}}d^{\{2\}}A^{(k+1,l+1)}] &=&0
\,,\nonumber\end{eqnarray} and that $\co$ commutes with $\gamma$. [Note that the linear combinations of

\noindent $ d^{\{1\}}A^{(k+1,l)}$ and $d^{\{2\}}A^{(k,l+1)}$ are schematic, we essentially mean two linearly independent combinations of these terms that satisfy the above relations.]
According to these relations, the following couples form contractible pairs $u^\ell \leftrightarrow v^\ell$:
\begin{eqnarray}
\co A^{(i,j)}\, &\leftrightarrow &\,\co \big[d^{\{1\}}A^{(i+1,j)}+ d^{\{2\}}A^{(i,j+1)}]\,)\nnn
\co\big[d^{\{1\}}A^{(i,j)}- d^{\{2\}}A^{(i-1,j+1)}]\,&\leftrightarrow &\,\co d^{\{1\}}d^{\{2\}}A^{(i,j+1)}\nnn
\co\big[d^{\{1\}}A^{(i+1,j-1)}- d^{\{2\}}A^{(i,j)}]\,&\leftrightarrow &\,\co d^{\{1\}}d^{\{2\}}A^{(i+1,j)}\nnn
\co\big[d^{\{1\}}A^{(i,j-1)}- d^{\{2\}}A^{(i-1,j)}]\,&\leftrightarrow &\,\co d^{\{1\}}d^{\{2\}}A^{(i,j)}\nn
\end{eqnarray}
Consequently, one can perform a change of variable within
the sets $[A^{(k,l)}]$, mixing $\co d^{\{1\}}A^{(k,l)} $ and $\co d^{\{2\}}A^{(k-1,l+1)}$,
so that the ghosts
$A^{(i,j)}$ in the interior and all their derivatives do not
appear in $H(\g)\,$.

\item[$\bullet$]\underline{Lowest corner}

\noindent On the one hand, we have $\g A_{[q,0]}^{(p-q,q)}=0\,$.
As the operator $\g$ introduces a derivative, $
A_{[q,0]}^{(p-q,q)}\,$ cannot be $\g$-exact. As a result, $A_{[q,0]}^{(p-q,q)}$ is a
$w^i$-variable and thence belongs to $H(\g)\,$. On the other hand,
we find $\pa_{\n}A_{\m_1\ldots\m_q}^{(p-q,q)}=$
$\g\big[A^{(p-q-1,q)}_{\n\m_1\ldots\m_q}+$
$(-)^{p-q}\frac{q}{p+1}A^{(p-q,q-1)}_{\m_1\ldots\m_q\vert\,\n}\big]\,$,
which implies that all the derivatives of $A^{(p-q,q)}$ do not
appear in $H(\g)\,$. 

\item[$\bullet$]\underline{Border}

\noindent  If a ghost $A^{(i,j)}$ stands on a border of Figure 1,
it means that either (i) its reducibility relation involves
only one ghost (see e.g. Fig. 2), or (ii) there exists only one
field whose reducibility relation involves $A^{(i,j)}\,$ (see
e.g. Fig. 3):
\begin{itemize}
\item[(i)] Suppose $A^{(i,j)}$ stands on the left-hand (lower)
edge of Figure 1. We have the relations \begin{eqnarray}\g
A^{(i,j)} \propto d^{\{2\}}A^{(i,j+1)}\,&,&\;
\g \big[d^{\{2\}}A^{(i,j+1)}\big]=0\,,\nonumber\\
\g \big[d^{\{1\}}A^{(i,j)}\big]\propto
d^{\{1\}}d^{\{2\}}A^{(i,j+1)}\,&,&\;
\g \big[d^{\{1\}}d^{\{2\}}A^{(i,j+1)}\big]= 0\,,\nonumber\\
\g A^{(i,j-1)} \propto d^{\{2\}}A^{(i,j)}\,&,&\;
\g \big[d^{\{2\}}A^{(i,j)}\big]=0\,,\nonumber
\end{eqnarray} so that the corresponding sets $[A^{(i,j)}]$ on the
left-hand edge do not contribute to $H(\g)$. We reach similar
conclusion if $A^{(i,j)}$ lies on the right-hand (lower) border
of Figure 1, substituting $d^{\{1\}}$ for $d^{\{2\}}$ when
necessary. 
\item[(ii)] Since, by assumption, $A^{(i,j)}$ does not
sit in a corner of Fig. 1 (but on the higher left-hand or 
right-hand border), its reducibility transformation involves two
ghosts, and we proceed as if it were in the interior. The only
difference is that $\co d^{\{1\}}d^{\{2\}}A^{(i,j)}$ will be equal
to either $\g \co d^{\{1\}}A^{(i,j-1)}$ or $\g \co
d^{\{2\}}A^{(i-1,j)}\,$, depending on whether the field above
$A^{(i,j)}$ is $A^{(i-1,j)}$ or $A^{(i,j-1)}\,$.
\end{itemize}

\item[$\bullet$]\underline{Left-hand corner}

\noindent In this case, the ghost $A^{(i,j)}$ is characterized by
a rectangular-shape Young diagram (it is the only one with this
property). Its reducibility transformation involves only one ghost
and there exists only one field whose reducibility transformation
involves $A^{(i,j)}\,$. Because of its symmetry properties,
$d^{\{2\}}A^{(i,j)}\sim d^{\{1\}}A^{(i,j)}\,$. Better,
$d^{\{2\}}$ is not well-defined on $A^{(i,j)}\,$, it is only
well-defined on $d^{\{1\}}A^{(i,j)}\,$. Therefore, the derivatives
$\pa_{\m_1\ldots\m_k}A^{(i,j)}$ decompose into $\co A^{(i,j)}\,$,
$\co d^{\{1\}}A^{(i,j)}\,$ and $\co
d^{\{1\}}d^{\{2\}}A^{(i,j)}\,$. The first set $\co A^{(i,j)}$ and the second set

\noindent $\co d^{\{1\}}A^{(i,j)}\,$ form
$u^{\ell}$-variables associated with $\co d^{\{2\}}A^{(i,j+1)}\,$
and  $\co d^{\{1\}}d^{\{2\}}A^{(i,j+1)}\,$ respectively.
The third one forms
$v^{\ell}$-variables with $\co d^{\{2\}}A^{(i-1,j)}\,$.

\item[$\bullet$]\underline{Top corner}

\noindent In the case where $A^{(i,j)}$ is the gauge field, we proceed exactly as in the
``Interior'' case, except that the variables $\co d^{\{1\}}d^{\{2\}}A^{(i,j)}=0$ are not
grouped with any other variables any longer. They constitute true $w^i$-variables and are
thus present in $H(\g)\,$.
Recalling the definition of the curvature $K\,$, we have
$\co d^{\{1\}}d^{\{2\}}A^{(i,j)}\propto [K]\,$.

\item[$\bullet$]\underline{Right-hand corner}

\noindent In this case, the field $A^{(i,j)}$ is the $p$-form ghost $A^{(0,q)}_{[p]}\,$.
We have the $(u,v)$-pairs
$(A^{(0,q)}, d^{\{1\}} A^{(1,q)})\,$,
$(\co d^{\{2\}}A^{(0,q)},\co d^{\{1\}}d^{\{2\}} A^{(1,q)})\,$ and \\
$( \co d^{\{1\}}A^{(0,q-1)} , \co d^{\{1\}}d^{\{2\}} A^{(0,q)} )\,$.
The derivative $d^{\{1\}}A_{[p]}^{(0,q)}$ $\propto$ $D^0_{[p+1]}$ is a
$w^i$-variable since
it is invariant and no other variable $\pa_{\m_1\ldots \m_k}A^{(i,j)}$ possesses the
same symmetry.
\end{itemize}
$\Box$

Let us recall (Section \ref{genthm}) that the polynomials $\a ([K],[A^*])$ in the curvature, the antifields and all their derivatives are called ``invariant polynomials''. 
Furthermore, let $\left\{\omega^I\left(A^{(p-q,q)},D^0\right)\right\}$ be
 a basis of the algebra of polynomials in the variables $A^{(p-q,q)}_{[\m_1 \dots \m_q]}$
 and $D^0_{[\m_0 \dots \m_{p}]}$. Any element of $H(\g)$ can be decomposed in
 this basis, hence for any $\g$-cocycle $\a$
 \be\gamma\a=0 \quad\Leftrightarrow\quad
 \a=\a_I([K],[\Phi^*])\;
 \omega^I\left(A^{(p-q,q)},D^0\right) + \g \b
 \label{gammaa}\ee where the $\a_I$ are invariant polynomials.
 Moreover, $\a_I\omega^I$ is $\g$-exact if and only if all the
 coefficients $\a_I$ are zero \be \a_I\omega^I=\gamma\b,\quad
 \Leftrightarrow\quad \a_I=0,\quad\mbox{for
 all}\,\,I.\quad\label{gammab}\ee 

\vspace{.2cm}

We will denote by ${\cal N}$ the algebra generated by all the ghosts
and the non-invariant derivatives of the field $\phi$.
The entire algebra of the fields and antifields is then generated by the invariant polynomials and the elements of ${\cal N}$.

\section{Invariant Poincar\'e lemma}
\label{InvariantPoincarelemma}

The space of {\it invariant} local forms is the space of (local)
forms that belong to $H(\gamma)$. The algebraic Poincar\'e lemma (Theorem \ref{d_cohomology})
 tells us that any closed form is exact\footnote{except for the constants, which are closed without being exact, and the topforms, which are closed but not necessarily exact.}.
However, if the form is furthermore invariant, it is not guaranteed
that the form is exact in the space of invariant forms. The
following lemma  tells us more about this important subtlety, in a
limited range of form degree.

\begin{lemma}[Invariant Poincar\'e lemma in form degree $k<p+1$]\label{invPoinclemma}
Let $\a^k$ be an invariant local $k$-form, $k<p+1\,$. 
$$\mbox{If}\quad d\a^k=0\,,\quad \mbox{then}
\quad\a^k=Q(K^{q+1}_{\m_1 \ldots \m_{p+1}})+d\b^{k-1}\,,$$ where
$Q$ is a polynomial in the $(q+1)$-forms $$K^{q+1}_{\m_1 \ldots
\m_{p+1}}\equiv K_{\m_1 \ldots \m_{p+1} \vert\, \n_1 \ldots
\n_{q+1}} dx^{\n_1} \ldots d x^{\n_{q+1}}\,,$$ and $\b^{k-1}$ is
an invariant local form. 

A closed invariant local form of form-degree $k<n$ and
of strictly positive antifield number is always exact in the space
of invariant local forms.
\end{lemma}

\noindent The proof is directly inspired from the one given in
\cite{Henneaux:1996ws} (Theorem 6).

\subsection{Beginning of the proof of the invariant Poincar\'e\\ lemma}
\label{Beginningoftheproof}

The second statement of the lemma (\ie  the case $antifield(\a^k)\neq
0$) is part of a general theorem (see e.g. \cite{Dubois-Violette:1991is}). It will not be reviewed here. Let us stress that it holds for any form-degree except the maximal degree $n$. 

We will thus assume that $antifield(\a^k) =0$, and prove the first
part of Lemma \ref{invPoinclemma} by induction:
\begin{description}
    \item[Induction basis:] For $k=0$, the invariant Poincar\'e lemma is trivially satisfied: $d \a^0=0$ implies that $\a^0$ is a constant by the usual Poincar\'e lemma.
    \item[Induction hypothesis:] The lemma 
    holds in form degree $k^\prime$ such that $0\leq k^\prime <k\,.$
    \item[Induction step:] We will prove in the sequel that under the induction
hypothesis, the lemma  holds in form degree
$k$. 

Because $d\a^k=0$ and $\gamma\a^k=0$, we can build a descent as
follows\begin{eqnarray}
d\a^k=0 \Rightarrow \a^k&=&da^{k-1,0} \label{form320asub}\\
0&=&\gamma a^{k-1,0} + da^{k-2,1}\label{form321asub}\\
& \vdots& \nonumber\\
0&=&\gamma a^{k-j,j-1} + da^{k-j-1,j}\label{form323asub} \\
0&=&\gamma a^{k-j-1,j}\,,\label{form324asub}
\end{eqnarray}
where $a^{r,i}$ is a $r$-form of pureghost number $i\,$.
The pureghost number of  $a^{r,i}$ lies in the range
$0\leq i\leq k-1\,$.
Of course, since we assume $k<p+1\,$, we have $i<p\,$.
The descent stops at Eq.(\ref{form324asub}) either because $k-j-1=0$ or
because $a^{k-j-1,j}$ is invariant.
The case $j=0$ is trivial since it gives immediately $\a^k=d\b^{k-1}\,$,
where $\b^{k-1}\equiv a^{k-1,0}$ is invariant. Accordingly, we assume from now
on that $j>0\,$.

Since we are dealing with a descent, it is helpful to introduce one of its
building blocks, which is the purpose of the next subsection. We will
complete the induction step in Section \ref{endprfindstp}.
\end{description}

\subsection{A descent of $\g$ modulo $d$}\label{descentogmodulod}
\label{descentgammamodd}

Let us define the following differential forms built up from the
ghosts
$$D^l_{\m_1 \ldots \m_{p+1}} \equiv (-)^{l(q+1)+q}\pa^{}_{[\m_1}
A^{(0,q-l)}_{\m_2 \ldots \m_{p+1} ] \vert\, \n_1 \ldots \n_{l}}
dx^{\n_1} \ldots d x^{\n_l}\,,$$ for $0 \leq  l \leq  q\,$.
It is easy to show that these fields verify the following descent:
\bqn 
\g (D^0_{\m_1 \ldots \m_{p+1}})&=&0  \,, \label{desc1} \\
\g (D^{l+1}_{\m_1 \ldots \m_{p+1}})+d D^l_{\m_1 \ldots \m_{p+1}}&= &0 \;,\quad\quad 0 \leq  l \leq  q-1\,,\nonumber \\
d D^q_{\m_1 \ldots \m_{p+1}}&=&K^{q+1}_{\m_1 \ldots \m_{p+1}}\,.
\label{desc2}
\eqn
It is convenient to introduce the inhomogeneous form
$$D_{\m_1\ldots \m_{p+1}}=\sum_{l=0}^{q}D^l_{\m_1 \ldots \m_{p+1}} $$
because it satisfies a so-called ``Russian formula'' \be (\g +
d)D_{\m_1 \ldots \m_{p+1}}=K^{q+1}_{\m_1 \ldots \m_{p+1}}\,,
\label{russianf} \ee which is a compact way of writing the descent
(\ref{desc1})--(\ref{desc2}). \vspace*{.2cm}

Let $\o_{(s,m)}$ be a homogeneous polynomial of degree  $s$ in $K$ and of degree $m$ in $D$. Its decomposition is
$$\o_{(s,m)}(K,D)=\o^{s(q+1)+mq,0}+...+\o^{s(q+1)+j,mq-j}+...+\o^{s(q+1),mq}$$ where $\o^{s(q+1)+j,mq-j}$ has form degree $s(q+1)+j$
and pureghost number $mq-j$.  Due to Eq.(\ref{russianf}), the
polynomial satisfies \bqn (\g + d)\o_{(s,m)}=K^{q+1}_{\m_1 \ldots
\m_{p+1}} \frac{\partial^L \o_{(s,m)}}{\partial D_{\m_1 \ldots
\m_{p+1}}}\,,\label{rhsdd}\eqn the form degree decomposition of
which leads to the descent 
\bqn
\g (\o^{s(q+1),mq})&=& 0\,,\nonumber \\
\g (\o^{s(q+1)+j+1,mq-j-1})+d \o^{s(q+1)+j,mq-j} &= & 0\, ,\quad0
\leq  j \leq q-1\nonumber
\\
\g (\o^{s(q+1)+q+1,(m-1)q-1}) + \,d \o^{s(q+1)+q,(m-1)q} &= &
K_{\m_1 \ldots \m_{p+1}}^{q+1}\Big[\frac{\partial^L \o}{\partial
D_{\m_1 \ldots \m_{p+1}}}\Big]^{s(q+1),(m-1)q} \label{rhsd} \eqn
where $[\frac{\partial \o}{\partial D}]^{s(q+1),(m-1)q}$ denotes
the component of form degree $s(q+1)$ and pureghost equal to
$(m-1)q$ of the derivative $\frac{\partial \o}{\partial D}$. 
This
component is the homogeneous polynomial of degree $m-1$ in the
variable $D^0$,  $$\Big[\frac{\partial \o}{\partial D_{\m_1
\ldots \m_{p+1}}}\Big]^{s(q+1),(m-1)q}=\frac{\partial \o}{\partial
D_{\m_1 \ldots \m_{p+1}}}\vert_{D=D^0}\,.$$ The
right-hand side of Eq.(\ref{rhsd}) vanishes if and only if the
right-hand side of Eq.(\ref{rhsdd}) does.

Two cases arise depending on whether the r.h.s. of Eq.(\ref{rhsdd}) vanishes or not.
\begin{itemize}
    \item The r.h.s. of Eq.(\ref{rhsdd}) vanishes: then the descent is said not to be obstructed in any strictly positive pureghost number and goes all the way down
    to the bottom equations\begin{eqnarray}
\g (\o^{s(q+1)+mq,0})+d \o^{s(q+1)+mq+1,1} &= & 0\, ,\quad0
\leq j \leq q-1\nonumber
\\ d(\o^{s(q+1)+mq,0})&=& 0\,.\nonumber
\end{eqnarray}
    \item The r.h.s. of Eq.(\ref{rhsdd}) is not zero : then the descent is  obstructed
    after $q$ steps. 
It is not possible to find an $\tilde{\o}^{s(q+1)+q+1,(m-1)q-1} $ such that $$\g (\tilde{\o}^{s(q+1)+q+1,(m-1)q-1} )+ \,d \o^{s(q+1)+q,(m-1)q} = 0\,,$$
because the r.h.s. of Eq.(\ref{rhsd}) is  an element of
    $H(\gamma)$. This element is
    called the {\it obstruction} to the descent. One also says   that this obstruction
    cannot be lifted more than $q$ times, and $\o^{s(q+1),mq}$ is
    the top of the ladder (in this case it must be an element of $H(\gamma)$).
\end{itemize}
This covers the general type of ladder (descent as well as lift)
that do not contain the $p\,$-th generation ghost $A^{(p-q,q)}$.

\subsection{End of the proof of the invariant Poincar\'e lemma}
\label{endprfindstp}

As $j<p$, Theorem \ref{Hgamma} implies that the equation
(\ref{form324asub}) has nontrivial solutions only when $j=mq$ for
some integer $m$
\begin{equation}
a^{k-mq-1,mq}=\sum_I\a_I^{k-mq-1}\,\o_I^{0,mq}\,, \label{blbl}
\end{equation}
up to some $\gamma$-exact term. The $\a_I^{k-mq-1}$'s are invariant
forms, and $\{\o_I^{0,mq}\}$ is a basis of polynomials of degree
$m$ in the variable $D^0$. The ghost $A^{(p-q,q)}$ are absent
since the pureghost number is $j=mq< p$.

The equation (\ref{form323asub}) implies $d\a_I^{k-mq-1}=0$. Together with the induction hypothesis, this implies
\begin{equation}
\a_I^{k-mq-1}=P_I(K^{q+1}_{\m_1 \ldots
\m_{p+1}})+d\b^{k-j-2}\quad\,,\label{blblbl}
\end{equation}where the polynomials $P_I$ of order $s$ are present
iff $k-mq-1=s(q+1)$. Inserting the expression (\ref{blblbl}) into Eq.(\ref{blbl}) we
find that, up to trivial redefinitions,
$a^{k-j-1,j}$ is a polynomial in $K^{q+1}_{\m_1 \ldots \m_{p+1}}$
and $D^0_{\,\m_1 \ldots \m_{p+1}}$.

From the analysis performed in Section \ref{descentogmodulod},
we know the two types of lifts that such an $a^{k-j-1,j}$ can belong to.
In the first case, $ a^{k-j-1,j}$ can be lifted up to form degree zero
but the resulting $a^k$ vanishes.
The second type of lift is obstructed after $q$
steps. Therefore, since $j=mq\,$,  $a^{k-j-1,j}$  belongs to a descent of
type (\ref{form320asub})--(\ref{form324asub}) only if $j=q\,$.
 Without loss of generality we can thus take
$a^{k-q-1}_q=P(K^{q+1}_{\m_1 \ldots \m_{p+1}},D^0)$ where $P$ is
a homogeneous polynomial with a linear dependence in $D^0$ (since
$m=1$). In such a case, it can be lifted up to Eq.(\ref{form320asub}). Furthermore, because $a^{k-1,0}$ is defined
up to an invariant form $\b^{k-1,0}$ by the equation
(\ref{form321asub}), the term $da^{k-1,0}$ of Eq.(\ref{form320asub})
must be equal to the sum
$$da^{k-1,0}=\underbrace{P(K^{q+1},K^{q+1})}_{\equiv Q(K^{q+1}_{\m_1 \ldots \m_{p+1}})}+d\b^{k-1,0}$$
of a homogeneous polynomial $Q$ in $K^{q+1}$ (the lift of the
bottom) and a form $d$-exact in the invariants. \quad \qedsymbol

\section{General property of $H(\g \vert d)$}

 The cohomological space $H(\g \vert d)$ is the space of
 equivalence classes of forms $a$ such that $\g a+db=0$, identified
 by the relation $a\sim a'$ $\Leftrightarrow$ $a'=a+\g c+df$. We
 shall need properties of $H(\g \vert d)$ in strictly positive
 antifield number.  

 The second part of Lemma \ref{invPoinclemma}, in the particular case were one deals with $d$-closed
 invariant forms that involve no ghosts (one considers only
 invariant polynomials), has the following useful consequence on
 general $\g$-mod-$d$-cocycles with $antifield >0\,$, but possibly 
 {\it pureghost} $\neq 0\,$.

 \vspace{.3cm} \noindent {\bf{Consequence of Lemma \ref{invPoinclemma}}}
 \vspace{.1cm}

 {\it{If $a$ has strictly positive antifield number (and involves
 possibly the ghosts), the equation $\gamma a + d b = 0$ is
 equivalent, up to trivial redefinitions, to $\gamma a = 0.$ That
 is,
 \begin{equation}
 \left.
 \begin{array}{c}
 \gamma a + d b = 0, \\
 antigh(a)>0
 \end{array}
 \right\} \Leftrightarrow \left\{
 \begin{array}{c}
 \gamma a' = 0\,, \\
 a'=a+dc
 \end{array}\right.\,.\label{blip}
 \end{equation}
 Thus, in antifield number $>0$, one can always choose
 representatives of $H(\g \vert d)$ that are strictly annihilated
 by $\g$}}. For a proof, see \cite{Barnich:1994db,Barnich:1994mt} or the proof of a similar statement in the spin-3 case (Section \ref{invPlemma}).

\section{Cohomology of $\d$ modulo $d\,$: $H^n_k(\d \vert\, d)$}
\label{Characteristiccohomology}

In this section, we compute the cohomology of $\d$ modulo $d$ in
top form-degree and antifield number $k$, for $k\geq q\,$.
We will also restrict ourselves to $k>1\,$. The group $H^n_1(\d \vert\, d)$ describes the infinitely many conserved currents and will not be studied here.
\vspace*{.2cm}

Let us first recall that by the  general theorem \ref{9.1} of Section \ref{genthm}, since the theory at hand has  reducibility order $p-1$, \bqn
H_k^n(\d \vert\, d)=0\; for\; k>p+1\,. \label{usefll}\eqn

The computation of the cohomology groups $H_k^n(\d \vert\, d)$
for $q \leq k\leq p+1$ follows closely the procedure  used for
$p$-forms in \cite{Henneaux:1996ws}. It relies on the following
proposition and theorem:

\begin{proposition}\label{bilin}
Any solution of $\d a^n+d b^{n-1} =0$ that is at least bilinear in the antifields is necessarily
trivial.
\end{proposition}
This is a trivial rewriting of Theorem \ref{quadr}.

\begin{theorem}\label{pplusun}
A complete set of representatives of $H^n_{p+1}(\d \vert\, d)$ is
given by the antifields $C^{* \, n}_{p+1\,\m_1 \ldots \m_q}$, \ie 
\bqn
 \d a^n_{p+1}+d a^{n-1}_{p}=0
 \;\Rightarrow \; a^n_{p+1}=\l^{\m_{[q]}}C^{* \, n}_{p+1\,\m_{[q]}} +
\d b_{p+2}^n+d b_{p+1}^{n-1}\,,
 \nonumber
 \eqn
where the $\l^{[\m_1 \dots \m_q]}$ are constants.
\end{theorem}
Note that representatives with an explicit $x$-dependence are not considered in the latter theorem, 
because they would not lead to Poincar\'e-invariant deformations.  

\noindent{ \bfseries{Proof :}} 
{ \underline{Candidates}:} any polynomial of antifield
number $p+1$ can be written \bqn a_{p+1}^n=\Lambda^{[\m_1 \dots
\m_q]}C^{* \, n}_{p+1\,[\m_1 \dots \m_q]} +\m_{p+1}^n+\d
b_{p+2}^n+d b_{p+1}^{n-1}\,, \nonumber \eqn where $\Lambda$ does
not involve the antifields and where $\m_{p+1}^n$ is at least
quadratic in the antifields. The cocycle condition $ \d
a^n_{p+1}+d a^{n-1}_{p}=0$ then implies \bqn -\Lambda^{[\m_1 \dots
\m_q]}d C^{* \, n-1}_{p\,[\m_1 \dots \m_q]}+ \d(\m_{p+1}^n +d
b_{p+1}^{n-1})=0\,. \nonumber \eqn By taking the Euler-Lagrange
derivative of this equation with respect to

\noindent $C^{*}_{p\,[\m_1 \dots
\m_q]\vert\, \n}$, one gets the weak equation
$\pa^{\n}\Lambda^{[\m_1 \dots \m_q]}\approx 0\,$. Considering
$\n$ as a form index, one sees that $\Lambda$ belongs to
$H_0^0(d\vert\, \d)$. The isomorphism $H_0^0(d\vert\, \d)/
\mathbb{R} \cong H^n_{n}(\d\vert\, d)$ (see \cite{Barnich:1994db})
combined with the knowledge of $H^n_{n}(\d\vert\, d)\cong 0$ (by
Eq.\bref{usefll}) implies $\Lambda^{[\m_1 \dots
\m_q]}=\l^{[\m_1 \dots \m_q]}+\d \n_1^{[\m_1 \dots \m_q]}$ where
$\l^{[\m_1 \dots \m_q]}$ is a constant. The term

\noindent $\d \n_1^{[\m_1
\dots \m_q]}C^{* \, n}_{p+1\,[\m_1 \dots \m_q]} $ can be rewritten as a term at least bilinear in the antifields up to a $\d$-exact
term.  Inserting $a_{p+1}^n=\l^{[\m_1 \dots \m_q]}C^{* \,
n}_{p+1\,\m_1 \dots \m_q}+\m_{p+1}^n+\d b_{p+2}^n+d b_{p+1}^{n-1}$
into the cocycle condition, we see that $\m_{p+1}^n$ has to be a
solution of $\d \m_{p+1}^n+d b^{n-1}=0$ and is therefore
trivial by Proposition \ref{bilin}. \vspace*{.2cm}

\underline{Nontriviality}: It remains to show that the
cocycles $a^n_{p+1}=\l C^{*\,n}_{p+1}$ are nontrivial. Indeed one
can prove that  $\l C^{*\,n}_{p+1}=\d u_{p+2}^n + d v_{p+1}^{n-1}$
implies that
 $\l C^{*\,n}_{p+1}$ vanishes. It is straightforward when $u_{p+2}^n$ and $v_{p+1}^{n-1}$
do not depend explicitly on $x$: $\d$ and $d$ bring in a
derivative while $\l C^{*\,n}_{p+1}$ does not contain any. If $u$
and $v$ depend explicitly on $x$, one must expand them and the
equation $\l C^{*\,n}_{p+1}=\d u_{p+2}^n + d v_{p+1}^{n-1}$
according to the number of derivatives of the fields and
antifields to reach the conclusion. Explicitly, $u_{p+2}^n
=u_{p+2,\,0}^n + \ldots +u_{p+2,\,l}^n$ and $
v_{p+1}^{n-1}=v_{p+1,\,0}^{n-1}+ \ldots +v_{p+1,\,s}^{n-1}$. If
$s>l$, the equation in degree $s+1$ reads $0= d'
v_{p+1,\,s}^{n-1}$ where $d'$ does not differentiate with respect
to the explicit dependence in $x$. This in turn implies that
$v_{p+1,\,s}^{n-1}= d' \tilde{v}_{p+1,\,s-1}^{n-1}$ and can be
removed by redefining $v_{p+1}^{n-1}$: $v_{p+1}^{n-1} \rightarrow
v_{p+1}^{n-1}-d \tilde{v}_{p+1,\,s-1}^{n-1}$. If $l>s$,  the
equation in degree $l+1$ is $0=\d u_{p+2,\,l}^n$ and implies,
together with the acyclicity of $\d$, that one can remove
$u_{p+2,\,l}^n$ by a trivial redefinition of $u_{p+2}^n\,$. If
$l=s>0$, the equation in degree $l+1$ reads $0= \d u_{p+2,\,l}^n+
d' v_{p+1,\,l}^{n-1}\,$. Since there is no cohomology in antifield
number $p+2$, this implies that $u_{p+2,\,l}^n=\d
\bar{u}_{p+3,\,l-1}^n + d'\tilde{u}_{p+2,\,l-1}^{n-1}$ and can be
removed by trivial redefinitions: $u_{p+2}^n \rightarrow u_{p+2}^n
-\d \bar{u}_{p+3,\,l-1}^n $ and $v_{p+1}^{n-1}\rightarrow
v_{p+1}^{n-1}- d \tilde{u}_{p+2,\,l-1}^{n-1}\,$. Repeating the
steps above, one can remove all $u_{p+2,\,l}^n$ and
$v_{p+1,\,s}^{n-1}$ for $l,\,s>0\,$. One is left with $\l
C^{*\,n}_{p+1}=\d u_{p+2,\,0}^n + d' v_{p+1,\,0}^{n-1}\,$. The
derivative argument used in the case without explicit
$x$-dependence now leads to the desired conclusion.
\quad
\qedsymbol

\begin{theorem}\label{BKzero}
The cohomology groups $H_k^n(\d \vert\, d)$ ($k>1$) vanish
unless $k=n-r(n-p-1)$ for some strictly positive integer $r\,$.
Furthermore, for those values of $k\,$,  $H_k^n(\d \vert\, d)$ 
has at most one nontrivial class.
\end{theorem} 
\noindent{ \bfseries{Proof :}} 
We already know that $H_k^n(\d \vert\,
d)$ vanishes for $k>p+1$ and that $H_{p+1}^n(\d \vert\, d)$ has one nontrivial class. Let us assume that the theorem has been proved for
all $k$'s strictly greater than $K$ (with $K<p+1$) and  extend it to $K$.
Without loss of generality we can assume that the cocycles of
$H_K^n(\d \vert\, d)$ take the form (up to trivial terms)
$a_{K}=\l^{\m_1 \ldots \m_{p+1-K}\vert\, \n_1 \ldots \n_q}C^*_{K\;
\n_1 \ldots \n_q \vert\, \m_1 \ldots \m_{p+1-K}}+ \m $, where $\l$
does not involve the antifields and $\m$ is at least bilinear in
the antifields. Taking the Euler-Lagrange derivative of the
cocycle condition with respect to $C^*_{K-1}$ implies that
$\l^{p+1-K}_{\n_1 \ldots \n_q}\equiv \l_{\m_1 \ldots
\m_{p+1-K}\vert\, \n_1 \ldots \n_q}dx^{\m_1} \ldots
dx^{\m_{p+1-K}}$ defines an element of $H_0^{p+1-K}(d \vert\,
\d)$. If $\l$ is $d$-trivial modulo $\d$, then it is
straightforward to check that $\l  C^{*\,n-p-1+K}_{K}$ is trivial
or bilinear in the antifields. Using the isomorphism
$H_0^{p+1-K}(d \vert\, \d) \cong H_{n-p-1+K}^{n}(\d \vert\, d)$,
we see that $\l$ must be trivial unless $ n-p-1+K=n-r(n-p-1)\,$, in which case $H_{n-p-1+K}^{n}(\d \vert\, d)$ has one nontrivial class. Since $K=n-(r+1)(n-p-1)$ is also of the required form, the
theorem extends to $K$. \quad \qedsymbol

\begin{theorem} \label{BK}
Let $r$ be a strictly positive integer. A complete set of
representatives of $H_k^n(\d \vert\, d)$ ($k=n-r(n-p-1)\geq  q
$) is given by the terms of form-degree $n$ in the expansion of all possible homogeneous
polynomials $P(\tilde{H})$ of degree $r$ in $\tilde{H}$ (or equivalently $P(\tilde{\cal H})$ of degree $r$ in $\tilde{\cal H}$). 
\end{theorem}
\noindent{ \bfseries{Proof :}} 
It is obvious from the definition of $\tilde{H}$ and from 
Eq.(\ref{htilde}) that the term of form-degree $n$ in $P^{(r)}(\tilde{H})$
has the right antifield number and is a cocycle of $H_k^n(\d
\vert\, d)$. 
Furthermore, as $\tilde{\cal H}=\tilde{H}+ d( \ldots)$, $P^{(r)}(\tilde{\cal H})$ belongs to the same cohomology class as $P^{(r)}(\tilde{H})$ and can as well be chosen as a representative of this class.
To prove the theorem, it is then enough, by Theorem
\ref{BKzero}, to prove that the cocycle $P^{(r)}(\tilde{H})\vert \,_k^n$ is nontrivial. The proof
is by induction:  we know the theorem to be true for $r=1$ by
Theorem \ref{pplusun}, supposing that the theorem is true for $r-1$,
(\ie  $[P^{(r-1)}(\tilde{H})]^n_{k+n-p-1}$ is not trivial in
$H^n_{k+n-p-1}(\d\vert d)$) we prove that $[P^{(r)}(\tilde{H})]^n_k$
is not trivial either.
\vspace{.2cm}

Let us assume that
$[P^{(r)}(\tilde{H})]^n_k$ is trivial: $[P^{(r)}(\tilde{H})]^n_k= \d (u_{k+1} d^nx) + d v_k^{n-1}$.
We  take the Euler-Lagrange derivative of this equation with respect to $C^{*}_{k,\m_{[q]}\vert \n_{[p+1-k]}}$.
For $k> q$, it reads:
\bqn
\a_{\m_{[q]} \vert\, \nu_{[p+1-k]}} =(-)^k \d(Z_{1\; \m_{[q]} \vert\, \nu_{[p+1-k]}})-Z_{0 \;\m_{[q]} \vert\, [\nu_{[p-k]},\n_{p+1-k}]} \,, \label{eul}
\eqn
where
\bqn
\a_{\m_{[q]} \vert\, \nu_{[p+1-k]}} d^n x &\equiv &\frac{\d^L [P^{(r)}(\tilde{H})]^n_k}{\d C_k^{*\; \m_{[q]} \vert\,\nu_{[p+1-k]}}}\,, \nonumber \\
Z_{k+1-j \; \m_{[q]} \vert\, \nu_{[p+1-j]}} &\equiv& \frac{\d^L u_{k+1}}{\d C_j^{*\; \m_{[q]}
 \vert\, \nu_{[p+1-j]}}} \,, \, \;{\rm for} \;  j =k, k+1 \,.\nonumber
\eqn
For $k=q$, there is an additional term:
\bqn
\a_{\m_{[q]} \vert\, \nu_{[p+1-q]}}& =&(-)^q \d(Z_{1\; \m_{[q]} \vert\, \nu_{[p+1-q]}})
\nnn
&&-(Z_{0 \;\m_{[q]} \vert\, [\nu_{[p-q]},\n_{p+1-q}]} - Z_{0 \;[\m_{[q]} \vert\, \nu_{[p-q]},\n_{p+1-q}]} )\,.\label{eul2}
\eqn
The origin of the additional term lies in the fact that
$C_q^{*\; \m_{[q]} \vert\,  \nu_{[p+1-q]}}$
does not possess all the irreducible components of $[q] \otimes [p+1-q]\,$: the completely antisymmetric component $[p+1]$ is missing. Taking the  Euler-Lagrange derivative with respect to this field thus involves projecting out this component.

We will first solve the equation (\ref{eul}) for $k>q$, then come back to  Eq.(\ref{eul2}) for $k=q$.
\vspace{.2cm}

Explicit computation of  $\a_{\m_{[q]} \vert\, \nu_{[p+1-k]}} $ for $k>q$ yields:
\bqn
\a_{\m_{[q]} \vert\, \nu_{[p+1-k]}}
=
[\tilde{H}^{\r^1_{[q]}}]_{0,\,\s^1_{[n-p-1]}}
\ldots [\tilde{H}^{\r^{r-1}_{[q]}}]_{0,\,\s^{r-1}_{[n-p-1]}}
a_{\m_{[q]} \vert\r^1_{[q]} \vert \ldots \vert\r^{r-1}_{[q]}}\d^{[\s^1_{[n-p-1]}\ldots \s^{r-1}_{[n-p-1]}]}_{\nu_{[p+1-k]}} 
\nonumber  \,,
\eqn
where $a$ is a constant tensor and the notation  $[A]_{k,\,\n_{[p]}}$ means the coefficient $A_{k,\,\n_{[p]}}$, with antifield number $k$, of the $p$-form component of $A=\sum_{k,l} A_{k,\,\n_{[l]}}dx^{\n_1} \ldots dx^{\n_{l}}$.
Considering the indices $\nu_{[p+1-k]}$ as form indices, Eq.(\ref{eul}) reads:
\bqn
\a^{p+1-k}_{\m_{[q]}}&=& [\tilde{H}^{\r^1_{[q]}}]_0^{n-p-1} \ldots [\tilde{H}^{\r^{r-1}_{[q]}}]_0^{n-p-1}a_{\m_{[q]} \vert \r^1_{[q]} \vert \ldots \vert\r^{r-1}_{[q]}}\nnn
&=&\Big[\prod_{i=1}^{(r-1)}\tilde{H}^{\r^{i}_{[q]}}\Big]^{p+1-k}_{0}
a_{\m_{[q]} \vert\r^1_{[q]} \vert \ldots \vert\r^{r-1}_{[q]}}
\nonumber \\
&=&
(-)^k \d(Z^{p+1-k}_{1\; \m_{[q]} })+(-)^{p-k+1} d\,Z^{p-k}_{0 \;\m_{[q]} }\,.\nonumber
\eqn
The latter equation is equivalent to $$ \Big[\prod_{i=1}^{(r-1)}\tilde{H}^{\r^{i}_{[q]}}\Big]^{n}_{n-p-1+k} a_{\m_{[q]} \vert \r^1_{[q]} \vert \ldots \vert\r^{r-1}_{[q]}}=\d (\ldots) + d (\ldots)\,,$$
which contradicts the induction hypothesis. The assumption that
$[P^{(r)}(\tilde{H})]^n_k$ is trivial is thus wrong, which proves the theorem for $k>q$.
\vspace{.2cm}

The  philosophy of the
 resolution of Eq.(\ref{eul2}) for $k=q$  goes as follows \cite{Boulanger:2004rx}: first, one has to constrain the last term of Eq.(\ref{eul2}) in order to get an equation similar to the equation
(\ref{eul}) treated previously,
then one solves this equation in the same way as for $k>q$.

Let us constrain the last term of Eq.(\ref{eul2}). Eq.(\ref{eul2}) and explicit computation of $\a_{\m_{[q]} \vert\, \nu_{[p+1-k]}}$ imply
\bqn
\pa_{[\n_{p+1-q}}\a_{\m_{[q]} \vert\,
\n_{[p-q]}] \l}&=&(-)^q \d(\pa_{[\n_{p+1-q}}Z_{1 \; \m_{[q]} \vert\,
\n_{[p-q]}] \l} ) - b \,\pa_{[\n_{p+1-q}}Z_{0 \; \m_{[q]}
\vert\, \nu_{[p-q]}],\l}\hspace{1.2cm}
\nonumber \\
&\approx &\!\!b \,\pa_{\l} (
[\tilde{H}^{\r^1_{[q]}}]_{0,\,\s^1_{[n-p-1]}} \ldots [\tilde{H}^{\r^{r-1}_{[q]} }]_{0 ,\,\s^{r-1}_{[n-p-1]}} \d^{[\s^1_{[n-p-1]}\ldots \s^{r-1}_{[n-p-1]}]}_{[\nu_{[p+1-k]}}\nnn
&&\hspace{7.3cm}\times a_{\m_{[q]} ]\vert\r^1_{[q]} \vert \ldots \vert\r^{r-1}_{[q]}})
\nonumber
\eqn where
$b=\frac{q}{(p+1)(p+1-q)} $.
By the isomorphism $H_0^0(d\vert \d)/\mathbb{R}\cong H^n_{n}(\d\vert d)\cong 0\,$, the latter equation implies
\bqn
Z_{0 \; [\m_{[q]}
\vert\, \nu_{[p-q]},\n_{p+1-q}]}\approx -[\tilde{H}^{\r^1_{[q]}}]_{0,\,\s^1_{[n-p-1]}} \ldots [\tilde{H}^{\r^{r-1}_{[q]} }]_{0,\,\s^{r-1}_{[n-p-1]}}\hspace{3.5cm}\nnn
\times a_{\m_{[q]} \vert\r^1_{[q]} \vert \ldots \vert\r^{r-1}_{[q]}}\d^{[\s^1_{[n-p-1]}\ldots \s^{r-1}_{[n-p-1]}]}_{\nu_{[p+1-k]}}\nonumber
\eqn
(the constant solutions are removed by considering the equation in polynomial degree $r-1$ in the fields and antifields.).
Inserting this expression for 

\noindent $Z_{0 \; [\m_{[q]}
\vert\, \nu_{[p-q]},\n_{p+1-q}]}$ into Eq.(\ref{eul2}) and redefining $Z_1$ in a suitable way yields

\noindent Eq.(\ref{eul}) for $k=q$. The remaining of the proof is then the same as for $k>q$. \quad \qedsymbol

\vspace{.2cm}

These theorems give us a complete description of all the
cohomology group

\noindent $H_k^n(\d\vert\, d)$ for $k\geq  q $ (with $k>1\,$).

\section{Invariant cohomology of $\d$ modulo $d$
}
\label{Invariantcharacteristiccohomology}

In this section, we compute the set of invariant  solutions $a^n_k$ ($k \geq  q$) of the
equation $\d a^n_k+d b^{n-1}_{k-1}=0$, up to trivial terms
$a^n_k=\d b^n_{k+1}+d c^{n-1}_k$, where $b^n_{k+1}$ and
$c^{n-1}_k$ are invariant. This space of solutions is the
invariant  cohomology of $\d$ modulo $d$, $H_k^{inv}(\d \vert\, d)$. 
We first  compute representatives of all the  cohomology classes of $H_k^{inv}(\d \vert\, d)$, then we sort out the cocycles without explicit $x$-dependence.

\begin{theorem} \label{cohoinva}
For $k\geq  q$, a complete set of invariant solutions  of the equation $\d a^n_k+db^{n-1}_{k-1}=0$ is given by the polynomials in the curvature $K^{q+1}$ and in $\tilde{\cal H}$ (modulo trivial solutions): 
$$\d a^n_k+d b^{n-1}_{k-1}=0 \Rightarrow a^n_k= P(K^{q+1},\tilde{\cal H})\vert\,_k^n +\d \m_{k+1}^n+d \n_k^{n-1} \,,$$
where $\m_{k+1}^n$ and $\n_k^{n-1}$ are invariant forms.
\end{theorem}
\noindent{ \bfseries{Proof :}} 
From the previous section, we know that for $k\geq  q$ the
general solution of the equation $\d a^n_k+d
b^{n-1}_{k-1}=0$ is $a^n_k=Q(\tilde{\cal H})\vert\,^n_k+\d m^n_{k+1}+d
n^{n-1}_k$ where $Q(\tilde{\cal H})$ is a homogeneous polynomials of
degree $r$ in $\tilde{\cal H}$ (it exists only when $k=n-r(n-p-1)$).
Note  that $m^n_{k+1}$ and $n^{n-1}_k$ are not necessarily
invariant. However, one can prove the following theorem (the lengthy proof of which is provided in the appendix \ref{append}):

\begin{theorem}\label{cohoinv}\label{CH}
Let $\a^n_k$ be an invariant polynomial ($k\geq  q$). If $\a_k^n = \d
m_{k+1}^n + d n_k^{n-1} $, then
$$\a_k^n = R^{(s,r)}(K^{q+1},\tilde{\cal H})\vert\,_k^n+\d \m_{k+1}^n + d \n_k^{n-1}\,,$$
where  $R^{(s,r)}(K^{q+1},\tilde{\cal H})$ is a polynomial of degree $s$ in $K^{q+1}$ and $r$ in $\tilde{\cal H}$, such that the strictly positive integers $s,r$ satisfy
 $n=r(n-p-1)+k+s(q+1)$ and $\m_{k+1}^n$ and $\n_k^{n-1}$ are invariant forms.
\end{theorem}

\noindent As $a^n_k$ and $Q(\tilde{\cal H})\vert\,^n_k$ are invariant, this theorem implies that
$$a^n_k=P^{(s,r)}(K^{q+1},\tilde{\cal H})\vert\,_k^n+\d \m_{k+1}^n + d \n_k^{n-1}\,,$$
 where $P^{(s,r)}(K^{q+1},\tilde{\cal H})$ is a polynomial of non-negative degree $s$ in $K^{q+1}$ and of strictly positive degree $r$ in $\tilde{\cal H}$. Note that the polynomials of non-vanishing degree in $K^{q+1}$ are trivial in $H_k^{n}(\d \vert\, d)$ but not necessarily in $H_k^{n\, inv}(\d \vert\, d)$. \quad \qedsymbol

\vspace{.2cm}

Part of the  solutions found in Theorem \ref{cohoinva} depend explicitely on the coordinate $x$, because $\tilde{\cal H}\vert \,_0$ does.  Therefore the question arises
whether there exist other representatives of the same
nontrivial equivalence class $[P^{(s,r)}(K^{q+1},\tilde{\cal H})\vert\,^n_k]\in
H^{n\, inv}_k(\d\mid d)$ that \textit{do not} depend explicitly on $x$. The answer is
negative when $r>1$.  In other words, we can prove the general theorem:

\begin{theorem}\label{quad}
When $r>1$, there is no nontrivial invariant cocycle in the equivalence class $[P^{(s,r)}(K^{q+1},\tilde{\cal H})\vert\,^n_k]$ $\in H^{n\, inv}_k(\d\mid d)$ without explicit $x$-dependence. 
\end{theorem}

To do so, we first prove the following lemma:
\begin{lemma}Let $P(K^{q+1},\tilde{\cal H})$ be a homogeneous polynomial of order $s$ in the curvature $K^{q+1}$ and $r$ in $\tilde{\cal H}$.
If $r\geq  2$, then the component $P(K^{q+1},\tilde{\cal H})|^n_k$ always contain terms of
order $r-1$($\neq 0$) in $\tilde{\cal H}\vert \,_0 $. \end{lemma} 
\noindent{ \bfseries{Proof :}} Indeed,
$P(K^{q+1},\tilde{\cal H})$ can be freely expanded in terms of $\tilde{\cal H}\vert \,_0 $ and
the undifferentiated antifield forms. The Grassmann parity is the
same for all terms in the expansion of $\tilde{\cal H}$, therefore the
expansion is the binomial expansion up to the overall coefficient
of the homogeneous polynomial and up to relative signs obtained
when reordering all terms. Hence, the component
$P(K^{q+1},\tilde{\cal H})|^n_k$ always contains a term that is a product of
$(r-1)$ $\tilde{\cal H}\vert \,_0^{ n-p-1}$'s, a single antifield $C^{*\,n-p-1+k}_k$ and $s$ curvatures,
which possesses the correct degrees as can be checked
straightforwardly. \quad\qedsymbol

\noindent {\bf Proof of Theorem \ref{quad}:} \hspace{.5cm} Let us
assume that there exists a non-vanishing invariant $x$-independent
representative $\a^{n\,,\,inv}_{k}$ of the equivalence class\\
$[P^{(s,r)}(K^{q+1},\tilde{\cal H})|^n_k]\in H^{n \, inv}_k(\d\mid d)$, \ie \bqn
P^{(s,r)}(K^{q+1},\tilde{\cal H})|^n_k+\d\rho^{n}_{k+1}+d\sigma^{n-1}_{k}=\a^{n\,,\,inv}_{k}\,,\label{previousequ}\eqn
where $\rho^{n}_{k+1}$ and $\sigma^{n-1}_{k}$ are invariant and allowed to
depend explicitly on $x$.

We define the descent map $f:\a_m^r\rightarrow \a_{m-1}^{r-1}$
such that $\d \a_m^r+d \a_{m-1}^{r-1}=0$, for $r\leq n$. This
map is well-defined on equivalence classes of $H^{inv}(\d\vert d)$ when $m>1$ and
preserves the $x$-independence of a
representative. Hence, going down $k-1$ steps, it
is clear that the equation (\ref{previousequ}) implies:
$$
P^{(s,r)}(K^{q+1},\tilde{\cal H})|^{n-k+1}_1+\d\rho^{n-k+1}_2+d\sigma^{n-k}_1=\a^{n-k+1\,,\,inv}_1\,,$$
with $\a^{n-k+1\,,\,inv}_1\neq 0$.

We can decompose this equation in the polynomial degree in the
fields, antifields, and all their derivatives. Since $\d$ and $d$
are linear operators, they preserve this degree; therefore \be
P^{(s,r)}(K^{q+1},\tilde{\cal H})|^{n-k+1}_{1,\,r+s}+\d\rho^{n-k+1}_{2,\,r+s}+d\sigma^{n-k}_{1,\,r+s}
=\a^{n-k+1\,,\,inv}_{1,\,r+s}\,,\label{decompose}\ee where $r+s$
denotes the polynomial degree. The homogeneous polynomial

\noindent
$\a^{n-k+1\,,\,inv}_{1,\,r+s}$ of polynomial degree $r+s$ is linear in
the antifields of antifield number equal to one, and depends on
the fields only through the curvature.

Finally, we introduce the number operator $N$ defined by \bqn
N\,&=&\,r\,\,\partial_{\r_1}\ldots\partial_{\r_r}\phi_{\m_1\ldots\m_p\,|\,\n_1\ldots\n_q}
\,\,\frac{\partial}{\partial
(\partial_{\r_1}\ldots\partial_{\r_r}\phi_{\m_1\ldots\m_p\,|\,\n_1\ldots\n_q})}\nn\\
&&+\,(r+1)\,\partial_{\r_1}\ldots\partial_{\r_r}\Phi_A^*\,\,\frac{\partial}{\partial
(\partial_{\r_1}\ldots\partial_{\r_r}\Phi_A^*)}
- x^{\m} \,\,\frac{\partial}{\partial x^{\m}}\nonumber 
\eqn
 where
$\{\Phi_A^*\}$ denotes the set of all antifields. It follows
immediately that $\d$ and $d$ are homogeneous of degree one and
the degree of $\tilde{\cal H}$ is also equal to one,
$$N(\d)=N(d)=1=N(\tilde{\cal H})\,.$$
Therefore, the decomposition in $N$-degree of the equation
(\ref{decompose}) reads in $N$-degree  equal to $m=r+2s$, \be
P^{(s,r)}(K^{q+1},\tilde{\cal H})|^{n-k+1}_{1,\,r+s}+\d\rho^{n-k+1}_{2,\,r+s,\,r+2s-1}+d\sigma^{n-k}_{1,\,r+s,\,r+2s-1}
=\a^{n-k+1\,,\,inv}_{1,\,r+s,\,r+2s}\label{endproof}\ee and, in $N$--degree
 equal to $m>r+2s$,
$$\d\rho^{n-k+1}_{2,\,r+s,\,m-1}+d\sigma^{n-k}_{1,\,r+s,\,m-1}
=\a^{n-k+1\,,\,inv}_{1,\,r+s,\,m}\,.$$ The component
$\a^{n-k+1\,,\,inv}_{1,\,r+s,\,r+2s}$ of $N$-degree  equal to $r+2s$ is $x$-independent, depends linearly on the (possibly differentiated) antifield of antifield number 1, and is of order $r+s-1$ in the (possibly differentiated) curvatures. Direct counting shows that there is no polynomial of $N$-degree equal to $r+2s$ satisfying these requirements when $r\geq  2$. Indeed, one would have $N\geq 2r+2s-1\,$, which is compatible with $N=r+2s$ only for $r\leq 1$.
 Thus for $r\geq  2$ the component
$\a^{n-k+1\,,\,inv}_{1,\,r+s,\,r+s}$ vanishes, and then the equation
(\ref{endproof}) implies that $P^{(s,r)}(K^{q+1},\tilde{\cal H})|^{n-k+1}_{1,\,r+s}$ is trivial (and even vanishes when $s=0$, by Theorem \ref{BK}).

In conclusion, if $P(K^{q+1},\tilde{\cal H})$ is a polynomial that is quadratic or more in $\tilde{\cal H}$,
then there exists no nontrivial invariant representative without explicit $x$-dependence in the cohomology
class $[P(K^{q+1},\tilde{\cal H})]$ of $H^{inv}(\d\vert d)$. 
\quad \qedsymbol

\vspace{.2cm}

This leads us to the following theorem:
\begin{theorem}
\label{thm8.1}
The invariant solutions $a^n_k$  ($k\geq  q$) of the equation $\d a^n_k+d
b^{n-1}_{k-1}=0$ without explicit $x$-dependence are all trivial
in $H_k^{inv}(\d \vert\, d)$ unless $k=p+1-s(q+1)$ for
some non-negative  integer $s$. For those values of $k$, the
 nontrivial representatives are given by  polynomials that
are linear in $C^{*\; n-p-1+k}_k$ and of order $s$ in $K^{q+1}$. 
\end{theorem}
\noindent{ \bfseries{Proof :}} 
By Theorem \ref{cohoinva}, invariant solutions of the equation $\d a^n_k+d
b^{n-1}_{k-1}=0$ are polynomials in $K^{q+1}$ and $\tilde{\cal H}$ modulo trivial terms. When the polynomial is quadratic or more in $\tilde{\cal H}$,  then Theorem \ref{quad} states that there is no representative without explicit $x$-dependence in its cohomology class, which implies that it should be rejected. The remaining solutions are the polynomials linear in  $\tilde{\cal H}\vert \,_k=C^{*\; n-p-1+k}_k$ and of arbitrary order in $K^{q+1}$. They are invariant and $x$-independent, they thus belong to the set of  looked-for solutions.
\quad \qedsymbol

\section{Self-interactions}
\label{self-interactions}

The proof is given for a single $[p,q]$-field $\phi$ but
extends trivially to a set $\{\phi^a\}$ containing a finite
number $n$ of them (with fixed $p$ and $q$) by writing some
internal index $a=1,\ldots,N$ everywhere. 
\vspace*{.2cm}

It was shown in Section \ref{cons} that   the
first-order nontrivial consistent local interactions are in
one-to-one correspondence with elements $a$ of the cohomology
$H^{n,0}(s \vert\, d)$ of the BRST-differential $s$ modulo the
total derivative $d\,$, in maximum form-degree $n$ and in ghost
number $0\,$. 
Let us recall
(Section \ref{ss:psga}) that (i) the antifield-independent piece 
is the deformation of the Lagrangian; (ii) the terms linear in the
ghosts contain the information about the deformation of the
reducibility conditions; (iii) the other terms give the
information about the deformation of the gauge algebra.

The general procedure to compute $H^{n,0}(s \vert\, d)$ has been explained in
Section \ref{ss:coh}. One can check that the assumptions stated in the latter section are satisfied by the theory we are dealing with. Indeed, the BRST-differential splits as the sum of the differentials $\g$ and $\d$  given in Section \ref{BRSTdifferential} ; the property (\ref{blip1}) is the consequence of Lemma \ref{invPoinclemma}, \ie  (\ref{blip}) ; finally, one defines the action of the differential $D$ as giving zero except for
$$
D A^{(0,q)}_{\m_1\ldots \m_{p}}=dx^{\m_0}\pa_{[\m_0}A^{(0,q)}_{\m_1\ldots \m_{p}]}=(-)^q dx^{\m_0}D^0_{\m_0\ldots \m_{p}}\,,
$$
and the $D$-degree  is the number of $D^0_{\m_0\ldots \m_{p}}$. This number is obviously bounded at given pureghost number.

\hfill

Let us summarize the computation of Section \ref{ss:coh} . 
A solution $a$ of $sa+db=0$ can be decomposed according to the antifield number as $a=a_0+a_1 + \ldots + a_k$, where $a_i$ 
has antifield number $i$ and  satisfies the descent
 \bqn
\d a_1 + \g a_0 + d b_0 &=&0 \,,\nonumber \\
\d a_2 + \g a_1 + d b_1 &=&0 \,,\nonumber \\
&\vdots &\nonumber \\
\d a_k + \g a_{k-1} + d b_{k-1}&=&0 \,,\nonumber \\
\g a_k&=&0 \label{descente}\,. \eqn 
The last equation of this descent  implies that $a_k=\a_J\,\o^J$ 
where $\a_J$ is an invariant
polynomial and $\o^J$ is a polynomial in the  ghosts of $H(\g)$:
$A^{(p-q,q)}_{\m_{[q]}}$ and $D^0_{\m_{[p+1]}}$. Inserting
this expression for $a_k$ into the second equation from the bottom  leads
to the result that $\a_J$ should be an element of
$H^{n,\,inv}_k(\d\vert\, d)$ \footnote{To be precise, the last statement applies to the component of $\a_J$ of lowest $D$-degree.}. Furthermore, if $\a_J$ is trivial in
this group, then $a_k$ can be removed by trivial redefinitions.
The vanishing of $H^{n,\,inv}_k(\d\vert\, d)$ is thus a sufficient
condition to remove the component $a_k$ from $a$. It is however
not  a necessary condition, as we will see in the sequel.

\subsection{Computation of $a_k$ for $k>1$}

Nontrivial interactions correspond to nontrivial elements
 of $H^{n,\,inv}_k(\d\vert\, d)$. The requirement that
the Lagrangian should be translation-invariant implies
that we can restrict ourselves to $x$-independent elements of
this group. By Theorem \ref{thm8.1},
$H^{n,\,inv}_k(\d\vert\, d)$ contains nontrivial
$x$-independent elements only if $k=p+1-s(q+1)$ for some
non-negative integer $s$. The form of the nontrivial elements is then

\noindent $ \a_k^n=C_k^{*\,n-p-1+k} (K^{q+1})^s \,.$ In order to be
(possibly) nontrivial, $a_k$ must thus be a polynomial linear in
$C_k^{*\,n-p-1+k}$, of order $s$ in the curvature $K^{q+1}$ and of
appropriate orders in the ghosts $A^{(p-q,q)}_{\m_{[q]}}$ and
$D^0_{\m_{[p+1]}}$.

As $a_k$ has ghost number zero, the antifield number of $a_k$  should match its pureghost number. Consequently,  as the ghosts $A^{(p-q,q)}_{\m_{[q]}}$ and $D^0_{\m_{[p+1]}}$ have $ pureghost$ $=p$ and $q$ respectively, the equation $k=rp+mq$ should be satisfied for some positive integers $r$ and $m$.
If there is no couple of integers $r,m$ to match $k$, then no $a_k$ satisfying the  equations of the descent (\ref{descente}) can be constructed and $a_k$ thus vanishes.

In the sequel, we will suppose that $r$ and $m$
satisfying $k=rp+mq$ can be found and classify the different cases
according to the value of $r$ and $m$: (i) $r\geq  2$, (ii)
$r=1$, (iii)  $r=0$, $m>1$, and (iv) $r=0$, $m=1$. We will show
that the corresponding candidates $a_k$  are either obstructed in
the lift to $a_0$ or that they are trivial, except in the case (iv).
In that case, $a_k$ can be lifted but $a_0$ depends explicitly
on $x$ and contains more than two derivatives.

\paragraph{(i) Candidates with $r\geq  2$ :}

The constraints $k\leq p+1$ and $k=rp+mq$ have no solutions\footnote{There is
a solution in the case previously considered in  \cite{Boulanger:2000rq}, where $p=q=1$, $r=2$.
The latter solution gives rise to Einstein's theory of gravity.}.

\paragraph{(ii) Candidates with ${r=1}$ :}

The conditions $k=mq+p\leq p+1$ are only satisfied for $q=1=m$. As shown in a particular case and guessed in general in \cite{Bekaert:2002uh}, the lift of these candidates is obstructed after one step without any additionnal assumption.

 Let us be more explicit.
Given the constraints on $r$, $q$ qnd $m$, one has $k=p+1$ and $s=0$. The candidate thus reads  $$a_{p+1}^n= C_{p+1}^{*\,n}{}_\m A^{(p-1,1)}_\n D^0_{\r_{[p+1]}}f^{\m\vert\n\vert\r_{[p+1]}}\,,$$ where $f$ is some covariantly constant tensor that contracts the indices, \ie it is build out of metrics and Levi-Civita densities. Since $p>1$ and $n>p+2 $ by assumption, $f$ must be the Levi-Civita density: $f^{\m\vert\n\vert\r_{[p+1]}}\sim\ve^{\m\n\r_{[p+1]}} \,$ and the space-time dimension must be  $n=p+3$.
One can easily lift $a_{p+1}^n$ a first time. The lift $a_p^{n-1}$ is  of the form 
$$a_p^{n-1}\sim C_{p}^{*\,n-1}{}_\m \Big( A^{(p-1,1)}_\n D^1_{\r_{[p+1]}} + [A_{\n\s}^{(p-2,1)}+A^{(p-1,0)}_{\n\s}]dx^\s D^0_{\r_{[p+1]}}\Big)\ve^{\m\n\r_{[p+1]}}\,,$$
up to some  signs and factors irrelevant for our argument.

However, there is  an obstruction to the construction of $a_{p-1}^{n-2}\,$.  
Let us first assume that $p>2\,$.
Using $dD^1=K^2\,$, one computes that $\d a_p^{n-1}$ is proportional to $ C_{p-1}^{*\,n-2}{}_\m  A^{(p-1,1)}_\n K^2_{\r_{[p+1]}}\ve^{\m\n\r_{[p+1]}}\,,$ modulo $d$- and $\g$-coboundaries. This term is not $\g$-exact modulo $d\,$.\footnote{This is easily seen by a reasoning similar to the one used at the end of Section \ref{defun}.} The whole candidate must thus vanish.

In the case $p=2\,$, the same obstruction is present, as well as another one. Indeed, the $\d$-variation of the second term of $a_p^{n-1}$ now involves the nontrivial term $ C_{p-1}^{*\,n-2}{}_\m  D^0_{\n\s\t} dx^\s dx^\t D^0_{\r_{[3]}}\ve^{\m\n\r_{[3]}}\,$. Obviously, it does not cancel the first obstruction, so the conclusion stays the same.

\paragraph{(iii) Candidates with $r=0$, ${m>1}$ :}

For a nontrivial candidate to exist at $k=mq$, Theorem \ref{thm8.1} tells us that $p$ and $q$ should
satisfy the relation $p+1=mq+s(q+1)$  for  some positive or null
integer $s$. The candidate then  has the form \bqn a^{n}_{mq} =
 C_{mq \, \n_{[q]}}^{*\,n-p-1+mq}\,
\o^{\n_{[q]}}_{(s,m)}(K,D^0)\,, \nonumber \eqn where 
 $\o_{(s,m)}$ is a polynomial of order $s$ in the curvature form and of order $m$ in the ghost $D^0 $ (see Section
\ref{descentgammamodd} for further details about this $\o$ and the ones that 
appear later in the descent).
 \vspace*{.2cm}

We will  show that these candidates are either trivial or that there is an obstruction to lift them up to $a_0^n$ after $q$ steps.
\vspace*{.2cm}

It is straightforward to check that, for $1\leq  j \leq
q$, the terms
$$a^{n}_{mq-j} = C_{mq-j}^{*\, n-p-1+mq-j} \o^{s(q+1)+j,\,mq-j}$$
 satisfy the descent equations, 
since, as $m>1$, all antifields $C_{mq-j}^{*\, n-p-1+mq-j}$ are
invariant. The set of summed indices $\n_{[q]}$ is implicit as
well as the homogeneity degree of the generating polynomials
$\o_{(s,m)}$. We can thus lift $a^{n}_{mq}$ up to $a^n_{(m-1)q}$.
As $m>1$, this is not yet $a_0\,$.

However, unless $a^n_{mq}$ is trivial, there is no $a^{n}_{(m-1)q-1}$ such that\bqn\g
(a^{n}_{(m-1)q-1}) +\d a^{n}_{(m-1)q} +d
\b^{n-1}_{(m-1)q-1}=0\,.\label{lookfor}\eqn Indeed, we have 
\bqn \d a^{n}_{(m-1)q}&=&-\g(C_{(m-1)q-1}^{*\,n-(s+1)(q+1)}\,
\o^{(s+1)(q+1),\,(m-1)q-1})
\nonumber \\
&&+(-)^{n-mq}\,C_{(m-1)q-1}^{*\,n-(s+1)(q+1)}
K^{q+1}\Big[\frac{\partial^L \o}{\partial D}\Big]^{s(q+1),(m-1)q}
\,. \nonumber \eqn  Without loss of generality, we can suppose
that
$$a^n_{(m-1)q-1} = C^{*\,n-(s+1)(q+1)}_{(m-1)q-1} \,\bar{a}^{(s+1)(q+1)}_0
+ \bar{a}^n_{(m-1)q-1}\,, $$ where there is an implicit summation
over all possible coefficients $\bar{a}^{(s+1)(q+1)}_0$, and most
importantly the two $\bar{a}$'s \textit{ do not}\footnote{This is
not true in the case
--- excluded in this paper --- where $p=q=1$ and $m=2$\,: since
$C^{*}_{(m-1)q-1}\equiv C^*_0 $ has antifield number zero, the antifield number counting  does not forbid that the $\bar{a}$'s
  depend on $C^*_0 $. Candidates arising in this way are treated in
\cite{Boulanger:2000ni} and give rise to a consistent deformation
of Fierz-Pauli's theory in $n=3$.} depend on $C^{*}_{(m-1)q-1} $.
Taking the Euler-Lagrange derivative of Eq.(\ref{lookfor}) with
respect to $C^{*}_{(m-1)q-1} $ yields \bqn \g
(\bar{a}^{(s+1)(q+1)}_{0 }-\o^{(s+1)(q+1),\,(m-1)q-1}) \propto K^{q+1}\Big[\frac{\partial^L \o}{\partial
D}\Big]^{s(q+1),(m-1)q} \,. \nonumber \eqn    The product of
nontrivial elements of $H(\g)$ in the r.h.s. is not $\g$-exact
and constitutes an obstruction to the lift of the candidate,
unless it vanishes. 
The latter happens only when the polynomial
$\o_{(s,m)}$ can be expressed as $$\o^{\n_{[q]}}_{(s,m)}(K,D)=K^{q+1\,\m_{[p+1]}}\frac{\partial^L \tilde{\o}^{\n_{[q]}}_{(s-1,m+1)}(K,D)}{\partial D^{ \m_{[p+1]}}}\,,$$ for some
polynomial $\tilde{\o}^{\n_{[q]}}_{(s-1,m+1)}(K,D)$ of order $s-1$ in $K^{q+1}$ and $m+1$ in $D$.
 However, in this case,  $a^n_{mq}$ can be
removed by the trivial redefinition  $$a^n\rightarrow a^n+ s (\tilde{H}_{\n_{[q]}}\tilde{\o}^{\n_{[q]}}_{(s-1,m+1)} \vert ^n)\,.$$

This completes the proof that these candidates are either trivial
or that their lift is obstructed. As a consequence, they do not
lead to  consistent interactions and can be  rejected. Let us
stress that no extra assumptions are needed to get this result. In the particular case $q=1$, this had already been  guessed  but not been proved in \cite{Bekaert:2002uh}.

\paragraph{(iv) Candidates with $r=0\,$, $m=1$ :}
These candidates exist only when the condition $p+2=(s+1)(q+1)$ is
satisfied, for some strictly positive integer $s\,$. It is
useful for the analysis to write the indices explicitly:
 \bqn
a^n_{q}&=&g^{\n_{[q]}\parallel\,\m^1_{[p+1]}\vert\, \ldots \vert\,
\m^{s+1}_{[p+1]}} \,C_{q\,\n_{[q]}}^{*\,n-p-1+q}  \left(
\prod_{i=1}^sK^{q+1}_{\m^i_{[p+1]}}\right)
D^{0}_{\m^{s+1}_{[p+1]}} \,, \nonumber \eqn where $g$ is a
constant tensor.

We can split the analysis into two cases: (i)  $g \rightarrow
(-)^q g$  under the exchange $\m^s_{[p+1]} \leftrightarrow
\m^{s+1}_{[p+1]}$, and (ii)  $g \rightarrow (-)^{q+1} g$ under the
same transformation.

In the case (i), $a_q^n$ can be removed by adding the trivial term
$s\, m^n$ where $m^n=\sum_{j=q}^{2q}m^n_j$ and $$
m^n_j=(-)^{n-q}\frac{1}{2} \,
g^{\n_{[q]}\parallel\,\m^1_{[p+1]}\vert\, \ldots \vert\,
\m^{s+1}_{[p+1]}} \,\, C^{*\,n-p-1+j}_{j\;\n{[q]}}\left(
\prod_{i=1}^{s-1}K^{q+1}_{\m^i_{[p+1]}}\right)  \Big[D_{\m_{[p+1]}^s}
D_{\m_{[p+1]}^{s+1}}\Big]^{2q+1-j}\,. $$ This construction
does not work in the case (ii) where the symmetry of $g$  makes
$m^n$ vanish.

In the case (ii), the candidate $a_q^n$ can be lifted up to $a_0^n$:
\bqn a^n_0\propto \,
f^{\s_{[p+1]}\parallel\,\m^1_{[p+1]}\vert\, \ldots \vert\,
\m^{s+1}_{[p+1]}} _{\t_{[n-p-q-1]}} x^{\t_1} dx^{\t_2} \ldots
dx^{\t_{n-p-q-1}}\,K^{q+1}_{\s_{[p+1]}}\,\left(
\prod_{i=1}^sK^{q+1}_{\m^i_{[p+1]}}\right) D^q_{\m^{s+1}_{[p+1]}}
\,,\nonumber \eqn where the constant tensor $f$ is defined by
$$f_{\t_{[n-p-q-1]}}^{\s_{[p+1]}\parallel\,\m^1_{[p+1]}\vert\, \ldots
\vert\, \m^{s+1}_{[p+1]}}\equiv
g^{\n_{[q]}\parallel\,\m^1_{[p+1]}\vert\, \ldots \vert\,
\m^{s+1}_{[p+1]}}\,\,\epsilon^{\s_{[p+1]}}_{\quad\,\,\,\,\,\,\n_{[q]}\t_{[n-p-q-1]}}\,.$$Let
us first note that this deformation does not affect the gauge
algebra, since it is linear in the ghosts.

The Lagrangian deformation $a_0^n$ depends explicitly on $x$,
which is not a contradiction with translation invariance of the
physical theory if the $x$-dependence of the Lagrangian can be
removed by adding a total derivative and/or a $\d$-exact term. If
it were the case, $a_0^n$ would have the form $a_0^n=xG (\ldots) +
x^{\a}d (\ldots)_{\a}$. We have no proof that $a_0^n$
does not have this form, but it is not obvious and we think it
very unlikely. In any case, this deformation  is ruled out if one requires
 that the deformation of the Lagrangian contains at
most two derivatives. \vspace*{.2cm}

So far, we have considered all the possible deformations that involve terms $a_k$ with $k\geq 2\,$ and we have checked whether they have a Lagrangian counterpart. We now turn to the deformations that stop at antifield number one or zero.

\subsection{Computation of $a_1$}

The term $a_1$ vanishes without any further assumption
 when $q>1\,$. Indeed, when $q>1\,$,  the vanishing of the
cohomology of $\g$ in {\it puregh $1$} implies that there is
no nontrivial $a_1\,$.

This is not true when $q=1$, as there are some nontrivial
cocycles with pureghost number equal to one. However, it can
be shown \cite{Bekaert:2002uh} that any nontrivial $a_1^n$ leads
to a deformation of the Lagrangian with at least four derivatives.

\subsection{Computation of $a_0$}

This leaves us with the problem of solving the equation
$\g a_0^n+ d\, b_0^{n-1} = 0$ for $a_0^n\,$.
Such solutions correspond to
deformations of the Lagrangian that are invariant up to a total
derivative. 
 Their Euler-Lagrange derivatives $\frac{\d a_0}{\d C}$ must be gauge invariant and must satisfy Bianchi identities of the type  
 (\ref{Noether}) (because of the gauge invariance of $\int a_0$).
 Asking that $a_0$ should not contain more than two derivatives, we obtain that  $\frac{\d a_0}{\d C}$ must be at most linear in the curvature $K\,$. These three conditions together completely constrain $a_0\,$ and have only two Lorentz-invariant solutions.
 The first one is a cosmological-constant-like term that exists only when $p=q$: 
 \begin{eqnarray}
a_0=\Lambda \h_{\m_1 \n_1} \ldots \h_{\m_p \n_p}C^{\m_1 \ldots \m_p \vert \n_1 \ldots \n_p}\,.
\end{eqnarray}
 The second one, where $\frac{\d a_0}{\d C}$ are linear in the curvature $K$,
 is the free Lagrangian itself  \cite{deMedeiros:2003dc}. 
 
 So we conclude that, apart from a cosmological-constant-like term, the deformation only changes the coefficient of the free Lagrangian and is not essential.

\subsection{Results and discussion}

We have investigated
in flat space and under the assumptions of locality and
Poincar\'e invariance
the possibility of introducing interactions consistently.

\vspace{.2cm}

We have shown that there is no consistent smooth deformation of the
free theory for $[p,q]$-type tensor gauge fields with $p>1$ that
modifies the gauge algebra. The algebra thus always remains Abelian, which is unlike the case $p=q=1$ of linearized gravity, since the latter can be consistently deformed into the non-Abelian Einstein theory.

This result can be compared to a similar result for vector fields and $p$-forms. The Maxwell theory of the electromagnetic field can be deformed into non-Abelian Yang-Mills theories, while there are no non-Abelian theories for $p$-forms ($p>1\,$) \cite{Henneaux:1997ha}.

\vspace{.2cm}

The constraint on the deformations that modify the gauge transformations but leave them Abelian is very restrictive as well.
Indeed, for $q>1$, there exists no such deformation when there is no positive integer $r$ such that
$p+2=(r+1)(q+1)$. In that case, there might exist a consistent deformation of the gauge transformations but it is not obvious whether the corresponding deformation of the Lagrangian is invariant under translations or not. 
For $q=1$, there is no strong constraint.
In all cases, the deformations lead to Lagrangians that have at least four derivatives.

One can again compare this result with the corresponding result for $p$-forms.
It is interesting to notice that the potential deformation for $q>1$ has the same structure as the Chapline-Manton deformation of theories with several $p$-forms (see Appendix \ref{chapline}). 
However, in the $[p,q]$-case, the ghost number zero element of $H$ is not gauge invariant as it is for $p$-forms, and it is not known whether there is a gauge invariant element without explicit $x$-dependence in its equivalence class in $H(\d\vert d)\,$. This is the reason for the doubt on the invariance under translations of the candidate.

\vspace{.2cm}

One can also consider interactions that do not modify the gauge transformations. If one excludes deformations that
involve more than two derivatives in the Lagrangian, 
one finds only a cosmological constant-like term for
$p=q$.

No complete analysis has been done for the case where more derivatives are allowed. 
One can however say that any polynomial in the curvature is an acceptable deformation. Furthermore,
analogues of Chern-Simons terms also exist, like the term 
$$a_0= \pa_{[\m_1}\phi_{\m_2 \ldots \m_{p+1}]\vert [\n_1 \ldots \n_q,\n_{q+1}]}
\pa^{[\m_1}\phi^{\m_2\ldots \m_{p+1}]}{}_{\vert\n_{q+2}\ldots \n_{2q+1}}\, dx^{\n_1}\ldots dx^{\n_{2q+1}}$$ in $n=2q+1\,$ and with $q$ odd.

If one introduces other fields, then new possibilities arise. For example, one can couple $[p,q]$-fields to $p'$-forms by a generalization of the Chapline-Manton interaction (Appendix \ref{chapline}). The gauge transformations of the $p'$-form are deformed by this interaction, but not those of the $[p,q]$-field.



%% file: spin3.tex
\chapter{Interactions for spin-3 fields}
\label{spin3}


In this chapter, the problem of introducing consistent interactions among spin-3 gauge fields \cite{Bekaert:2005jf, Boulanger:2005br} is  analysed in Minkowski space-time $\mathbb{R}^{n-1,1}$  
($n\geq 3$) using BRST-cohomological methods. 
Under the assumptions of locality and Poincar\'e  invariance, all the   
perturbative, consistent deformations of the Abelian gauge algebra are determined, together 
with the corresponding deformations of the quadratic action, at first order in the
deformation parameter. Conditions for the consistency of the algebra at second order are examined as well.

Following the cohomological procedure, we first classify all the 
possible first-order deformations  of the spin-3 gauge algebra. 
Then, we investigate whether these algebra-deforming terms give rise to consistent first-order 
vertices. The parity-preserving and the parity-breaking terms are considered separately. In both cases, two  deformations are found  that make the algebra non-Abelian.  All these algebra-deforming terms lead to nontrivial deformations of the quadratic Lagrangian, modulo some constraints on the structure constants. 

When parity invariance is demanded, on top of the covariant cubic vertex of Berends, Burgers and van Dam \cite{Berends:1984wp}, a 
cubic vertex is found which corresponds to a non-Abelian gauge algebra related to an  internal, non-commutative, invariant-normed algebra (like in Yang-Mills's theories). 
This new cubic vertex brings in five derivatives of the field: it is of the form
$\cl_1\sim g_{[abc]}(h^a \pa^2 h^b \pa^3 h^c + h^a \pa h^b \pa^4 h^c)$. At second order, the Berends-Burgers-van Dam vertex is ruled out by a first test of consistency, which the five-derivative vertex passes.

In the parity-breaking case, non-Abelian deformations of the spin-3 algebra exist in space-time dimensions $n=3$ and $n=5\,$, and lead to consistent vertices.   
The first one, in dimension $n=3$, is defined for spin-3 gauge  fields that take value in an 
internal, anticommutative, invariant-normed algebra $\ca$, while the second one is defined in 
a space-time of dimension $n=5$ for fields that take value in a commutative, invariant-normed 
internal algebra $\cb$. 
However, as we demonstrate, consistency conditions at second order in the coupling imply that
the algebras $\ca$ and $\cb$ must also be nilpotent of order three and associative, 
respectively. 
In turn, this means that the $n=3$ parity-breaking deformation is trivial while 
the algebra $\cb$ is a direct sum of one-dimensional ideals --- 
provided the metrics which define the norms in $\ca$ and $\cb$ are positive-definite, which is 
required by the positivity of energy.       
Essentially, this signifies that we may consider only one \emph{single} self-interacting 
spin-3 gauge field in the $n=5$ case, similarly to what happens in 
Einstein gravity \cite{Boulanger:2000rq}.

\vspace{.1cm}

The chapter is organized as follows. In Section
\ref{sec:FreeTheory}, we review the free theory of massless spin-3
gauge fields represented by completely symmetric rank-$3$ tensors. The
sections \ref{sec:BRSTSettings} to \ref{Invariantcharacteristiccohomology3} gather together the main BRST
results needed for the exhaustive treatment of the interaction problem:
The BRST spectrum of the theory is presented in Section
\ref{BRSTspectrum3}. Some cohomological results have already been
obtained in \cite{Bekaert:2005ka}, such as the cohomology $H^*(\g)$
of the gauge differential $\gamma$ and the so called
characteristic cohomology $H_k^n(\d\vert d)$ in antifield number
$k\geq 2$. We recall the content of these groups in Sections
\ref{cohogamma2} and \ref{Characteristiccohomology3}.  Section \ref{invPlemma}
is devoted to the invariant Poincar\'e Lemma and to $H(\g\vert d)\,$. The
calculation of the invariant characteristic cohomology $H_k^{n}(\d
\vert d,H(\g))$ constitutes the core of the BRST analysis and is
achieved in Section \ref{Invariantcharacteristiccohomology3}. Several
technicalities related to Schouten identities left to
 the appendix \ref{schouten}.  The
self-interaction question is finally answered in Sections
\ref{interactions} and  \ref{sec:def}, for  parity-invariant 
and parity-breaking deformations respectively.
To conclude, we summarize the results and discuss them in Section \ref{conclusions}.
  
\vspace{.1cm}

Let us stress that the computations of the cohomology groups are not merely trivial generalizations of the corresponding computations for spin two. Indeed, an important feature of spin-3 fields, which is absent from the spin-2 case, is the tracelessness condition
on the gauge parameter. Quadratic non-local actions \cite{Francia:2002aa,Francia:2002pt} have been proposed in order to get rid of this trace 
constraint, but we do not
discuss the non-local formulation here because an important hypothesis of the BRST procedure is locality.
\footnote{Notice that by introducing a pure gauge field (sometimes refered to as ``compensator''), it is possible to write a local (but higher-derivative) action for spin-3 \cite{Francia:2002aa,Francia:2002pt} that is invariant under 
unconstrained gauge transformations.
Very recently, this action was generalized to the arbitrary spin-$s$ case by further adding an auxiliary field \cite{Francia:2005bu} (see also \cite{Pashnev:1998ti} for an older non ``minimal'' version of it).}

\section{Free theory}
\label{sec:FreeTheory}

The local action for a collection
$\{h^a_{\m\n\r}\}$ of $N$ non-interacting completely symmetric
massless spin-3 gauge fields in flat space-time is
\cite{Fronsdal:1978rb} (see Chapter \ref{appendixA})
\begin{eqnarray}
S_0[h^a_{\m\n\r}] = \sum_{a=1}^N \int d^n x
                &\big[& -\frac{1}{2}\,\pa_{\s}h^a_{\m\n\r}\pa^{\s}h^{a\m\n\r} +
                        \frac{3}{2}\,\pa^{\m}h^a_{\m\r\s}\pa_{\n}h^{a\n\r\s} +
          \nonumber \\
               && \frac{3}{2}\,\pa_{\m}h^a_{\n}\pa^{\m}h^{a\n} +
                  \frac{3}{4}\,\pa_{\m}h^{a\m}\pa_{\n}h^{a\n} -
                        3 \,\pa_{\m}h^a_{\n}\pa_{\r}h^{a\r\m\n} \,\,\,\big]\,,\
\label{freeaction}
\end{eqnarray}
where $h^a_{\m}:=\eta^{\n\r}h^a_{\m\n\r}\,$.
The Latin indices are internal indices taking $N$ values. They are raised
and lowered with the Kronecker delta's $\d^{ab}$ and $\d_{ab}$.
The Greek indices are space-time indices taking $n$ values, which are
lowered (resp. raised) with the ``mostly plus'' Minkowski metric $\eta_{\m\n}$ (resp. $\eta^{\m\n}$).

The action (\ref{freeaction}) is invariant under the gauge
transformations
\begin{eqnarray}
        \delta_{\l}h^a_{\m\n\r} = 3 \,\pa^{}_{(\m}\l^a_{\n\r)}\,,\quad \eta^{\m\n}\l^a_{\m\n}\equiv 0\,,
\label{gaugetransfo3}
\end{eqnarray}
where the gauge parameters $\l^a_{\n\r}$ are symmetric and traceless.
Curved (resp. square) brackets on space-time indices denote strength-one complete symmetrization
(resp. antisymmetrization) of the indices.
The gauge transformations (\ref{gaugetransfo3}) are Abelian and irreducible.

The field equations read
\begin{eqnarray}
        \frac{\d S_0}{\d h^a_{\m\n\r}} \equiv G^{\m\n\r}_a = 0\,,
        \label{eom}
\end{eqnarray}
where
\begin{eqnarray}
G^{a}_{\m\n\r}:=F^{a}_{\m\n\r} -\frac{3}{2}
\eta^{}_{(\m\n}F^a_{\r)} \label{einstein}
\end{eqnarray}
is the ``Einstein'' tensor and $F^{a}_{\m\n\r}$ the Fronsdal (or
``Ricci'') tensor
\begin{eqnarray}
        F^{a}_{\m\n\r} :=
        \Box h^{a}_{\m\n\r} - 3 \,\pa^{\s}\pa^{}_{(\m}h^a_{\n\r)\s} + 3 \, \pa^{}_{(\m}\pa^{}_{\n}h_{\r)}^a\,.
\label{Fronsdal}
\end{eqnarray} We denote $F_{\m}=\h^{\n\r}F_{\m\n\r}\,$. 
The Fronsdal tensor is gauge invariant thanks to the tracelessness
of the gauge parameters. Because the action is invariant under the gauge transformations (\ref{gaugetransfo3}),
$$0=\d_{\l}S_0[h^a_{\m\n\r}]= -3\, \sum_{a=1}^N \int d^n x \,
\Big[\pa^{\r}G^{a}_{\m\n\r}-\frac{1}{n}\,\eta_{\m\n}\pa^{\r}G^{a}_{\r}\Big]\,\l^a_{\m\n}\;,
$$ 
where $G^{a}_{\r} := \eta^{\m\n} G^{a}_{\m\n\r}$, the
Einstein tensor $G^{a}_{\m\n\r}$ satisfies the Noether identities
\begin{eqnarray}
        \pa^{\r}G^{a}_{\m\n\r}-\frac{1}{n}\,\eta_{\m\n}\pa^{\r}G^{a}_{\r}\equiv 0\;.
\label{Noether3}
\end{eqnarray}
These identities have the symmetries of the gauge parameters $\l^a_{\m\n}\,$; in other words,
the l.h.s. of Eq.(\ref{Noether3}) is symmetric and traceless.

The gauge symmetries enable one to get rid of some components of
$h^a_{\m\n\r}\,$, leaving it on-shell with $N^n_3$ independent
physical components, where $N^n_3$ is the dimension of the
irreducible representation of the ``little group'' $O(n-2)$
($n\geq 3$) corresponding to a completely symmetric rank $3$
traceless tensor in dimension $n-2$. One has
$N^n_3=\frac{n^3-3n^2-4n+12}{6}\,$. Of course, $N^{4}_3=2$ for the
two helicity states $\pm 3\,$ in dimension $n=4\,$. Note also that
there is no propagating physical degree of freedom in $n=3$ since
$N^{3}_3=0\,.$ This means that the theory in $n=3$ is topological.

An important object is the curvature (or ``Riemann'') tensor
\cite{Weinberg:1965rz,deWit:1979pe,Damour:1987vm}
\begin{eqnarray}
        K^a_{\a\m|\b\n|\g\r}:= 8 \pa^{}_{[\g} \pa^{}_{[\b}\pa^{}_{[\a}h^a_{\m]\n]\r]}\nn
\end{eqnarray}
which is antisymmetric in $\a\m\,$, $\b\n\,$, $\g\r$ and invariant under gauge transformations
(\ref{gaugetransfo3}), where the gauge parameters $\lambda^a_{\m\n}$ are however {\textit{not}} necessarily traceless.

Its importance, apart from gauge invariance with unconstrained gauge parameters, stems from the fact that the field equations (\ref{eom}) are equivalent\footnote{As usual in field theory, 
we work in the space $\cs$ of $C^{\infty}$ functions that, together with all their derivatives, decrease to zero at infinity faster 
than any 
negative power of the coordinates. In particular, polynomials in $x^{\m}$ are forbidden.} 
to the following equations
\begin{eqnarray}
        \eta^{\a\b} K^a_{\a\m|\b\n|\g\r} = 0\,.\label{TrK}
\end{eqnarray}
This was proved in the work \cite{Bekaert:2003az,Bekaert:2003zq} by combining various former
results \cite{Damour:1987vm,Dubois-Violette:2001jk,DVH,Francia:2002aa,Francia:2002pt}.

There exists another field equation for completely symmetric gauge fields in the 
unconstrained approach, which also involves the curvature tensor but is 
non-local \cite{Francia:2002aa} (see also \cite{Francia:2002pt}). 
The equivalence between both unconstrained field equations was proved in 
\cite{Bekaert:2003az}.
One of the advantages of the non-local field equation of \cite{Francia:2002aa} is 
that it can be derived from an action principle. 
The equation (\ref{TrK}) is obtained from the general 
$n$-dimensional bosonic mixed symmetry case \cite{Bekaert:2003az} by specifying 
to a completely symmetric rank-3 gauge field and is \cite{Bekaert:2003zq} a 
generalization of Bargmann-Wigner's equations in $n=4$ \cite{Bargmann:1948ck}. 
However, it cannot be directly obtained from an action principle. 
For a recent work in direct relation to \cite{Francia:2002aa,Francia:2002pt}, 
see \cite{Francia:2005bu}.
 
Notice that when $n=3$, the equation (\ref{TrK}) implies that the curvature 
vanishes on-shell, which reflects the ``topological'' nature of the theory in the 
corresponding dimension. 
This is similar to what happens in 3-dimensional Einstein gravity, where the vacuum 
field equations $R_{\m\n}:=R^{\a}_{~\,\m\a\n}\approx 0$ imply that the Riemann tensor 
$R^{\a}_{~\,\m\b\n}$ is zero on-shell. 
The latter property derives from the fact that the conformally-invariant Weyl tensor 
identically vanishes in dimension $3\,$, allowing the Riemann tensor to be expressed 
entirely in terms of the Ricci tensor $R_{\m\n}\,$. 
Those properties are a consequence of a general theorem (see \cite{Hamermesch} p. 394) 
which states that a tensor transforming in an irreducible representation of $O(n)$ identically vanishes if the corresponding Young diagram is such that the sum of the lengths of the first two columns exceeds $n\,$.    

Accordingly, in dimension $n=3$ the curvature tensor $K^a_{\a\m|\b\n|\g\r}$ can be 
written \cite{Damour:1987vm} as 
\begin{eqnarray}
        K^a_{\a\m|\b\n|\g\r} \equiv \frac{4}{3}( S^a_{\a\m|[\b[\g}\h_{\r]\n]} + S^a_{\b\n|[\g[\a}
 \h_{\m]\r]}+ S^a_{\g\r|[\a[\b} \h_{\n]\m]} )\,,
\label{curvD3}
\end{eqnarray}
where the tensor $S^a_{\a\m|\n\r}$ is defined, in dimension $n=3$, by
 $$S^a_{\a\m|\n\r}=2\pa_{[\a} F^a_{\m]\n\r}-\frac{3}{2}\Big[ 2\pa_{[\a} F_{\m]}^{a} 
 \,\h_{\n\r} - \pa_{\r} F_{[\a}^{a}\,\h_{\m]\n} - \pa_{\n} F_{[\a}^{a}\,\h_{\m]\r} 
 +  \pa_{\a} F_{(\n}^{a}\,\h_{\r)\m} - \pa_{\m} F_{(\n}^{a}\,\h_{\r)\a}\Big]\,.$$
It is antisymmetric in its first two indices and symmetric in its last two indices. 
For the expression of $S^a_{\a\m|\n\r}$ in arbitrary dimension
$n\geq 1\,$, see \cite{Damour:1987vm} where the curvature tensor 
$K^a_{\a\m|\b\n|\g\r}$ is decomposed under the (pseudo-)orthogonal group $O(n-1,1)\,$. 
The latter reference contains a very careful analysis of the structure 
of Fronsdal's spin-3 gauge theory, as well as an interesting ``topologically massive''  
spin-3 theory in dimension $n=3\,$.   

%
\section{BRST construction}
\label{sec:BRSTSettings}
%
\label{BRSTspectrum3}
%
According to the general rules of the BRST-antifield formalism (Section \ref{s:faf}), a Grassmann-odd ghost
$C_{\m\n}^a$ is introduced, which accompanies each Grassmann-even gauge parameter $\l_{\m\n}^a$.
In particular, it possesses the same algebraic symmetries as $\l_{\m\n}^a$: it is symmetric and
traceless in its space-time indices.
Then, to each field and ghost of the spectrum, a corresponding
antifield  is added, with the same algebraic symmetries but
the opposite Grassmann parity. A $\mathbb{Z}$-grading called {\textit{ghost number}} ($gh$) is associated with the BRST differential $s$, while the {\textit{antifield number}} ($antifield$)
of the antifield $Z^*$ associated with the field (or ghost) $Z$
is given by $antifield(Z^*)\equiv gh(Z)+1\,$.
More precisely, in the theory under consideration, the spectrum of fields (including ghosts)
and antifields together with their respective ghost and antifield numbers is given by
\begin{itemize}
\item the fields $h^a_{\m\n\r}\,$, with ghost number $0$ and antifield number $0$;
\item the ghosts $C^a_{\m\n}\,$, with ghost number $1$ and antifield number $0$;
\item the antifields $h^{*\m\n\r}_a\,$, with ghost number $-1$ and antifield number $1$;
\item the antifields $C^{*\m\n}_a\,$, with ghost number $-2$ and antifield number $2\,$.
\end{itemize}
The BRST differential $s$ of  the free theory (\ref{freeaction}), (\ref{gaugetransfo3}) is generated by the functional
\begin{eqnarray}
W_0 = S_0 [h^a] \;+ \int d^nx\; ( 3\, h^{*\m\n\r}_a \, \pa_{\m} C_{\n\r}^a )\,. \nonumber
\end{eqnarray}
More precisely, $W_0$ is the generator of the BRST differential $s$ of the free theory through
\begin{eqnarray}
        s A = (W_0, A)\, \nonumber
\end{eqnarray}
where the antibracket $(~,~)$ has been defined by Eq.(\ref{antibracket def}).
The functional $W_0$ is a solution of the \emph{master equation}
\begin{eqnarray}
        (W_0,W_0)=0\,.
\end{eqnarray}
In the theory at hand, the BRST-differential $s$ decomposes into 
\be
\label{diffbrst}
s=\g + \d \,.
\ee
The first piece $\g\,$, the differential along the gauge orbits, is associated with another
grading called \textit{pureghost number} ($pureghost$) and increases it by one unit, whereas the Koszul-Tate differential $\d$ decreases the antifield  number by one unit.
The differential $s$ increases the ghost number by one unit.
Furthermore, the ghost, antifield and pureghost gradings are not
independent. We have the relation
\begin{eqnarray}
        gh = pureghost - antifield \,.
\end{eqnarray}

The pureghost number, antifield number, ghost number and Grassmann parity of the various fields are displayed in Table \ref{table13}.

\begin{table}[!ht]
\centering
\begin{tabular}{|c|c|c|c|c|}
\hline Z  & $puregh(Z)$  & $antifield(Z)$  & $gh(Z)$  & parity (mod $2$)\\ \hline
$h^a_{\m\n\r}$   &$0$  & $0$  &$0$ &$0$ \\
$C^a_{\m\n}$ & $1$ & $0$ & $1$ & $1$ \\
$h^{*\m\n\r}_a$ & $0$& $1$ & $-1$ & $1$ \\
$C^{*\m\n}_a$ & $0$ & $2$ & $-2$ & $0$ \\
\hline
\end{tabular}
\caption{\it pureghost number, antifield number, ghost number
and parity of the (anti)fields.\label{table13}}
\end{table}
The action of the differentials  $\delta$ and $\gamma$ gives zero on all the
fields of the formalism except in the few following cases:
\begin{eqnarray}
\d h^{*\m\n\r}_a &=&G^{\m\n\r}_a\,,
\nonumber \\
\d C^{*\m\n}_a &=& -3 ( \pa_{\r}h^{*\m\n\r}_a - \frac{1}{n}\eta^{\m\n}\pa_{\r}h^{*\r}_a )\,,
\nonumber \\
\gamma h^a_{\m\n\r} &=& 3\,\pa^{}_{(\m}C_{\n\r)}^a\,.
\nonumber 
\end{eqnarray}
Let us draw attention on the right-hand side of the second equation. It is built from the Noether identities \bref{Noether3} for the equations of motion by replacing the latter by the antifield $h^{*\m\n\r}_a\,$. It thus exhibits the tracelessness property of the gauge parameter.

\section{Cohomology of $\g$}
\label{cohogamma2}

In the context of local free theories in Minkowski space for
massless spin-$s$ gauge fields represented by completely symmetric
(and double traceless when $s>3$) rank $s$ tensors, the groups
$H^*(\g)$ have recently been calculated \cite{Bekaert:2005ka}.
We only recall the latter results in the special case
$s=3$ and introduce some new notations.
\begin{prop}\label{Hgamma3} The cohomology of $\g$ is
isomorphic to the space of functions depending on
\begin{itemize}
  \item the antifields $h^{*\m\n\r}_a$, $C^{*\m\n}_a$ and their derivatives, denoted by
  $[\Phi^{*i}]\,$,
  \item the curvature and its derivatives $[K^a_{\a\m|\b\n|\g\r}]\,$,
  \item the symmetrized derivatives $\pa^{}_{(\a_1}\ldots\pa^{}_{\a_k}F^a_{\m\n\r)}$ of the Fronsdal tensor,
  \item the ghosts $C_{\m\n}^a$ and the traceless parts of $\pa^{}_{[\a}C_{\m]\n}^a$ and
  $\pa^{}_{[\a}C_{\m][\n,\b]}^a$.
\end{itemize} 
Thus, identifying with zero any $\gamma$-exact term in $H(\gamma)$, we have   
$$ \g f=0 $$
if and only if $$f=
f\left([\Phi^{*i}],[K^a_{\a\m|\b\n|\g\r}],\{F^a_{\m\n\r}\},
                               C_{\m\n}^a, \widehat{T}^a_{\a\m\vert\n}, \widehat{U}^a_{\a\m\vert\b\n}
     \right)$$
where $\{F^a_{\m\n\r}\}$ stands for the completely symmetrized
derivatives $\pa^{}_{(\a_1}\ldots\pa^{}_{\a_k}F^a_{\m\n\r)}$ of
the Fronsdal tensor, while $\widehat{T}^a_{\a\m\vert\n}$ denotes
the traceless part of $T^a_{\a\m\vert\n}:=\pa^{}_{[\a}C_{\m]\n}^a$ and $\widehat{U}^a_{\a\m\vert\b\n}$ the
traceless part of ${U}^a_{\a\m\vert\b\n}:=\pa^{}_{[\a}C_{\m][\n,\b]}^a\,$.
\end{prop}

This proposition provides the possibility of writing down the
most general gauge-invariant interaction terms. Such
higher-derivative Born-Infeld-like Lagrangians were already
considered in \cite{Damour:1987fp}. These deformations are
consistent to all orders but they do not deform the gauge
transformations (\ref{gaugetransfo3}). Also notice that any
function involving the Fronsdal tensor or its derivatives corresponds to
a field redefinition since it is proportional to the equations of motion
(cf. (\ref{trivialinteractions})). \vspace{.3cm}

Let $\{\o^I\}$ be a basis of the space of polynomials in the
$C_{\m\n}^a$, $\widehat{T}^a_{\a\m\vert\n}$ and $\widehat{U}^a_{\a\m\vert\b\n}$
(since these variables anticommute, this space is finite-dimensional).
If a local form $a$ is $\gamma$-closed, we have
\begin{eqnarray}
        \g a = 0 \quad\Rightarrow\quad a \,=\,
        \a_J([\Phi^{i*}],[K],\{F\})\,
        \o^J(C_{\m\n}^a,\widehat{T}^a_{\a\m\vert\n},\widehat{U}^a_{\a\m\vert\b\n}) + \g b\,,
\end{eqnarray}
If $a$ has a fixed, finite ghost number, then $a$ can only contain
a finite number of antifields. Moreover, since the
{\textit{local}} form $a$ possesses a finite number of
derivatives, we find that the $\a_J$ are polynomials. Such 
polynomials $\a_J([\Phi^{i*}],[K],\{F\})$ are called  
{\textit{invariant polynomials}}.\vspace{.3cm}

\noindent {\textbf{Remark 1}:} Because of the Damour-Deser identity
\cite{Damour:1987vm}
$$\eta^{\a\b}K_{\a\m|\b\n|\g\r}\equiv 2\, \pa_{[\g}F_{\r]\m\n}\,,$$ the derivatives of the
Fronsdal tensor are not all independent  of the curvature tensor
$K$. This is why, in Proposition \ref{Hgamma3}, the completely
symmetrized derivatives of $F$ appear, together with all the
derivatives of the curvature $K$. However, from now on, we will
assume that every time the trace $\eta^{\a\b}K_{\a\m|\b\n|\g\r}$
appears, we substitute $2\pa_{[\g}F_{\r]\m\n}$ for it. With this convention, we can write
$\a_J([\Phi^{i*}],[K],[F])$ instead of the unconvenient notation
$\a_J([\Phi^{i*}],[K],\{F\})$.

\vspace{.3cm}

\noindent {\textbf{Remark 2}:} 
Proposition \ref{Hgamma3} must be slightly modified in the special $n=3$ case. 
As we said in the section \ref{sec:FreeTheory}, the curvature tensor $K$ can be expressed in terms of 
the first partial derivatives of the Fronsdal tensor, see Eq.(\ref{curvD3}).  
Moreover, the ghost variable $\widehat{U}^a_{\a\m\vert\b\n}$ identically vanishes 
because it possesses the symmetry of the Weyl tensor. 
Thus, in dimension $n=3$ we have 
\begin{eqnarray}
        \g a = 0 \quad\Rightarrow\quad a \,=\,
        \a_J([\Phi^{i*}],[F])\,
        \o^J(C_{\m\n}^a,\widehat{T}^a_{\a\m\vert\n}) + \g b\,.
\end{eqnarray} 
Another simplifying property in $n=3$ is that the variable $\widehat{T}^a_{\a\m\vert\n}$
can be replaced by its dual 
\begin{eqnarray}
\widetilde{T}^a_{\a\b}:=\ve^{\m\n}_{~~\a}\widehat{T}^a_{\m\n\vert\b} 
\quad (\widehat{T}^a_{\m\n\vert\r}=-\frac{1}{2}\ve_{\m\n}^{~~\a}\widetilde{T}^a_{\a\r})
\label{tdual} 
\end{eqnarray}
which is readily seen to be symmetric and traceless, as a consequence of the symmetries 
of $\widehat{T}^a_{\a\m\vert\n}\,$;  
\begin{eqnarray}
        \widetilde{T}^a_{\a\b}=\widetilde{T}^a_{\b\a}\,, \quad
\h^{\a\b}\widetilde{T}^a_{\a\b}= 0\,.
\label{proptdual}
\end{eqnarray}

\noindent {\textbf{Remark 3}:} It is possible to make a link with
the variables occurring in the frame-like first-order formulation
of free massless spin-3 fields in Minkowski space-time
\cite{Vasiliev:1980as} (see Section \ref{vasidescr}). In this context, the spin-3 field is represented
off-shell by a frame-like object $e_{\m|ab}$, symmetric and
traceless in the internal indices $(a,b)$. The spin-3 connection
$\o_{\m|b|a_1a_2}$ is traceless in the internal latin indices,
symmetric in ($a_1,a_2$) and obeys $\o_{\m|(b|a_1a_2)}\equiv 0$. The
gauge transformations are $\d e_{\m|ab} = \pa_{\m} \xi_{ab} +
\a_{\m|ab}$, $\d \o_{\m|b|a_1a_2} = \pa_{\m}\a_{b|a_1a_2} +
\Sigma_{\m|b|a_1a_2}$, where the parameter $\xi_{ab}$ is symmetric and
traceless in $(a,b)$, the generalized Lorentz parameter
$\a_{\m|ab}$ is completely traceless, symmetric in ($a,b$) and
satisfies the identity $\a_{(\m|ab)}\equiv 0$, so that it belongs
to the $o(n-1,1)$-irreducible module labeled by the Young tableau
{\footnotesize{\begin{picture}(22,14)(0,0)
\multiframe(1,4)(10.5,0){2}(10,10){$a$}{$b$}
\multiframe(1,-6.5)(10.5,0){1}(10,10){$\m$}
\end{picture}}}. Finally, the parameter $\Sigma_{\m|a|bc}$ transforms in the
$o(n-1,1)$ irreducible representation associated with the Young
tableau {\footnotesize{
\begin{picture}(25,14)(0,0)
\multiframe(1,4)(10.5,0){2}(10,10){$b$}{$c$}
\multiframe(1,-6.5)(10.5,0){2}(10,10){$a$}{$\m$}
\end{picture}}},
in the manifestly symmetric convention. By choosing the
generalized Lorentz parameter appropriately, it is possible to
work in the gauge where the frame-field $e_{\m|ab}$ is completely
symmetric, $e_{\m|ab}=e_{(\m|ab)}\equiv h_{\m a b}$. Then, it is
still possible to perform a gauge transformation with parameters
$\a_{\m|ab}$ and $\xi_{ab}$, provided the traceless component of
$\pa_{[\m}\xi_{a]b}$ be equal to $-\a_{[\m|a]b}$. The traceless
component of $\pa_{[\m}\xi_{a]b}$ is nothing but the variable
$\widehat{T}_{\m\a\vert\b}$ in the BRST conventions. Furthermore,
in the 1.5 formalism where the connection is still present in the
action, but viewed as a function of $e_{\m|a_1a_2}$, consistency
with the ``symmetric gauge'' $e_{\m|ab}=e_{(\m|ab)}\equiv h_{\m a
b}$ implies that the traceless component of the second derivative
$\pa_{[a}\xi_{b][c,\mu]}$ be entirely determined by
$\Sigma_{\m|b|ac}$. The traceless component of
$\pa_{[a}\xi_{b][c,\mu]}$ is the variable
$\widehat{U}_{\a\b\vert\c\m}$ in the BRST language. The relations
$\widehat{T}_{\m\a\vert\b}\longleftrightarrow \a_{\m|ab}$ and
$\widehat{U}_{\a\b\vert\c\m}\longleftrightarrow \Sigma_{\m|b|ac}$
are now manifest (note that we work in the manifestly
antisymmetric convention, as opposed to the choice made in
\cite{Vasiliev:1980as}). The variables
$\{C_{\m\n},\widehat{T}_{\m\a\vert\b},\widehat{U}_{\a\b\vert\c\m}\}\in
H(\gamma)$ in the ghost sector are in one-to-one correspondence
with the gauge parameters
$\{\xi_{\m\n},\a_{\m|ab},\Sigma_{\m|b|ac}\}$ of the first-order
formalism \cite{Vasiliev:1980as}.

%
\section{Invariant Poincar\'e lemma and property of $H(\g \vert d)$}
\label{invPlemma}

We shall need several standard results on the cohomology of $d$ in
the space of invariant polynomials.
\begin{prop}\label{2.2}
In form degree less than $n$ and in antifield number strictly greater than $0$,
the cohomology of $d$ is trivial in the space of invariant
polynomials.
That is to say, if $\a$ is an invariant polynomial, the equation
$d \a = 0$ with $antifield(\a) > 0$ implies
$ \a = d \b$ where $\b$ is also an invariant polynomial.
\end{prop}
\noindent The latter property is called {\it Invariant Poincar\'e Lemma}; it is rather generic for gauge theories
(see e.g.  \cite{Boulanger:2000rq} for a proof), as well as
the following:

\begin{prop}\label{csq}
If $a$ has strictly positive antifield number, then the equation
$\gamma a + d b = 0$ is equivalent, up to trivial redefinitions,
to $\gamma a = 0$. More precisely, one can always add $d$-exact
terms to $a$ and get a cocycle $a' := a  + d c$ of $\gamma$, such
that $\g a'= 0$.
\end{prop}
\vspace*{.2cm}

\noindent{\bfseries{Proof}}: Along the lines of 
\cite{Boulanger:2000rq}, we consider the descent associated with
$\gamma a + d b = 0$: from this equation, one infers, by using the
properties $\gamma^2 = 0$, $\gamma d + d \gamma = 0$ and the
triviality of the cohomology of $d$, that $\gamma b + dc = 0$ for
some $c$.  Going on in the same way, we build a ``descent''
\begin{eqnarray}
\gamma a + d b &=& 0\nn\\\gamma b + dc &=& 0\nn\\\gamma c + de &=& 0\,,\nn\\ &\vdots&\label{descent}\\
\gamma m + dn &=& 0\,,\nn\\ \gamma n &=& 0\,.\nn
\end{eqnarray}
in which each successive equation has one less unit of
form-degree. The descent ends with $\gamma n = 0$ either because
$n$ is a zero-form, or because one stops earlier with a
$\gamma$-closed term. Now, because $n$ is $\gamma$-closed, one
has, up to trivial, irrelevant terms, $n = \a_J \omega^J$.
Inserting this into the previous equation in the descent yields
\be 
d (\a_J) \omega^J \pm \a_J d \omega^J + \gamma m = 0 .
\label{keya3} 
\ee 
In order to analyse this equation, we introduce a new differential.
\vspace{2mm}

\noindent \textbf{Definition (differential $D$)}: The action of the differential $D$ on
$h^a_{\mu \nu\rho}$, $h^{*\mu \nu\r}_a$, $C^{*\m\n}_a$ and all
their derivatives is the same as the action of the total
derivative $d$, but its action on the ghosts is given by :
\begin{eqnarray}
D C^a_{\m\n} &=& {\frac {4}{3}} \, d x^{\a}\, {\widehat{T}}^a_{\a(\mu\vert\nu )}\,,
\nonumber \\
D  {\widehat{T}}^a_{\m\a\vert\b} &=& d x^{\r} \, {\widehat{U}}^a_{\m\a\vert\r\b}\,,
\nonumber \\
D(\partial _{( \rho } C^a_{\mu \nu)}) &=& 0\,,
\nonumber \\
D(\partial _{\rho _1 \ldots \rho _t} C_{\mu\n}^a) &=& 0 ~ {\rm \ if \
}~ t\geq 2 .
\end{eqnarray}
The above definitions follow from
\begin{eqnarray}
        \pa_{\a}C^a_{\m\n} &=& \frac{1}{3}(\gamma h^a_{\a\m\n})+\frac{4}{3}T^a_{\a(\m\vert\n)}\,,
        \nonumber \\
        \pa_{\r}T_{\m\a\vert\b} &=& -\frac{1}{2}\,\g(\pa_{[\a}h_{\m]\b\r})+U_{\m\a\vert\r\b}\,,
        \nonumber \\
        \pa_{\r}U_{\m\a\vert\n\b} &=& \frac{1}{3} \g (\pa_{[\m}h_{\a]\r[\b,\n]})\,.
\end{eqnarray}
The operator $D$ thus coincides with $d$ up to $\gamma$-exact terms.

It follows from the definitions that $D\omega^J = A^J{}_I
\omega^I$ for some constant matrix $A^J{}_I$ that involves $dx^\m$
only. One can rewrite Eq.(\ref{keya3}) as \be \underbrace{d
(\a_J) \omega^J \pm \a_J D \omega^J}_{=(d\a_J\,\pm\,\a_I
A^I{}_J)\omega^J} + \gamma m' = 0 \ee which implies, \be d (\a_J)
\omega^J \pm \a_J D \omega^J = 0 \label{keya4} \ee since a term of
the form $\b _J \omega^J $ (with $\b _J$ invariant) is $\g$-exact
if and only if it is zero. It is also convenient to introduce a
new grading.
\vspace{2mm}

\noindent\textbf{Definition ($D$-degree)}: The number of
${\widehat{T}}_{\a\m|\n}$'s plus twice the number of
${\widehat{U}}_{\a\m|\b\n}$'s is called the $D$-degree. It is
bounded because there is a finite number of
${\widehat{T}}_{\a\m|\n}$'s and ${\widehat{U}}_{\a\m|\b\n}$'s,
which are anticommuting.
 The operator $D$ splits as the
sum of an operator $D_1$ that raises the $D$-degree by one unit,
and an operator $D_0$ that leaves it unchanged. $D_0$ has the
same action as $d$ on $h_{\mu \nu\r}$, $h^{*\mu \nu\r}$,
$C^{*\alpha\b}$ and all their derivatives, and gives $0$ when
acting on the ghosts. $D_1$ gives $0$ when acting on all the
variables but the ghosts on which it reproduces the action of $D$.

Let us expand Eq.(\ref{keya3}) according to the $D$-degree. At lowest
order, we get \be d \a_{J_0} = 0 \ee where $J_0$ labels the
$\omega^J$ that contain no derivative of the ghosts ($D\omega^J
= D_1 \omega^J $ contains at least one derivative). This equation
implies, according to Proposition \ref{2.2}, that $\a_{J_0} = d \b
_{J_0}$ where $\b _{J_0}$ is an invariant polynomial. Accordingly,
one can write \be \a_{J_0}\omega^{J_0} = d(\b _{J_0} \omega^{J_0})
\mp \b _{J_0} D\omega^{J_0} + \hbox{ $\g$-exact terms}  . \ee The
term $\b _{J_0} D\omega^{J_0}$ has $D$-degree equal to $1$. Thus,
by adding trivial terms to the last term $n$($=\a_J\o^J$) in
the descent  (\ref{descent}), we can assume that it does not
contain any term of $D$-degree $0$. One can then successively
remove the terms of $D$-degree $1$, $D$-degree $2$, etc, until one
gets $n = 0$. One then repeats the argument for $m$ and the
previous terms in the descent (\ref{descent}) until one gets
$b = 0$, \ie , $\g a = 0$, as requested.\quad \qedsymbol

\section{Cohomology of $\d$ modulo $d\,$: $H^n_k(\d \vert\, d)$}
\label{Characteristiccohomology3}

In this section, we review the local Koszul-Tate cohomology
groups in top form-degree and antifield numbers $k\geq 2\,$.
The group $H^D_1(\d \vert\, d)$ describes the infinitely many
conserved currents and will not be studied here.
\vspace*{.2cm}

Let us first recall that by the general theorem \ref{9.1}, since the 
free spin-3 theory has no reducibility,
\begin{eqnarray}
H_p^n(\d \vert\, d)=0\; for\; p>2\,.  \label{usefll3}
\end{eqnarray}
We are thus left with the computation of
$H_2^n(\d \vert\, d)\,$. The cohomology $H_2^n(\d \vert\, d)$ is given by the following
theorem.

\begin{prop}\label{H2}
A complete set of representatives of $H^n_2(\d\vert d)$ is given by the antifields
$C_a^{*\m\n}$, up to explicitly $x$-dependent terms. In detail,
$$
        \left.
        \begin{array}{ll}
         \delta a^n_2 + d b^{n-1}_1 = 0\,,
        \nonumber \\
         \quad a^n_2 \sim  a^n_2 + \delta c^n_3 + d c^{n-1}_2
         \end{array}\right\}
          \quad \Longleftrightarrow \quad
        \left\{ \begin{array}{ll}
        a^n_2 = L^a_{\m\n}(x)C_a^{*\m\n}d^n x + \delta b^n_3 + d b_2^{n-1}\,,
\nonumber \\
L^a_{\m\n}(x) = \lambda^a_{\m\n} + A^a_{\m\n\vert\r}x^{\r} + B^a_{\m\n\vert\r\s}x^{\r}x^{\s}\,.
\end{array}\right.
$$
The constant tensor $\lambda^a_{\m\n}$ is symmetric and traceless
in the indices $\m\n$, and so are the constant tensors
$A^a_{\m\n\vert\r}$ and $B^a_{\m\n\vert\r\s}$. Moreover, the
tensors $A^a_{\m\n\vert\r}$ and $B^a_{\m\n\vert\r\s}$ transform in
the irreducible representations of $GL(n,\mathbb{R})$ labeled
by the Young tableaux
\begin{picture}(22,16)(0,0)
\multiframe(1,4)(10.5,0){2}(10,10){$\m$}{$\n$}
\multiframe(1,-6.5)(10.5,0){1}(10,10){$\r$}
\end{picture}
and
\begin{picture}(25,16)(0,0)
\multiframe(1,4)(10.5,0){2}(10,10){$\m$}{$\n$}
\multiframe(1,-6.5)(10.5,0){2}(10,10){$\r$}{$\s$}
\end{picture}, meaning that
\begin{eqnarray}
        A^a_{\m\n\vert\r} &=& A^a_{\n\m\vert\r}\,,\quad A^a_{(\m\n\vert\r)}\equiv 0\,,
        \nonumber \\
        B^a_{\m\n\vert\r\s} &=& B^a_{\n\m\vert\r\s} = B^a_{\m\n\vert\s\r}\,, \quad
        B^a_{(\m\n\vert\r)\s} = 0\,.
\end{eqnarray}
Together with the tracelessness constraints on the constant
tensors $A^a_{\m\n\vert\r}$ and

\noindent $B^a_{\m\n\vert\r\s}\,,$ the
$Gl(n,\mathbb{R})$ irreducibility conditions written here above
imply that the tensors $\lambda^a_{\m\n}$, $A^a_{\m\n\vert\r}$ and
$B^a_{\m\n\vert\r\s}$ respectively transform in the
irreducible representations of $O(n-1,1)$ labeled by the Young
tableaux
\begin{picture}(24,16)(0,0)
\multiframe(1,-1)(10.5,0){2}(10,10){$\m$}{$\n$}
\end{picture}, \begin{picture}(22,16)(0,0)
\multiframe(1,4)(10.5,0){2}(10,10){$\m$}{$\n$}
\multiframe(1,-6.5)(10.5,0){1}(10,10){$\r$}
\end{picture}
and
\begin{picture}(25,16)(0,0)
\multiframe(1,4)(10.5,0){2}(10,10){$\m$}{$\n$}
\multiframe(1,-6.5)(10.5,0){2}(10,10){$\r$}{$\s$}
\end{picture}.
\end{prop}
\noindent The proof of Proposition \ref{H2} in the general
spin-$s$ case has been given in  \cite{Bekaert:2005ka} (see also \cite{Barnich:2005bn}).
The spin-3 case under consideration was already written in 
\cite{Nazim}.

\section{Invariant cohomology of $\d$ modulo $d$ 
}
\label{Invariantcharacteristiccohomology3}

We have studied above the cohomology of $\delta$ modulo $d$ in the
space of arbitary local functions of the fields $h^a_{\m \n\r}$,
the antifields $\Phi^{*i}$, and their derivatives.  One can also
study $H^n_k(\delta \vert d)$ in the space of invariant
polynomials in these variables, which involve $h^a_{\m\n\r}$ and
its derivatives only through the curvature $K$, the Fronsdal
tensor $F$, and their derivatives (as well as the antifields and
their derivatives). The above theorems remain unchanged in this
space, {\emph{i.e.}} $H_k^{n,inv}(\d \vert\, d)\cong 0$ for $k>2\,$.
This very nontrivial property is crucial for the computation of 
$H^{n,0}(s \vert\, d)$ and is a consequence of
\begin{theorem}\label{2.6}
Assume that the invariant polynomial $a_{k}^{p}$
($p =$ form-degree, $k =$ antifield number) is $\delta$-trivial modulo $d$,
\begin{eqnarray}
a_{k}^{p} = \delta \mu_{k+1}^{p} + d \mu_{k}^{p-1} ~ ~ (k \geq 2).
\label{2.37}
\end{eqnarray}
Then, one can always choose $\mu_{k+1}^{p}$ and $\mu_{k}^{p-1}$ to be
invariant.
\end{theorem}
To prove the theorem, we need the following lemma, a proof of which can be found e.g. in
\cite{Boulanger:2000rq}.
\begin{lemma} \label{l2.1}
If $a $ is an invariant polynomial that is $\delta$-exact, $a = \d b$,
then, $a $ is $\delta$-exact in the space of invariant polynomials.
That is, one can take $b$ to be also invariant.
\end{lemma}

The next three subsections are devoted to the proof of Theorem \ref{2.6}. As the proof for the space-time dimension $n=3$ is slightly different, we first consider the general case $n>3$ and afterwards the particular case $n=3$.

%
\subsection{Propagation of the invariance in form degree}
%
We first derive a chain of
equations with the same structure as Eq.(\ref{2.37})
\cite{Barnich:1994mt}. Acting with $d$ on Eq.(\ref{2.37}), we get $d
a_{k}^{p} = - \d d \m^{p}_{k+1}$.  Using the lemma and the fact
that $d a_{k}^{p}$ is invariant, we can also write $da_{k}^{p}=
-\d a_{k+1}^{p+1}$ with $a_{k+1}^{p+1}$ invariant. Substituting
this into $d a_{k}^{p} = - \d d \m^{p}_{k+1}$, we get $\d \left[
a_{k+1}^{p+1}-d \m_{k+1}^{p} \right]=0$. As $H(\d)$ is trivial in
antifield number $>0$, this yields \be a_{k+1}^{p+1}=\d
\m^{p+1}_{k+2}+d\m^{p}_{k+1} \ee which has the same structure as
Eq.(\ref{2.37}). We can then repeat the same operations, until we
reach form-degree $n$, \be a^{n}_{k+n-p}=\d \m^{n}_{k+n-p+1}+ d
\m^{n-1}_{k+n-p}. \ee

Similarly, one can go down in form-degree. Acting with $\d$ on
Eq.(\ref{2.37}), one gets $\d a^{p}_{k}=-d (\d \m^{p-1}_{k})$. If the
antifield number $k-1$ of $\d a^{p}_{k}$ is greater than or equal
to one (\ie , $k>1$), one can rewrite, thanks to Proposition
\ref{2.2}, $\d a^{p}_{k}=-d a^{p-1}_{k-1}$ where $a^{p-1}_{k-1}$
is invariant. (If $k=1$ we cannot go down and the bottom of the
chain is Eq.(\ref{2.37}) with $k=1$, namely
$a_1^p=\d\m_2^p+d\m_1^{p-1}$.) Consequently $d \left[
a^{p-1}_{k-1}-\d \m^{p-1}_{k} \right]=0$ and, as before, we deduce
another equation similar to Eq.(\ref{2.37}) : \be a^{p-1}_{k-1}=\d
\m^{p-1}_{k}+d\m^{p-1}_{k-1}. \ee Applying $\d$ on this equation
the descent continues. This descent stops at form degree zero or
antifield number one, whichever is reached first, \ie , \bqn
&{\rm{either}}&~~a^{0}_{k-p}=\d \m^{0}_{k-p+1}
\nonumber \\
&{\rm{or}}&~~a^{p-k+1}_{1}=\d \m^{p-k+1}_{2}+d \m^{p-k}_{1}.
\eqn
Putting all these observations together we can write the entire descent as
\bqn
a^{n}_{k+n-p}  &=& \d \m^{n}_{k+n-p+1}+d \m^{n-1}_{k+n-p}
\nonumber \\
& \vdots &
\nonumber \\
a^{p+1}_{k+1}  &=& \d \m^{p+1}_{k+2}+d \m^{p}_{k+1}
\nonumber \\
a^{p}_{k}  &=& \d \m^{p}_{k+1}+d \m^{p-1}_{k}
\nonumber \\
a^{p-1}_{k-1}  &=& \d \m^{p-1}_{k}+d \m^{p-2}_{k-1}
\nonumber \\
& \vdots &
\nonumber \\
{\rm{either}}~~a^{0}_{k-p}&=&\d \m^{0}_{k-p+1}
\nonumber \\
{\rm{or}}~~a^{p-k+1}_{1}&=&\d \m^{p-k+1}_{2}+d \m^{p-k}_{1}
\eqn
where all the $a^{p \pm i}_{k \pm i}$ are invariants.

Let us show that when one of the $\m$'s in
the chain is invariant,
we can actually choose all the other $\m$'s in such a way that they
share this property. In other words, the invariance property propagates up and down in the ladder.
Let us thus assume that $\m^{c-1}_{b}$ is
invariant. This $\m^{c-1}_{b}$ appears in
two equations of the descent :
\bqn
a^{c}_{b} &=& \d \m^{c}_{b+1}+d \m^{c-1}_{b},
\nonumber \\
a^{c-1}_{b-1} &=& \d \m^{c-1}_{b}+ d \m^{c-2}_{b-1} \eqn (if we
are at the bottom or at the top, $\m^{c-1}_{b}$ occurs in only one
equation, and one should just proceed from that one). The first
equation tells us that $ \d \m^{c}_{b+1}$ is invariant. Thanks to
Lemma \ref{l2.1} we can choose $\m^{c}_{b+1}$ to be invariant.
Looking at the second equation, we see that $ d \m^{c-2}_{b-1}$ is
invariant and by virtue of Proposition \ref{2.2}, $\m^{c-2}_{b-1}$
can be chosen to be invariant since the antifield number $b$ is
positive. These two $\m$'s appear each one in two different
equations of the chain, where we can apply the same reasoning. The
invariance property propagates then to all the $\m$'s.
Consequently, it is enough to prove the theorem in form degree
$n$.

\subsection{Top form-degree}

Two cases may be distinguished depending on whether the antifield number $k$ is greater than $n$ or not.\vspace{.3cm}

In the first case, one can prove the following lemma:
\begin{lemma} \label{l2.2}
If $a^n_k$ is of antifield number $k>n$, then the ``$\m$''s in Eq.(\ref{2.37}) can
be taken to be invariant.
\end{lemma}
\noindent{\bf{Proof for $k>n$}} : If $k>n$, the last equation of the
descent is $a^{0}_{k-n}=\d \m^{0}_{k-n+1}$. We can, using Lemma
\ref{l2.1}, choose $\m^{0}_{k-n+1}$ invariant, and so, all the
$\m$'s can be chosen to have the same property.
\quad \qedsymbol
\vspace{.3cm}

It remains therefore to prove Theorem \ref{2.6} in the case where
the antifield number satisfies $k\leq n$.
Rewriting the top
equation (\ie Eq.(\ref{2.37}) with $p=n$) in dual notation, we have
\be a_k=\d b_{k+1}+\pa_{\r}j^{\r}_{k},~ (k\geq 2).
\label{2.44} \ee We will work by induction on the antifield
number, showing that if the property expressed in Theorem \ref{2.6} is true for $k+1$ (with
$k>1$), then it is true for $k$. As we already know that it is
true in the case $k>n$, the theorem will be proved.\vspace{.4cm}

\noindent{\bf{Inductive proof for $k\leq n$}} : The proof follows the lines of
 \cite{Barnich:1994mt} and decomposes into three parts. First, all Euler-Lagrange derivatives of Eq.(\ref{2.44}) are computed. Second, the Euler-Lagrange (E.L.) derivative of an invariant quantity is also invariant. This property is used to express the E.L. derivatives of $a_k$ in terms of invariants only. Third, the homotopy formula is used to reconstruct $a_k$ from his E.L. derivatives. \vspace{1.5mm}

{\bf (i)} Let us take the E.L.  derivatives of Eq.(\ref{2.44}). Since the E.L.
derivatives with respect to the $C^*_{\a}$ commute with $\d$, we
get first :
\begin{eqnarray}
\frac{\d^L a_k}{\d C^*_{\a\b}} =\d Z^{\a\b}_{k-1}
\label{2.45}
\end{eqnarray}
with
$Z^{\a\b}_{k-1}=\frac{\d^L b_{k+1}}{\d C^*_{\a\b}}$. For the E.L.
derivatives of $b_{k+1}$ with respect to $h^*_{\m\n\r}$ we obtain,
after a direct computation,
\begin{eqnarray}
\frac{\d^L a_k}{\d h^*_{\m\n\r}}=-\d X^{\m\n\r}_k + 3\pa^{(\m}Z^{\n\r)}_{k-1}.
\label{2.46}
\end{eqnarray}
where $X^{\m\n\r}_{k}=\frac{\d^L b_{k+1}}{\d h^*_{\m\n\r} }$.
Finally, let us compute the E.L. derivatives of $a_k$ with respect to the fields.
We get :
\begin{eqnarray}
\frac{\d^L a_k}{\d h_{\m\n\r}}=\d Y^{\m\n\r}_{k+1} + {\cg}^{\m\n\r\vert\a\b\g}
X_{\a\b\g\vert k}
\label{2.47}
\end{eqnarray}
where $Y^{\m\n\r}_{k+1}=\frac{\d^L b_{k+1}}{\d h_{\m\n\r}}$ and
${\cg}^{\m\n\r\vert\a\b\g}(\partial)$ is the second-order self-adjoint differential operator
appearing in the equations of motion (\ref{eom}): $$G^{\m\n\r}={\cg}^{\m\n\r\vert\a\b\g}\,h_{\a\b\g}\,.$$
The hermiticity of $\cg$ implies
${\cg}^{\m\n\r\vert\a\b\g}={\cg}^{\a\b\g\vert\m\n\r}$.\vspace{2mm}

{\bf (ii)} The E.L. derivatives of an invariant object are invariant. Thus,
$\frac{\d^L a_k}{\d C^*_{\a\b}}$ is invariant.  Therefore, by Lemma \ref{l2.1}
and Eq.(\ref{2.45}), we have also
\begin{eqnarray}
\frac{\d^L a_k}{\d C^*_{\a\b}} =\d Z'^{\a\b}_{k-1}
\label{2.45'}
\end{eqnarray}
for some invariant $Z'^{\a\b}_{k-1}$.
Indeed, let us write the decomposition 
$Z^{\a\b}_{k-1} = Z'^{\a\b}_{k-1} + {\tilde{Z}}^{\a\b}_{k-1}$,
where ${\tilde{Z}}^{\a\b}_{k-1}$ is obtained from ${Z}^{\a\b}_{k-1}$ by setting to zero all
the terms that belong only to $H(\g)$. The latter operation clearly commutes
with taking the $\d$ of something, so that Eq.(\ref{2.45}) gives  $0 = \d {\tilde{Z}}^{\a\b}_{k-1}$ which,
by the acyclicity of $\d$, yields ${\tilde{Z}}^{\a\b}_{k-1}=\d \s_k^{\a\b}$ where
$\s_k^{\a\b}$ can be chosen to be traceless.
Substituting $\d \s_k^{\a\b} + Z'^{\a\b}_{k-1}$ for $Z^{\a\b}_{k-1}$ in Eq.(\ref{2.45})
gives Eq.(\ref{2.45'}).

Similarly, one easily verifies that
\begin{eqnarray}
\frac{\d^L a_k}{\d h^*_{\m\n\r}}=-\d X'^{\m\n\r}_k + 3\pa^{(\m}Z'^{\n\r)}_{k-1}\,,
\label{2.46'}
\end{eqnarray}
where $X^{\m\n\r}_k = X'^{\m\n\r}_k + 3\pa^{(\m}\s^{\n\r)}_{k} + \d \r_{k+1}^{\m\n\r}$.
Finally, using ${\cg}^{\m\n\r}{}_{\a\b\g}\,\pa^{(\a}\s^{\b\g)}{}_k=0$ due to
the gauge invariance of the equations of motion ($\s_{\a\b}$ has been taken traceless), we find
\begin{eqnarray}
\frac{\d^L a_k}{\d h_{\m\n\r}} = \d Y'^{\m\n\r}_{k+1}+{\cg}^{\m\n\r}{}_{\a\b\g}
{X'}^{\a\b\g}_k
\label{2.47'}
\end{eqnarray}
for the invariants $X'^{\m\n\r}_k$ and $Y'^{\m\n\r}_{k+1}$.
Before ending the argument by making use of the homotopy formula,
it is necessary to know more about the invariant $Y'^{\m\n\r}_{k+1}$.

Since $a_k$ is invariant, it depends on the fields only
through the curvature $K$, the Fronsdal tensor and their derivatives.
(We remind the reader of  our convention of Section \ref{cohogamma2} to substitute $2\pa_{[\g}F_{\r]\m\n}$
for $\eta^{\a\b}K_{\a\m|\b\n|\g\r}$ everywhere.)
We then express the Fronsdal tensor in terms of the
Einstein tensor (\ref{einstein}): $F_{\m\n\r} = G_{\m\n\r} - \frac{3}{n}\eta_{(\m\n}G_{\r)}$, so that we can write $a_k = a_k([\Phi^{*i}],[K],[G])$\,, where $[G]$ denotes the
Einstein tensor and its derivatives.
We can thus write
\begin{eqnarray}
\frac{\d^L a_k}{\d h_{\m\n\r}} = {\cg}^{\m\n\r}{}_{\a\b\g}
{A'}^{\a\b\g}_k + \pa_{\a}\pa_{\b}\pa_{\g}{M'}^{\a\m\vert\b\n\vert\g\r}_k
\label{2.49}
\end{eqnarray}
where $${A'}^{\a\b\g}_k\propto\frac{\d a_k}{\d G_{\a\b\g}}$$ and $${M'}_k^{\a\m\vert\b\n\vert\g\r}\propto {\frac{\d a_k}{\d K_{\a\m\vert\b\n\vert\g\r}}}$$ are both invariant and respectively have the same symmetry properties as the ``Einstein'' and ``Riemann'' tensors.

Combining Eq.(\ref{2.47'}) with Eq.(\ref{2.49}) gives
\begin{eqnarray}
\d Y'^{\m\n\r}_{k+1} = \pa_\a\pa_\b\pa_\g{M'}_k^{\a\m\vert\b\n\vert\g\r}
                       + {\cg}^{\m\n\r}{}_{\a\b\g} {B'}^{\a\b\g}_k
\label{2.50}
\end{eqnarray}
with ${B'}^{\a\b\g}_k:={A'}^{\a\b\g}_k-{X'}^{\a\b\g}_k$.
Now, only the first term on the right-hand side of Eq.(\ref{2.50}) is divergence-free,
$\pa_{\m}(\pa_{\a\b\g}{M'}_k^{\a\m\vert\b\n\vert\g\r})\equiv0$,
not the second one which instead obeys
a relation analogous to the Noether identities (\ref{Noether3}).\footnote{This is were the computation for spin 3 starts to diverge from the computation for lower spins.
In the latter case, the second term on the right-hand side of Eq.(\ref{2.50}) is also divergenceless. For spins higher than two, only the traceless part of its divergence vanishes, which complicates the subsequent calculations.
}
 As a result, we have
$\d\Big[\pa_{\m}({Y'}^{\m\n\r}_{k+1}-\frac{1}{n}\eta^{\n\r}{Y'}_{k+1}^{\m})\Big]=0\,$,
where ${Y'}_{k+1}^{\m}\equiv \eta_{\n\r}{Y'}_{k+1}^{\m\n\r}\,$.
By Lemma \ref{l2.1}, we deduce
\begin{eqnarray}
        \pa_{\m}({Y'}^{\m\n\r}_{k+1}-\frac{1}{n}\eta^{\n\r}{Y'}_{k+1}^{\m})+\d {F'}_{k+2}^{\n\r}=0
        \,, \label{truc}
\end{eqnarray}
where ${F'}_{k+2}^{\n\r}$ is invariant and can be chosen symmetric and traceless.
Eq.(\ref{truc}) determines a cocycle of $H^{n-1}_{k+1}(d\vert\d)$, for given
 $\n$ and $\r$. Using the general isomorphisms
 $H^{n-1}_{k+1}(d\vert\d)\cong H^{n}_{k+2}(\d\vert d)\cong 0$ ($k\geq 1$)
 \cite{Barnich:1994db} gives
\begin{eqnarray}
        {Y'}^{\m\n\r}_{k+1}-\frac{1}{n}\eta^{\n\r}{Y'}_{k+1}^{\m}=
        \pa_{\a}T_{k+1}^{\a\m\vert\n\r} + \delta P^{\m\n\r}_{k+2}
        \,, \label{truc2}
\end{eqnarray}
where both $T^{\a\m\vert\n\r}_{k+1}$ and $P^{\m\n\r}_{k+2}$ are
invariant by the induction hypothesis. Moreover,
$T^{\a\m\vert\n\r}_{k+1}$ is antisymmetric in its first two
indices. The tensors $T^{\a\m\vert\n\r}_{k+1}$ and
$P^{\m\n\r}_{k+2}$ are both symmetric-traceless in $(\n, \r)$.
This results easily from taking the trace of Eq.(\ref{truc2}) with
$\eta_{\n\r}$ and using the general isomorphisms
$H^{n-2}_{k+1}(d\vert\d)\cong H^{n-1}_{k+2}(\d\vert d)\cong
H^{n}_{k+3}(\d\vert d)\cong 0$ \cite{Barnich:1994db} which hold since $k$ is positive. From Eq.(\ref{truc2}) we obtain
\begin{eqnarray}
        {Y'}^{\m\n\r}_{k+1} =
        \pa_{\a} [ T_{k+1}^{\a\m\vert\n\r}+\frac{1}{n-1}\eta^{\n\r}T_{k+1}^{\a\vert\m} ]
        + \delta [ P^{\m\n\r}_{k+2} + \frac{1}{n-1}\eta^{\n\r}P^{\m}_{k+2} ]\,,
         \label{truc3}
\end{eqnarray}
where $T_{k+1}^{\a\vert\m}\equiv \eta_{\n\r}T_{k+1}^{\a\n\vert\r\m}$ and
$P_{k+2}^{\m}\equiv \eta_{\n\r}P_{k+2}^{\n\r\m}\,$.
Since $Y'^{\m\n\r}_{k+1}$ is symmetric in $\m$ and $\n$, we have also
$\pa_{\a}[T_{k+1}^{\a[\m\vert\n]\r}+\frac{1}{n-1}T_{k+1}^{\a\vert[\m}\eta^{\n]\r}]$
$+\;\delta [ P^{[\m\n]\r}_{k+2} + \frac{1}{n-1}\eta^{\r[\n}P^{\m]}_{k+2} ]=0\,$.
The triviality of $H^{n}_{k+2}(d \vert \d)$ ($k>0$) implies again that
$(P^{[\m\n]\r}_{k+2} + \frac{1}{n-1}\eta^{\r[\n}P^{\m]}_{k+2})$ and
$(T_{k+1}^{\a[\m\vert\n]\r}+\frac{1}{n-1}T_{k+1}^{\a\vert[\m}\eta^{\n]\r})$ are trivial,
in particular,
\begin{eqnarray}
T_{k+1}^{\a[\m\vert\n]\r}+\frac{1}{n-1}T_{k+1}^{\a\vert[\m}\eta^{\n]\r}
=\pa_{\b}S^{\b\a\vert\m\n\vert\r}_{k+1}+\d Q^{\a\m\n\r}_{k+2}
\label{derStoT}
\end{eqnarray}
where $S^{\b\a\vert\m\n\vert\r}_{k+1}$ is antisymmetric in ($ \b, \a$) and ($\m,\n$).
Moreover, it is traceless in $\m, \n, \r\,$ as the left hand side of the above equation
shows.
The induction assumption allows us to choose
$S^{\b\a\vert\m\n\vert\r}_{k+1}$ and $Q^{\a\m\n\r}_{k+2}$ invariant.
We now project both sides of Eq.(\ref{derStoT}) on the symmetries of the Weyl tensor.
For example, denoting by $W^{\b\vert\m\n\vert\a\r}_{k+1}$ the projection
${\cw}^{\m\;\n\;\a\;\r\;}_{\m'\n'\a'\r'}S^{\b\a'\vert\m'\n'\vert\r'}_{k+1}$ of
$S^{\b\a\vert\m\n\vert\r}_{k+1}$, we have
\begin{eqnarray}
        W^{\b\vert\m\n\vert\a\r}_{k+1} &=& W^{\b\vert\a\r\vert\m\n}_{k+1}
        = - W^{\b\vert\n\m\vert\a\r}_{k+1} = - W^{\b\vert\m\n\vert\r\a}_{k+1}\,,
        \nonumber \\
        W^{\b\vert\m[\n\vert\a\r]}_{k+1} &=& 0\,,\quad \eta_{\m\a}W^{\b\vert\m\n\vert\a\r}_{k+1}=0\,.
        \nonumber
\end{eqnarray}
As a consequence of the symmetries of $T_{k+1}^{\a\m\vert\n\r}$, the projection of Eq.(\ref{derStoT}) on the symmetries of the Weyl tensor gives
\begin{eqnarray}
        0 = \pa_{\b}W^{\b\vert\m\n\vert\a\r}_{k+1}+ \d (\dots)
\label{derW}
\end{eqnarray}
where we do not write the (invariant) $\d$-exact terms explicitly because they play no role in
what follows.
Eq.(\ref{derW}) determines, for given $(\m, \n, \a, \r)$, a cocycle of
$H^{n-1}_{k+1}(d\vert\d,H(\gamma))$.  Using again the isomorphisms \cite{Barnich:1994db}
$H^{n-1}_{k+1}(d\vert\d)\cong H^{n}_{k+2}(\d\vert d)\cong 0$ ($k\geq 1$) and
the induction hypothesis, we find
\begin{eqnarray}
        W^{\b\vert\m\n\vert\a\r}_{k+1} = \pa_{\g}\phi^{\g\b\vert\m\n\vert\a\r}_{k+1} +
        \d (\dots)
        \label{Wintermsofphi}
\end{eqnarray}
where $\phi^{\g\b\vert\m\n\vert\a\r}_{k+1}$ is invariant, antisymmetric in $(\g, \b)$ and
possesses the symmetries of the Weyl tensor in its last four indices. The $\d$-exact
term is invariant as well.
Then, projecting the invariant tensor $4\,\phi^{\g\b\vert\m\n\vert\a\r}_{k+1}$ on the
symmetries of the curvature tensor $K^{\g\b\vert\m\n\vert\a\r}$ and calling the
result $\Psi^{\g\b\vert\m\n\vert\a\r}_{k+1}$ which is of course invariant, we find after some rather lengthy algebra (which takes no time using \emph{Ricci} \cite{Lee})
\be
        {Y'}^{\m\n\r}_{k+1} = \pa_{\a}\pa_{\b}\pa_{\c}\Psi^{\a\m\vert\b\n\vert\c\r}_{k+1}
                              + {\cg}^{\m\n\r}{}_{\a\b\g}\widehat{X}^{\a\b\g}{}_{k+1}+\d(\ldots)\,,
                             \label{result1} \ee
with
\be
        \widehat{X}_{\a\b\g\vert k+1} := \frac{2}{n-2}\cy^{\s\t\r}_{\a\b\c}\Big(
        -S^{\m}_{~\;\s|\m\t|\r\;k+1}
        +\frac{1}{n}\eta_{\s\t}[S_{\m\n|~~\;|\r\;k+1}^{~~~\m\n}+
        S_{\m\n|~\;\r|~~k+1}^{~~~\m~~\n}]\Big)
        \label{result2}
\ee
where $\cy^{\s\t\r}_{\a\b\c}=\cy^{(\s\t\r)}_{(\a\b\c)}$ projects on completely symmetric  rank-3 tensors.\vspace{2mm}

{\bf (iii)} We can now complete the argument. The homotopy formula
\begin{eqnarray}
a_k = \int^{1}_{0}dt\,\left[C^*_{\a\b}\frac{\d^L a_k}{\d C^*_{\a\b}}+
h^*_{\m\n\r}\frac{\d^L a_k}{\d h^*_{\m\n\r}}+
h_{\m\n\r}\frac{\d^L a_k}{\d h_{\m\n\r}}\right](th\,,\,th^*\,,\,tC^*)
\label{homotopy}
\end{eqnarray}
enables one to reconstruct $a_k$ from its E.L. derivatives. Inserting the expressions (\ref{2.45'})-(\ref{2.47'}) for
these E.L. derivatives, we get
\begin{eqnarray}
a_k=\d\Big(\int^{1}_{0}dt\,[C^*_{\a\b}Z'^{\a\b}_{k-1}
+h^*_{\m\n\r}X'^{\m\n\r}_{k}+h_{\m\n\r}Y'^{\m\n\r}_{k+1}](t)\,\Big)+\pa_{\r}k^{\r}.\label{invhf}
\end{eqnarray}
The first two terms in the argument of $\d$ are manifestly invariant. To prove that the third term can be assumed to be invariant in Eq.(\ref{invhf}) without loss of generality, we use Eq.(\ref{result1}) to find that
$$h_{\m\n\r}\,Y'^{\m\n\r}_{k+1}=-\Psi^{\a\m\vert\b\n\vert\c\r}_{k+1}K_{\a\m\vert\b\n\vert\c\r}
+G_{\a\b\g}\widehat{X}^{\a\b\g}{}_{k+1}+\pa_\r \ell^\r+\d(\ldots)\,,$$
where we integrated by part thrice to get the first term of the r.h.s. while
the hermiticity of ${\cg}^{\m\n\r\vert\a\b\g}$ was used to obtain the second term.

We are left with $a_k = \d \m_{k+1} +
\pa_{\r}\n^{\r}_k\,$, where $\m_{k+1}$ is invariant. That
$\n^{\r}_{k}$ can now be chosen invariant is straightforward.
Acting with $\g$ on the last equation yields $\pa_{\r} (\g
{\n}^{\r}_{k}) =0\,$. By the Poincar\'e lemma, $\g {\n}^{\r}_{k} =
\pa_{\s} (\t_k^{[\r \s]})\,$. Furthermore, Proposition \ref{csq}
on $H(\g\vert\, d)$ for positive antifield number $k$ implies that
one can redefine $\n^{\r}_{k}$ by the addition of trivial
$d$-exact terms such that one can assume $\g {\n}^{\r}_{k}=0\,$.
As the pureghost number of ${\n}^{\r}_{k}$ vanishes, the last
equation implies that $\n^{\r}_{k}$ is an invariant
polynomial. 

This ends the proof for $n>3$. \quad \qedsymbol

\subsection{Special case $n=3$}

Let us point out the place where 
the proof of Theorem \ref{2.6} must be adapted to $n=3\,$  \cite{Bekaert:2005jf}. 
It is when one makes use of the projector on the symmetries of the Weyl tensor. Above, the equations (\ref{truc3}) and (\ref{derStoT}) 
 are used to obtain (\ref{result1}) and (\ref{result2}). During this procedure, one had to project 
$\pa_{\b}S^{\b\a\vert\m\n\vert\r}_{k+1}$
on the symmetries of the Weyl tensor. 
In dimension $3$, this gives zero identically. 

If we denote by $W^{\b\vert\m\n\vert\a\r}_{k+1}$ the projection
${\cw}^{\m\;\n\;\a\;\r\;}_{\m'\n'\a'\r'}S^{\b\a'\vert\m'\n'\vert\r'}_{k+1}$ of
$S^{\b\a\vert\m\n\vert\r}_{k+1}$ on the symmetries of the Weyl tensor, we have of course 
$W^{\b\vert\m\n\vert\a\r}_{k+1}=0$.  
Then, obviously
$$        0 = \frac{2}{3}\,\pa_{\a}\pa_{\b}\Big[
        W^{\m\vert\a\n\vert\b\r}_{k+1}+W^{\m\vert\a\r\vert\b\n}_{k+1}+
        W^{\n\vert\a\m\vert\b\r}_{k+1}+W^{\n\vert\a\r\vert\b\m}_{k+1}+
        W^{\r\vert\a\m\vert\b\n}_{k+1}+W^{\r\vert\a\n\vert\b\m}_{k+1}
        \Big].
$$
Substituting for $W^{\m\vert\a\n\vert\b\r}_{k+1}$ its expression in terms of 
$S^{\a\b\vert\g\d\vert\r}_{k+1}$ and using Eqs.(\ref{truc3}) and (\ref{derStoT})
we find 
$0 = {Y}^{\m\n\r}_{k+1}-{\cg}^{\m\n\r}{}_{\a\b\g}\widehat{X}^{\a\b\g}{}_{k+1}+\d(\ldots)\,$, 
where $\widehat{X}^{\a\b\g}{}_{k+1}$ is still given by Eq.(\ref{result2}). 
The result (\ref{result1}) is thus recovered except for the first $\Psi\,$-term. 
This is linked to the fact that, in $n=3$, an invariant polynomial depends on the field 
$h_{\m\n\r}$ only through the Fronsdal tensor $F^{\m\n\r}$,  
see Eq.(\ref{curvD3}). The Eqs.(\ref{2.49}) and (\ref{2.50})  are changed accordingly. 
The proof then proceeds as in the general case $n>3$, where one sets $\Psi$ to zero.  \quad \qedsymbol
%

\section{Parity-invariant self-interactions}
\label{interactions}

As explained in Section \ref{cons}, nontrivial
consistent interactions are in one-to-one correspondance with
elements of $H^{n,0}(s\vert d)$, {\it i.e. }  solutions $a$ of the
equation \be s a+ d b =0\,, \label{topeq}\ee with form-degree $n$
and ghost number zero, modulo the equivalence relation
$$a\sim a+sp+dq\,.$$
Moreover, one can quite generally expand $a$ according to the antifield
number, as \be a=a_0+a_1+a_2+ \ldots a_k\,,\label{antighdec}\ee
where $a_i$ has antifield number $i$.  The expansion stops at some
finite value of the antifield number by locality, as was proved in
\cite{Barnich:1994mt}.
Let us recall (see also Section \ref{ss:psga}) the meaning of the various
components of $a$ in this expansion. The antifield-independent
piece $a_0$ is the deformation of the Lagrangian; $a_1$, which is
linear in the antifields $h^{*\m \n\r}$, contains the information
about the deformation of the gauge symmetries, given by the
coefficients of $h^{*\m \n\r}$; $a_2$ contains the information
about the deformation of the gauge algebra (the term $C^{*} C C$
gives the deformation of the structure functions appearing in the
commutator of two gauge transformations, while the term $h^* h^* C
C$ gives the on-shell closure terms); and the $a_k$ ($k>2$) give
the informations about the deformation of the higher-order
structure functions and the reducibility conditions.

Using the cohomological theorems of the previous sections and the
reasoning of Section \ref{ss:coh}, one can remove all components of $a$ with antifield
number greater than 2. Indeed, the properties required to use the analysis of Section  \ref{ss:coh} are satisfied: (i) is just Eq.\bref{diffbrst}, (ii) is Proposition \ref{csq}, and
(iii) is true since there are only a finite number of ghosts in $H(\g)$ at given pureghost number (see Proposition \ref{Hgamma3}).
Then the key point in the analysis is that the invariant characteristic
cohomology $H^{n,inv}_k(\delta \vert d)$ controls the obstructions to
the removal of the term $a_k$ from $a$ and that all 
$H^{n,inv}_k(\delta \vert d)$ vanish for $k>2$ by \ref{usefll3} and
Theorem \ref{2.6}. 
This proves the first part of the following theorem:
\begin{theorem}\label{antigh2}Let $a$ be a local topform that is a 
nontrivial solution of the equation (\ref{topeq}). 
Without loss of generality,  one can assume
that the decomposition (\ref{antighdec}) stops at antifield number
two, \ie  \be a=a_0+a_1+a_2\,.\label{defdecomps}\ee

If the last term $a_2$ is  parity and Poincar\'e-invariant, then
it can always be written as the sum of \be
a_2^2=f^a{}_{bc}\,C^{*\m\n}_{a} (T^b_{\m\a\vert
\b}T^c_{\n\a\vert \b} -2T^b_{\m\a\vert
\b}T^c_{\n\b\vert \a} + \frac{3}{2}\,
C^{b\,\a\b}U^c_{\m\a\vert \n\b})\, d^nx\label{a22}\ee
and \be a^4_2=g^{a}{}_{bc}\,C^{*\m\n}_{a}U^b_{\m\a\vert
\b \l}U^c_{\n\a\vert \b\l}\,d^nx\,,\label{a42}\ee where
$f^a{}_{bc}$ and $g^a{}_{bc}$ are some arbitrary constant tensors
that are antisymmetric under the exchange of $b$ and $c$. 
Furthermore $a_2^4$ vanishes when $n=3,4\,$.
\end{theorem}
\noindent This most general parity and Poincar\'e invariant
expression for $a_2$ is computed in Section \ref{subsec}.

Let us note that the two components of $a_2$ do not contain the
same number of derivatives: $a^2_2$ and $a^4_2$ contain
respectively two and four derivatives. This implies that $a^2_2$
and $a^4_2$ lead to Lagrangian vertices with resp. three and five
derivatives. The first kind of deformation (three derivatives) was
studied in \cite{Berends:1984wp}, however the case with five
derivatives has never explicitly been considered before in flat space-time analyses.

Another consequence of the different number of derivatives in
$a^2_2$ and $a^4_2$ is that the descents associated with both
terms can be studied separately. Indeed, the operators appearing
in the descent equations to be solved by $a_2$, $a_1$ and $a_0$ (see Eqs.(\ref{first})-(\ref{minute}) in the next subsection) are all
homogeneous with respect to the number of derivatives, which means
that one can split $a$ into eigenfunctions of the operator
counting the  number of derivatives and solve the equations
separately for each of them. After the proof of Theorem \ref{antigh2} in Section \ref{subsec}, when we conpute the gauge transformations and the vertices associated with the deformations of the algebra, 
we thus split the
analysis: the descent starting from $a^2_2$ is analysed in Section
\ref{defun}, while the descent associated with  $a^4_2$ is treated
in Section \ref{defdeux}.

\subsection{Most general term in antifield number
two}\label{subsec}

As has been shown in Section \ref{ss:coh}, similarly to the finiteness of the decomposition of $a$, Eq.(\ref{antighdec}), one can assume that the antifield number decomposition of $b$ is finite. Furthermore, since $a$ stops at antifield number $2$, without loss of generality one has $b=b_0+b_1\,.$
Inserting the expansions of $a$ and $b$ into Eq.(\ref{topeq}) and
decomposing $s$ as $s=\d+\g$ yields \bqn
\g a_0+\d a_1 +d b_0=0 \,,\label{first}\\
\g a_1+\d a_2 +d b_1=0 \,, \label{second}\\
\g a_2=0\,. \label{minute} \eqn 
The general solution of
Eq.(\ref{minute}) is given by Proposition \ref{Hgamma3}.
The latter implies that, modulo trivial terms, $a_2$ has the form 
$a_2=\a_I \o^I$, where $\a_I $ is an invariant polynomial, depending thus on the field $\phi$, the antifields and all their derivatives, while the
$\{\o^I\}$ provide a basis of the polynomials in $C_{\m\n},
\widehat{T}_{\m\n\r}, \widehat{U}_{\m\n\r\s}$ (see Section
\ref{cohogamma2}). Let us stress that, as $a_2$ has ghost number
zero and antifield number two, $\o^I$ must have ghost number two.

 The further
constraints on $a_2$ follow from the results obtained in Sections
\ref{invPlemma}-\ref{Invariantcharacteristiccohomology3}, applied
to the equation (\ref{second}).

Acting with $\g$ on Eq.(\ref{second}) and using the
triviality of $d$, one gets that $b_1$ should also be an element
of $H(\g)$, {\it i.e.}, modulo trivial terms, $b_1=\b_I \o^I$,
where the $\b_I$ are invariant polynomials.

Let us further expand $a_2$ and $b_1$ according to the $D$-degree
defined in the proof of Proposition \ref{csq} in Section
\ref{invPlemma}. The $D$-degree is related to the differential $D$ and counts the number of $\widehat{T}$'s plus twice the number of $\widehat{U}$'s. One has 
\bqn a_2=\sum_{i=0}^M a_2^i= \sum_{i=0}^M
\a_{I_i} \o^{I_i}\,, \hspace{.5cm} b_1=\sum_{i=0}^{M}
b_1^i=\sum_{i=0}^{M} \b_{I_i} \o^{I_i}\,,\nonumber \eqn 
where $a_2^i$, $b_1^i$ and $\o^{I_i}$ have $D$-degree $i$ and $M$ is the maximal $D$-degree in pureghost number two. 
Since the action of the differential $D$ is the same as the action of the exterior derivative $d$ modulo $\g$-exact terms, the
equation (\ref{second}) reads
$$\sum_i \d [  \a_{I_i} \o^{I_i}]+\sum_i D [\b_{I_i} \o^{I_i}]= \g(\ldots)\,,$$
or equivalently, remembering that $ D \o^{I_i}=A^{I_i}_{I_{i+1}}\o^{I_{i+1}}\,$,
$$\sum_i \d [  \a_{I_i} ]\o^{I_i}+\sum_i d [\b_{I_i}] \o^{I_i} \pm \sum_i \b_{I_{i}} A^{I_i}_{I_{i+1}}\o^{I_{i+1}}= \g(\ldots)\,,$$
where the $\pm$ sign is fixed by the parity of $\b_{I_{i}} \,$.
This implies
\bqn 
\d [ \a_{I_i} ]+d [\b_{I_i} ]\pm \b_{I_{i-1}}
A^{I_{i-1}}_{I_{i}}= 0\, \label{third}
\eqn 
for each $D$-degree
$i$, as the elements of the set $\{\o^I\}$ are linearly
independent nontrivial elements of $H(\g)$.

We now analyse this equation for each $D$-degree.
\vspace*{2mm}

\noindent\textbf{\underline{D-degree decomposition}}:

\begin{itemize}
  \item \textbf{degree zero :} In $D$-degree 0, the  equation reads $\d [ \a_{I_0} ]+ d
[\b_{I_0} ]=0$, which implies that $\a_{I_0}$ belongs to
$H_2(\d\vert d)$. In antifield number 2, this group has nontrivial
elements given by Proposition \ref{H2}, which are  proportional to
$C^{*\m\n}_{a}$ . The requirement of translation-invariance
restricts the coefficient of $C^{*\m\n}_{a}$  to be constant.
Indeed, it can be shown \cite{Henneaux:1992ig} that if the Lagrangian
deformation $a_0$ is invariant under translations, then so are the
other components of $a$. On the other hand, in $D$-degree 0 and
ghost number 2, we have  $\o^{I_0}=C^b_{\m\r}C^{c}_{\n\s}$. To get
a parity and Lorentz-invariant $a^0_2$, $\o^{I_0}$  must be
completed by multiplication with  $C^{*\m\n}_{a}$ and some
parity-invariant and covariantly constant tensor, {\it i.e.} a
product of $\eta_{\m\n}$'s. The only $a^0_2$ that can be thus
built  is $a_2^0= C^{*\m\n}_{a} C^b_{\m\r}C^{c\r}_{\n}f^{a}_{bc}
d^n x $, where $f^{a}_{bc}$ is some constant tensor that
parametrizes the deformation. From this expression, one computes
that $b^0_1=\b_{I_{0}}\o^{I_0}=-3\, (h^{*\m\n\a}_{a}-\frac{1}{n}\eta^{\m\n} h_a^{*\a})
C^b_{\m\r}C^{c\r}_{\n}f^{a}_{bc}  *  (dx_\a)\,,$ where $*(dx_\a)=\frac{1}{(n-1)!}\e_{\a \m_1 \ldots \m_{n-1}} dx^{\m_{1}}\ldots dx^{\m_{n-1}}$.

  \item \textbf{degree one :} We now analyse Eq.(\ref{third}) in $D$-degree 1, which reads
\be \label{blabla3} \d [ \a_{I_1} ]+d [\b_{I_1} ]+\b_{I_{0}}
A^{I_{0}}_{I_{1}}= 0\,.\ee The last term can be read off $
\b_{I_{0}} A^{I_{0}}_{I_{1}}\o^{I_1}\propto
(h^{*\m\n\a}_{a}-\frac{1}{n}\eta^{\m\n} h_a^{*\a}) f^{a}_{bc}
d^{n} x\ \widehat{T}^b_{\a (\m\vert\r)}C^{c\r}_{\n}\,,$  and
should be $\d$-exact modulo $d$ for a solution of Eq.(\ref{blabla3})
to exist. However, the coefficient of $\widehat{T}^b_{\a
(\m\vert\r)}C^{c\r}_{\n}$ is not $\d$-exact modulo $d$. This is
easily seen in the space of $x$-independent functions, as both
$\d$ and $d$ bring in one derivative while the coefficient
contains none. As $\b_{I_{1}}$ is allowed to depend  explicitely
on $x^\m$, the argument is actually slightly more complicated: one must
expand $\b_{I_{1}}$ according to the number of derivatives of the
fields in order to reach the conclusion. The detailed argument can
be found in the proof of Theorem 7.3 in 
\cite{Henneaux:1996ws}. As $\b_{I_{0}} A^{I_{0}}_{I_{1}}$  is not
$\d$-exact modulo $d$, it must vanish if Eq.(\ref{blabla3}) is to be
satisfied. This implies that $f^{a}_{bc} $ vanishes, so that
$a^0_2=0$ and $b^0_1=0\,.$ One thus gets that $\a_{I_1} $ is an
element of $H_2(\d\vert d)$. However, there is no way to complete
it in a Poincar\'e-invariant way because the only $\o^{I_1}$ is
$\o^{I_1}=\widehat{T}^b_{\m \n\vert\r}C^{c}_{\a \b}$, which has an
odd number of Lorentz indices, while $\a_{I_1} \propto
C^{*\m\n}_{a} $ has an even number of them. Thus $a^1_2=0=b_1^1$.
  \item \textbf{degree two :} The equation (\ref{third}) in $D$-degree 2 is then $\d [ \a_{I_2}
]+d [\b_{I_2} ]=0$, which implies that $\a_{I_2}$ belongs to
$H_2(\d\vert d)$. One finds, most generally when $n>3$, that 
\bqn
a^2_2&= &C^{*\m\n}_{a} (\widehat{T}^b_{\m\a\vert \b}\widehat{T}^c_{\n\a\vert \b} f^a_{[bc]}+\widehat{T}^b_{\m\a\vert \b}\widehat{T}^c_{\n\b\vert \a} g^a_{[bc]}+ C^{b\,\a\b}\widehat{U}^c_{\m\a\vert \n\b}k^a_{bc})d^nx\,,\hspace{1.5cm} \label{candi2} \\
 b_1^2 &= &-3\,(h^{*\m\n\r}_{a}-\frac{1}{n}\eta^{\m\n} h_a^{*\r})\nnn
&&\times
(\widehat{T}^b_{\m\a\vert \b}\widehat{T}^c_{\n\a\vert \b}
f^a_{[bc]}+\widehat{T}^b_{\m\a\vert \b}\widehat{T}^c_{\n\b\vert
\a} g^a_{[bc]}+ C^{b\,\a\b}\widehat{U}^c_{\m\a\vert
\n\b}k^a_{bc})*(dx_\r)\,, \nonumber \eqn where $f^a_{[bc]}$,
$g^a_{[bc]}$ and $k^a_{bc}$ are three a priori independent
constant tensors. When $n=3$, there are linear dependences that slightly modify the analysis for this candidate, this case  will be treated at the end of the proof.
  \item \textbf{degree three :} Now, in the equation for $a^3_2$, we have
 \bqn\b_{I_{2}} A^{I_{2}}_{I_{3}}\o^{I_3}\propto \Big[ h^{* \m\n\r}_a \widehat{U}^b_{\m\a\vert \r \b} \widehat{T}_{\n}^{c\ \a\vert \b}( f^a_{[bc]}+g^a_{[bc]}-\frac{2}{3}k^a_{cb})\hspace{1.5cm}\nnn
-\frac{1}{n}h^{* \r}_a \widehat{U}^b_{\m\a\vert \r \b} \widehat{T}_{\m}^{c\ \a\vert \b}( f^a_{[bc]}+\frac{1}{2}g^a_{[bc]})\Big]\, d^nx\,,\nn
\eqn
which implies, when $n>3$, that $g^a_{[bc]}=-2 \,f^a_{[bc]}$ and
$k^a_{bc}=\frac{3}{2}\, f^a_{[bc]}\,,$ since the coefficients of $
\widehat{U}^b_{\m\a\vert \r \b} \widehat{T}_{\n}^{c\ \a\vert \b}$
and $\widehat{U}^b_{\m\a\vert \r \b} \widehat{T}_{\m}^{c\ \a\vert
\b}  $ are not  $\d$-exact modulo $d\,.$ 
We thus obtained the component (\ref{a22}) of $a_2$, which is the expression $a_2^2$ found here modulo trivial terms. 
Provided that the above conditions are satisfied,
$\a_{I_3}$ must be in $H_2(\d\vert d)$. But no
Poincar\'e-invariant $a^3_2$ can be built because
$\o^{I_3}=\widehat{T}^b_{\m\a\vert \b}\widehat{U}^c_{\n\r\vert
\s\t}$ has an odd number of Lorentz indices, so $a^3_2=0$.
  \item \textbf{degree four :} 
  Repeating the same arguments for $a^4_2$, one gets $$a^4_2=g^{a}{}_{bc}\,C^{*\m\n}_{a}
  \widehat{U}^b_{\m\a\vert\b \l}\widehat{U}^c_{\n\a\vert \b\l}d^nx$$
 and $b^4_1 =-3\,(h^{*\m\n\r}_{a}-\frac{1}{n}\eta^{\m\n}
h_a^{*\r})\widehat{U}^b_{\m\a\vert \b \l}\widehat{U}^c_{\n\a\vert
\b\l} g^{a}_{bc}*\big( dx_\r\big)\,,$ for some constant structure
function $g^{a}_{bc}$.  It is important to notice that $a_2^4$ vanishes in dimension less than five
because of the Schouten identity $$0\equiv C^{*\n_1}_{\m_1} \widehat{U}_{\m_2 \m_3\vert}^{b~~~~\n_2 \n_3} \widehat{U}_{\m_4 \m_5\vert}^{c~~~~\n_4 \n_5}\d^{[\m_1}_{[\n_1} \ldots \d^{\m_5]}_{\n_5]} \propto C^{*\m\n}\widehat{U}^b_{\m\a\vert\b \l}\widehat{U}^c_{\n\a\vert \b\l}\,.$$
No condition is imposed on $g^{a}_{bc}$ by
equations in higher $D$-degree because $D_1 b^4_1 =0$. 
We now obtained the component (\ref{a42}).
  \item \textbf{degree higher than four :} Finally, there are no $a^i_2$ for $i>4$ because there is no ghost
combination $\o^{I_i}$ of ghost number two and $D$-degree higher
than four.
\end{itemize}
\noindent Summarizing, we have almost proved the second part of Theorem
\ref{antigh2}: it remains to show that the component of $D$-degree two, $a_2^2$, in space-time dimension $n=3$ can be chosen with the same form as in the other dimensions.
So let us return to the analysis of Eq.\bref{third} in $D$-degree two when $n=3$. One can again write the most general  $a_2^2$ as (\ref{candi2}).
However
the second term  is linearly dependent on the first one and the 
last one vanishes, because of Schouten identities. These identities are due to the fact 
that one cannot antisymmetrize over more indices than the number of space-time dimensions;  
they read $0= C^{*\n_1}_{\m_1}\widehat{T}^{b~~~\n_2}_{\m_2 \m_3\vert}\widehat{T}^{c 
\n_3\n_4\vert}_{~~~~~~\m_4}\d^{[\m_1}_{[\n_1} \ldots \d^{\m_4]}_{\n_4]} \propto  C^{*\m\n} 
(2\widehat{T}^b_{\m\a\vert \b}\widehat{T}^c_{\n\a\vert \b} -\widehat{T}^b_{\m\a\vert 
\b}\widehat{T}^c_{\n\b\vert \a} )\,$, $0=C^{*\n_1}_{\m_1} C^{\n_2}_{\m_2} 
\widehat{U}_{\m_3 \m_4\vert}^{~~~~\n_3 \n_4}\d^{[\m_1}_{[\n_1} \ldots \d^{\m_4]}_{\n_4]} 
\propto C^{*\m\n}C^{\a\b}\widehat{U}_{\m\a\vert \n\b}\,.$  We can however also take the 
above form for $a_2^2$ in $n=3$, keeping in mind that in this case $g^a_{[bc]}$ and 
$k^a_{bc}$ are arbitrary, provided $g^a_{[bc]} \neq -\frac{1}{2}f^a_{[bc]}$ so that 
$a_2^2$ is nonvanishing.
In $D$-degree 3, $\b_{I_{2}} A^{I_{2}}_{I_{3}}\o^{I_3}$ now vanishes by Schouten identities. 
We 
can then use the arbitrarity of $g^a_{[bc]}$ and $k^a_{bc}$  to impose 
the above conditions and have the same result as in higher dimensions.

This completes the proof  of Theorem
\ref{antigh2}.

\subsection{Berends--Burgers--van Dam's deformation}
\label{defun}

In this section, we consider the deformation related to $a^2_2$
given by (\ref{a22}). As explained above, $a_2=a^2_2$  must
now be completed into a solution $a$ of $s a+db=0$ by adding terms
with lower antifield number. The complete solution $a$ provides
then the first-order deformation term $W_1=\int a$ of an
interacting theory. The next step is to check that higher-order
terms  $W_2$, $W_3$, etc. can be built to get the full interacting
theory.

In the case considered here, we show that a first-order interaction term $W_1$ can be constructed; however, there is an obstruction to the existence of $W_2$, which prevents its completion into a consistent interacting theory.

\subsubsection{Existence of a first-order deformation}

In this section, the descent equations (\ref{first}) and (\ref{second}), {\it i.e.} $ \g a_0+ \d a_1 +d b_0=0$ and  $\g a_1+ \d a_2 +d b_1=0$,  are solved for $a_1$ and $a_0$.

The latter of these  equations admits the particular solution
\bqn
a_1^p&=&-\frac{3}{2}\,\Big[\,(h^{*\m\n\r}_{a}-\frac{1}{n}\eta^{\m\n} h_a^{*\r})\nnn
&&\hspace{2cm}\times \Big(2 \pa_{[\m}h^b_{\a]\b\r}(T^c_{\n\a\vert \b} -2 T^c_{\n\b\vert \a} )+ h^b_{\a\b\r}U^c_{\m\a\vert \n\b}-3 C^{b\,\a\b}\pa_{[\n}h^c_{\b]\r[\a,\m]}\Big) \nonumber \\
&&\hspace{4cm}+ \frac{1}{n} h^{*\r}_a T^b_{\r\a\vert \b} (\pa_\s h^{c\,
\s\a\b}-\pa^\a h^{c\,\b}-\pa^\b h^{c\,\a})\,\Big]\,f^a_{bc}\,d^nx
\,.\nonumber \eqn To this particular solution, one must add the
general solution $\bar{a}_1$ of $\g \bar{a}_1+d b_1=0\,,$ or
equivalently (by Proposition \ref{csq}) of $\g \bar{a}_1=0$. In
ghost number zero, antifield number one and with two derivatives,
this solution is, modulo trivial $\d$-, $\g$- and $d$-exact terms,
$$\bar{a}_1=h^{*\,a}_{\m\n\r} G^{b\,\m\n}_{\s} C^{c\,\r\s}l^1_{(ab)c}
+h^{*a}_{\m}G^b_{\n}C^{c\,\m\n}l^2_{(ab)c}+h^{*a\,\m}G^b_{\m\n\r}C^{c\,\n\r}l^3_{abc}\,,$$
where $l^1_{(ab)c}$, $l^2_{(ab)c}$ and $l^3_{abc}$ are some arbitrary constants. For future convenience, we also add to $a^p_1+\bar{a}_1$ the trivial term $\g b_1$ where
\bqn
b_1=
f^a_{bc} h^{*}_{a\m\n\r} (-\textstyle{\frac{3}{2}}h^{b\m\s\t}\pa^\n h^{c\r}_{~\s\t}
 -2h^{b\m\s\t}\pa_\s h^{c\n\r}_{~~\t}+3 h^{b \m} \pa^\n h^{c\r}
 -3h^{b}_{\s}\pa^\m h^{c\n\r\s}\hspace{1cm}
\nnn+2h^{b}_{\s}\pa^\s h^{c\m\n\r})
\nnn
+f_{abc} h^{*a}_{\m}(\textstyle{2} h^{b\m\n\r}\pa_\n h^{c}_\r -h^{b\m\n\r}\pa^\s h^{c}_{\n\r\s}                                                       
 +\textstyle{3} h^{b\m}\pa^\s h^c_{\s}-\textstyle{\frac{1}{2}}h^{b}_{\n\r\s}\pa^{\m}h^{c \n\r\s}   
+\textstyle{6} h^{b}_{\n}\pa_\r h^{c\m\n\r})
\,.\nonumber
\eqn
In short, up to trivial terms, the most general $a_1$, solution of $\g a_1+ \d a_2 +d b_1=0$, is $a_1=a^p_1+\bar{a}_1+\g b_1\,.$

The next step is to find $a_0$ such that $\g a_0+ \d a_1+d
b_0=0\,.$ A cumbersome but straightforward computation shows that
necessary (and, as we will see, sufficient) conditions for a
solution $a_0$ to exist are (i) $f^a_{[bc]}$ is totally
antisymmetric, or more precisely 
$\d_{ad}f^d_{[bc]}=f_{[abc]}$,
(ii) $l^1_{(ab)c}=l^2_{(ab)c}=0$ and (iii)
$l^3_{abc}=-\frac{9}{8}f_{[abc]}\,.$ This computation follows the
lines of an argument developped in \cite{Boulanger:2000rq}, which
considers the most general $a_0$ and matches the coefficients of
the terms with the structure $C h'h'$, where $h'$ denotes the
trace of $h$. 
In three and four dimensions, one must take into account that some of these terms are related by Schouten identities (see Appendix \ref{schouten} for a definition); however, this does not change the conclusions.
Once  the conditions (i) to (iii) are satisfied, one can
explicitly build the solution $a_0$, which  corresponds to the
spin-3 vertex found in \cite{Berends:1984wp}, in which the
structure function $f_{abc}$ has been replaced by
$-\frac{3}{8}f_{abc}\,.$ The explicit deformation $a_0$ of the Lagrangian
will be given shortly for completeness. 
It is unique up to solutions $\bar{a}_0$ of the homogeneous equation 
$\g \bar{a}_0+d b_0=0\,.$

We have thus proved by a new method that the spin-3 vertex of
\cite{Berends:1984wp} is the only consistent nontrivial first-order deformation of the free spin-3 theory with at
most\footnote{The developments above prove the three-derivatives
case. For less derivatives, it follows from above that $a_2=0$,
which implies that  $\g a_1= 0$ by Eq.(\ref{second}); however there
is no such parity and Poincar\'e-invariant $a_1$ with less than
two derivatives, so $a_1=0$ as well.} three derivatives in the
Lagrangian, modulo deformations $\bar{a}_0$ of the latter that are
gauge-invariant up to a total derivative, {\it i.e.} such that $\g
\bar{a}_0+ d b_0=0\,.$ However, as is known from
\cite{Berends:1984rq}, this deformation cannot be completed to all
orders, as is proved again below.

\subsubsection{Explicit first-order vertex and gauge transformation}
\label{azeroun}

For completeness, we provide here the explicit first-order vertex and gauge transformation of the Berends--Burgers--van Dam cubic interaction.

The deformation of the vertex is 
 $$\int a_0\,=\,f_{[abc]}\,S^{abc}\,\,;\quad
S^{abc}[h^d_{\m\n\r}]\,=\, -\frac{3}{8}\,\int{\cal
L}_{BBvD}^{abc}\,d^n x\;,$$ 
\hfill

where  
\bqn &{\cal
L}_{BBvD}^{abc}&= -\frac{3}{2} \,h^{a \a} h^{b
\b,\,\c}h^c_{\b,\,\a\g} +3\,h^{a \a,\,\b} h^{b \c}h^c_{\g,\,\a\b}
+6 \,h^{a \a\b\g,\,\d} h^{b }_{\a}h^c_{\b,\,\g\d}\nnn &&
+\frac{1}{2}\,h^{a \a} h^{b
\b\c\d,\,\e}h^c_{\b\g\d,\,\a\e} +h^{a
\a}_{~~,\,\a\b} h^{b }_{\g\d\e}h^{c\g\d\e,\,\b} +h^{a \a,\,\b}
h^{b \g\d\e}h^{c}_{\g\d\e,\,\a\b} \nnn &&
-3\,h^{a}_{ \a\b\g} h^{b
\a\b}_{~~~\d,\,\e}h^{c \d,\,\g\e} -3\,h^{a}_{ \a\b\g} h^{b
\a\b\d,\,\g\e}h^{c}_{ \d,\,\e} +3\,h^{a}_{
\a\b\g,\,\d} h^{b \a\b\e}h^{c~,\,\g\d}_{ \e} \nnn&&
+3\,h^{a~~~,\,\g\d}_{
\a\b\g} h^{b \a\b\e}h^{c}_{ \e,\,\d} -\frac{9}{4}\,h^{a}_{
\a,\,\b\g} h^{b \b}h^{c\g ,\,\a} -\frac{1}{4}\,h^{a}_{ \a,\,\b}
h^{b \b,\,\g}h^{c~,\,\a}_{\g }\nonumber \\&& -3\,h^{a}_{ \a\b\g}
h^{b\d,\,\a}h^{c~,\,\b \g}_{ \d} -\frac{3}{2}\,h^{a ~,\,\a}_{ \a}
h^{b \b,\,\g}h^{c~~~,\,\d}_{\b\g\d} +3\,h^{a}_{ \a}
h^{b}_{\b,\,\g}h^{c~\b\g,\,\a\d}_{ \d} \nnn &&
+\frac{3}{2}\,h^{a
~,\,\a\b}_{ \a} h^{b \g,\,\d}h^{c}_{\b\g\d}
+3\,h^{a}_{ \a,\,\b} h^{b}_{\g,\,\d}h^{c\b\g\d,\,\a}
-\frac{3}{2}\,h^{a }_{ \a} h^{b
~~~,\,\b}_{\b\g\d}h^{c~\g\d,\,\a\e}_\e \nnn &&
-6\,h^{a
~~~,\,\a\d}_{\a\b\g} h^{b \b,\,\e}h^{c}_{\d\e}{}^\g
 +6\,h^{a ~~~,\,\a\d}_{\a\b\g} h^{b \b}h^{c~~\g,\,\e}_{\d\e}
-2\,h^{a}_{ \a\b\g,\,\d} h^{b
~\a\d,\,\e}_{\l}h_\e^{c~\l\b,\g}\nnn &&
 +h^{a}_{ \a\b\g} h^{b
~~~,\,\a}_{\d\e\l}h^{c~\d\e\l,\,\b\g}
 -3\,h^{a}_{ \a\b\g}{}^{,\,\a} h^{b
~\b\g,\,\e}_{\d}h^{c}_{\e\l}{}^{\d,\,\l}\nnn && +3\,h^{a~~~,\,\a\d}_{
\a\b\g} h^{b \b\g\e,\,\l}h^{c}_{\e\d\l} +6\,h^{a}_{ \a\b\g,\,\d}
h^{b \a\b\e,\,\l}h^{c}_{\e\l}{}^{\d,\,\g}\,, \nonumber \eqn
where we remind that indices after a coma denote partial derivatives.

The first-order deformation of the gauge transformations is given by
 $$\delta^1_{\l}h^a_{\m\n\r} =f^a{}_{bc}\,\Phi^{bc}_{\m\n\r}\,,$$
where $ \Phi^{bc}_{\m\n\r} $ is the completely symmetric component of 
\begin{eqnarray}
        {\cal \phi}^{bc}_{\m\n\r} &= &
6\, h^{b\s}\l^c_{\m\s,\n\r}
-3\, h^{b\s}\l^c_{\m\n,\r\s}
+6\, h^b_{\m,\n}\l^{c\, ~,\s}_{\s\r}
-6\, h^b_{\m}\l^{c~~~\;\s}_{\s\n,\r}
\nonumber\\ &&
-\frac{15}{4} h^b_{\m\s\t,\n}\l^{c\s,\t}_\r
+\frac{31}{4} h^{b\s\t}_{\m}\l^c_{\n\s,\t\r}
+\frac{9}{4} h^b_{\m\n\s,\r\t}\l^{c\s\t}
-\frac{11}{2} h_{\m\n}^{b\; ~\s,\t}\l^c_{\s(\t,\r)}
\nonumber\\ &&-6\, h^b_{\m\n\s,\r}\l^{c\s\t}_{~~~,\t}
-\frac{3}{4} h^b_{\m\s\t,\n}\l^{c\s\t}_{~~~,\r}
-\frac{9}{8} h^b_{\m\s\t,\n\r}\l^{c\s\t}
+\frac{9}{8} h^{b\s\t}_{\m}\l^c_{\s\t,\n\r}
\nonumber\\ &&-\frac{1}{2} h^b_{\m\n\s,\t}\l^{c\t,\s}_\r
+\frac{13}{8} h^{b\s\t}_\m\l^c_{\n\r,\s\t}
+4\, h^b_{\m\n\r,\s}\l^{c\s\t}_{~~~,\t}
-\frac{9}{8} h^{b~~~,\s\t}_{\m\n\r}\l^c_{\s\t}
\nonumber\\
&
\!\!\!\!+\eta_{\m\n}\Big(&
\frac{9}{4} (h^{b\s,\t}_{~~~\t}\l^c_{\r\s}- h^b_{\s,\r\t}\l^{c\s\t}
- h^{b~~~,\h\s}_{\h\s\t}\l^{c\t}_{\r})
+\frac{9}{8} (h^{b\, ,\s\t}_\s\l^c_{\r\t}+ h^{b\h\s\t}_{~~~~,\h\r}\l^c_{\s\t})
\nonumber\\ &&
\!\!\!\!\!\!\!\!\!+6\,( h^{b\,,\s}_{\s}\l^{c~\,,\t}_{\r\t}- h^{b\s}\l^{c~~~\t}_{\s\t,\r}
- h^{b\s}\l^{c~~~\t}_{\s\r,\t}- h^b_{\s}\l^{c\,~,\s\t}_{\r\t}
- h^b_{\r}\l^{c\,~,\s\t}_{\s\t}
+2\, h^{b~~\,,\s}_{\r\s\t}\l^{c\t,\h}_{\h})
\nonumber\\ &&
\!\!\!\!\!\!+\frac{3}{2} (h^{b\h\s\t}\l^c_{\s\t,\h\r}
-h^b_{\h\s\t,\r}\l^{c\, \s\t,\h})
+ (1-\frac{3}{4n})(2\, h^{b\s,\t}\l^c_{\s\t,\r}-\,h^{b\s\t,\h}_\h \l^c_{\s\t,\r})
\nonumber\\ &&
\!\!\!\!\!\!+(2+\frac{3}{4n})( h^b_{\s,\t}\l_\r^{c\s,\t}+ h^b_{\s,\t}\l_\r^{c\t,\s}
                                        - h_{\s}^{b\t\h,\s}\l^c_{\r\t,\h}- h^b_{\r\s\t}\l_{\h}^{c\s,\t\h}
+ \frac{1}{2}h_\r^{b\s\t}\l^{c~~~\h}_{\s\t,\h})
\nonumber\\ &&
\!\!\!+\frac{9}{8}(1-\frac{1}{n})(- h^{b~~~~\h}_{\r\s\t,\h}\l^{c\s\t}
+2\, h_{\r\s}^{b~\t,\h\s}\l^c_{\h\t}
- h_\r^{b\,,\s\t}\l^c_{\s\t})
\Big)\,.
\nonumber\end{eqnarray}
This expression is equivalent to that of \cite{Berends:1984wp} modulo field redefinitions.

\subsubsection{Obstruction for the second-order deformation}
\label{obstr}

In the previous subsections, we have constructed a first-order deformation $W_1=\int \,\ll(a_0+a_1+a_2\rr)$ of the free functional $W_0\,.$
As explained in Section \ref{cons}, a consistent second-order deformation $W_2$ must satisfy the condition (\ref{deformation3}), \ie  \be (W_1,W_1)=-2 s W_2\,.\label{order2} \ee
Expanding $(W_1,W_1)$ according to the antifield number, one
finds
$$(W_1,W_1) = \int d^n x \,(\a_0 + \a_1 + \a_2 )\,,$$
where the term of antifield number two $\a_2$ comes from
the antibracket of $a_2$ with itself.

If one also expands $W_2$ according to the antifield number, one
gets from Eq.(\ref{order2}) the following condition on $\a_2$ (it is
easy to see that the expansion of $W_2$ can be assumed to stop at
antifield number three, $W_2 = \int d^n x (c_0 + c_1 + c_2 + c_3)$
and that $c_3$ may be assumed to be invariant, $\g c_3 = 0$) \be
\a_2= -2(\g c_2+\d c_3)+\pa_\m b^\m_2\,. \ee 
Explicitly, \bqn
\a_2&=\frac{1}{2} f^{}_{abc}f^c_{~de} C^{*a}_{\m\n} &\Big(
-4\widehat{T}^{b \m\a\vert \b}\widehat{T}^{d\n\r\vert
\s}\widehat{U}^e_{\a\r\vert \b\s} +5\widehat{T}^{b \m\a\vert
\b}\widehat{T}^{d\n\r\vert \s}\widehat{U}^e_{\a\s\vert \b\r}
\nnn
&&\!\!\!-3\widehat{T}^{b \m\a\vert \b}\widehat{T}^{d}_{\a\r\vert
\s}\widehat{U}_{\hspace{20pt}\b}^{e\s\n\vert \r}
+\widehat{T}^{b \m\a\vert \b}\widehat{T}^{d}_{\b\r\vert \s}\widehat{U}_{\hspace{20pt}\a}^{e\r\n\vert \s}
+\widehat{T}^{b \m\a\vert \b}\widehat{T}^{d}_{\b\r\vert \s}\widehat{U}_{\hspace{20pt}\a}^{e\s\n\vert \r}
\nnn &&
-\frac{3}{2} \widehat{U}^{b \m\a\vert \n\b}\widehat{T}^d_{\a\r\vert \s}\widehat{T}^{e~\r\vert \s}_{~\b}
+3\widehat{U}^{b \m\a\vert \n\b}\widehat{T}^d_{\a\r\vert \s}\widehat{T}^{e~\s\vert \r}_{~\b}
\nnn &&
+\frac{9}{4}\widehat{U}^{b \m\a\vert \n\b}C^{d\r\s}\widehat{U}^{e }_{\a\s\vert \b\r}
+\frac{3}{2}C^b_{\a\b}\widehat{U}^{d\r\m\vert \s\a}\widehat{U}^{e~\n\vert \b}_{~\r\hspace{12pt}\s}
\nonumber \\
&&
-\frac{3}{4}C^b_{\a\b}\widehat{U}^{d\r\m\vert \s\a}\widehat{U}^{e~\n\vert \b}_{~\s\hspace{12pt}\r}
+\frac{3}{4}C^{b\a\b}\widehat{U}^{d }_{\r\a\vert \s\b}\widehat{U}^{e\r\m\vert \s\n}\Big)+\g(\ldots)\,.
\nonumber
\eqn
It is impossible to get an expression with three ghosts, one $C^{*}$
and no fields, by acting with $\d$ on $c_3$, so we
can assume without loss of generality that $c_3$ vanishes,
which implies that  $\a_2$ should be $\g$--exact modulo total derivatives.

However, $\a_2$ is not a  mod-$d$ $\g$-coboundary unless it vanishes. 
Indeed, suppose we have $$\a_2 = \g(u) +  \pa_\m k^\m  \,.$$ 
Both $u$ and $k^\m$ have antifield number two and we can restrict
ourselves to their components linear in $C^*\,$ without loss of
generality (so that the gauge algebra closes off-shell at second order).
We can also assume that $u$ contains $C^*$
undifferentiated, since derivatives can be removed through
integration by parts. As the Euler derivative of a divergence is
zero, we can reformulate the question as to whether the following
identity holds,
$$ \frac{\d^{L}\a_2 }{\d  C^{* a}_{\m\n}}
 = \frac{\d^{L}(\g u)}{\d  C^{* a}_{\m\n}}=-\g\Big( \frac{\partial^{L}u}{\partial  C^{* a}_{\m\n}}\Big) \,.$$
since $\g C^{*}=0$ and $ C^{* }$ appears undifferentiated in $u$. On the other hand, $
\frac{\d^{L}\a_2 }{\d  C^{* a}_{\m\n}} $ is a sum of nontrivial
elements of $H(\g)$; it can be $\g$-exact only if it vanishes. 
Consequently, a necessary condition
for the closure of the gauge transformations ($c_2$ may be assumed to be linear in the antifields) is $\a_2=0$.

Finally, $\a_2$ vanishes if and only if
either $n=3$, since $\widehat{U}^{a }_{\m\n\vert \r\s} $ vanishes identically in this dimension because of its symmetry,
or  
$f_{abc}f^c_{~de} =0\,$ (nilpotency of the algebra). The latter condition implies the vanishing of $f_{abc}$ 
(by Lemma \ref{lemalg}), and thus of the whole deformation candidate.
So, the deformation is obstructed at second order when $n>3\,$.

Let us note th  at originally, in the work \cite{Berends:1984rq}, the
obstruction to this first-order deformation appeared
under the weaker form $f^{}_{abc}f^c_{~de} =f^{}_{adc}f^c_{~be}\,$ 
(associativity). It was also obtained by demanding the closure of the algebra of gauge transformations at second order in the deformation parameter.

\subsection{Five-derivative deformation}
\label{defdeux}
\label{azerodeux}

We now consider the deformation related to $a_2=a^4_2$,
written in Equation (\ref{a42}).
In this case, the general solution $a_1$ of $\g a_1+ \d a_2 +d
b_1=0$ is, modulo trivial terms, \be a_1=-2\,
(h^{*\m\n\r}_a-\frac{1}{n}
\eta^{\m\n}h^{*\r}_a)\pa^{}_{[\m}h^b_{\a]\r [\b,\l]}
U^c_{\n\a\vert \b\l} g^{a}_{[bc]}\, d^nx +\bar{a}_1 \,,
\label{a1max} \ee where $\bar{a}_1 $ is an arbitrary element of
$H(\g)$ .

When the structure constant is completely antisymmetric in its indices, $\d_{ad}g^d_{[bc]}$ $=g_{[abc]}\,$, a Lagrangian deformation $a_0$ such that $ \g a_0+ \d a_1 +d b_0=0$ can be computed. Its expression is quite long and is given later in this section.
We used the symbolic manipulation program FORM \cite{form} for its computation.

 This nontrivial first-order deformation of the free theory had not been found in the previous spin-three analyses, which is related to the assumption usually made that the Lagrangian deformation should contain at most three derivatives, while it contains five of them in this case.
However, it would be very interesting to see whether the cubic vertex 
could be related to the flat space limit of
the higher-spin vertices of the second reference of \cite{Fradkin:1987ks}.
At first order in the deformation parameter, it is possible to take
some flat space-time limit of the $(A)dS_n$ higher-spin cubic
vertices. An appropriate flat limit must be taken: the
dimensionless coupling constant $g$ of the full higher-spin gauge theory
should go to zero in a way which compensates the non-analyticity 
$\sim 1/\Lambda^m$ 
in the cosmological constant $\Lambda$ of the cubic vertices,
\emph{i.e.} such that the ratio $g/\Lambda^m$ is finite.
The spin-3 vertices could then be recovered in such a limit
from the action of \cite{Lopatin:1987hz} by substituting the linearized
spin-3 field strengths for the nonlinear ones at quadratic order and
replacing the auxiliary and extra connections by their expressions in
terms of the spin-3 gauge field obtained by solving the linearized
torsion-like constraints, as explained in
\cite{Vasiliev:2004qz,Sagnotti:2005,Fradkin:1987ks} (and references
therein). Such a relation would provide a geometric meaning for the
complicated expression of the five-derivative vertex. 

The next step is to find the second-order components of the deformation. Similarly to the previous case, it can easily be  checked that we can assume $c_3=0$. However, no  obstruction arises from the constraint $\a_2\equiv (a_2,a_2)=-2 \g c_2+\pa_m k^\m$. If this candidate for an interacting theory is obstructed, the obstructions arise at some later stage, {\it i.e.} beyond the (possibly on-shell) closure of the gauge transformations.

For completeness, one should check whether $ \g a_0+ \d a_1 +d b_0=0$
admits a solution $a_0$ when the structure constant
$g^d_{~bc}=g^d_{~[bc]}$ is not completely antisymmetric but has
the ``hook'' symmetry property $\d^{}_{d[a} g^d{}_{bc]}=0$.
However, the computations involved are very cumbersome and we were
not able to reach any conclusion about the existence of such an
$a_0$.


\vspace{.2cm}

We now give the deformation $a_0$ related to the
element $a^4_2$ with completely
antisymmetric structure constants. It satisfies the equation $\g
a_0 + \d a_1+ d b_0=0$ for $a_1$ defined by Eq.(\ref{a1max}), in
which $\bar{a}_1=0$. The deformation is $ \int a_0\,=\,g^{[abc]}\,T_{abc}\,\,;\quad T_{abc}[h^d_{\m\n\r}]\,=\, \frac{1}{2}\,\int{\cal L}^{}_{abc}\,d^n x\;,$
where

{\small

\bqn &{\cal L}^{}_{abc}\,= 
h_{a}^{\m\n\r}\,\,\Big(&-{\textstyle \frac{7}{4}}\, \pa_{\m\n}
h_{b}^{\l\s\t}\pa_{\r\s\t}h_{c\l} 
-{\textstyle \frac{1}{4}}\, \pa_{\m\n}
h_{b}^{\l\s\t}\pa_{\r\h}\pa^\h h_{c\l\s\t} 
-{\textstyle \frac{1}{2}}\,
\pa_{\m\n} h_{b}^{\l}\pa_{\r\l \s}h_{c}^{\s}  \nonumber\\&&
-{\textstyle \frac{3}{4}}\,
\pa_{\m\n} h_{b}^{\l}\pa_{\r\s\t}h_{c\l}^{~~\s\t}
-{\textstyle \frac{5}{3}}\, \pa_{\m} h_{b}^{\l\s\t}\pa_{\n\r\l\h}h_{c\s\t}^{\h}
+{\textstyle \frac{1}{2}}\, \pa_{\m} h_{b}^{\l\s\t}\pa_{\n\r\h}\pa^\h
h_{c\l\s\t}  \nonumber\\&&
+{\textstyle \frac{2}{3}}\, \pa_{\m}
h_{b}^{\l}\pa_{\n\r\s\t}h_{c\l}^{~~\s\t} 
-{\textstyle \frac{4}{3}}\, \pa_{\m}
h_{b}^{\l}\pa_{\n\r\s}\pa^\s h_{c\l} 
+{\textstyle \frac{5}{4}}\,
\pa_{\s\t} h_{b}^{\s\t\l}\pa_{\m\n\r} h_{c\l} \nonumber\\&&
-{\textstyle \frac{5}{3}}\,
\pa_{\s\t} h_{b}^{\s\l\h}\pa_{\m\n\r} h_{c\l\h}^{\t}
+{\textstyle \frac{3}{4}}\, \pa_{\s}\pa^\s h_{b}^{\l\h\t}\pa_{\m\n\r}
h_{c\l\h\t} 
+{\textstyle \frac{1}{2}}\, \pa_{\s\t} h_{b}^{\s}\pa_{\m\n\r}
h_{c}^{\t}\nonumber\\&& 
+{\textstyle \frac{23}{12}}\, \pa_{\s\t}
h_{b}^{\l}\pa_{\m\n\r} h_{c\l}^{~~\s\t} 
-{\textstyle \frac{4}{3}}\,
\pa_{\s}\pa^{\s} h_{b}^{\l}\pa_{\m\n\r} h_{c\l} 
-{\textstyle \frac{51}{16}}\,
\pa_{\m\n}h_{b\r}\pa_{\s\t}\pa^\s h_{c}^{\t}\nonumber\\&& 
-{\textstyle \frac{11}{8}}\,
\pa_{\m} h_{b\n}^{~~\s\t}\pa_{\r\s\t\l} h_{c}^{\l}
+{\textstyle \frac{5}{4}}\, \pa_{\m} h_{b\n \s\t}\pa_{\r\l\h}\pa^\t
h_{c}^{\s\l\h} 
-{\textstyle \frac{3}{8}}\, \pa_{\m} h_{b\n
\s\t}\pa_{\r\l}\pa^{\l\t} h_{c}^{\s} \nonumber\\&&
+{\textstyle \frac{9}{4}}\, \pa_{\m}
h_{b\n\s\t}\pa_{\r\l\h}\pa^\h h_{c}^{\s\t\l} 
-{\textstyle \frac{1}{12}}\,
\pa_{\m} h_{b\n}\pa_{\r\l \s\t} h_{c}^{\l\s\t}
-{\textstyle \frac{3}{2}}\, \pa_{\m} h_{b\n}\pa_{\r\l \s}\pa^\s h_{c}^{\l}\nonumber\\&&
-{\textstyle \frac{11}{16}}\, \pa_{\l} h_{b\m}^{~\s\t}\pa_{\n\r \s\t}
h_{c}^{\l} 
-{\textstyle \frac{1}{4}}\, \pa_{\l \h} h_{b\m\s\t}\pa_{\n\r }\pa^\t
h_{c}^{\l\h\s} 
+{\textstyle \frac{3}{4}}\, \pa_{\l}\pa^{ \l}
h_{b\m\s}^{\t}\pa_{\n\r \t} h_{c}^{\s}\nonumber\\&& 
+{\textstyle \frac{7}{4}}\,
\pa_{\h\l} h_{b\m}\pa_{\n\r}\pa^\h h_{c}^{~\l} 
-{\textstyle \frac{19}{16}}\,
\pa_{\h} \pa^\h h_{b\m}\pa_{\n\r\l}h_{c}^{~\l} 
+{\textstyle \frac{11}{4}}\,
\pa_{\m\l} h_{b\n}^{~~\l\s}\pa_{\s \t\h} h_{c\r}^{~~\t\h}\nonumber\\&&
+{\textstyle \frac{3}{4}}\, \pa_{\m} h_{b\n \s\t}\pa^{\s \t \l\h}
h_{c\r\l\h}
 +{\textstyle \frac{7}{8}}\, \pa_{\m} h_{b\n
\s\t}\pa^{\s \t \l}\pa_{\l} h_{c\r} 
+{\textstyle \frac{3}{2}}\, \pa_{\m} h_{b\n
\s\t}\pa^{\s \l}\pa_{\l\h} h_{c\r}^{~~\t\h} \nonumber\\&&
-\, \pa_{\m} h_{b\n
\s\t}\pa^{ \l\h}\pa_{\l\h} h_{c\r}^{~~\s\t}
 +\, \pa_{\m} h_{b\n
}\pa_{ \l}\pa^{\l\s\t} h_{c\r\s\t} 
+{\textstyle \frac{7}{4}}\,
\pa^{\s} h_{b\m \s\t}\pa^{\t \l\h}\pa_{\n} h_{c\r\l\h}\nonumber\\&&
-{\textstyle \frac{9}{8}}\, \pa^{\s} h_{b\m \s\t}\pa^{\t \l}\pa_{\n\l} h_{c\r}
+{\textstyle \frac{1}{4}}\, \pa^{\l} h_{b\m }^{~~\s\t}\pa_{\n\s\t\h}
h_{c\r\l}^{\h} 
-{\textstyle \frac{3}{4}}\, \pa^{\l} h_{b\m
}^{~~\s\t}\pa_{\n\s\t\l} h_{c\r} \nonumber\\&&
+2\, \pa^{\l\t}
h_{b\m \l\s}\pa_{\n\t\h} h_{c\r}^{~~\s\h}
 -{\textstyle \frac{1}{4}}\, \pa_{\t}
h_{b\m \l\s}\pa_{\n\h}\pa^{\l\h} h_{c\r}^{~~\s\t} 
+{\textstyle \frac{3}{4}}\,
\pa^{\t} h^\l_{b\m \s}\pa_{\n\l\t\h} h_{c\r}^{~~\s\h} \nonumber\\&&
+ \pa^{\l}
h_{b\m \s\t}\pa_{\n\l\h}\pa^\h h_{c\r}^{~~\s\t}
\!-{\textstyle \frac{1}{4}} \pa^{\s\t} h_{b\m \s\t}\pa_{\h}\pa^{\h\l}
h_{c\n\r\l} 
\!-{\textstyle \frac{3}{4}} \pa^{\s} h_{b\m
\s\t}\pa_{\h}\pa^{\t\h\l} h_{c\n\r\l} \nonumber\\&&
+{\textstyle \frac{3}{4}}\, \pa^{\l}
h_{b\m \s\t}\pa_{\h}\pa^{\s\t\h} h_{c\n\r\l} 
+{\textstyle \frac{3}{2}}\,\pa_{\l} h_{b\m \s\t}\pa^{\l\s\t\h} h_{c\n\r\h}
-{\textstyle \frac{1}{4}}\, \pa^{\l} h_{b\m }\pa_{\s\t}\pa^{\s\t} h_{c\n\r\l}\nonumber \\&&
+{\textstyle \frac{3}{4}}\, \pa^{\l} h_{b\m\l\h }\pa_{\s\t}\pa^{\s\t}
h_{c\n\r}^{\h} 
+{\textstyle \frac{3}{2}}\, \pa_{\s\t} h_{b\m\l\h }\pa^{\l\s\t}
h_{c\n\r}^{\h}
+{\textstyle \frac{1}{3}}\, \pa_{\m} h_{b\n\r\l }\pa^{\l \s\t\h}h_{c\s\t\h} \nonumber \\&&
-{\textstyle \frac{15}{4}}\, \pa_{\m} h_{b\n\r\l}\pa^{\l \s\t}\pa_\s h_{c\t} 
-{\textstyle \frac{11}{4}}\, \pa_{\m} h_{b\n\r\l}\pa^{ \s\t\h}\pa_\s h_{c\t\h}^{\l} 
+{\textstyle \frac{1}{2}}\, \pa_{\m}
h_{b\n\r\l }\pa^{ \s\t}\pa_{\s\t} h_{c}^{\l}
\nonumber\\&& 
+{\textstyle \frac{1}{2}}\,
\pa_{\h} h_{b\m\n\l }\pa^{ \l}\pa_{\r\s\t}
h_{c}^{\h\s\t} 
-{\textstyle \frac{1}{2}}\, \pa_{\h} h_{b\m\n\l}\pa^{ \l\s}\pa_{\r\s} h_{c}^{\h}
 -\, \pa_{\s} h_{b\m\n\l }\pa^{
\l\s}\pa_{\r\h} h_{c}^{\h} \nonumber\\&&
-{\textstyle \frac{3}{4}}\, \pa^{ \h}\pa_{\h}
h_{b\m\n\l }\pa_{\r\s\t} h_{c}^{~\l \s\t} 
+{\textstyle \frac{1}{2}}\, \pa^{
\s\t} h_{b\m\n\l }\pa_{\r\s\t} h_{c}^{\l }
+{\textstyle \frac{7}{4}}\, \pa^{\l} h_{b\m\n\l }\pa_{\h\s\t}\pa^\h h_{c\r}^{~~\s\t} 
\nonumber\\&&
-{\textstyle \frac{1}{4}}\, \pa^{\l} h_{b\m\n\l }\pa_{\s\t}\pa^{\s\t}
h_{c\r} 
-{\textstyle \frac{3}{2}}\, \pa^{\h} h_{b\m\n\l }\pa_{\s}\pa^{\l\s\t}
h_{c\r\h\t} 
-2\,\pa_{\h} h_{b\m\n\l }\pa^{\h\l\s\t}h_{c\r\s\t}
\nonumber\\&& 
+{\textstyle \frac{1}{2}}\, \pa_{\h} h_{b\m\n\l
}\pa^{\h\l\s}\pa_\s h_{c\r} +{\textstyle \frac{1}{4}}\, \pa_{\h} h_{b\m\n\l
}\pa^{\s\t}\pa_{\s\t} h_{c\r}^{~~\h\l} +{\textstyle \frac{1}{2}}\, \pa_{\h}
h_{b\m\n\l }\pa^{\h\s\t}\pa_{\s} h_{c\r\t}^{\l}\nonumber\\&& -{\textstyle \frac{1}{4}}\,
\pa_{\h} h_{b\m\n\r }\pa_{\l\s\t}\pa^{\l} h_{c}^{\h\s
\t} -{\textstyle \frac{3}{8}}\, \pa_{\h} h_{b\m\n\r
}\pa_{\l\s}\pa^{\l\s} h_{c}^{\h} -{\textstyle \frac{1}{2}}\, \pa_{\h}
h_{b\m\n\r }\pa^{\h\l} \pa_{\l\s}h_{c}^{\s}\nonumber\\
&&
-{\textstyle \frac{27}{16}}\,
\pa_{\m\n} h_{b \l}\pa^{\l\s\t} h_{c\r\s\t} +{\textstyle \frac{15}{16}}\,
\pa_{\m\n} h_{b \l}\pa^{\l\s}\pa_\s h_{c\r}
-{\textstyle \frac{1}{8}}\, \pa_{\m\n} h_{b \l}\pa^{\s}\pa_{\s\h}
h_{c\r}^{~~\l\h} \nonumber
\eqn
\newpage
\bqn
&&+{\textstyle \frac{1}{3}}\, \pa_{\m} h_{b}^{
\l\s\t}\pa_{\n\l\s\t} h_{c\r} +{\textstyle \frac{1}{2}}\, \pa_{\m\l} h_{b}^{
\l}\pa_{\n\s}\pa^\s h_{c\r} -{\textstyle \frac{33}{16}}\, \pa_{\m} h_{b}^{
\l}\pa_{\n\l\s\t}h_{c\r}^{~~\s\t}\nonumber\\&& -{\textstyle \frac{23}{4}}\,
\pa_{\m}\pa^\s h_{b}^{ \l}\pa_{\n\l\s}h_{c\r} +{\textstyle \frac{5}{8}}\,
\pa_{\m}h_{b}^{ \l}\pa_{\n\s}\pa^{\s\t} h_{c\r\l\t} -3\,
\pa_{\m}h_{b}^{ \l\s\t}\pa_{\n\l\h}\pa^{\h} h_{c\r\s\t}\nonumber\\
&&
-{\textstyle \frac{1}{4}}\, \pa_{\l}h_{b}^{ \l\s\t}\pa_{\m\n\s\t}
h_{c\r} -{\textstyle \frac{3}{2}}\,
\pa^{\l\s}h_{b\l}\pa_{\m\n}\pa^\t h_{c\r\s\t} +{\textstyle \frac{11}{4}}\,
\pa^{\l\s}h_{b\l}\pa_{\m\n\s} h_{c\r} \nonumber\\&&-{\textstyle \frac{15}{16}}\,
\pa^{\s\t}h_{b}^{\l}\pa_{\m\n\l} h_{c\r\s\t} +{\textstyle \frac{43}{16}}\,
\pa^{\s}\pa_\s h_{b}^{\l}\pa_{\m\n\l} h_{c\r}
-{\textstyle \frac{11}{4}}\, \pa^{\s\t}h_{b}^{\l}\pa_{\m\n\s} h_{c\r\l\t}\nonumber\\&&
+{\textstyle \frac{19}{8}}\, \pa^{\s}\pa_\s h_{b}^{\l}\pa_{\m\n\t}
h_{c\r\l}^{\t} +{\textstyle \frac{9}{4}}\, \pa_{\h\l}h_{b}^{\h\s\t}\pa_{\m\n\s}
h_{c\r\t}^{\l} +{\textstyle \frac{3}{4}}\, \pa_{\h}h_{b}^{\l\s\t}\pa_{\m\n\s\t}
h_{c\r\l}^{\h}\nonumber\\&& +{\textstyle \frac{15}{4}}\,
\pa_{\l}h_{b}^{\l\s\t}\pa_{\m\n\h}\pa^\h h_{c\r\s\t} -3\,
\pa^{\h}h_{b}^{\l\s\t}\pa_{\m\n\h\l} h_{c\r\s\t} -{\textstyle \frac{1}{2}}\,
\pa_{\m}h_{b\l\s\t}\pa^{\l\s\t\h} h_{c\n\r\h}\nonumber\\&&  -{\textstyle \frac{19}{4}}\,
\pa_{\m}h_{b\l}\pa^{\l\s\h}\pa_\s h_{c\n\r\h}
+{\textstyle \frac{1}{2}}\, \pa_{\m}h_{b}^{\l}\pa^{\s\t}\pa_{\s\t} h_{c\n\r\l}
-{\textstyle \frac{5}{2}}\, \pa_{\m}\pa^\h h_{b}^{\l\s\t}\pa_{\h\s\t}
h_{c\n\r\l}\nonumber\\&& -{\textstyle \frac{21}{4}}\, \pa_{\m}h_{b}^{\l\s\t}\pa^\h
\pa_{\h\s\t} h_{c\n\r\l} +{\textstyle \frac{1}{6}}\, \pa_{\l}h_{b}^{\l\s\t}
\pa_{\m\s\t}\pa^\h h_{c\n\r\h} -{\textstyle \frac{1}{2}}\,
\pa^{\h}h_{b}^{\l\s\t} \pa_{\m\l\s\t} h_{c\n\r\h} \nonumber\\&&-5\,
\pa^{\l\h}h_{b\l} \pa_{\m\h\s} h_{c\n\r}^{\s} -{\textstyle \frac{1}{2}}\,
\pa^{\s\h}h_{b}^{\l} \pa_{\m\h\l} h_{c\n\r\s} -{\textstyle \frac{9}{2}}\,
\pa^{\h}\pa_\h h_{b}^{\l} \pa_{\m\l}\pa^\s
h_{c\n\r\s}\nonumber\\&& -{\textstyle \frac{1}{2}}\, \pa^{\s}\pa_\s h_{b}^{\l}
\pa_{\m\h}\pa^\h h_{c\n\r\l} +\, \pa^{\s\t} h_{b}^{\l}
\pa_{\m\s\t} h_{c\n\r\l} -{\textstyle \frac{11}{2}}\, \pa^\h \pa_\s
h_{b}^{\l\s\t} \pa_{\m\t\h} h_{c\n\r\l}\nonumber\\&& +{\textstyle \frac{3}{4}}\, \pa^\h
\pa_\h h_{b}^{\l\s\t} \pa_{\m\s\t} h_{c\n\r\l}
-{\textstyle \frac{1}{4}}\, \pa_\l h_{b}^{\l\s\t} \pa_{\s\t\h}\pa^\h
h_{c\m\n\r} +{\textstyle \frac{1}{2}}\, \pa^\h h_{b}^{\l\s\t} \pa_{\l\s\t\h}
h_{c\m\n\r}\nonumber\\&& -{\textstyle \frac{7}{8}}\,  \pa_\l h_{b}^{\l} \pa_{\s\t}\pa^{\s\t}
h_{c\m\n\r} -{\textstyle \frac{7}{4}}\,  \pa^{\s\t} h_{b}^{\l}
\pa_{\l\s\t}h_{c\m\n\r}\,\Big)\nonumber\\
\nonumber \\
&+\,\,h_{a}^{\m}\,\,\Big(&
{\textstyle \frac{1}{2}}\,
 \pa_{\m} h_{b}^{\l\s\t} \pa_{\l\s\t
\r}h_{c}^{\r} -{\textstyle \frac{13}{16}}\,   \pa_{\m} h_{b}^{\s\t\l} \pa_{\s\t
\n\r}h_{c\l}^{~~\n\r} +{\textstyle \frac{9}{16}}\,
 \pa_{\m} h_{b}^{\s\t\l} \pa_{\s\t \n}\pa^\n h_{c\l}\nonumber\\&&
+{\textstyle \frac{1}{2}}\,  \pa_{\m\l}  h_{b}^{\l\n\r} \pa^{\s\t}\pa_\s
h_{c \n\r\t} -{\textstyle \frac{3}{4}}\,  \pa_{\m} h_{b}^{\l\n\r}
\pa_{\l\s} \pa^{\s\t}h_{c \n\r\t} +\,   \pa_{\m} h_{b}^{\l\n\r}
\pa^{\s\t}\pa_{\s\t} h_{c \l\n\r}\nonumber\\&& +{\textstyle \frac{1}{2}}\, \pa_{\m}  h_{b\l}
\pa^{\l\n\r\s} h_{c \n\r\s} -{\textstyle \frac{1}{2}}\,
 \pa_{\m} h_{b}^{\l} \pa_{\l\n\r}\pa^\n h_{c}^{ \r}
+{\textstyle \frac{3}{16}}\,  \pa_{\m}  h_{b}^{\l} \pa^{\n\r\s}\pa_\s
h_{c\l\n\r}\nonumber\\&& +{\textstyle \frac{1}{4}}\,   \pa_{\m} h_{b}^{\l}
\pa^{\r\s}\pa_{\r\s} h_{c\l}-\,  \pa_{\l} h_{b}^{\l\r\s}
\pa_{\m\r\s\t} h_{c}^{\t} +{\textstyle \frac{1}{2}}\, \pa_{\t} h_{b}^{\l\r\s}
\pa_{\m\l\r\s} h_{c}^{\t}\nonumber\\&&  +{\textstyle \frac{23}{16}}\,
 \pa_{\l}  h_{b}^{\l\n\r} \pa_{\m\n\s\t}
h_{c\r}^{~~\s\t} -{\textstyle \frac{3}{4}}\,  \pa_{\l} h_{b}^{\l\n\r}
\pa_{\m\n\s}\pa^\s h_{c\r} -{\textstyle \frac{5}{8}}\,
 \pa^{\l}  h_{b}^{\n\r\s} \pa_{\m\n\r\t} h_{c\l\s}^{\t}\nonumber\\&&
+{\textstyle \frac{25}{4}}\,   \pa_{\l} h_{b}^{\l\n\r} \pa_{\m\s}\pa^{\s\t}
h_{c\n\r\t} +\,  \pa^{\h}  h_{b}^{\l\n\r}
\pa_{\m\l\s}\pa^{\s} h_{c\h\n\r} -6\,  \pa^{\h} h_{b}^{\l\n\r}
\pa_{\m\h\l\s} h_{c\n\r}^\s\nonumber\\&& -\,   \pa^{\h} h_{b}^{\l\n\r}
\pa_{\m\h\s} \pa^{\s} h_{c\l\n\r} -{\textstyle \frac{1}{4}}\,
 \pa^{\l\n} h_{b\l} \pa_{\m\n\r} h_{c}^{\r}
-{\textstyle \frac{1}{2}}\,   \pa_{\n}  h_{b}^{\l} \pa_{\m\l\r\s}
h_{c}^{\n\r\s}\nonumber\\&& +{\textstyle \frac{1}{4}}\,  \pa^{\r\n} h_{b}^{\l}
\pa_{\m\l\r}h_{c\n} -{\textstyle \frac{1}{2}}\,  \pa^{\r}\pa_{\r} h_{b}^{\l}
\pa_{\m\l\n}h_{c}^{\n} +{\textstyle \frac{5}{4}}\, \pa^{\r}\pa_{\r}  h_{b}^{\l}
\pa_{\m\n}\pa^\n h_{c\l}\nonumber\\&& -\,
 \pa^{\n\r}  h_{b}^{\l} \pa_{\m\n\r} h_{c\l}
-{\textstyle \frac{5}{12}}\,   \pa_{\l} h_{b}^{\l\n\r} \pa_{\n\r\s\t}
h_{c\m}^{~~\s\t} +{\textstyle \frac{1}{3}}\,  \pa_{\l} h_{b}^{\l\n\r}
\pa_{\n\r\s}\pa^\s h_{c\m} \nonumber\\&&+{\textstyle \frac{2}{3}}\,
 \pa_{\s} h_{b}^{\l\n\r} \pa_{\l\n\r\t}h_{c\m}^{~~\s\t}
-\,   \pa^{\s} h_{b}^{\l\n\r} \pa_{\l\n\r\s}h_{c\m}
+{\textstyle \frac{9}{16}}\,  \pa_{\l}  h_{b}^{\l} \pa_{\n\r\s}\pa^\n
h_{c\m}^{~~\r\s}\nonumber\\&& +{\textstyle \frac{1}{8}}\,   \pa_{\l}  h_{b}^{\l}
\pa_{\n\r}\pa^{\n\r} h_{c\m} -{\textstyle \frac{3}{8}}\,   \pa_{\n} h_{b}^{\l}
\pa_{\l\r\s}\pa^{\s} h_{c\m}^{~~\n\r} +{\textstyle \frac{3}{8}}\,
 \pa^{\n}  h_{b}^{\l} \pa_{\l\n\r\s} h_{c\m}^{~~\r\s}\nonumber\\&&
-{\textstyle \frac{1}{4}}\,   \pa^{\n}  h_{b}^{\l} \pa^{\r\s} \pa_{\r\s}
h_{c\m\n\l} +{\textstyle \frac{1}{4}}\,   \pa^{\n} h_{b}^{\l} \pa^{\r\s}
\pa_{\n\r} h_{c\m\l\s} -{\textstyle \frac{1}{8}}\, \pa_{\l} h_{b}^{\l\n\r}
\pa^{\s\t} \pa_{\n\s} h_{c\m\r\t}\nonumber\\&& -{\textstyle \frac{3}{4}}\,
 \pa^{\l} h_{b}^{\n\r\s} \pa_{\n\r\t}
\pa^{\t}h_{c\m\l\s} +2\,   \pa^{\l} h_{b}^{\n\r\s} \pa_{\l\n\r\t}
h_{c\m\s}^{\t} -{\textstyle \frac{1}{4}}\,  \pa_{\l} h_{b}^{\l\r\s} \pa^{\n\t}
\pa_{\n\t} h_{c\m\r\s}\nonumber\\&& +{\textstyle \frac{1}{2}}\,
 \pa_{\n} h_{b}^{\l\r\s} \pa^{\n\t} \pa_{\l\t}
h_{c\m\r\s} +{\textstyle \frac{1}{4}}\,  \pa^{\n}  h_{b\m \n\r}
\pa^{\r\l\s\t} h_{c\l\s\t} -{\textstyle \frac{1}{2}}\,   \pa^{\l} h_{b\m \n\r}
\pa^{\n\r\s\t} h_{c\l\s\t}\nonumber\\&& +{\textstyle \frac{3}{16}}\,  \pa^{\l} h_{b\m \n\r}
\pa^{\n\r\s} \pa_\s h_{c\l} -{\textstyle \frac{3}{4}}\,  \pa_{\n} h_{b\m \r\s}
\pa^{\n\r\s \l} h_{c\l} +{\textstyle \frac{9}{4}}\, \pa^{\n\l}
h_{b\m \n\r}  \pa^{\s\t }\pa_\s h_{c\l\t}^{\r}\nonumber\\&& +{\textstyle \frac{3}{2}}\,
 \pa^{\n\l}  h_{b\m \n\r} \pa^{\s\t } \pa_\l
h_{c\s\t}^{\r} +{\textstyle \frac{7}{8}}\,   \pa^{\n}  h_{b\m \n\r} \pa^{\l\s\t
} \pa_\l h_{c\s\t}^{\r} -{\textstyle \frac{1}{2}}\, \pa^{\n} h_{b\m \n\r}
\pa_{\s\t} \pa^{\s\t } h_{c}^{\r}\nonumber\\&& +{\textstyle \frac{1}{2}}\,
 \pa_{\l}  h_{b\m \n\r} \pa_{\s\t} \pa^{\n\t }
h_{c}^{\l\r\s} +{\textstyle \frac{5}{4}}\,  \pa_{\l} h_{b\m \n\r}\pa^{\n\l }
\pa_{\s\t} h_{c}^{\r\s\t} +{\textstyle \frac{1}{2}}\, \pa_{\l} h_{b\m
\n\r}\pa^{\n\l\s } \pa_{\s}   h_{c}^{\r}\nonumber\\&& +{\textstyle \frac{1}{4}}\,
 \pa_{\l}  h_{b\m \n\r}\pa_{\s \t} \pa^{\s\t}
h_{c}^{\l\n\r} -{\textstyle \frac{1}{2}}\,  \pa_{\l} h_{b\m
\n\r}\pa_{\s \t} \pa^{\l\t} h_{c}^{\n\r\s} +{\textstyle \frac{1}{2}}\, \pa_{\l}
h_{b\m }\pa_{\n\r\s } \pa^{\s} h_{c}^{\l\n\r}\nonumber\\&& +{\textstyle \frac{1}{6}}\,
 \pa^{\l}  h_{b\m }\pa_{\l\n\r\s } h_{c}^{\n\r\s}
+{\textstyle \frac{1}{8}}\,   \pa_{\l}  h_{b\m }\pa_{\n\r } \pa^{\n\r }
h_{c}^{\l} -{\textstyle \frac{1}{4}}\,  \pa^{\l} h_{b\m
}\pa_{\l\n\r } \pa^{\n} h_{c}^{\r}\,\Big) \,. \nonumber \eqn

}
\hfill

%
\section{Parity-breaking self-interactions}
\label{sec:def}
%
In this section, we first compute all possible parity-breaking and Poincar\'e-invariant first-order deformations of the Abelian spin-3 gauge algebra. We find that such deformations exist in three and five dimensions.  We then proceed separately for $n=3$ and $n=5$. We analyse the corresponding first-order deformations 
of the quadratic Lagrangian and find that they both exist. Then, consistency conditions 
at second order are obtained which make the $n=3$ deformation trivial and which 
constrain the $n=5$ deformation to involve only one \emph{single} gauge field. 
%
\subsection{Most general term in antifield number two}
%
The first part of Theorem \ref{antigh2} is stil true for parity-breaking deformations, as the property of parity-invariance is not needed to prove it. If one allows for parity-breaking interactions, the second part must be completed by the following statement:
\begin{theorem}\label{defa2}
Let $a=a_0+a_1+a_2$ be a local topform that is a nontrivial solution of the
equation $sa+db=0\,$. 
If the last term $a_2$ is parity-breaking and Poincar\'e invariant, then it is trivial except
in three and five dimensions. In those cases,  modulo trivial terms, it can be written respectively
\begin{eqnarray} \label{deftrois} a_2=f^a_{~[bc]} \e^{\m\n\r}C^{* \a\b}_a C^b_{\m\a}
\pa_{[\n}C^c_{\r]\vert \b}d^3x\,\end{eqnarray}
and
\begin{eqnarray} a_2 = g^a_{~(bc)}\ve^{\m\n\r\s\t}C^{*\a}_{a~\;\m}
\pa_{[\n}C_{\r]}~^{\!\!\!\!\!\!\!b\,\b} \pa_{\a[\s}C^c_{\t]\b}d^5 x\,.
\label{a2n5}
\end{eqnarray}
The structure constants $f^a_{~[bc]}$ define an internal, 
anticommutative algebra $\ca$ while the structure constants  $g^a_{~(bc)}$ define an 
internal, commutative algebra $\cb\,$.  

\end{theorem}
\noindent{ \bfseries{Proof :}} 
The proof differs from the corresponding proof in the parity-invariant case by new terms arising in the $D$-degree decomposition of $a_2$. We refer to Section \ref{interactions} for  the beginning of the proof and turn immediately to the resolution of Eq.(\ref{third}), \ie
\begin{eqnarray} \d\a_{I_i}+d\b_{I_i}\pm\b_{I_{i-1}}
A^{I_{i-1}}_{I_{i}}= 0\, \label{third3}\end{eqnarray} for each $D$-degree
$i$. The results depend on the dimension, so we split the analysis into the cases $n=3$, $n=4$, $n=5$ and $n>5$.

\vspace{3mm}

\noindent\textbf{\underline{D-degree decomposition}}:

\vspace{3mm}

{\bfseries{Dimension 3}}
\begin{itemize}
  \item {\bf degree zero} : In $D$-degree 0, the  equation  (\ref{third3}) reads $\d\a_{I_0}+d\b_{I_0}=0$, which implies that $\a_{I_0}$ belongs to
$H_2(\d\vert d)$. 
In antifield number 2, this group has nontrivial
elements given by Proposition \ref{H2}, which are  proportional to
$C^{*\m\n}_{a}$ . The requirement of translation-invariance
restricts the coefficient of $C^{*\m\n}_{a}$  to be constant. On the other hand, in $D$-degree 0 and
ghost number 2, we have  $\o^{I_0}=C^b_{\m\r}C^{c}_{\n\s}$. To get
a parity-breaking but Lorentz-invariant $a^0_2$, a scalar quantity must be build 
by contracting $\o^{I_0}$,  $C^{*\m\n}_{a}$, the tensor $\ve^{\m\n\r}$ and a product of
$\eta_{\m\n}$'s. This cannot be done because there is an odd number of indices, so $a^0_2$ vanishes: $a^0_2=0$. One can then also choose $b^0_1=0$.

  \item {\bf degree one} : We now analyse Eq.(\ref{third3}) in $D$-degree 1. It reads 
  $ \d\a_{I_1} +d\b_{I_1}= 0\,$ and implies that $\a_{I_1} $ is an
  element of $H_2(\d\vert d)$. 
  Therefore the only parity-breaking and Poincar\'e-invariant $a^1_2$ that can be built is
 
\noindent  $a^1_2=f^a_{~bc}\ve^{\m\n\r} C_a^{*\a\b}C^{b}_{\a\m} T^{c}_{ \n\r\vert \b}d^3x\,$. 
  Indeed, it should have the structure $\ve C^*C\widehat{T}$ (or $\ve C^*CT$, up to trivial 
  terms), contracted with $\h$'s. In an equivalent way, it must have the structure 
  $C^*C\widetilde{T}\,$, contracted with $\h$'s, where the variable $\widetilde{T}$ has been 
  introduced in Eq.(\ref{tdual}). Due to the symmetry properties (\ref{proptdual}) of 
  $\widetilde{T}\,$ which are the same as the symmetries of $C^a_{\m\n}$ and 
  $C^{*\m\n}_{a}$, there is only one way of contracting $\widetilde{T}\,$, 
  $C$ and $C^*$ together: $f^a_{~bc}C^{*\m\n}_{a}C_{\m}^{b\,\r}\widetilde{T}^c_{\n\r}\,$. 
  No Schouten identity  (see Appendix \ref{schouten}) can come into play because of the 
  number and the 
  symmetry of the fields composing 
  $f^a_{~bc}C^{*\m\n}_{a}C_{\m}^{b\,\r}\widetilde{T}^c_{\n\r}\,$. The latter term is 
  proportional to  
  $a^1_2=f^a_{~bc}\ve^{\m\n\r} C_a^{*\a\b}C^{b}_{\a\m} T^{c}_{ \n\r\vert \b}d^3x\,$, up to 
  trivial terms. 
  One can now easily compute that $b^1_1=- 3\,f^a_{~bc}\ve^{\m\n\r} 
  (h_a^{*\a\b\l}-\frac{1}{3}\eta^{\a\b}
  h_a^{*\l})C^{b}_{\a\m}\widehat{T}^{c}_{\n\r\vert \b}\frac{1}{2}\ve_{\l\s\t}dx^\s  dx^\t\,$.

  \item {\bf degree two} : The equation (\ref{third3}) in $D$-degree 2 is 
  $\d\a_{I_2}+d\b_{I_2}-\b_{I_{1}}A^{I_{1}}_{I_{2}}=0$, with
  \begin{eqnarray} 
  -\b_{I_{1}}A^{I_{1}}_{I_{2}}\o^{I_2}&=&3\,f^a_{~bc}\ve^{\m\n\r} 
  (h_a^{*\a\b\l}-\frac{1}{3}\eta^{\a\b}
  h_a^{*\l})(\frac{4}{3} \widehat{T}^{b}_{\h (\a\vert\m)} \widehat{T}^{c}_{ \n\r\vert \b})
  \frac{1}{2} \ve_{\l\s\t} dx^\h dx^\s  dx^\t   \nonumber \\
  &=&2\ f^a_{~(bc)}\ve^{\m\n\r} 
  (h_a^{*\a\b\l}-\frac{2}{3}\eta^{\a\b}
  h_a^{*\l}) \widehat{T}^{b}_{\l \m\vert\a} \widehat{T}^{c}_{ \n\r\vert \b}  \ d^3 x   
  \,.
  \nonumber 
  \end{eqnarray}
The latter equality holds up to irrelevant trivial $\g$-exact terms. It is obtained 
by using the fact that there  are only two linearly independent scalars having the structure 
$\ve h^* \widehat{T}\widehat{T}\,$. They are 
$\ve^{\m\n\r}h^{*\a\b\g}\widehat{T}_{\m\n|\a}\widehat{T}_{\r\b|\g}$ and 
$\ve^{\m\n\r}h^{*\a}\widehat{T}_{\m\n|}^{~~\b}\widehat{T}_{\r\b|\a}\,$. 
To prove this, it is again easier to use the dual variable $\widetilde{T}$ instead of $\widehat{T}$. 
One finds that the linearly independent terms with the structure 
$\ve h^* \widetilde{T}\widetilde{T}$ are
$f^a_{~(bc)}\ve^{\m\n\r}h^{*a}_{~\m}{\widetilde{T}}_{\n}^{b\,\a}{\widetilde{T}}^c_{\r\a}$ 
and 
$f^a_{~(bc)}\ve^{\m\n\r}h^{*a\,\a\b}_{~\m}{\widetilde{T}}^{b}_{\n\a}{\widetilde{T}}^c_{\r\b}\,;$
 they are proportional to  
$f^a_{~(bc)}\ve^{\m\n\r}h_a^{*\a\b\g}\widehat{T}^b_{\m\n|\a}\widehat{T}^c_{\r\b|\g}$ and 
$f^a_{~(bc)}\ve^{\m\n\r}h^{*\a}_a\widehat{T}_{\m\n|}^{b~\;\b}\widehat{T}^c_{\r\b|\a}\,$.

Since the expression for $\b_{I_{1}}A^{I_{1}}_{I_{2}}$ is not $\d$-exact modulo $d\,,$ it must vanish: $f_{abc}=f_{a[bc]}\,.$ One then gets 
that $\a_{I_2}$ belongs to
$H_2(\d\vert d)$.  However, no such parity-breaking and Poincar\'e-invariant $a^2_2$ can be formed in $D$-degree 2, so $a_2^2=0=b_1^2\,.$ 
  
  \item {\bf degree higher than two} : Finally, there are no $a^i_2$ for $i>2$. Indeed, there is no ghost
combination $\o^{I_i}$ of ghost number two and $D$-degree higher
than two, because $\widehat{U}$ identically vanishes when $n=3$.
\end{itemize}

\vspace{2mm}
{\bfseries{Dimension 4}}

There is no nontrivial deformation of the gauge algebra in dimension 4.
\begin{itemize}

\item {\bf degree zero} : The equation (\ref{third3})  reads $\d \a_{I_0}+d \b_{I_0}=0$. It implies that $\a_{I_0}$ belongs to $H_2^4{(\d|d)}$, which means that $\a_{I_0}$ is of the form $k^a_{~bc}\ve^{\m\n\r\s}C_a^{*\a\b}d^4x$ where $k^a_{~bc}$ are some constants. 
It is obvious that all contractions of $\a_{I_0}$ with two undifferentiated ghosts $C$ in a Lorentz-invariant way identically vanish. One can thus choose $a_2^0=0$ and $b_1^0=0$.

\item {\bf degree one} : The equation in $D$-degree 1 reads $\d \a_{I_1}+d \b_{I_1}=0$. The nontrivial part of $\a_{I_1}$ has the same form as in $D$-degree 0. It is however impossible to build a nontrivial Lorentz-invariant $a_2^1$ because $\o_{I_1} \sim CT$ has an odd number of indices. So $a_2^1=0$ and $b_1^1=0$.

\item {\bf degree two} : In $D$-degree 2, the equation $\d \a_{I_2}+d \b_{I_2}=0$ must be studied. Once again, one has $\a_{I_2}=k^a_{~bc}\ve^{\m\n\r\s}C_a^{*\a\b}d^4x$. 
There are two sets of $\o_{I_2}$'s : $\widehat{T}^b_{\m\n\vert \a}\widehat{T}^c_{\r\s \vert \b}$ and $C^b_{\a\b}\widehat{U}^c_{\m\n|\r\s}\,.$ A priori there are three different ways to contract the indices of terms with the structure $\ve C^*\widehat{T}\widehat{T}$, but because of Schouten identities (see Appendix \ref{ectt}) only two of them are independent, with some symmetry constraints on the structure functions. No Schouten identities exist for terms with the structure $\ve C^*C\widehat{U}$. The general form of $a_2^2$ is thus, modulo trivial terms, 
$$\begin{array}{rcl}a_2^2&=&\stackrel{(1)}{k^a}_{[bc]}\ve^{\m\n\r\s}\ C_a^{*\a\b}\
   \widehat{T}^{b}_{\m\n \vert \a}\ \widehat{T}_{\r\s \vert \b}^c d^4 x+\stackrel{(2)}{k^a}_{(bc)}\ve^{\m\n\r\s}\
C^{*\a}_{a~\,\m}\ \widehat{T}^{b \, ~ \b}_{ \n\r\vert}\ \widehat{T}_{\s\a\vert \b}^c d^4 x\\&&+\stackrel{(3)}{k^a}_{bc}\ve^{\m\n\r\s}\
C^{*}_{a\m\a}\ C^b_{\n\b}\ \widehat{U}^{c\ \ \a\b}_{\r\s|}d^4 x\,,\end{array}$$
and $b_1^2$ is given by
$$\begin{array}{rcl}
b_1^2&=&
-3 \ \ve^{\m\n\r\s} \Big[ ( h_a^{*\l\a\b}-\frac{1}{4} 
h_a^{*\l}\h^{\a\b})\stackrel{(1)}{k^a}_{[bc]}
    \widehat{T}^{b}_{\m\n \vert \a}\ \widehat{T}_{\r\s \vert \b}^c \\
&&
+(h^{*\l\a}_{a~\;\;\m}-\frac{1}{4} h_a^{*\l}\d^\a_{\m})(\stackrel{(2)}{k^a}_{(bc)}
\widehat{T}^{b \, ~ \b}_{ \n\r\vert}\ \widehat{T}_{\s\a\vert \b}^c +\stackrel{(3)}
{k^a}_{bc}\ C^b_{\n\b}\ \widehat{U}^{c\ \ \ \b}_{\r\s\vert \a})
\Big]\nnn
&&\hspace{8cm}\frac{1}{3!}\ve_{\l \r\s\t}dx^\r dx^\s dx^\t
\,.\end{array}$$

\item {\bf degree three} : Eq.(\ref{third3}) now reads $\d \a_{I_3}+d \b_{I_3}+\b_{I_2}A_{I_3}^{I_2}=0$ , with
$$\begin{array}{rcl}\b_{I_2}A_{I_3}^{I_2}\o^{I_3}
&=&-\displaystyle\frac{3}{2}\stackrel{(1)}{k^a}_{[bc]}\ve^{\m\n\r\s}h_a^{*\l}
 \widehat{T}^{b ~~\a}_{\m\n\vert}\widehat{U}^c_{\l\a|\r\s}d^4 x\\
&&-3\stackrel{(2)}{k^a}_{(bc)}\ve^{\m\n\r\s}h_{a~~\m}^{*\a\l}\Big(\widehat{T}^{b~~\b}_{\n\a\vert}\widehat{U}^c_{\l\b|\r\s}-\widehat{T}^{b~~\b}_{\n\r\vert}
\widehat{U}^c_{\l\b|\s\a}\Big)d^4 x\\
&&+4\stackrel{(3)}{k^a}_{bc}\ve^{\m\n\r\s}h_{a~~\m}^{*\a\l}\widehat{T}^b_{\l(\b\vert \n)}\widehat{U}^{c\quad\b}_{\r\s|\a}d^4 x\\
&=&\Big( -\frac{3}{2}(\stackrel{(1)}{k^a}_{[bc]}+\stackrel{(2)}{k^a}_{(bc)})
\ve^{\b\g\r\s}h_a^{*\m}\h^{\a\n}\\
&&\quad
-(6\stackrel{(2)}{k^a}_{(bc)}+4\stackrel{(3)}{k^a}_{bc})\ve^{\m\n\l\b}
h^{*\g\r}_{a~~\l} \h^{\a\s}
\Big)
\,\widehat{T}^b_{\b\g\vert \a}\widehat{U}^c_{\m\n|\r\s}d^4x
\end{array} $$ 
The latter equality is obtained using Schouten identities 
(see Appendix \ref{ehtu}). 
It is obvious that the coefficient of $\o^{I_3}=\widehat{T}^b_{\b\g\vert \a}\widehat{U}^c_{\m\n|\r\s}$ cannot be $\d$-exact modulo $d$ unless it is zero. This implies that $\stackrel{(1)}{k^a}_{[bc]}=\stackrel{(2)}{k^a}_{(bc)}=\stackrel{(3)}{k^a}_{bc}=0$.
So $a_2^2$ is trivial and can be set to zero, as well as $b_1^2$. One now has $\d \a_{I_3}+d \b_{I_3}=0$, which has  the usual solution for $\a_{I_3}$, but there is no nontrivial Lorentz-invariant $a_2^3$ because there is an odd number of indices to be contracted.

\item {\bf degree higher than three} : Eq.(\ref{third3}) is $\d \a_{I_4}+d \b_{I_4}=0$, thus $\a_{I_4}$ is of the form $l^a_{~bc}\ve^{\m\n\r\s}C_a^{*\a\b}d^4x$. 
There are two different ways to contract the indices : $\ve^{\m\n\r\s}C_a^{*\a\b}\widehat{U}^b_{\m\n|\a\g}\widehat{U}_{\r\s|\b}^{c\ \ \ \g}$ and $\ve^{\m\n\r\s}C^{*}_{a\m\a}\widehat{U}^b_{\n\r|\b\g}\widehat{U}_\s^{c\a|\b\g}$, but  both functions vanish because of Schouten identities (see Appendix \ref{ecuu}). Thus $a_2^4=0$ and $b_2^4=0$. No candidates $a^i_2$ of ghost number two exist in $D$-degree higher than four because there is no appropriate $\o^{I_i}$.

\end{itemize}

\vspace{2mm}
{\bfseries{Dimension 5}}
\begin{itemize}

\item {\bf degree zero} : In $D$-degree 0, the equation (\ref{third3}) reads $\d \a_{I_0}+d\b_{I_0}=0\,,$ 
which means that $\a_{I_0}$ belongs to $H^{5}_2(\d|d)$.
However, $a^0_2$ cannot be build with such an $\a_{I_0}$ because the latter has an odd number of indices while $\o^{I_0}$  has an even one. So, $\a_{I_0}$ and
$\b_{I_0}$ can be chosen to vanish. 

\item {\bf degree one} : In $D$-degree 1, the equation becomes $\d \a_{I_1}+d\b_{I_1}=0$, so $\a_{I_1}$ belongs to $H^{5}_2(\d|d)$. However, it is impossible to build a non-vanishing Lorentz-invariant $a_2^1$ because  in a product $C^*C\widehat{T}$ there are not enough indices that can be antisymmetrised to be contracted with the Levi-Civita density.
So $\a_{I_1}$ and $\b_{I_1}$ can be set to zero.

\item {\bf degree two} : The equation (\ref{third3}) reads $\d\a_{I_2}+d\b_{I_2}=0$. Once again, there is no way to build a Lorentz-invariant $a_2^2$ because of the odd number of indices. So $\a_{I_2}=0$ and $\b_{I_2}=0$.

\item {\bf degree three} : In $D$-degree 3,
the equation is $\d \a_{I_3}+d\b_{I_3}=0$, so $\a_{I_3}\in H^{5}_2(\d|d)$. This gives rise to an $a_2$ of the form "$g\ve\, C^*\widehat{T}\widehat{U}d^5 x$". There is only one nontrivial Lorentz-invariant object of this form : $$a_2=g^a_{~bc}\ve^{\m\n\r\s\t}C^{*}_{a\m\a}\widehat{T}^{b}_{\n\r\vert\b}
\widehat{U}^{c\a\b\vert}_{~~~~\,\s\t}d^5 x\;.$$ It is equal to (\ref{a2n5}) modulo a $\g$-exact term.
One has 
$$\!\!b_1^3=\!\b_{I_3}\o^{I_3}=\!\!-3g^a_{~bc}\ve^{\m\n\r\s\t}
(h^{*~~\; \l}_{a\m\a}-\textstyle{\frac{1}{5}}\eta_{\m\a}h_a^{*\l})\widehat{T}^b_{\n\r\vert\b}
\widehat{U}^{c\a\b}_{\ \ \ \ |\s\t}\textstyle{\frac{1}{4!}}\ve_{\l \g\d\h \xi}dx^\g dx^\d dx^\h 
dx^{\xi}\,.$$

\item {\bf degree four} : The equation (\ref{third3}) reads 
$\d \a_{I_4}+d\b_{I_4}-\b_{I_3}A^{I_3}_{I_4}$, with $$\b_{I_3}A^{I_3}_{I_4}\o^{I_4}=-3g^a_{~[bc]}\ve^{\m\n\r\s\t}h^{*\a\l}_{a~~\;\m}
\widehat{U}^{b}_{\l\b|\n\r}\widehat{U}^{c\ \b}_{\ \a\ \ |\s\t}d^5 x$$ 
The coefficient of $\o^{I_4}\sim\widehat{U}\widehat{U}$ cannot be
$\d$-exact modulo $d$ unless it vanishes, which implies that
$g^a_{~bc}=g^a_{~(bc)}$. One is left with the equation $\d
\a_{I_4}+d\b_{I_4}=0$, but once again it has no
Lorentz-invariant solution because of the odd number of indices to be contracted. So $\a_{I_4}=0$ and $\b_{I_4}=0$.
\item {\bf degree higher than four}: There is again no $a^i_2$ for $i>4$, for the same reasons as in four dimensions.
\end{itemize}

{\bfseries{Dimension $\mathbf{n>5}$}}

No new $a_2$ arises because it is impossible to build a non-vanishing parity-breaking term by contracting an element of $H^n_2(\d\vert d)$, {\it i.e.}  $C^{*\m\n}$, two ghosts from the set $\{ C^{\m\n}, \widehat{T}^{\m\n\vert \r}, \widehat{U}^{\m\n\vert \r\s} \}$, an epsilon-tensor $\ve^{\m_1 \ldots \m_n}$ and metrics $\h_{\m\n}$.

\vspace{2mm}
Let us finally notice that throughout this proof we have acted as if
 $\a_I$'s trivial in $H^n_2(\d\vert d)$ lead to trivial $a_2$'s. The correct statement
is that trivial $a_2$'s correspond to $\a_I$'s trivial in $H^n_2(\d\vert d,H(\g))$ (see Section \ref{cons} for more details). However, both statements are equivalent in this case, since both groups are isomorphic (Theorem  \ref{2.6}).

This ends the proof of Theorem \ref{defa2}.\quad \qedsymbol


\subsection{Deformation in 3 dimensions}
\label{dim3peu}

In the previous section, we determined that the only nontrivial first-order deformation of the free theory in three dimensions deforms the gauge algebra by the term (\ref{deftrois}). 
We now check that this deformation can be consistently lifted and leads to a consistent first-order deformation of the Lagrangian. 
However, we then show that obstructions arise at second order, {\it i.e.} that 
one cannot construct a corresponding consistent second-order deformation unless the whole 
deformation vanishes. 

\subsubsection{First-order deformation}
\label{fodn3}
A consistent first-order deformation exists if one can solve Eq.(\ref{first}) for $a_0$, where $a_1$ is obtained from Eq.(\ref{second}). 
The existence of  a solution $a_1$ to Eq.(\ref{second}) with $a_2=a_2^1$ is a consequence of the analysis of the previous section. Indeed, the $a_2$'s of Theorem \ref{defa2} are those that admit an $a_1$ in Eq.(\ref{second}).
Explicitely,  $a_1$ reads, modulo trivial terms, \begin{eqnarray}
a_1=f^a_{~[bc]}\e^{\m\n\r} \Big[ 3\, (h_a^{*\a\b\l}-\frac{1}{3} \h^{\a\b} h_a^{* \l})\, (\frac{1}{3}h^{b}_{\a\m\l}T^{c}_{ \n\r\vert \b}+\frac{1}{2} C^{b}_{\a\m} \pa_{[\r}h^{c}_{ \n] \b\l})
\nonumber \\
+\frac{1}{3}h_a^{*\l} T^b_{\l\n\vert \m}h_\r^c +h^*_{a\m} C^{b\,\a}_{\,\n}(-\frac{1}{2} \pa^\l h^c_{\l\a\r}+\pa_{(\a}h^c_{\r)})\Big] d^3 x\,.\nonumber 
\end{eqnarray}

On the contrary, a new condition has to be imposed on the structure function for the existence of an $a_0$ satisfying Eq.(\ref{first}).  Indeed a necessary condition for $a_0$ to  exist is that $\d_{ad}f^d_{~[bc]}=f_{[abc]}$, 
which means that the corresponding internal anticommutative algebra $\ca$ is endowed with an invariant norm. The internal metric we use is $\d_{ab}$, which is positive-definite.  
The condition is also sufficient and $a_0$ reads, modulo trivial terms,
\begin{eqnarray}
a_0=f_{[abc]}\e^{\m\n\r}\Big[
\frac{1}{4} \pa_\m h^a_{\n\a\b}\pa^\a h^{b\b}h^{c}_\r
+\frac{1}{4} \pa_\m h^a_{\n\a\b}\pa^\a h^{b\b\g\d}h^{c}_{\r\g\d}
-\frac{5}{4} \pa_\m h^a_{\n\a\b}\pa^\a h^{b\g}h^{c\b}_{\r\g}\nonumber \\
-\frac{3}{8} \pa_\m h^a_{\n}\pa^\a h^{b}_{\a}h^{c}_{\r}
+\frac{1}{4}\pa_\m h^{a\a\b}_{\n}\pa^\g h^{b}_{\g}h^{c}_{\r\a\b}
-\pa_\m h^{a}_{\n}\pa^\g h^{b}_{\a\b\g}h_{\r}^{c\a\b}\nonumber \\
+\frac{1}{2} \pa_\m h^{a}_{\n\a\b}\pa^\g h^{b}_{\a\g\d}h_{\r}^{c\b\d}
+2 \pa_\m h^{a}_{\n}\pa^\b h^{b\g}h_{\r}^{c\b\g}
-\frac{1}{4}\pa_\m h^{a}_{\n\a\b}\pa^\g h^{b\a\b\d}h_{\r\g\d}^{c}\nonumber \\
-\frac{1}{4}\pa_\m h^{a}_{\n\a\b}\pa^\g h^{b\b}h_{\r\g}^{c~\a}
-\frac{5}{8}\pa_\m h^{a}_{\n}\pa_\r h^{b\b}h_{\b}^{c}
+\frac{7}{8}\pa_\m h^{a}_{\n\a\b}\pa_\r h^{b\a\b\g}h_{\g}^{c}\nonumber \\
+\frac{1}{4}\pa_\m h^{a}_{\n\a\b}\pa_\g h_\r^{b\a\g}h^{c\b}
+\frac{1}{4}\pa_\m h^{a}_{\n}\pa^\a h_{\r\a\b}^{b}h^{c\b}
-\frac{1}{4}\pa_\m h^{a}_{\n\a\b}\pa^\g h_{\r\g\d}^{b}h^{c\a\b\d}\nonumber \\
-\frac{1}{8}\pa_\m h^{a}_{\n}\pa^\a h_{\r}^{b}h^{c}_{\a}
-\frac{1}{8}\pa_\m h^{a}_{\n\a\b}\pa^\g h_{\r}^{b\a\b}h^{c}_{\g}
\Big]d^3x\,.\nonumber 
\end{eqnarray}
To prove these statements about $a_0$, one writes the most general $a_0$ with two derivatives, that is Poincar\'e-invariant but breaks the parity symmetry. One  inserts this $a_0$ into the equation to solve, {\it i.e.} $\d a_1+\g a_0=d b_0$, and computes the $\d$ and $\g$ operations. One takes an Euler-Lagrange derivative of the equation with respect to the ghost, which removes the total derivative  $d b_0$. The equation becomes $\frac{\d}{\d C_{\a\b}}(\d a_1+\g a_0)=0$, which we multiply by $C_{\a\b}$. The terms of the equation have the structure $\ve C \pa^3h h$ or $\ve C \pa^2h \pa h$.
One  expresses them as linear combinations of a set of linearly independent quantities, which is not obvious as there are Schouten identities relating them (see Appendix \ref{cpahh}). One can finally solve the equation for the arbitrary coefficients in $a_0$, yielding 
the above results.

\subsubsection{Second-order deformation}
Once the first-order deformation $W_1=\int (a_0+a_1+a_2)$ of the free theory is determined, the next step 
is to investigate whether a corresponding second-order deformation $W_2$ exists. 
This second-order deformation of the master equation is constrained to obey $  s W_2 = -\frac{1}{2} (W_1,W_1)\,,$ (see Section \ref{cons}).
Expanding both sides according to the antighost number yields several conditions.
The maximal antighost number condition reads 
\begin{eqnarray}
-\frac{1}{2}(a_2,a_2) = \g c_2 + \d c_3 + d f_2  
\nonumber
\end{eqnarray}
where we have taken $W_2 = \int d^3x \ (c_0+c_1+c_2+c_3)$ and $antigh(c_i)=i\,$. 
It is easy to see that the expansion of $W_2$ can indeed be assumed to stop at antighost number  
$3$ and that $c_3$ may be assumed to be invariant.
 
\hfill

The calculation of $(a_2,a_2)$, where 
        $a_2 = f^a_{~[bc]} \varepsilon^{\mu\nu\rho} C^{*\alpha\beta}_a \, 
        C^{b}_{\mu\alpha} \partial_{{\nu}}C^c_{\rho\beta}\,$, 
gives

\begin{eqnarray}
        (a_2,a_2) \!\!\!&=&\! \!\!2 \frac{\d^R a_2}{\d C^{*\a\b}_a}\frac{\d^L a_2}{\d C_{\a\b}^a}
  \nonumber \\
  &=&\!\!\!\g\mu + d\n + 2f^a_{~bc}f_{ead}\ve^{\m\n\r}\ve_{\a\l\t}
  \Big[\,
  \frac{1}{2}\,C^{*e\s\x}C_{\m}^{b\,\a}{\widehat{T}}^c_{\n\r|\s}
  {\widehat{T}}^{d\l\t|}_{~~~~~\x}
  \nonumber \\
 \! &\!\!\! \! \! +&\!\!\!\!\!\!\! 
  \frac{1}{2}\,C^{*e\s\x}C_{\m\s}^{b}{\widehat{T}}_{\n\r|}^{c~~\a}
  {\widehat{T}}^{d\l\t|}_{~~~~~\x}
-\frac{1}{3}\,C^{*e\a\x}C_{\m}^{b\s}{\widehat{T}}^c_{\n\r|\s}
  {\widehat{T}}^{d\l\t|}_{~~~~~\x}-
  \frac{2}{3}\,C^{*e\s\x}{\widehat{T}}^{b\l}_{~~(\m|\a)}
  {\widehat{T}}^c_{\n\r|\s}C_{\x}^{d\,\t}
  \nonumber \\
 \! &\!\!\!\! \! -&\!\!\!\!\!\!
  \frac{2}{3}\,C^{*e\s\x}{\widehat{T}}^{b\l}_{~~(\m|\s)}
  {\widehat{T}}^{c~~\a}_{\n\r|}C_{\x}^{d\,\t}
+\frac{4}{9}\,C^{*e\a\x}{\widehat{T}}^{b\l}_{~~(\m|\s)}
  {\widehat{T}}^{c~~\s}_{\n\r|}C_{\x}^{d\,\t}\,
  \Big]\,.
\end{eqnarray}

It is impossible to get an expression with three ghosts, one $C^{*}$
and no field, by acting with $\d$ on $c_3\,$.  We 
can thus assume without loss of generality that $c_3$ vanishes,
which implies that  $(a_2,a_2)$ should be $\g\,$-exact modulo total derivatives.

The use of the variable $\widetilde{T}_{\a\b}:=\ve^{\m\n}_{~~\a}\widehat{T}_{\m\n|\b}$
instead of $\widehat{T}_{\m\n|\r}(=-\frac{1}{2}\ve^{\a}_{~\m\n}\widetilde{T}_{\a\r})$ 
simplifies the calculations. We find, after expanding the products of $\ve$-densities,  
\begin{eqnarray}
        (a_2,a_2)&=&\g\mu + d\n + f^a_{~bc}f_{ead}C^{*e\s\t}
  \Big[\,
    C^{b\m\a}{\widetilde{T}}^c_{\m\s}{\widetilde{T}}^d_{\a\t}    
  + C^{b\m}_{~~\s}{\widetilde{T}}^c_{\m\a}{\widetilde{T}}^{d\,\a}_{\;\t}
  \nonumber \\  
  && - \frac{2}{3}\,C^{b\m\a}{\widetilde{T}}^c_{\m\a}{\widetilde{T}}^d_{\s\t}   
  + C^{d\m}_{~~\s}{\widetilde{T}}^b_{\m\a}{\widetilde{T}}^{c\,\a}_{\;\t}  
  - \frac{1}{3}\,C^d_{\s\t}{\widetilde{T}}^{b\a\m}{\widetilde{T}}^c_{\a\m} 
  \Big]\,.
  \label{inter1}
\end{eqnarray}
We then use the only possible Schouten identity 
\begin{eqnarray}
        0 &\equiv& C^{*e\,\t}_{~[\s}C^{b\,\m}_{\;\a}{\widetilde{T}}^{c\,\s}_{\;\m}
        {\widetilde{T}}^{d\,\a}_{\;\t]}\nonumber \\
        &=& \frac{1}{24}\Big[
         - C^{*e\s\t}C^{b\m\a}{\widetilde{T}}^c_{\s\t}{\widetilde{T}}^d_{\m\a}
         + 2 C^{*e\s\t}C^{b\m\a}{\widetilde{T}}^c_{\s\m}{\widetilde{T}}^d_{\a\t}
         + 2 C^{*e\s\t}C^b_{\s\m}{\widetilde{T}}^c_{\t\a}{\widetilde{T}}^{d\,\a\m}
         \nonumber \\
         &&- C^{*e\s\t}C^b_{\s\t}{\widetilde{T}}^c_{\m\n}{\widetilde{T}}^{d\,\m\n}
         - C^{*e\s\t}C^{b\m\n}{\widetilde{T}}^c_{\m\n}{\widetilde{T}}^d_{\s\t}
   + 2 C^{*e\s\t}C_{\s}^{b\,\m}{\widetilde{T}}^c_{\m\a}{\widetilde{T}}^{d\,\a}_{~\t}
        \Big] \label{Sca2a2}
\end{eqnarray}
in order to substitute in Eq.(\ref{inter1}) the expression of $C^{*e\s\t}C^{b\m\a}{\widetilde{T}}^c_{\m\s}{\widetilde{T}}^d_{\a\t}$ in terms of 
the other summands appearing in Eq.(\ref{Sca2a2}).  
Consequently, the following expression for $(a_2,a_2)_{a.b.}$ contains only linearly independent terms: 
\begin{eqnarray}
        (a_2,a_2)\!\!\! &=&  \g\mu + d\n + C^{*e\s\t}\Big[
         {\textstyle\frac{1}{2}}f^a_{~bc}f_{dea}C^{b\m\a}{\widetilde{T}}^c_{\s\t}
   {\widetilde{T}}^d_{\m\a}
                +{\textstyle \frac{1}{6}}f^a_{~bc}f_{dea}C^{b\m\a}{\widetilde{T}}^d_{\s\t}
                {\widetilde{T}}^{c}_{\m\a}
        \nonumber \\
        &&\!\!\!\!\!\! + {\textstyle{2}} f^a_{~c(b}f_{d)ea}C^{b\,\m}_{\s}{\widetilde{T}}^c_{\t\a}
        {\widetilde{T}}^{d\,\a}_{~\m}
        +{\textstyle \frac{1}{2}}f^a_{~b[c}f_{d]ea}C^{b}_{\s\t}{\widetilde{T}}^c_{\m\a}
        {\widetilde{T}}^{d\m\a}
        +{\textstyle \frac{1}{3}}f^a_{~bc}f_{dea}C^{d}_{\s\t}{\widetilde{T}}^c_{\m\a}
        {\widetilde{T}}^{b\m\a}
        \Big]\nonumber\,,
\end{eqnarray}
where we used that the structure constants of $\ca$ obey 
$f_{abc}\equiv \d_{ad}f^d_{~bc}=f_{[abc]}$. 

Therefore, the above expression is a $\g\,$-cobounday modulo $d$ if and only if 
 $f^a_{~bc}f_{dea}=0$, meaning that the internal algebra $\ca$ is nilpotent of order three. 
 In turn, this implies\footnote{The internal metric $\d_{ab}$ being Euclidean, the 
condition $ f^a_{~bc}f_{aef}\equiv \d_{ad} f^a_{~bc}f^d_{~ef} = 0$ can be seen as 
expressing the vanishing of the norm of a vector in Euclidean space (fix $e=b$ and $f=c$), 
leading to $f^a_{~bc}=0$. } that $f^a_{~bc}=0$ and the deformation is trivial.


\subsection{Deformation in 5 dimensions}

Let us perform the same analysis for the candidate in five dimensions.

\subsubsection{First-order deformation}
\label{fodn5}
First, $a_1$ must be computed from $a_2$ (given by (\ref{a2n5})), using the equation $\d a_2+\g a_1+d b_1=0\,$: 
$$\begin{array}{rcl}\d
a_2&=&-3g^a_{~(bc)}\ve^{\m\n\r\s\t}\pa_\l
h^{*\a\l}_{a~~\m}\pa_{[\n}C^b_{\r]\b}\pa_{\a[\s}C^{c\b}_{\t]}d^5
x\\&=&-d
b_1+3g^a_{~(bc)}\ve^{\m\n\r\s\t}h^{*\a\l}_{a~~\m}\,\Big[\pa_{\l[\n}C^b_{\r]\b}\pa_{\a[\s}C^{c\b}_{\t]}
+\pa_{[\n}C^b_{\r]\b}\pa_{\l\a[\s}C^{c\b}_{\t]}\Big] d^5
x\end{array}$$
We recall that it is a consequence of Theorem \ref{defa2} that $g^a_{~bc}$
is symmetric in its lower indices, thereby defining a commutative algebra. 
Therefore the first term  between square bracket vanishes because of the symmetries of 
the structure constants $g^a_{~bc}$ of the internal commutative algebra $\cb\,$.  
We finally obtain, modulo trivial terms,
$$a_1=\frac{3}{2}g^a_{~(bc)}\ve^{\m\n\r\s\t}h^{*\a\l}_{a~~\,\m}\pa_{[\n}^{\ }C^{b\
\b}_{\r]}\left[\pa_{\b[\s}h^c_{\t]\l\a}-2\pa_{\l[\s}h^c_{\t]\a\b}\right]d^5
x\,.$$
The element $a_1$ gives the first order deformation of the gauge transformations. 
By using the definition of the generalized de Wit--Freedman connections 
\cite{deWit:1979pe}, we get the following simple expression for $a_1$: 
\begin{eqnarray}
        a_1=g^a_{~(bc)}\ve^{\m\n\r\s\t}h^{*\a\b}_{a~~\,\m}\pa_{[\n}^{\ }C^{b\ \l}_{\r]} 
        {\Gamma}^c_{\l[\s;\t]\a\b}  d^5x\,,
        \label{a1dWF}
\end{eqnarray}
where ${\Gamma}^c_{\l\s;\t\a\b}$ is the second spin-3 connection 
\begin{eqnarray}
        {\Gamma}^c_{\l\s;\t\a\b} =3\, \pa_{(\t}\pa_{\a}h^c_{\b)\l\s} 
        + \pa_{\l}\pa_{\s}h^c_{\t\a\b}  - \frac{3}{2}\,\big(\pa_{\l}\pa_{(\t}h^c_{\a\b)\s} 
        +\pa_{\s}\pa_{(\t}h^c_{\a\b)\l}\big)\nonumber
\end{eqnarray}
transforming under a gauge transformation $\d_{\l}h^a_{\m\n\r}=3\,\pa_{(\m}\l^a_{\n\r)}$ 
according to 
\begin{eqnarray}
        \d_{\l} {\Gamma}^c_{\r\s;\t\a\b} = 3\, \pa_{\t} \pa_{\a} \pa_{\b}\l^c_{\r\s}\,. \nonumber
\end{eqnarray}
The expression (\ref{a1dWF}) for $a_1$ implies that the deformed gauge transformations are 
\begin{eqnarray}
        \stackrel{(1)}{{\d_{\l}}} h^a_{\m\a\b} = 3\, \pa_{\m}\l^a_{\a\b} + g^a_{~(bc)} \,
        \ve_{\m}^{~\,\n\r\s\t}\,{\Gamma}^b_{\g\n;\r\a\b}\,\pa_{\s}^{\ }\l^{c\ \g}_{\;\t}\,,
\label{defogt}
\end{eqnarray}
where the right-hand side must be completely symmetrized over the indices $(\m\a\b)\,$.       

The cubic deformation of the free Lagrangian, $a_0$, is obtained from $a_1$
by solving the top equation $\d a_1+\g a_0+d b_0=0$. 

Again, we consider the most general cubic
expression involving four derivatives and apply $\g$ to it, then we
compute $\d a_1$. We take the Euler-Lagrange derivative with respect to $C_{\a\b}$ of the sum of the two expressions, and multiply  by $C_{\a\b}$ to get  a sum of terms of the form $\ve C\pa^4 h \pa h$ or $\ve C\pa^3 h\pa^2 h$. These are not related by Schouten identities and are therefore independent; all coefficients of the obtained equation thus have  to vanish. When solving this system of equations,
we find that $g_{abc}\equiv \d_{ad}g^d_{~bc}$ must be completely symmetric. 
In other words, the corresponding internal commutative algebra $\cb$ 
possesses an invariant norm. As for the algebra $\ca$ of the $n=3$ case, 
the positivity of energy requirement
imposes a positive-definite internal metric with respect to which the norm is defined.  

Finally, we obtain the following solution for $a_0$:
\begin{eqnarray}\nonumber
a_0=\!{\textstyle \frac{3}{2}}g_{(abc)}\ve^{\m\n\r\s\t}\! \left\{\!-\frac{1}{8}\pa_\m\Box
h^a_\n\pa_\r h^b_\s h^c_\t +\frac{1}{2}\pa^3_{\m\a\b} h^a_\n\pa_\r
h^{b\a\b}_\s h^c_\t+\frac{1}{4}\pa_\m\Box h^{a\a\b}_\n\pa_\r
h^b_{\s\a\b}h^c_\t\right.
\\\nonumber+\frac{3}{8}\pa_\m\Box
h^a_\n\pa_\r h^{b\a\b}_\s h^c_{\t\a\b}-\frac{1}{2}\pa_\m\Box
h^{a\a\b}_\n\pa_\r h^b_{\s\a\g} h^{c\ \g}_{\t\b}-\frac{1}{2}\pa^{3\a\b}_\m h^a_\n\pa_\r h^b_{\s\a\g} h^{c\ \g}_{\t\b}
 \\
\nonumber -\frac{1}{2}\pa^{3\a\b}_\m h^a_{\n\a\g}\pa_\r h^b_\s h^{c\
\g}_{\t\b}-\frac{1}{4}\pa^{3\a\b}_\m h^a_{\n\a\b}\pa_\r h^{b\g\d}_\s
h^c_{\t\g\d}-\frac{1}{2}\pa^{3\a\b}_\m h^a_{\n\g\d}\pa_\r h^b_{\s\a\b}
h^{c\g\d}_\t
 \\ \left. +\pa^{3\a\b}_\m h^a_{\n\b\g}\pa_\r
h^{b\g\d}_\s
h^c_{\t\a\d}+\frac{1}{2}\pa^2_{\m\a}h_\n^{a\a\b}\pa_\r^{2\g}h^b_\s
h^c_{\t\b\g}-\pa^2_{\m\a}h_{\n\b\g}^a\pa_{\r\d}^2h^{b\a\b}_\s
h^{c\g\d}_\t \right\}d^5 x\,.\nonumber \end{eqnarray}

\subsubsection{Second-order deformation}

The next step is the equation at order 2 : $(W_1,W_1)=-2sW_2$. In particular,  its antighost 2 component reads $(a_2,a_2)=\d c_3+\g c_2+d f_2\label{a2a21}\,.$ The left-hand side is directly computed from Eq.(\ref{a2n5}) : 
$$\begin{array}{rcl} \displaystyle (a_2,a_2) \!\!&=&\!\!-g^a_{bc}g_{dea}^{}\ve^{\bar{\m}\bar{\n}\bar{\r}\bar{\s}\bar{\t}}\ve_\m^{\ \,\n\r\s\t}\d^{(\m}_{\bar{\t}} \d^{\a)}_\d \Big[4\pa_{\bar{\m}}C_{\bar{\n}}^{*d\g}\pa_{\g\bar{\r}}C_{\bar{\s}}^{e\d}+2\pa_{\g\bar{\m}}C_{\bar{\n}}^{*d\g}\pa_{\bar{\r}}C_{\bar{\s}}^{e\d} \Big]\quad\\
&&\hspace{9cm}\times\ \pa_\n^{\ }C_\r^{b\b}\pa_{\a\s}C_{\t\b}^{c} \\&=& \!\!\displaystyle -12 g^a_{b[c}g_{d]ea}^{} C^{*b\a\b}\widehat{U}_{\ \a}^{c\ \g|\m\n}\widehat{U}_{\b\g}^{d\ \,|\r\s}\widehat{U}^e_{\m\n|\r\s}
+ \g c_2 + \pa_{\m}j^{\m}_2\,. \end{array}$$
The first term appearing in the right-hand side of the
above equation is a nontrivial element of $H(\g\vert d)$ . 
Its vanishing implies  that the structure constants $g_{(abc)}$ of the 
commutative invariant-normed algebra $\cb$ 
must obey the associativity relation $g_{~\;b[c}^a g_{d]ea}^{\ }=0$. As for the spin-2 deformation problem (see \cite{Boulanger:2000rq}, Sections 5.4 and 6), this means that, modulo 
redefinitions of the fields, there is no cross-interaction between different kinds of spin-3 gauge fields provided the internal metric in $\cb$ is positive-definite --- which is demanded 
by the positivity of energy. 
The cubic vertex $a_0$ can thus be written as a sum of independent self-interacting vertices, one for each field $h_{\m\n\r}^a\,$, $a=1,\ldots,N\,$. 
Without loss of generality, we may drop the internal index $a$ and consider only one \emph{single} self-interacting spin-3 gauge field $h_{\m\n\r}\,$.

\section{Results and discussion}
\label{conclusions}

In this chapter we carefully analysed the problem of introducing
consistent interactions among a countable collection of
spin-3 gauge fields in flat space-time of arbitrary dimension
$n\geq 3\,$. For this purpose we used the powerful BRST
cohomological deformation techniques, in order to be as exhaustive
as possible.
Let us underline that most of the cohomologies that we computed for the intermediate steps
are interesting for their one sake. For example, the cohomology of $\d$ modulo $d$ provides a complete list of the conserved forms.

The results proved in Sections \ref{interactions}
and \ref{sec:def} constitute strong yes-go and no-go theorems
that generalize previous works on spin-3 self-interactions.
We summarize them in this section, considering separately the parity-invariant and parity-breaking deformations.
We also provide the explicit first-order gauge transformations.

Let us first recall the results for parity-invariant deformations of the gauge algebra and transformations.
\begin{theorem}\label{galgebradefs}Let $h^a_{\m\n\r}$ be
a collection of spin-3 gauge fields ($a=1,\ldots,N$) described by
the local and quadratic action of Fronsdal.

At first order in some smooth deformation parameter, the
nontrivial consistent local deformations of the (Abelian) gauge
algebra that are invariant under parity and Poincar\'e
transformations, may always be assumed to be closed off-shell and are
in one-to-one correspondence with the structure constant tensors
$$C^a{}_{bc}=-C^a{}_{cb}$$ of an anticommutative internal algebra, that may be taken
as deformation parameters.

Moreover, the most general gauge transformations deforming the
gauge algebra at first order in $C=(f,g)$ are equal to
\begin{eqnarray}
        &\delta_{\l}h^a_{\m\n\r} &= 3
        \,\pa^{}_{(\m}\l^a_{\n\r)}+f^a{}_{bc}\,\Phi^{bc}_{\m\n\r}+\,g^a{}_{bc}\,(\Psi^{bc}_{\m\n\r}-\frac{1}{n}\,\eta^{}_{(\m\n}\Psi^{bc}_{\r)})+\co(C^2)\,,\quad\quad
\label{defgtransfo}\end{eqnarray}up to gauge transformations
that either are trivial or do not deform the gauge algebra at
first order, where $\Phi^{bc}_{\m\n\r}$ and $\Psi^{bc}_{\m\n\r}$
are bilinear local functions of the gauge field
$h^a_{\m\n\r}$ and the traceless gauge parameter
$\l^a_{\m\n}$. 
The expression for $\Phi$ is lengthy and has been given in Section \ref{azeroun},
while 
\be
\Psi^{bc}_{\m\n\r}=-\frac{1}{ 3}\,\eta^{\a\b}\partial^{}_{[\m}h^b_{\a]\n[\s,\t]}\partial^{}_{[\r}\l^{c\,\,\s,\t}_{\b]}+\mbox{perms}\,,\label {Psi}
\ee
where a coma denotes a partial derivative\footnote{For example
$\Phi^i_{,\,\a}\equiv\pa_{\a}\Phi^i$.} and ``perms'' stands for the sum of terms obtained via all nontrivial permutations of the indices $\m\,,\n\,,\r\,$ from the first term of the r.h.s.

\noindent The structure constant tensors $f^a{}_{bc}$ and
$g^a{}_{bc}$ are some arbitrary constant tensors that are
antisymmetric in the indices $bc$. In mass units, the coupling constant $f^a{}_{bc}$ has dimension $-n/2$ and
$g^a_{bc}$ has dimension $-2-n/2$.

Both of these deformations exist in any dimension $n \geq 5$. In the cases $n=3,4\,$, the structure constant tensor $g^a{}_{bc}$ vanishes. 
\end{theorem}

In the parity-breaking case, one finds the following deformations of the gauge algebra and transformations:
\begin{theorem}\label{galgebradefsbis}Let $h^a_{\m\n\r}$ be
a collection of spin-3 gauge fields ($a=1,\ldots,N$) described by
the local and quadratic action of Fronsdal.

At first order in some smooth deformation parameter, the
nontrivial consistent local deformations of the (Abelian) gauge
algebra that are invariant under  Poincar\'e
transformations but not under parity transformations, may always be assumed to be closed off-shell and  exist only in 3 or in 5 space-time dimensions. 
They are
in one-to-one correspondence with the structure constant tensors
$f^a{}_{bc}=-f^a{}_{cb}$ of an anticommutative internal algebra in three dimensions and with 
the structure constant tensors
$g^a{}_{bc}=g^a{}_{cb}$ of  commutative internal algebra in five dimensions.

Moreover, the most general gauge transformations deforming the
gauge algebra at first order are equal to
\begin{eqnarray}
        &\delta_{\l}h^a_{\m\n\r} &=
\d^3_n \, f^a{}_{bc}\,\Big(\Psi^{bc}_{\m\n\r}-\frac{1}{n}\,\eta^{}_{(\m\n}\Psi^{bc}_{\r)}
+\eta^{}_{(\m\n}\Phi^{bc}_{\r)}\Big)
+\d^5_n \, g^a{}_{bc}\, \Omega^{bc}_{\m\n\r}
\,,
\label{defgtransfobis}\end{eqnarray}
up to gauge transformations
that either are trivial or do not deform the gauge algebra at
first order, where $\Psi^{bc}_{\m\n\r}\,,$  $\Phi^{bc}_{\r}$ and 
$\Omega^{bc}_{\m\n\r}$ are given by
\begin{eqnarray}
\Psi^{bc}_{\m\n\r}&=&\ve^{\a\b\g}(\frac{1}{3}
h^b_{\m\n\a}\pa_{[\b}\l^c_{\g]\r}-\frac{1}{2} \l^b_{\m\a} \pa_{[\b}h^c_{\g]\n\r})\,+ \, perms\nnn
\Phi^{bc}_{\r}&=&\ve^{\a\b\g}[-\frac{1}{3}\pa_{[\r}\l^b_{\a]\b}h^c_\g +\eta_{\r\a} \l^{b\s}_\b (-\frac{1}{2}\pa^{\l}h^c_{\g\s\l}+\pa_{(\s}h^c_{\g)})]
\nnn
\Omega^{bc}_{\m\n\r}&=&\frac{1}{3}\,\ve_\m{}^{\a\b\g\l}
 \pa_{[\a}\l^b_{\b]\s}\Gamma^{c\s}{}_{[\g,\l]\n\r}
+ \, perms
\end{eqnarray}
and ``perms'' stands for the sum of terms obtained via all nontrivial permutations of the indices $\m\,,\n\,,\r\,$ of the r.h.s. 
\end{theorem}

\noindent
Let us make two remarks. Firstly, without imposing any restriction on the maximal number of derivatives (as was implicit in most former works)
we prove that the allowed possibilities are extremely restricted.

Secondly, the first parity-invariant deformation of the gauge symmetries (corresponding to the coefficients $f^a{}_{bc}$)  corresponds to the first-order interaction of Berends--Burgers--van Dam \cite{Berends:1984wp}, while the other deformations 
had not been explicitely found in previous analyses of spin-3 self-interactions (involving no other type of fields).
An intriguing question is whether these gauge algebra deformations
can be obtained from an appropriate flat space-time limit of the 
$(A)dS_n$ higher-spin algebras containing a finite-dimensional 
non-Abelian internal subalgebra (studied in details by Vasiliev and 
collaborators \cite{Vasiliev:2004cm}).
An indication that this might be the case is provided by the 
deformation of the gauge transformations (\ref{defgtransfo}) involving 
the tensor $\Psi^{ab}_{\m\n\r}$. 
The presence of the term $\partial^{}_{[\m}h^b_{\a]\n[\s,\t]}$ in (\ref{Psi})
is reminiscent of the second frame-like connection (see e.g. the second reference 
of \cite{Sagnotti:2005}). They both involve two derivatives of the spin-3 field and
have the $gl(n)$-symmetry corresponding to the Young diagram 
{\footnotesize{
\begin{picture}(35,14)(0,0)
\multiframe(1,4)(10.5,0){3}(10,10){$$}{$$}{$ $}
\multiframe(1,-6.5)(10.5,0){2}(10,10){$$}{$$}
\end{picture}}}. 
More comments in that direction are given in Sections \ref{cohogamma2} and \ref{defdeux}. 

\vspace{.2cm}

An important physical question is whether or not these first-order
gauge symmetry deformations possess some Lagrangian counterpart,
\textit{i.e.} if there exist vertices that are invariant under
(\ref{defgtransfo}) and \bref{defgtransfobis} at first order in the deformation parameters. The following theorem
provides a sufficient condition for that in the parity-invariant case:
\begin{theorem}\label{cvertices}
Let  the constant tensor $C_{abc}=(f_{abc},g_{abc})$ be
completely antisymmetric, where $C_{abc}:=\d_{ad}C^d{}_{bc}\,$. Then,

$\bullet$ The quadratic local action (\ref{freeaction})  admits a
first-order consistent deformation \be
\cs[h^a_{\m\n\r}]\,=\,\cs_0\,+\,f_{abc}\,S^{abc}\,+\,g_{abc}\,T^{abc}\,+\,\co(C^2)\,,\label{defcubact}\ee
which is gauge invariant under the deformed gauge transformations
(\ref{defgtransfo}) at first order in the deformation parameters.
Furthermore, this antisymmetry condition on the tensor $f^a{}_{bc}$ is necessary for
the existence of the corresponding deformation of the action.

$\bullet$ The vertices in the first-order
deformations are determined uniquely by the structure constants $f_{abc}$ and
$g_{abc}$, modulo vertices that do not
deform the gauge algebra. The corresponding local functionals
$S^{abc}[h^d_{\m\n\r}]$ and $T^{abc}[h^d_{\m\n\r}]$ are cubic in
the gauge field and respectively contain three and five
derivatives. Actually, there are no other nontrivial
consistent vertices containing at most three derivatives that
deform the gauge transformation at first order.

$\bullet$ At second order in $C$, the deformation of the
gauge algebra can be assumed to close off-shell without loss of generality,
but it is obstructed if and only if $f_{abc}\neq 0\,$.  
\end{theorem}

\noindent The first-order covariant cubic deformation
$S^{bc}{}_a[h^d_{\m\n\r}]$ is the Berends--Burgers--van Dam vertex
\cite{Berends:1984wp} (reviewed for completeness in Section
\ref{azeroun}) while the other cubic deformation
$T^{bc}{}_a[h^d_{\m\n\r}]$ is written in Section
\ref{azerodeux}. 
The antisymmetry condition $g_{abc}=g_{[abc]}$ on the structure
constant of the second deformation is only sufficient for the existence of a consistent vertex at first order.
It would be interesting to establish whether a constant tensor
$g^a_{~[bc]}$ with the ``hook'' symmetries $\d_{d[a} g^d{}_{bc]}=0$
might not also give rise to a consistent first-order vertex. If
this first-order non-Abelian deformation turned out to exist, then
there would be no other one, under the assumptions stated above.

It is possible to provide a more intrinsic characterization of the
conditions on the constant tensors. Let $\cal A$
be an {\it anticommutative} algebra of dimension $N$ with a basis
$\{T_a\}$ . Its multiplication law $\ast:{\cal A}^2\rightarrow \cal
A$ obeys $a\ast b=-b\ast a$ for any $a,b\in\cal A$, which is equivalent to
the fact that the structure constant tensor $C^a{}_{bc}$ defined
by $T_b\ast T_c=C^a{}_{bc}\,T_a$ is antisymmetric in the covariant
indices: $C^a{}_{bc}=-C^a{}_{cb}$. Moreover, let us assume that the
algebra $\cal A$ is a Euclidean space, {\it i.e.} it is endowed
with a scalar product $\langle\,\,,\,\rangle:{\cal A}^2\rightarrow
\mathbb R$ with respect to which the basis $\{T_a\}$ is
orthonormal, $\langle\,T_a\,,\,T_b\,\rangle=\d_{ab}$.
For an anticommutative algebra, the scalar
product is said to be {\it invariant} (under the left or right
multiplication) if and only if
$\langle\,a\ast b\,,\,c\,\rangle=\langle\,a\,,\,b\ast c\,\rangle$ for any
$a,b,c\in \cal A\,$, and the latter
property is equivalent to the complete antisymmetry of the
trilinear form
$$C:{\cal A}^3\rightarrow{\mathbb R}:(a,b,c)\mapsto
C(a,b,c)=\langle\,a\,,\,b\ast c\,\rangle$$ or, in
components, to the complete antisymmetry property of the covariant
tensor $C_{abc}:=\d_{ad}\,C^d{}_{bc}$.

The gauge algebra
inferred from the Berends--Burgers--van Dam vertex is inconsistent
at second order \cite{Bengtsson:1983bp,Berends:1984rq} and no
corresponding quartic interaction can be constructed
\cite{Bengtsson:1986bz}.
Originally, consistency of the Berends--Burgers--van Dam deformation at second order was shown to require that $f^d{}_{ec}f^e{}_{ab}=f^d{}_{ae}f^e{}_{bc}$ \cite{Berends:1984rq}, which means that the corresponding internal algebra is associative $(a\ast b)\ast c=a\ast (b\ast c)$.
In Section \ref{obstr}, we actually obtain a stronger condition from consistency: $f^d{}_{ec}f^e{}_{ab}=0$, {\it i.e.} the internal algebra is nilpotent of order three: $(a\ast b)\ast c=0$. In any case, to derive that the Berends--Burgers--van Dam vertex is inconsistent at order two, one may use the following well-known lemma
\begin{lemma}\label{lemalg}
If an anticommutative algebra endowed with an invariant
scalar product is associative, then the product of any two elements is zero (in other words, the algebra is nilpotent of order two).
\end{lemma}
\noindent{ \bfseries{Proof :}} 
 Under the hypotheses of Lemma \ref{lemalg}, one gets  $\langle\,a\ast b\,,\,b\ast a\,\rangle=\langle\,a\,,\,b\ast (b\ast a)\,\rangle=\langle\,a\,,\,(b\ast b)\ast a\,\rangle=0$ which implies $a\ast b=0$ for any $a,b\in \cal A$. \quad \qedsymbol

\vspace{.2cm}

An exciting result is that the second
deformation corresponding to $g_{abc}=g_{[abc]}$ passes the gauge
{\em{algebra}} consistency requirement where the vertex of
Berends, Burgers and van Dam fails. It would be very interesting to investigate
whether there exist second-order gauge {\em{transformations}} that are
consistent at this order and  whether the deformation of the Lagrangian could then be extended to higher
orders in the deformation parameter.
Unfortunately, the lengthy nature of the five-derivative cubic
vertex makes further analysis very tedious.

\vspace{.2cm}

Let us now turn to the existence of first-order Lagrangians for the deformations that do not preserve the parity invariance.

\begin{theorem}\label{cverticesbis}
 The quadratic local action (\ref{freeaction})  admits a
first-order consistent parity-breaking deformation \be
\cs[h^a_{\m\n\r}]\,=\,\cs_0\,+\,\d^3_n f_{[abc]}\,U^{abc}\,+\,\d^5_n\, g_{(abc)}\,V^{abc}\,+\,\co(f^2,g^2)\,,\label{defcubactbis}\ee
which is gauge invariant under the deformed gauge transformations
(\ref{defgtransfobis}) at first order in the deformation parameters.
Furthermore, the complete antisymmetry and symmetry conditions on the tensors $f_{[abc]}:= \d_{ad}f^d{}_{bc}$ and $g_{(abc)}:= \d_{ad}g^d{}_{bc}$ are necessary for
the existence of the corresponding deformation of the action. The explicit expressions of the latter can be found in Sections \ref{fodn3} and \ref{fodn5} respectively.

$\bullet$ The vertices in the first-order
deformations are determined uniquely by the structure constants $f_{[abc]}$ and
$g_{(abc)}$, modulo vertices that do not
deform the gauge algebra. The corresponding local functionals
$U^{abc}[h^d_{\m\n\r}]$ and $V^{abc}[h^d_{\m\n\r}]$ are cubic in
the gauge field and respectively contain two and four
derivatives. 

$\bullet$ At second order in $f$ and $g$, the deformation of the
gauge algebra can be assumed to close off-shell without loss of generality,
but it is obstructed if and only if $f_{abc}\neq 0\,$. Furthermore, the algebra associated with $g$ must be associative.  
\end{theorem}

By relaxing the parity invariance requirement, one thus obtains two more consistent non-Abelian first-order deformations that lead to a cubic vertex in the Lagrangian. 
The first one, defined in $n=3$, involves a multiplet of gauge fields  
$h^a_{\m\n\r}$ taking values in an internal, anticommutative, invariant-normed 
algebra $\ca\,$. 
The fields of the second one, living in a space-time of dimension $n=5$, take value in an 
internal, commutative, invariant-normed algebra $\cb\,$. 
Taking the metrics which define the inner product in $\ca$ and $\cb$ positive-definite 
(which is required for the positivity of energy), the $n=3$ candidate gives rise to inconsistencies when continued at perturbation order two, whereas the $n=5$ one passes the 
 test and can be assumed to involve only \emph{one} kind of 
self-interacting spin-3 gauge field $h_{\m\n\r}$, bearing no internal ``color'' index.

Remarkably, the cubic vertex of the $n=5$ deformation is rather simple. 
Furthermore, the Abelian gauge transformations are deformed by the addition of a term involving the second de Wit--Freedman connection in a straightforward way, cf. 
Eq.(\ref{defogt}). 
The relevance of this second generalized Christoffel symbol in relation to a hypothetical spin-3 covariant derivative was already stressed in \cite{Bengtsson:1983bp}.  

It is interesting to compare the results of the present spin-3 analysis with those found 
in the spin-2 case first studied in \cite{Boulanger:2000ni}. 
There, two parity-breaking first-order consistent non-Abelian deformations of 
Fierz-Pauli theory were obtained, also living in dimensions $n=3$ and $n=5$.
The massless spin-2 fields in the first case bear a color index, the internal algebra 
$\widetilde{\ca}$ 
being commutative and further endowed with an invariant scalar product. 
In the second, $n=5$ case, the fields take value in an anticommutative, invariant-normed 
internal algebra $\widetilde{\cb}$. 
It was further shown in \cite{Boulanger:2000ni} that the $n=3$ first-order consistent 
deformation could be continued to \emph{all} orders in powers of the coupling constant, the 
resulting full interacting theory being explicitly written down \footnote{Since the 
deformation is consistent, starting from $n=3$ Fierz-Pauli, the complete $n=3$ 
interacting theory of \cite{Boulanger:2000ni} describes no propagating physical 
degree of freedom. 
On the contrary, the topologically massive theory in \cite{Deser:1981wh,Deser:1982vy} 
describes a massive graviton with \emph{one} propagating degree of freedom (and not 
\emph{two}, as 
was erroneously typed in \cite{Boulanger:2000ni}.}. 
However, it was not determined in \cite{Boulanger:2000ni} whether the $n=5$ candidate could 
be continued to all orders in the coupling constant. Very interestingly, 
this problem was later solved in \cite{Anco:2003pf}, where a consistency condition was 
obtained at second order in the deformation parameter, \emph{viz} the algebra 
$\widetilde{\cb}$ must be nilpotent of order three. 
Demanding positivity of energy and using  
the results of \cite{Boulanger:2000ni}, the latter nilpotency condition implies      
that there is actually no $n=5$ deformation at all: the structure constant of the 
internal algebra $\cb$ must vanish \cite{Anco:2003pf}. 
Stated differently, the $n=5$ first-order deformation 
candidate of \cite{Boulanger:2000ni} was shown to be inconsistent \cite{Anco:2003pf} when continued at second order in powers of the coupling constant, in analogy with the spin-3 
first-order deformation written in \cite{Berends:1984wp}.   

In the present spin-3 case, the situation is somehow the opposite.
Namely, it is the $n=3$ deformation which shows inconsistencies when going to second order, 
whereas the $n=5$ deformation passes the first test. Also, in the $n=3$ case the fields take 
values in an anticommutative, invariant-normed internal algebra $\ca$ whereas the fields 
in the $n=5$ case take value in a commutative, invariant-normed algebra $\cb\,$. 
However, the associativity condition deduced from a second-order consistency condition is 
obtained for the latter case, which implies that the algebra $\cb$ is a direct sum of 
one-dimensional ideals. 
We summarize the previous discussion in Table \ref{table133}. 
\begin{table}[!ht]
\centering
\begin{tabular}{|c||c|c|}
\hline 
     & $s=2$                          & $s=3$ \\ \hline \hline
$n=3$& $\widetilde{\ca}$ commutative  & $\ca$ anticommutative,\\
     & and invariant-normed           &  invariant-normed and \\
     &                                & nilpotent of order $3$ \\
 \hline
$n=5$ & $\widetilde{\cb}$ anticommutative, & $\cb$ commutative,\\
      & invariant-normed and               & invariant-normed and\\ 
      &  nilpotent of order $3$            & associative \\
\hline
\end{tabular}
\caption{\it Internal algebras for the parity-breaking first-order 
deformations of spin-2 and spin-3 free gauge theories.\label{table133}}
\end{table}

It would be of course very interesting to investigate further the $n=5$ deformation 
exhibited here, since if the deformation can be consistently continued to all orders 
in powers of the coupling constant, this would give the first consistent interacting 
Lagrangian for a single higher-spin gauge field.

\vspace{.2cm}

It would also be of interest to enlarge the set of fields to spin 2, 3 and 4 and see if this allows to remove some previous obstructions at order two. A hint that this might be sufficient comes from the fact that the commutator of two spin-3 generators produces spin-2 and spin-4 generators for the bosonic higher-spin algebra of  \cite{Vasiliev:2004qz}.

\vspace{.2cm}

Let us finally comment on the Abelian interactions of spin-3 fields. 
To constrain these interactions, one should compute the cohomology of $\d$ modulo $d$ in antighost number one, $H^n_1(\d\vert d)\,,$ which corresponds to the conserved currents. This has never been done, so no complete list of the Abelian interactions can be given.
Nevertheless, let us mention three kinds of interactions that involve spin-3 fields, without modifying their gauge transformations. The most obvious one is any polynomial in the curvature. Other possible deformations of the Lagrangian are related to Chern-Simons-like terms, e.g. in $n=3\,$,
$$a_0= K_{\m_1\m_2\vert \n_1\n_2\vert \r_1\r_2} \pa^{[\m_1}h^{\m_2]\r_3 [\n_1,\n_2]}dx^{\r_1}dx^{\r_2}dx_{\r_3}\;.$$
Finally, if one introduces $p$-forms, one can build Chapline-Manton-like interactions that couple them to the spin-3 fields. This generalization is presented in the Appendix \ref{chapline}. It leaves the gauge transformations of the spin-3 field unchanged while deforming those of the $p$-form.


%% file: conclusion.tex
\chapter*{Conclusions}
\addcontentsline{toc}{chapter}{Conclusions}
\markboth{}{}

In this thesis, we have studied two aspects of higher-spin gauge field theories: dualities and interactions.

\vspace{.2cm}

The first aspect is related to the presence of dualities, \ie  ``hidden'' 
symmetries among gauge field theories. We considered the question of whether two higher-spin theories corresponding to different irreducible representations of the Poincar\'e group can have the same physical content. Duality relations were already known at the level of the equations of motion and Bianchi identities, here we proved that these dualities hold also at the level of the action. As a consequence, the dual theories are formally equivalent.
Our main result is that the free theory of a completely symmetric gauge fields is dual at the level of the action to the  free theory of mixed-symmetry ``hook'' fields of the same spin, in specific dimensions. For example, in five space-time dimensions the spin-two theory of Pauli and Fierz is dual to the theory of a mixed-symmetry spin-two field written by Curtright.  

In four space-time dimensions the duality exchanges the electric and magnetic degrees of freedom of the field. This property led us to introduce external magnetic sources for higher-spin fields, thereby generalizing to arbitrary spin the work of Dirac on the coupling of magnetic monopoles to the electromagnetic field. Similarly to the quantization condition on the product of the electric and magnetic charges for electromagnetism, there is a quantization condition on the product of conserved ``electric'' and ``magnetic'' charges for higher spins.

\vspace{.2cm}

The second aspect of higher-spin gauge field theories that has been analysed in this thesis is the problem of interactions.
Self-interactions of exotic spin-two gauge fields have been studied, as well as self-interactions of completely symmetric spin-three fields. This was done in the BRST field-antifield formalism developped by Batalin and Vilkovisky, using the technique of consistent deformations of the master equation proposed by Barnich and Henneaux. 

For the exotic spin-two fields, we obtained a strong no-go result against the deformation of the Abelian algebra of gauge transformations. No Einstein-like theory thus exists for spin-two fields other than the graviton. 

On the other hand, in the spin-three case, we found two deformations of the gauge algebra that are consistent at first-order in the deformation parameter and fulfill some second-order consistency conditions.  An open question is whether they are related to the nonlinear equations written by Vasiliev \cite{Vasiliev:1990en,
Vasiliev:2004qz,Sagnotti:2005} in  the limit where the cosmological constant vanishes.
It would also be most interesting to investigate further whether they can be consistently continued to higher orders. They would then constitute the first consistent interactions of higher-spin gauge fields that do not involve an infinite tower of higher-spin fields. 


%% file: young.tex
\chapter{Young Tableaux}
\label{young}

In this appendix\footnote{This appendix is based on the introduction to Young tableaux of the second reference of \cite{Sagnotti:2005}. }, we introduce the Young diagrams and Young tableaux. Their importance stems from 
the fact that they completely
characterize the irreducible representations of ${gl}(M)$ and
$o(M)$. \vspace{.2cm}

A \textit{Young diagram} $[n_1,\, n_2, \ldots n_p]$ is a diagram
which consists of a finite number $p>0$ of columns of identical
squares. The lengths of the columns are  finite and do not increase:
$n_1\geq n_2\geq \ldots\geq n_p\geq 0$. The
Young diagram  $[n_1,\, n_2, \ldots n_p]$  is represented as follows:

\begin{picture}(80,88)(-20,0)
\multiframe(0,10)(0,10.5){7}(10,10){}{}{}{}{}{}{}\put(0,0){$n_1$}
\multiframe(10.5,31)(0,10.5){5}(10,10){}{}{}{}{}\put(12.5,21){$n_2$}
\multiframe(21,41.5)(0,10.5){1}(20,41.5){\vdots}\put(25,31.5){$\cdots$}
\multiframe(41.5,62.5)(0,10.5){2}(10,10){}{}\put(44.5,52.5){$n_{p-1}$}
\multiframe(52,73)(0,10.5){1}(10,10){}\put(55,62.5){$n_{p}$}
\end{picture}

\noindent
A {\it Young tableau} is a filled Young diagram, \ie it is constituted by a Young diagram and a set of values assigned to each box of the Young diagram. 

\vspace{.2cm}

Let us consider covariant tensors of ${gl}(M)$:
$A_{a\,b\,c\ldots}$ where $a,b,c, \ldots=1,2,\ldots M$. Simple
examples of these are the symmetric tensor $A^S_{a\vert b}$ such that
$A^S_{a\vert b}-A^S_{b\vert a}=0$, or the antisymmetric tensor $A^A_{a
b}$ such that $A^A_{ab}+A^A_{ba}=0$.

A complete set of covariant tensors irreducible under ${gl}(M)$ is
given by the tensors $A_{a^1_1 \ldots a^1_{n_1}\vert \,\ldots \,\vert\, a^p_1
\ldots a^p_{n_p} }$ ($n_i \geq n_{i+1}$) that are antisymmetric in each
set of indices $\{a^i_1 \ldots a^i_{n_{i}}\}$ with fixed $i$ and
that vanish when one antisymmetrizes the indices of a set $\{a^i_1
\ldots a^i_{n_{i}}\}$ with any index $a^j_l$ with $j>i$.
If one requires that the tensor be also irreducible under $o(M)$, then it must be traceless.\footnote{For proofs of these statements, we recommand the reference \cite{Hamermesch}. 
} The
properties of these irreducible tensors can be conveniently encoded
into Young tableaux. The Young diagram $[n_1,\, n_2, \ldots n_p]$
is associated with the tensor $A_{a^1_1 \ldots a^1_{n_1},\,\ldots
\,,\, a^p_1 \ldots a^p_{n_p} }$. Each box of the Young diagram is
related to an index of the tensor, boxes of the same column
corresponding to antisymmetric indices. So, in a natural way, the components of the tensor correspond to Young tableaux. Finally, the property that antisymmetrization over a set of indices and an additional index vanishes is translated into the rule that the antisymmetrization of
all the indices of a column with an index from any column to the right vanishes.
For example,  the irreducible tensors $ A^S_{a\vert b}$  and $A^A_{a
b}$  are associated with the Young tableaux
\begin{picture}(65,12)(-5,0)
\multiframe(8.5,0)(8.5,0){1}(8,8){\tiny b}
\multiframe(0,0)(8.5,0){1}(8,8){\tiny a} \put(23,0){and}
\multiframe(50,4)(8.5,0){1}(8,8){\tiny a}
\multiframe(50,-4.5)(8.5,0){1}(8,8){\tiny b}
\end{picture}, respectively.

   \vspace{.2cm}

In the notation developed here, the irreducible tensors are
manifestly antisymmetric in groups of indices. This is a convention: one
could as well choose to have manifestly symmetric groups of
indices of non-increasing length, corresponding to rows of the Young tableau. The irreducibility condition is then that the symmetrization of all indices of a row and an index of a lower row must vanish.  The choice of convention 
depends very much on the context, \ie the tensors at hand. In this thesis, we always use the antisymmetric convention.

\vspace{.2cm}

To end this introduction to Young diagrams, we give some
``multiplication rules'' of one or two box(es) with an arbitrary Young tableau.

Let us start with the tensor product of a vector (characterized by one
box) with an irreducible tensor under $gl(M)$ characterized by a
given Young tableau. It decomposes as the direct sum of irreducible
tensors under $gl(M)$ corresponding to all possible Young tableaux
obtained by adding one box to the initial Young tableau,  e.g.
\begin{center}
\begin{picture}(200,20)(30,0)
\multiframe(0,10)(10.5,0){1}(10,10){}
\multiframe(10.5,10)(10.5,0){1}(10,10){}
\multiframe(0,-0.5)(10.5,0){1}(10,10){} \put(27,11){$\otimes$}
\multiframe(43,10)(10.5,0){1}(10,10){$*$} \put(65,10){$\simeq$}
\multiframe(80,10)(10.5,0){1}(10,10){}
\multiframe(90.5,10)(10.5,0){1}(10,10){}
\multiframe(101,10)(10.5,0){1}(10,10){$*$}
\multiframe(80,-0.5)(10.5,0){1}(10,10){}
\put(120,10){$\oplus$} \multiframe(140,10)(10.5,0){1}(10,10){}
\multiframe(150.5,10)(10.5,0){1}(10,10){}
\multiframe(140,-0.5)(10.5,0){1}(10,10){}
\multiframe(150.5,-0.5)(10.5,0){1}(10,10){$*$}
\put(170,10){$\oplus$} \multiframe(190,10)(10.5,0){1}(10,10){}
\multiframe(200.5,10)(10.5,0){1}(10,10){}
\multiframe(190,-.5)(10.5,0){1}(10,10){}
\multiframe(190,-11)(10.5,0){1}(10,10){$*$}
\put(230,10){.}
\end{picture}
\end{center}
The decomposition of the tensor product of an antisymmetric two-form (characterized by one column of two boxes) with the same kind of tensors is computed in a similar way: one sums all the possible Young tableaux obtained by adding two boxes to the initial Young tableau, provided one never adds both boxes on the same line. 
E.g.
\begin{center}
\begin{picture}(200,30)(30,0)
\multiframe(0,20)(10.5,0){1}(10,10){}
\multiframe(10.5,20)(10.5,0){1}(10,10){}
\multiframe(0,9.5)(10.5,0){1}(10,10){} 
\put(27,21){$\otimes$}
\multiframe(43,9.5)(10.5,0){1}(10,10){$*$}
\multiframe(43,20)(10.5,0){1}(10,10){$*$} 
\put(65,20){$\simeq$}
\multiframe(80,20)(10.5,0){1}(10,10){}
\multiframe(90.5,20)(10.5,0){1}(10,10){}
\multiframe(101,20)(10.5,0){1}(10,10){$*$}
\multiframe(80,9.5)(10.5,0){1}(10,10){}
\multiframe(90.5,9.5)(10.5,0){1}(10,10){$*$}
\put(120,20){$\oplus$} 
\multiframe(140,20)(10.5,0){1}(10,10){}
\multiframe(150.5,20)(10.5,0){1}(10,10){}
\multiframe(140,9.5)(10.5,0){1}(10,10){}
\multiframe(150.5,9.5)(10.5,0){1}(10,10){$*$}
\multiframe(140,-1)(10.5,0){1}(10,10){$*$}
\put(170,20){$\oplus$} 
\multiframe(190,20)(10.5,0){1}(10,10){}
\multiframe(200.5,20)(10.5,0){1}(10,10){}
\multiframe(190,9.5)(10.5,0){1}(10,10){}
\multiframe(211,20)(10.5,0){1}(10,10){$*$}
\multiframe(190,-1)(10.5,0){1}(10,10){$*$}
\put(230,20){$\oplus$} 
\multiframe(250,20)(10.5,0){1}(10,10){}
\multiframe(260.5,20)(10.5,0){1}(10,10){}
\multiframe(250,9.5)(10.5,0){1}(10,10){}
\multiframe(250,-11.5)(10.5,0){1}(10,10){$*$}
\multiframe(250,-1)(10.5,0){1}(10,10){$*$}
\put(280,10){.}
\end{picture}
\end{center}
For the tensor product of a symmetric tensor with two indices (characterized by a two-box row), the two boxes added must belong to different columns: 
\begin{center}
\begin{picture}(200,30)(30,0)
\multiframe(0,20)(10.5,0){1}(10,10){}
\multiframe(10.5,20)(10.5,0){1}(10,10){}
\multiframe(0,9.5)(10.5,0){1}(10,10){} 
\put(27,21){$\otimes$}
\multiframe(53.5,20)(10.5,0){1}(10,10){$*$}
\multiframe(43,20)(10.5,0){1}(10,10){$*$} 
\put(75,20){$\simeq$}
\multiframe(90,20)(10.5,0){1}(10,10){}
\multiframe(100.5,20)(10.5,0){1}(10,10){}
\multiframe(111,20)(10.5,0){1}(10,10){$*$}
\multiframe(90,9.5)(10.5,0){1}(10,10){}
\multiframe(100.5,9.5)(10.5,0){1}(10,10){$*$}
\put(130,20){$\oplus$} 
\multiframe(150,20)(10.5,0){1}(10,10){}
\multiframe(160.5,20)(10.5,0){1}(10,10){}
\multiframe(150,9.5)(10.5,0){1}(10,10){}
\multiframe(160.5,9.5)(10.5,0){1}(10,10){$*$}
\multiframe(150,-1)(10.5,0){1}(10,10){$*$}
\put(180,20){$\oplus$} 
\multiframe(200,20)(10.5,0){1}(10,10){}
\multiframe(210.5,20)(10.5,0){1}(10,10){}
\multiframe(200,9.5)(10.5,0){1}(10,10){}
\multiframe(221,20)(10.5,0){1}(10,10){$*$}
\multiframe(200,-1)(10.5,0){1}(10,10){$*$}
\put(240,20){$\oplus$} 
\multiframe(260,20)(10.5,0){1}(10,10){}
\multiframe(270.5,20)(10.5,0){1}(10,10){}
\multiframe(260,9.5)(10.5,0){1}(10,10){}
\multiframe(281,20)(10.5,0){1}(10,10){$*$}
\multiframe(291.5,20)(10.5,0){1}(10,10){$*$}
\put(305,10){.}
\end{picture}
\end{center}

For the (pseudo)orthogonal algebras $o(M-N,N)$, the tensor product
of a vector (characterized by one box) with a traceless  tensor
characterized by a given Young tableau decomposes as the direct sum
of traceless tensors under $o(M-N,N)$ corresponding to all possible
Young tableaux obtained by adding or removing one box from the
initial Young tableau (a box can be removed as a result of
contraction of indices),  e.g.
\begin{center}
\begin{picture}(200,20)(30,0)
\multiframe(0,10)(10.5,0){1}(10,10){}
\multiframe(10.5,10)(10.5,0){1}(10,10){}
\multiframe(0,-0.5)(10.5,0){1}(10,10){} \put(27,11){$\otimes$}
\multiframe(43,10)(10.5,0){1}(10,10){} \put(64,10){$\simeq$}
\multiframe(80,10)(10.5,0){1}(10,10){}
\multiframe(90.5,10)(10.5,0){1}(10,10){}
\multiframe(101,10)(10.5,0){1}(10,10){}
\multiframe(80,-0.5)(10.5,0){1}(10,10){}
\put(120,10){$\oplus$} \multiframe(140,10)(10.5,0){1}(10,10){}
\multiframe(150.5,10)(10.5,0){1}(10,10){}
\multiframe(140,-0.5)(10.5,0){1}(10,10){}
\multiframe(150.5,-0.5)(10.5,0){1}(10,10){}
\put(170,10){$\oplus$} \multiframe(190,10)(10.5,0){1}(10,10){}
\multiframe(200.5,10)(10.5,0){1}(10,10){}
\multiframe(190,-.5)(10.5,0){1}(10,10){}
\multiframe(190,-11)(10.5,0){1}(10,10){}
\put(220,10){$\oplus$} \multiframe(240,10)(10.5,0){1}(10,10){}
\multiframe(240,-.5)(10.5,0){1}(10,10){}
\put(260,10){$\oplus$} \multiframe(280,10)(10.5,0){1}(10,10){}
\multiframe(290.5,10)(10.5,0){1}(10,10){}
\put(310,10){.}
\end{picture}
\end{center}


%% file: chapline.tex
\chapter{Chapline-Manton for exotic spin-2 fields and spin-s fields }
\label{chapline}

In this appendix, we 
generalize the Chapline-Manton interactions among $p$-forms to interactions that couple $[p,q]$-fields to $p'$-forms, as well as higher-spin gauge fields and $p$-forms. These interactions deform the gauge transformation for the $p$-forms and leave the gauge transformation of the higher-spin fields unchanged. 

\section{Chapline-Manton interaction}

Let us first introduce the usual Chapline-Manton interaction \cite{Chapline:1982ww}, which couples different kinds of $p$-forms.

One considers a $p$-form $A_{ \r_1 \dots \r_{p}}$ and a $q$-form $B_{ \r_1 \dots \r_{q}}\,$, which read in form notation $A^p=A_{ \r_1 \dots \r_{p}}\ dx^{\r_1} \ldots dx^{\r_{p}}$ and $B^q=B_{ \r_1 \dots \r_{q}}\ dx^{\r_1} \ldots dx^{\r_{q}}$.
Their respective field strengths are $F^{p+1}=d A^p$ and $H^{q+1}=dB^q$.
The dual $ ^*F^{n-p-1}$ of $F^{p+1}$ is defined by 
${}^*F^{n-p-1}=\frac{1}{(n-p-1)!}F_{ \r_1 \dots \r_{p+1}}\ \ve^{ \r_1 \dots \r_{n}}\ dx_{\r_{p+2}} \ldots dx_{\r_{n}}\,$.

\vspace{.2cm}

The action for the free theory describing these forms is 
$$\cs = \int\, (\,F^{p+1} {}^*F^{n-p-1} + H^{q+1} {}^*H^{n-q-1}\,)\,.$$
It is invariant under the gauge transformations
$$\d_{\Lambda} A^p= d \Lambda^{p-1}\;,\; \d_{\Omega} B^q=d\Omega^{q-1}\;.$$

The Chapline-Manton coupling  exists when $p$ and $q$ satisfy $p+1=q+k(q+1)$ for some positive integer $k$. (One can of course invert the role of $p$ and $q$.) It consists in the following deformation of the field strength $F^{p+1}\,$:
$$F^{p+1}\rightarrow \tilde{F}^{p+1}\equiv dA^p+ g\, B^q H^{q+1} \ldots H^{q+1}\,,$$
where there are $k$ factors $H^{q+1}\,,$ and $g$ is an arbitrary constant.
The interacting action is
$$ \cs = \int\, (\,\tilde{F}^{p+1} {}^*\tilde{F}^{n-p-1} + H^{q+1} {}^*H^{n-q-1}\,)\,,$$
which is invariant under the deformed gauge transformations
\bqn
\d_{\Lambda,\Omega} A^p&=& d\Lambda^{p-1}-g\, \Omega^{q-1} H^{q+1} \ldots H^{q+1}\nnn
\d_{\Lambda,\Omega} B^q&=&d\Omega^{q-1}\;.\nn
\eqn
Indeed, it is easy to check that the deformed field strength $\tilde{F}^{p+1} $ is invariant under this transformation.

\section{$[p,q]$-fields and $p'$-forms}

The Chapline-Manton-like interaction can be generalized to couple a $[p,q]$-field 

\noindent $\phi_{\m_1 \dots \m_{p}\vert \n_1 \dots \n_{q}}$ and a $r$-form $A_{ \r_1 \dots \r_{r}}$. In this case, $q$ and $r$ must be related by $r+1=q+k(q+1)$ for some strictly\footnote{The case $k=0$ is absent because there is no covariantly constant tensor $f$ with $p+1$ antisymmetric indices to contract the free indices of $D^q$ in \bref{cm}.}
 positive integer $k$.

The interacting Lagrangian is again obtained from the sum of the free Lagrangians for $\phi$ and $A$ by replacing the curvature of the $r$-form by a deformed curvature.
This deformed curvature $\tilde{F}^{r+1} \equiv \tilde{F}_{ \r_1 \dots \r_{r+1}}\ dx^{\r_1} \ldots dx^{\r_{r+1}} $ is now defined by 
\bqn
F^{r+1} \rightarrow \tilde{F}^{r+1}=d A^{r}+ K_{\m_{[p+1]}^1}^{q+1} \ldots K_{\m_{[p+1]}^k}^{q+1} D^q_{\r_{[p+1]}}f^{\m^1_{[p+1]}\vert \ldots \vert \m^k_{[p+1]} \vert \r_{[p+1]}}\,, \label{cm}
\eqn
where 
\bqn
D^{q}_{\r_{[p+1]}}&=& \pa_{[\r_1} \phi_{\r_2 \dots \r_{p+1}] \vert \n_1 \dots \n_{q}}dx^{\n_1} \ldots dx^{\n_{q}} \, ,\nnn
K_{\m_{[p+1]}}^{q+1}&=&\pa_{[\m_1}\phi_{\m_2 \ldots \m_{p+1}]\vert [\n_1 \ldots \n_q , \n_{q+1}]}dx^{\n_1} \ldots dx^{\n_{q+1}} \,,\nn
\eqn
 $f$ is a constant tensor such that\footnote{ When $f^{\m^1_{[p+1]}\vert \ldots \vert \m^k_{[p+1]} \vert \r_{[p+1]}}= (-)^{q}f^{\m^1_{[p+1]}\vert \ldots \vert \m^{k-1}_{[p+1]} \vert \r_{[p+1]}\vert \m^k_{[p+1]} }$ , the deformation of the curvature is a total derivative and can be removed by a redefinition of $A$.}
  $$f^{\m^1_{[p+1]}\vert \ldots \vert \m^k_{[p+1]} \vert \r_{[p+1]}}= (-)^{q+1}f^{\m^1_{[p+1]}\vert \ldots \vert \m^{k-1}_{[p+1]} \vert \r_{[p+1]}\vert \m^k_{[p+1]} }\,$$
and where we have used the short notation $\m_{[p]}$ to denote a collection of $p$ antisymmetric indices $ [\m_1 \dots \m_{p}]\, $.

The deformed curvature and thus the new Lagrangian are invariant under the deformed gauge transformation $\g$ defined by:
\bqn
\g A^{r}&= &d \Lambda^{r-1}+( -)^{q} K_{\m_{[p+1]}^1}^{q+1} \ldots K_{\m_{[p+1]}^k}^{q+1} D^{q-1}_{\r_{[p+1]}}f^{\m^1_{[p+1]}\vert \ldots \vert \m^k_{[p+1]} \vert \r_{[p+1]}}\,,
\nonumber \\
\g \phi_{\m_1 \dots \m_{p} \vert \n_1 \dots \n_{q}}&= &\pa_{[\m_1}A^{(1,0)}_{\m_2 \dots \m_{p}] \vert \n_1 \dots \n_{q}}\nonumber \\
&&+\,   A^{(0,1)}_{\m_1 \dots \m_{p} \vert [\n_1 \dots \n_{q-1},\n_{q}]}
+ \frac{p!}{(p-q+1)! q!}A^{(0,1)}_{\n_1 \dots \n_{q}[\m_{q+1} \dots \m_{p}\vert \m_1 \dots \m_{q-1},\m_{q}]} \,,\nn
\eqn
where $D^{q-1}_{\r_1 \ldots \r_{p+1}}= \pa_{[\r_1} A^{(0,1)}_{\r_2 \dots \r_{p+1}] \vert \n_1 \dots \n_{q-1}}dx^{\n_1} \ldots dx^{\n_{q-1}} \,$. (See Chapter \ref{spin2} for more details bout the undeformed spin-2 gauge transformation parameters).

\section{Higher-spin gauge fields and $p$-forms }

In a similar way, one can construct Chapline-Manton-like interactions coupling completely symmetric higher-spin gauge fields to $p$-forms with even $p=2k>0$.

The deformed lagrangian is the sum of the Fronsdal Lagrangian for the completely symmetric gauge field  $\phi_{(\m_1 \ldots \m_s)}$ and the free Lagrangian for the  $p$-form $A_{[\r_1\ldots \r_p]}$, where the curvature of the $p$-form has been replaced by a deformed curvature $\tilde{F}\,$.

We define 
\bqn
D^1_{\m^1_1 \m^1_2 \vert \ldots \vert \m^{s-1}_1 \m^{s-1}_2 }&= &
 \pa_{[\m^{s-1}_2} \ldots \pa_{[\m^2_2 }\pa_{[\m^1_2 } \phi_{\m^1_1 ]\m^1_2 ]\ldots \m^{s-1}_1] \n} dx^{\n} \nonumber \\
K^2_{\m^1_1 \m^1_2 \vert \ldots \vert \m^{s-1}_1 \m^{s-1}_2 }&=&d D^1_{\m^1_1 \m^1_2 \vert \ldots \vert \m^{s-1}_1 \m^{s-1}_2 }
\eqn
 where the antisymmetrizations in the r.h.s. are over the pairs  $[\m^i_1 \m^i_2]$. Note that $K^2$ is just the usual spin-$s$ curvature where two indices are considered as form-indices.
The deformed curvature for the $p$-form is then defined as follows:
\bqn
\tilde{F}^{p+1}\equiv d A^p + K^2 \ldots K^2 D^1 f
\eqn
where there are $k$ factors $K^2$, and the constant tensor $f$ contracts the free indices of the curvatures $K^2$ and $D^1$. In order for the deformation to be nontrivial, $f$ should be  symmetric under the exchange of the indices of $D^1$ with those of any $K^2$. Indeed, if $f$ is antisymmetric under this exchange, then the deformation of $F^{p+1}$ is a total derivative and can be removed by a redefinition of the field $A^p$. Of course, the interactions exist for a given $k$ only if an appropriate tensor $f$ can be found.

The new Lagrangian is invariant under the deformed gauge transformations
\bqn
\g A^p&= &d \Lambda^{p-1} + K^2 \ldots K^2 \Omega  f \nonumber \\
\g \phi_{\m_1 \ldots \m_s}&= &\pa_{(\m_1} \o_{\m_2 \ldots \m_s)}\nn
\eqn
where 
$\Omega_{\m^1_1 \m^1_2 \vert \ldots \vert \m^{s-1}_1 \m^{s-1}_2 }\equiv \pa_{[\m^{s-1}_2} \ldots \pa_{[\m^2_2 }\pa_{[\m^1_2 }\o_{\m^1_1 ]\m^2_1] \ldots \m^{s-1}_1 ]}$ .


%% file: Zino.tex
\chapter{First-order formulation of the free exotic spin-2 theory}
\label{zino}
\markboth{First-order formulation of the free exotic spin-2 theory}{First-order formulation of the free exotic spin-2 theory}

We consider a theory describing the free propagation of a gauge field
$\phi_{\m_1 \dots \m_{p} \vert \n_1 \dots \n_q }$, the 
symmetry properties of which are characterized by two columns of arbitrary
lengths $p$ and $q$, with $p \geq q$. These gauge fields thus obey  the
conditions
\bqn
&\phi_{\m_1 \dots \m_{p} \vert \n_1 \dots \n_q }
=\phi_{[\m_1 \dots \m_{p}] \vert \n_1 \dots \n_q
}=\phi_{\m_1 \dots \m_{p} \vert [\n_1 \dots \n_q]}\,,&
\nonumber \\
&\phi_{[\m_1 \dots \m_{p} \vert \n_1]\n_2 \dots \n_q }=0\,,&
\nonumber
\eqn

The action \bref{action} describing their free motion given in Section \ref{spin2} is of second order in the derivatives of the fields. As is shown in Section \ref{vasidescr}, higher-spin gauge field theories can be formulated either in a second-order formalism, or in a first-order one. This is also the case for spin-2 field theories. We review their first-order formulation in this appendix. In the particular case of a symmetric spin-2 field, the first-order formulation is simply the linearization of the formulation of gravity by Mac-Dowell and Mansouri \cite{MacDowell:1977jt}.  
The simple cases of $[2,1]$-, $[2,2]$- and $[3,1]$-fields have been written in
\cite{Zinoviev:2003ix}.
The first-order formulation of mixed symmetry fields has also been considered in $AdS$ in \cite{Alkalaev:2003qv}.

\hspace{.2cm}

The first-order theory is formulated in terms of the generalized vielbein\\
$e_{\m_1 \dots \m_{p} \vert \n_1 \dots \n_q }$ and of the generalized spin connection $\o_{\m_1 \dots \m_{q}\vert \n_1 \ldots \n_{p+1}}$, which are both antisymmetric in each of their sets of indices,
\bqn
e_{\m_1 \dots \m_{p} \vert \n_1 \dots \n_q }=e_{[\m_1 \dots \m_{p} ]\vert \n_1 \dots \n_q }
=e_{\m_1 \dots \m_{p} \vert[ \n_1 \dots \n_q] }\,,\nnn
\o_{\m_1 \dots \m_{q}\vert \n_1 \ldots \n_{p+1}}=\o_{[\m_1 \dots \m_{q}]\vert \n_1 \ldots \n_{p+1}}=\o_{\m_1 \dots \m_{q}\vert [\n_1 \ldots \n_{p+1}]}\,.\nn
\eqn
They satisfy no further identity. 

Let us  define $T_{\m_1 \dots \m_{p+1} \vert \n_1 \dots \n_q}$ by  $T_{\m_1 \dots \m_{p+1} \vert \n_1 \dots \n_q}=\pa_{[\m_1}e_{\m_2 \dots \m_{p+1}] \vert \n_1 \dots \n_q }\,.$   
The first-order Lagrangian then reads 
\bqn
\cl= \d^{[\r_1 \ldots \r_{q}\m_1 \ldots \m_{p+1} ]}_{[\t_1 \ldots \t_{q}\n_1 \ldots \n_{p+1}] } \o_{\r_1 \dots \r_{q}\vert}^{ \hspace{.9cm}\n_1 \ldots \n_{p+1}}
\left(  T_{\m_1 \ldots \m_{p+1}\vert}^{\hspace{1.2cm} \t_1 \dots \t_{q}}
-\frac{1}{2}\  \o_{[\m_1 \dots \m_{q}\vert \m_{q+1} \ldots \m_{p+1}]}^{\hspace{2.6cm}\t_1 \ldots \t_{q}}\right)\,.\nn
\eqn

As the Lagrangian depends on the vielbein only through its antisymmetrized derivative $T$, it is obviously invariant under the gauge transformation
$$ \d_{\xi} e_{\m_1 \dots \m_{p} \vert \n_1 \dots \n_q } = \pa_{[\m_1}\xi_{\m_2 \dots \m_{p}] \vert \n_1 \dots \n_q }\,,\;\d_{\xi}\o_{\m_1 \dots \m_{q}\vert \n_1 \ldots \n_{p+1}}=0\,,$$
with $\xi_{\m_1 \dots \m_{p-1} \vert \n_1 \dots \n_q }$ antisymmetric in its two sets of indices, $$\xi_{\m_1 \dots \m_{p-1} \vert \n_1 \dots \n_q }=\xi_{[\m_1 \dots \m_{p-1}]\vert \n_1 \dots \n_q }=\xi_{\m_1 \dots \m_{p-1} \vert [\n_1 \dots \n_q ]}\,.$$

The following gauge invariance of the action is less obvious:
\bqn
 \d_{\chi} e_{\m_1 \dots \m_{p} \vert \n_1 \dots \n_q } &= &\chi_{[\m_1 \dots \m_{q-1}\vert \m_{q+1} \dots \m_{p}]  \n_1 \dots \n_q}\;,\label{gaugino}\\
\d_{\chi} \o_{\m_1 \dots \m_{q} \vert \n_1 \dots \n_{p+1} }&=&\pa_{[\m_1} \chi_{\m_2 \dots \m_{q}] \vert \n_1 \dots \n_{p+1} }
\,,\nn \eqn
where $\chi_{\m_1 \dots \m_{q-1} \vert \n_1 \dots \n_{p+1} }$ is also antisymmetric in both sets of indices,
$$\chi_{\m_1 \dots \m_{q-1} \vert \n_1 \dots \n_{p+1} }=\chi_{[\m_1 \dots \m_{q-1}] \vert \n_1 \dots \n_{p+1} }=\chi_{\m_1 \dots \m_{q-1} \vert [\n_1 \dots \n_{p+1}] }\,.$$
To prove that the action is invariant under this transformation, one must notice that
$$ \d^{[\r_1 \ldots \r_{q}\m_1 \ldots \m_{p+1} ]}_{[\t_1 \ldots \t_{q}\n_1 \ldots \n_{p+1}] } 
\o_{\r_1 \dots \r_{q}\vert}^{1 \hspace{.7cm}\n_1 \ldots \n_{p+1}}
\o_{[\m_1 \dots \m_{q}\vert \m_{q+1} \ldots \m_{p+1}]}^{2\hspace{2.4cm}\t_1 \ldots \t_{q}}
$$
is symmetric for the exchange of $\o^1$ and $\o^2$. This can be checked by expanding the product of $\d$'s. The proof of the gauge invariance then follows rapidly.

\vspace{.2cm}

Let us now make contact with the second-order formulation.
The last symmetry property  can be used to derive an elegant expression of the equations of motion for $\o$, which reads
$$ T_{\m_1 \ldots \m_{p+1}\vert}^{\hspace{1.2cm} \t_1 \dots \t_{q}}
=  \o_{[\m_1 \dots \m_{q}\vert \m_{q+1} \ldots \m_{p+1}]}^{\hspace{2.7cm}\t_1 \ldots \t_{q}}\,.$$
They imply that one can express $\o$ in terms of derivatives of the vielbein, \ie that $\o$ is an auxilliary field. Indeed, all ireducible components of $\o$ are constrained by this equation. 
Inserting the expression $\o(e)$ into the action, one gets a two-derivative  action depending only on the vielbein $e\,.$ Furthermore, the analysis of the gauge invariance of this action shows that it depends only on the irreducible component of the vielbein that has the symmetry represented by the Young diagram $[p,q]\,.$ Indeed the invariance under the gauge transformation \bref{gaugino} implies that all other components are pure gauge.
Defining $\phi$ to be the irreducible component of the vielbein with symmetry $[p,q]\,,$ the action becomes the second-order action \bref{action}, up to some irrelevant overall constant factor.


%% file: spin2apenprim.tex

\chapter{Technical appendix}

\section{Proof of
Theorem \ref{cohoinv} }
\label{append}

We now give the complete (and lengthy) proof of
Theorem \ref{cohoinv}.
The proof is by induction and follows closely the steps of the
proof of similar theorems in the case of $1$-forms
\cite{Barnich:1994db,Barnich:1994mt},
$p$-forms\cite{Henneaux:1996ws} or gravity
\cite{Boulanger:2000rq}.

There is a general procedure to prove that the theorem \ref{cohoinv} holds for $k>n$,
that can be found e.g. in \cite{Boulanger:2000rq} and will not be repeated here.
We assume that the theorem  has been proved for any $k^{'}>k$, and show that it is
still valid for $k\,$.
\vspace*{.2cm}

The proof of the induction step is rather lengthy and is
decomposed into several steps:
\begin{itemize}
\item
the Euler-Lagrange derivatives of $a_k$ with respect to the fields
$\phi$ and $C^*_j$ ($1\leq j \leq p+1$) are computed in terms of
the Euler-Lagrange derivatives of $b_{k+1}$ (Section
\ref{sec7.1});
\item
it is shown that the Euler-Lagrange derivatives of $b_{k+1}$ can
be replaced by invariant quantities in the expression for the
Euler-Lagrange derivative of $a_k$ with the lowest antifield
number, up to some additionnal terms (Section \ref{sec7.2});
\item
 the previous step is extended to all the Euler-Lagrange derivatives of $a_k$
(Section \ref{sec7.3});
\item
the Euler-Lagrange derivative of $a_k$ with respect to the field
$\phi$ is reexpressed in terms of invariant quantities (Section
\ref{sec7.4});
\item
an homotopy formula is used to reconstruct $a_k$ from its
Euler-Lagrange derivatives (Section \ref{sec7.5}).
\end{itemize}

\subsection{Euler-Lagrange derivatives of $a_k$}
\label{sec7.1}

We define
\bqn
Z_{k+1-j\; \m_{[q]} \vert\, \nu_{[p+1-j]}} &=& \frac{\d^L b_{k+1}}{\d C_j^{* \; \m_{[q]}
 \vert\, \nu_{[p+1-j]}}} \,, \quad 1 \leq j \leq p+1 \,,\nonumber \\
Y_{k+1}^{\m_{[p]} \vert\,  \nu_{[q]}} &= &\frac{\d^L b_{k+1}}{\d
\phi_{\m_{[p]} \vert\,  \nu_{[q]}}} \,.\nonumber
\eqn
Then, the Euler-Lagrange derivatives of $a_k$ are given by
\bqn \frac{\d^L a_k}{\d C_{p+1}^{*\; \m_{[q]} }}
 &=&(-)^{p+1} \d Z_{k-p \; \m_{[q]}} \,,
 \label{vroum} \\
\frac{\d^L a_k}{\d C_j^{*\; \m_{[q]} \vert\, \nu_{[p+1-j]}}}
&=&(-)^j \d Z_{k+1-j \; \m_{[q]} \vert\, \nu_{[p+1-j]}} -Z_{k-j
\; \m_{[q]}\vert\, [\nu_{[p-j]},\n_{p+1-j}]}\;, 
\,q<j \leq p\,,
\nonumber\\
\frac{\d^L a_k}{\d C_j^{*\; \m_{[q]} \vert\, \nu_{[p+1-j]}}}
&=&(-)^j \d Z_{k+1-j \; \m_{[q]} \vert\, \nu_{[p+1-j]}} -Z_{k-j
\; \m_{[q]}\vert\, [\nu_{[p-j]},\n_{p+1-j}]} \vert\,_{sym\, of\,
C^{*}_j}\;,\nnn
&&\hspace{9cm}1\leq j \leq q \,,
\nonumber\\
\frac{\d^L a_k}{\d \phi^{\m_{[p]} \vert\,  \nu_{[q]}}} &=&\d
Y_{k+1 \; \m_{[p]} \vert\,  \nu_{[q]}} + \b D_{\m_{[p]} \vert\,
\nu_{[q]} \vert\, \r_{[p]} \vert\, \s_{[q]} }Z_k^{\s_{[q]}
\vert\,\r_{[p]} }\,, \label{EL}
\eqn
where $\b\equiv(-)^{(q+1)(p+\frac{q}{2})} \frac{(p+1)!}{q!(p-q+1)!}\,$, and
$D^{\m_{[p]}\vert \hspace{1.2cm}\s_{[q]} }_{\hspace{.6cm}\nu_{[q]} \vert\, \r_{[p]} \vert}\equiv \frac{1}{(p+1)!q!}\;\delta^{[ \s_{[q]}   \a \m_{[p]}
]}_{[\n_{[q]}\b  \r_{[p]}]} \pa_{\a}\pa^{ \b}$ is the
second-order self-adjoint differential operator defined by
$$G_{\m_{[p]} \vert\, \nu_{[q]}}\equiv D_{\m_{[p]} \vert\, \nu_{[q]} \vert\, \r_{[p]} \vert\, \s_{[q]}} C^{\r_{[p]} \vert\, \s_{[q]} }\,.$$

As in the proof of Theorem \ref{BK}, the projection on the symmetry of the indices of $C^*_j$ is needed when $j\leq q$, since in that case the variables
$C^*_j$ do not possess all the irreducible components of $[q] \otimes [p+1-j]\,$,
but only those where the length of the first column is smaller or equal to $p\,$. When $j>q$, the projection is trivial.

\subsection{Replacing $Z$ by an invariant in the Euler-Lagrange derivative of $a_k$ with
the lowest antifield number}
\label{sec7.2}


We should first note that, when $k<p+1\,$, some of the
Euler-Lagrange derivatives of $a_k$ vanish identically: indeed, as
there is no negative antifield-number field, $a_k$ cannot depend
on $C^{*}_j$ if $j>k$. Some terms on the r.h.s. of
Eqs.(\ref{vroum})-(\ref{EL}) also vanish:  $Z_{k+1-j}$ vanishes when
$j>k+1\,$. This implies that the $p+1-k$ top  equations of
the system
(\ref{vroum})-(\ref{EL}) are trivially satisfied: the $p-k$ first
equations involve only vanishing terms, and the $(p-k+1)$th
involves in addition the $\d$ of an antifield-zero term, which
also vanishes trivially.
The first nontrivial equation is then \bqn \frac{\d^L a_k}{\d
C^{*}_{k \; \m_{[q]} \vert\, \nu_{[p+1-k]}}} &=&(-)^k \d(Z_{1\;
\m_{[q]} \vert\, \nu_{[p+1-k]}} )-Z_{0 \; \m_{[q]} \vert\,
[\nu_{[p-k]},\n_{p+1-k}]} \vert\,_{sym\, of\, C^{*}_k} \,.\quad\quad
\label{ping}\eqn

Let us now define $[T^q_{\r_{[p+1]}}]_{\n_{[q]}}\equiv (-)^{q} \pa_{[\r_1}\phi_{\r_2 \ldots \r_{p+1}] \vert \n_{[q]}}$. We will prove the following lemma for $k\geq  q\,$:
\begin{lemma} \label{nontriv}
In the first nontrivial equation of the system
(\ref{vroum})-(\ref{EL}) (\ie Eq.(\ref{vroum}) when
$k\geq  p+1$ and Eq.(\ref{ping}) when $p+1 > k \geq  q$), 
 respectively $Z_{k-p}$ or $Z_1$  
satisfies
\bqn 
Z_{l\;\m_{[q]} \vert\, \nu_{[p+l-k]}}
&=&Z^{\prime}_{l\; \m_{[q]} \vert\, \nu_{[p+l-k]}}\label{basis}
\\
&+&(-)^{k-l}\d \b_{l+1\; \m_{[q]} \vert\, \nu_{[p+l-k]}} + \b_{l\; \m_{[q]}  \vert\,
[ \nu_{[p+l-k-1]},\nu_{p+l-k} ]}\vert\,_{sym\, of\, C^{*}_{k-l+1}} \nonumber \\
&+&A_l \Big[P^{(t)}_{\m_{[q]}}(\tilde{\cal H})+\frac{1}{s}T^q_{\r_{[p+1]}} \frac{\pa^L R^{(s,r)}_{\m_{[q]}}(K^{q+1},\tilde{\cal H}) }{\pa K^{q+1}_{\r_{[p+1]}}}\Big]_{l,\,\nu_{[p+l-k]}}\vert\,_{sym\, of\, C^{*}_{k-l+1}}  
\;,\nn
\eqn 
where $Z^{\prime}_l$ is invariant, the
$\b_l$'s are at least linear in ${\cal N}$ and possess the same
symmetry of indices as $Z_{l-1}\,$, $A_l\equiv(-)^{lp+p+1+\frac{l(l+1)}{2}}\,$,
$P^{(t)}$ is a polynomial of degree $t$ in $\tilde{\cal H}$ and $R^{(s,r)}$ is a polynomial of degree $s$ in $K^{q+1}$ and $r$ in $\tilde{\cal H}$. The polynomials are present only when $p-k=t(n-p-1)$ or $p+1-k=s(q+1)+r(n-p-1)$ respectively. 
\vspace{.1cm}

Moreover, when $p+1 > k \geq  q$,  the first nontrivial equation can be written
\bqn \frac{\d^L a_k}{\d
C^{*}_{k \; \m_{[q]} \vert\, \nu_{[p+1-k]}}}\!\! &=&(-)^{k} \d
Z_{1\,\m_{[q]}\vert\, \nu_{[p+1-k]}}^{\prime}-Z_{0\,\m_{[q]}\vert\, [\nu_{[p-k]},\n_{p+1-k}]}^{\prime}\vert\,_{sym\, of\, C^{*}_{k}}  \nonumber \\
&+&\!\!\!\!\Big([Q^{(m)}_{\m_{[q]}}(K^{q+1}) ]_{\nu_{[p+1-k]}}
+(-)^k [R^{(s,r)}_{\m_{[q]}}(K^{q+1},\tilde{\cal H}) ]_{0,\,\nu_{[p+1-k]}}\Big)\vert\,_{sym\, of\, C^{*}_{k}} \,, 
\nonumber \eqn
where $Z_{0}^{\prime}$ is an invariant  and 
$Q^{(m)}_{\m_{[q]}}(K^{q+1}) $ is a polynomial of degree $m$ in $K^{q+1}$, present only when $p+1-k=m(q+1)$.
\end {lemma}

The lemma will be proved 
now respectively for the cases
$k \geq  p+1\,$, $ q < k <p+1$ and $k=q\,$.

\vspace{.2cm}

\noindent 
\underline{\bf Proof of Lemma \ref{nontriv} for $k \geq  p+1$:}

\label{inductbasis1}

As $k-p>0\,$, there is  no trivially satisfied equation  and we
start with the top equation of the system (\ref{vroum})--(\ref{EL}).

The  lemma \ref{nontriv} is a direct consequence of the well-known  Lemma \ref{preliminary}
(see e.g. \cite{Boulanger:2000rq} ):
\begin{lemma}\label{preliminary}Let $\a$ be an invariant local form that is
$\d$-exact, \ie $\a=\d\b\,$. Then $\b=\b^\prime+\d\s\,$, where
$\b^\prime$ is invariant and we can assume without loss of
generality that $\s$ is at least linear in the variables of $\cal N\,$.
\end{lemma}

\noindent \underline{\bf Proof of Lemma \ref{nontriv} for $q <k< p+1$:}

The first nontrivial equation is (as $k>q\,$):
\bqn
\frac{\d^L a_k}{\d C^{*}_{k \; \m_{[q]} \vert\,
\nu_{[p+1-k]}}} &=&(-)^k \d(Z_{1\; \m_{[q]} \vert\, \nu_{[p+1-k]}}
)-Z_{0 \;\m_{[q]} \vert\, [\nu_{[p-k]},\n_{p+1-k}]} \,.
\label{pingg}
\eqn
We will first prove that $Z_1$ has the required
form, then we will prove the the first nontrivial equation can indeed be reexpressed as stated in Lemma \ref{nontriv}.

\paragraph{{ First part:}}
Defining
$\a_{0\;\m_{[q]}\vert \nu_{[p+1-k]}}\equiv\frac{\d^L a_q}{\d
C^{*}_{q \; \m_{[q]} \vert\, \nu_{[p+1-q]}}}$, the above equation can be
written  as \be \a_0^{p+1-k}=(-)^k \d (Z_1^{p+1-k})+ (-)^{p+1-k}
dZ_0^{p-k} \,,\label{alpha} \ee where we consider the indices $\nu_{[p+1-k]}$ as form-indices and omit to write the indices $\m_{[q]} $. Acting with $d$ on this equation
yields $d\a_0^{p+1-k}=(-)^{k+1} \d (dZ_1^{p+1-k})$. Due to Lemma
\ref{preliminary}, this implies that \be \a_1^{p+2-k} =
dZ_1^{p+1-k}+ \d Z_2^{p+2-k}\,,\label{beta}\ee for some invariant
$\a_1^{p+2-k}$ and some $Z_2^{p+2-k}$. These steps can be
reproduced to build a descent of equations ending with
\bqn\a_{n-p-1+k}^{n} = dZ_{n-p-1+k}^{n-1}+ \d
Z_{n-p+k}^{n}\,,\nonumber \eqn
where $\a_{n-p-1+k}^{n} $ is invariant.
As  
$n-p-1+k>k$, the induction hypothesis can be used and 
implies
$$\a_{n-p-1+k}^{n} =
dZ_{n-p-1+k}^{\prime\,\,n-1}+ \d Z_{n-p+k}^{\prime\,\,n}+[R(K^{q+1},\tilde{\cal H})]^n_{n-p-1+k}\,,$$
where $Z_{n-p+k}^{\prime\,\,n}$ and $Z_{n-p-1+k}^{\prime\,\,n-1}$ are invariant, and
$R(K^{q+1},\tilde{\cal H})$ is a polynomial of order $s$ in $K^{q+1}$ and $r$ in $\tilde{\cal H}$ (with $r,s>0$), present when $p+1-k=s(q+1)+r(n-p-1)$.
This equation can be lifted and implies that
$$\a_1^{p+2-k} = dZ_1^{\prime\, p+1-k}+ \d Z_2^{\prime \, p+2-k}+[R(K^{q+1},\tilde{\cal H})]^{p+2-k}_{1}\,,$$
for some invariant quantities $Z_1^{\prime\, p+1-k}$ and $Z_2^{\prime \,
p+2-k}$. 
Substracting the last equation from Eq.(\ref{beta}) yields
 $$
d\Big(Z_1^{p+1-k}-Z_1^{\prime\, p+1-k} -\frac{1}{s}T^q \Big[\frac{\pa^L R(K^{q+1},\tilde{\cal H})}{\pa K^{q+1}}\Big]_1^{p+1-k-q}\Big)+\d(\ldots)= 0\,.$$
As $H_1^{p+1-k}(d \vert\, \d)\cong H_{n-(p-k)}^{n} (\d \vert\,d)$, by Theorem \ref{BK} the solution of this equation is
\bqn Z_1^{p+1-k}\!=\!Z_1^{\prime\, p+1-k}\!
+\frac{1}{s}T^q \Big[\frac{\pa^L R(K^{q+1},\tilde{\cal H})}{\pa K^{q+1}}\Big]_1^{p+1-k-q}+d \b_1^{p-k}
 + \d \b_2^{ p+1-k} \quad\nnn+[P^{(t)}(\tilde{\cal H})]_1^{p+1-k}  \,,\nonumber\eqn
where the last term is present only when $p-k=t(n-p-1)$. 

This proves the first part of the induction basis, regarding $Z_1$.

\paragraph{Second part:}
We insert
the above result for $Z_1$ into Eq.(\ref{alpha}). Knowing that $\d(
[P(\tilde{\cal H})]_1^{p+1-k} )+ d (
[P(\tilde{\cal H})]_0^{p-k} )=0\,$
and defining $$W_0^{p-k}=(-)^{k+1} \Big((-)^{p} Z_0^{p-k}+ \d
\b_1^{p-k} + [P^{(t)}(\tilde{\cal H})]_0^{p-k} 
+\frac{1}{s}T^q \Big[\frac{\pa^L R(K^{q+1},\tilde{\cal H})}{\pa K^{q+1}}\Big]_0^{p-k-q} 
\Big)\,,$$  we get
\bqn
\a_0^{p+1-k}= (-)^{k} \d(Z_1^{\prime\,p+1-k})+d(W_0^{p-k})
+(-)^{k}[R(K^{q+1},\tilde{\cal H})]_0^{p-k}
\,. \nonumber
\eqn
Thus $d(W_0^{p-k})$ is an invariant and the invariant Poincar\'e
Lemma \ref{invPoinclemma} then states that
$$d(W_0^{p-k})=d(Z_0^{\prime\;p-k})+Q(K^{q+1}) $$
for some invariant $Z_0^{\prime\;p-k}$ and some polynomial in $K^{q+1}$, $Q(K^{q+1})$.
This straightforwardly implies 
\bqn \a_0^{p+1-k}=(-)^{k} \d
(Z_1^{\prime\,p+1-k})+d(Z_0^{\prime\;p-k})+Q(K^{q+1}) +(-)^{k}[R(K^{q+1},\tilde{\cal H})]_0^{p-k}\,, \nonumber \eqn
which completes the proof of Lemma \ref{nontriv} for $q < k <  p+1$.
 \quad \qedsymbol 

\vspace{.2cm}
\noindent \underline{\bf Proof of Lemma \ref{nontriv} for $k=q$:}
\label{inductbasis3}

The first nontrivial equation is \be \frac{\d^L a_q}{\d C^{*}_{q
\; \m_{[q]} \vert\, \nu_{[p+1-q]}}} =(-)^q \d(Z_{1 \;  \m_{[q]}
\vert\, \nu_{[p+1-q]}} ) -(Z_{0 \; \m_{[q]} \vert\,
[\nu_{[p-q]},\n_{p+1-q}]}  - Z_{0 \; [\m_{[q]} \vert\,
\nu_{[p-q]},\n_{p+1-q}]} )\,.\label{indbasq} \ee
 This equation is different from the equations treated in the previous cases because the operator acting on $Z_0$ cannot
be seen as a total derivative, since it involves the projection on a specific Young diagram.
The philosophy of the
 resolution of the latter problem goes as follows \cite{Boulanger:2004rx}:
\begin{itemize}
\item[(1)] one first constrains the last term of Eq.(\ref{indbasq}) to get an equation similar to Eq.(\ref{ping}) treated previously,
\item[(2)]  one solves it in the same way as for $q<k<p+1\,$.
\end{itemize}

We need the useful lemma \ref{lem1}  \cite{Boulanger:2004rx}.
\begin{lemma}\label{lem1}
If $\a_0^1$ is an invariant polynomial of antifield number 0 and
form degree 1 that satisfies 
\be
\alpha_{0 }^1 = \delta Z_{1}^1 +
d W_{0}^0 \,,\label{proj1biss} 
\ee
 then, for some invariant
polynomials ${Z'}_{1}^1$ and ${W'}_{0}^0\,$, 
\bqn
 Z_1^1 ={Z'}_1^1+\d \phi_2^1 + d \chi_{1}^0 \quad ,\label{une}\\
W_{0}^0= {W'}_{0}^0 + \d  \chi_1^0\,. \label{deux}
\eqn
\end{lemma}
\noindent {\bf{Proof:}}\hspace{.5cm}
Using standard techniques, one gets the following descent
\bqn
\a_1^2&= &\d Z_2^2 + d Z_1^1 \label{prems}\\
&\vdots &\nonumber \\
\a_{n-1}^n &= &\d Z_n^n + d Z_{n-1}^{n-1}\,, 
\nonumber  
\eqn
where all the $\a_{i-1}^i$ are invariant. 
As $n-1 \geq q+1$, by the induction hypothesis (\ie Theorem \ref{cohoinv} has been proved for $k>q$) we can choose $Z_n^n$ and $Z_{n-1}^{n-1}$ invariant. 
The invariance property propagates up until
$\a_1^2= \d {Z'}_2^2 + d {Z'}_1^1 $, where ${Z'}_2^2$ and ${Z'}_1^1$ have been chosen 
invariant.  
Substracting the latter equation from Eq.(\ref{prems}) and knowing that 
$H_1^1(\d\vert d) \cong H_n^n(\d\vert d)$ vanishes, we get Eq.(\ref{une}). 
Substituting  Eq.(\ref{une}) in Eq.(\ref{proj1biss}) and acting with $\g$, we find 
$d ( \g (W^0_0-\d \chi_{1}^0)) = 0\,$. Using the algebraic Poincar\'e lemma and the fact that 
there is no constant with positive pureghost number, this implies $\g (W^0_0-\d \chi_{1}^0)=0\,$, 
which in turn gives Eq.(\ref{deux}), as there exists no $\g$-exact term of pureghost number $0\,$.\hspace{1cm}\qedsymbol
\vspace*{.5cm}

As explained above, we now constrain the last term of
Eq.(\ref{indbasq}). The latter equation implies
\bqn \pa_{[\r}\a_{0\;\m_{[q]} \vert\,
\n_{[p-q]}]\nu_{p+1-q}}=(-)^q \d(\pa_{[\r}Z_{1 \; \m_{[q]} \vert\,
\n_{[p-q]}] \nu_{p+1-q}} ) - b \,\pa_{[\r}Z_{0 \; \m_{[q]}
\vert\, \nu_{[p-q]}],\n_{p+1-q}}\,, \nonumber \eqn where
$b\equiv\frac{q}{(p+1)(p+1-q)} $. Defining
\bqn
\tilde{\a}^1_{0\,[\r\m_{[q ]} \n_{[p-q]}]}&=&\pa_{[\r}\a_{0\;\m_{[q ]} \vert\, \n_{[p-q]}]
\nu_{p+1-q}} dx^{\n_{p+1-q}}\,, \nonumber \\
\tilde{Z}^1_{1\, [\r\m_{[q ]} \n_{[p-q]}]}&=&(-)^q \pa_{[\r}Z_{1 \; \m_{[q ]} \vert\,
\n_{[p-q]}] \nu_{p+1-q}} dx^{\n_{p+1-q}}\,, \nonumber \\
\tilde{W}^0_{0\,[\r\m_{[q ]} \n_{[p-q]}]}&=&- a\,\pa_{[\r}Z_{0 \;
\m_{[q ]} \vert\,  \nu_{[p-q]}]}\,, \nonumber
\eqn
and omitting to write the indices $[\r\m_{[q]}\n_{[p-q]}]$, the above equation reads $
\tilde{\a}^1_0=\d\tilde{Z}^1_1+ d\tilde{W}^0_0 \,$. Lemma
\ref{lem1} then implies that $\tilde{W}^0_0=I^{\prime\,0}_{0} + \d
m^0_1$ for some invariant $I^{\prime\,0}_{0}$. By the definition of
$\tilde{W}^0_0$, this statement is equivalent to \bqn
\pa_{[\r}Z_{0 \; \m_{[q]} \vert\,
\nu_{[p-q]}]}=I^{\prime}_{0\,[\m_{[q]} \nu_{[p-q]}\r]}+ \d
m_{1\,[\m_{[q]} \n_{[p-q]}\r]} \,.\nonumber \eqn Inserting this result into
Eq.(\ref{indbasq}) yields \bqn \a_{0\;\m_{[q]}\vert\,
\nu_{[p+1-q]}}-I^{\prime}_{0\,[\m_{[q]}  \nu_{[p+1-q]}]}=\d( (-)^q Z_{1
\; \m_{[q]} \vert\, \nu_{[p+1-q]}} +m_{1\,[\m_{[q]}
\nu_{[p+1-q]}]})\nnn
 -Z_{0 \; \m_{[q]}  \vert\,
[\nu_{[p-q]},\n_{p+1-q}]}\,.\nonumber \eqn 
This equation has the same
form as Eq.(\ref{pingg}) and can be solved in the same way to get the
following result: \bqn Z_{1 \; \m_{[q]}\vert\,
\n_{[p+1-q]}}&=&(-)^{q+1} m_{1 \; [\m_{[q]}\n_{[p+1-q]}]}+
Z^{\prime}_{1 \; \m_{[q]}\vert\, \n_{[p+1-q]}}
\nnn
&&+\b_{1\;\m_{[q]}\vert\, [ \nu_{[p-q]},\n_{p+1-q}]}
+\d \b_{2\;\m_{[q]}\vert\, \n_{[p+1-q]}}
\nonumber\\
&&+\frac{1}{s}\Big[T^q_{\r_{[p+1]} }\frac{\pa^L R_{\m_{[q]}}(K^{q+1},\tilde{\cal H})}{\pa K^{q+1}_{\r_{[p+1]} }}\Big]_{1,\, \n_{[p+1-q]}}+[P(\tilde{\cal H})]_{1,\,\n_{[p+1-k]}}\,,
\nonumber\\
\a_{0\;\m_{[q]}\vert\, \n_{[p+1-q]}}&=&I^{\prime}_{0\,[\m_{[q]}\vert\,
\n_{[p+1-q]}]}+(-)^q \d (Z^{\prime}_{1 \; \m_{[q]}\vert\,
\n_{[p+1-q]}})+Z_{0 \; \m_{[q]}\vert\,
[\nu_{[p-q]},\n_{p+1-q}]}^{\prime}
\nonumber \\
&&+ [Q_{\m_{[q]}}(K^{q+1})]_{\n_{[p+1-q]}}+(-)^{k}[R(K^{q+1},\tilde{\cal H})]_{0,\,\n_{[p+1-q]}}\,.\nonumber \eqn
Removing the completely antisymmetric parts of these equations
yields the desired result.\vspace*{.2cm}

This ends the proof of Lemma \ref{nontriv} for $k\geq  q\,$.\quad \qedsymbol

\subsection{Replacing all $ Z$ and $Y$ by invariants}
\label{sec7.3}

We will now prove the following lemma:
\begin{lemma} \label{yzinv}
The Euler-Lagrange derivatives of $a_k$ can be written
\bqn
\frac{\d^L a_k}{\d C_{p+1}^{* \;  \m_{[q]} }} &=&(-)^{p+1} \d(Z^{\prime}_{k-p \; \m_{[q]}  } )\,,
\nonumber\\
\frac{\d^L a_k}{\d C_j^{*\; \m_{[q]}  \vert\, \nu_{[p+1-j]}}} &=&(-)^j \d(Z^{\prime}_{k+1-j \;
\m_{[q]}  \vert\, \nu_{[p+1-j]}} )-Z^{\prime}_{k-j \; \m_{[q]}  \vert\, [ \nu_{[p-j]},\n_{p+1-j}]}\;, \nnn
&&\hspace{9cm} q<j \leq p \,,\nonumber\\
\frac{\d^L a_k}{\d C_j^{* \; \m_{[q]}  \vert\,  \nu_{[p+1-j]}}} &=&(-)^j \d(Z^{\prime}_{k+1-j \;
\m_{[q]} \vert\,  \nu_{[p+1-j]}} )-Z^{\prime}_{k-j \; \m_{[q]}  \vert\, [ \nu_{[p-j]},\n_{p+1-j}]}
\vert\,_{sym\, of\, C^{*}_j}\;,\nnn
&&\hspace{9cm}1\leq j \leq q \,,\nonumber\\
\frac{\d^L a_k}{\d \phi^{\m_{[q]}\vert\, \nu_{[q]}}} &=&\d
(Y^{\prime}_{k+1 \; \m_{[q]}\vert\, \nu_{[q]}} )+ \b D_{\m_{[q]}\vert\,
\nu_{[q]} \vert\,
 \r_{[p]} \vert\, \s_{[q]} }{Z'}_k^{\s_{[q]}
\vert\,\r_{[p]}  }\,, \nonumber\eqn where $Z_{l}^{\prime}$
($k-p\leq l \leq k$)  and $Y_{k+1}^{\prime}$ are invariant polynomials,
except in the following cases. When $k=p+1-m(q+1)$ for some strictly positive integer $m\,$, there is an additionnal term in the first nontrivial
equation:
\bqn
\frac{\d^L a_{k}}{\d C_k^{*\; \m_{[q]}\vert\,  \nu_{[p+1-k]}}}=(-)^k \d Z^{\prime}_{1\;
\m_{[q]}\vert\,  \nu_{[p+1-k]}} -Z^{\prime}_{0 \; \m_{[q]}\vert\, [\nu_{[p-k]},\n_{p+1-k}]}\nnn
+[Q_{\m_{[q]}}(K^{q+1})]_{\nu_{[p+1-k]}}\vert\,_{sym\, of\, C^{*}_k}
\,,\nonumber \eqn
where $Q$ is a polynomial of degree $m$ in $K^{q+1}$.
Furthermore, when $k=p+1-r(n-p-1)-s(q+1)$ for a couple of integer $r,s>0$, then there is an additional term in each Euler-Lagrange derivative:
\bqn
\frac{\d^L a_k}{\d C_j^{* \; \m_{[q]}  \vert\,  \nu_{[p+1-j]}}} &=&(-)^j \d(Z^{\prime}_{k+1-j \;
\m_{[q]} \vert\,  \nu_{[p+1-j]}} )-Z^{\prime}_{k-j \; \m_{[q]}  \vert\, [ \nu_{[p-j]},\n_{p+1-j}]}
\vert\,_{sym\, of\, C^{*}_j}\nonumber\\
&&+ (-)^{k+p+1}A_{k-j}[R_{\m_{[q]}}(K^{q+1},\tilde{\cal H})]_{k-j\;\nu_{[p+1-j]}}\vert\,_{sym\, of\, C^{*}_j}
\nonumber\\
\frac{\d^L a_k}{\d \phi^{\m_{[q]}\vert\, \nu_{[q]}}} &=&\d
(Y^{\prime}_{k+1 \; \m_{[q]}\vert\, \nu_{[q]}} )+ \b D_{\m_{[q]}\vert\,
\nu_{[q]} \vert\,
 \r_{[p]} \vert\, \s_{[q]} }{Z'}_k^{\s_{[q]}
\vert\,\r_{[p]}  }\nonumber\\
&&+A\,\d^{[\s_{[q]} \a \m_{[p]}\xi]}_{[\n_{[q]}\b \r_{[p+1]}]}
\pa_\a \pa^\b (x_{\xi} \,[R_{\s_{[q]}}(K^{q+1},\tilde{\cal H})]_k^{\r_{[p+1]}})\,,
\nonumber
\eqn
where $A=\b \frac{p+q+2}{(n-p-q-1)(p+1)!q!}A_k(-)^{p+k+1}$.
\end{lemma}

\noindent {\bf{Proof:}}\hspace{.5cm}
By Lemma \ref{nontriv}, we know that  the $Z$'s
involved in the first nontrivial equation satisfy Eq.(\ref{basis})
and that this equation has the required form. We
will proceed by induction and prove that when $Z_{k-j}$  (where $k-j\geq  1$) satisfies
Eq.(\ref{basis}), then the equation for $\frac{\d^L a_k}{\d C^{*}_j}$ also has the desired form and $Z_{k-j+1}$ also satisfies Eq.(\ref{basis}). 

Let us assume that $Z_{k-j }$ satisfies  Eq.(\ref{basis}) and
consider the following equation:
\bqn
\frac{\d^L a_k}{\d C^{*}_{j\; \m_{[q]}\vert\, \n_{[p+1-j]} }}=(-)^j \d(Z_{k+1-j}^{
\m_{[q]}\vert\, \n_{[p+1-j]}} )-Z_{k-j }^{ \m_{[q]} \vert\,
[\nu_{[p-j]},\n_{p+1-j}]}\vert\,_{sym\, of\,
C^{*}_j}\,. \label{xunk}
\eqn
Inserting  Eq.(\ref{basis}) for $Z_{k-j}$ into this equation yields
\bqn \!\!\!\!\!\!\!\!\!\!\!\!
\frac{\d^L a_k}{\d C^{*}_{j\; \m_{[q]}\vert\, \n_{[p+1-j]}}}=(-)^j \d \Big( Z_{k+1-j }^{ \m_{[q]}\vert\, \n_{[p+1-j]} }-\b_{k-j+1}^{\m_{[q]}\vert\, [ \nu_{[p-j]}, \nu_{p-j+1}]}\vert\,_{sym\, of\,C^{*}_j} \Big)\hspace{1.5cm}
\label{eqtot}
\\ 
\hspace{1.2cm}
+(-)^{k+p}a_{k-j}\d \Big[P^{\m_{[q]}}(\tilde{\cal H})+\frac{1}{s}T^q_{\r_{[p+1]}}\frac{\pa^L R^{\m_{[q]}}(K^{q+1},\tilde{\cal H})}{\pa K^{q+1}_{\r_{[p+1]}}}\Big]_{k-j+1}^{\n_{[p+1-j]} }\vert\,_{sym\, of\, C^{*}_j}
\nonumber \\
\hspace{1.2cm}-\Big(Z_{k-j }^{\prime\; \m_{[q]}\vert\, [\nu_{[p-j]},\n_{p+1-j}]}+(-)^{ p+k}A_{k-j}[R^{\m_{[q]}}(K^{q+1},\tilde{\cal H})]_{k-j}^{\n_{[p+1-j]}}\Big)\vert\,_{sym\, of\, C^{*}_j} 
\nonumber
\,.
\eqn
Note that one can omit to
project on the symmetries of $C^*_{j+1}$ when inserting
Eq.(\ref{basis}) into  Eq.(\ref{xunk}). Indeed the Young components that
are removed by this projection would be removed later anyway by
the projection on the symmetries of $C^*_j\,$.

Defining the invariant
\bqn Z_{k+1-j}^{\prime\; \m_{[q]}\vert\, \n_{[p+1-j]} }\equiv Z_{k+1-j }^{
\m_{[q]}\vert\, \n_{[p+1-j]} }\vert\,_{{\cal N}=0}\hspace{8.cm} \nonumber \\
+(-)^{k+p+j}A_{k-j} \Big[P^{\m_{[q]}}(\tilde{\cal H})+{\textstyle \frac{1}{s}}T^q_{\r_{[p+1]}}\frac{\pa^L R^{\m_{[q]}}(K^{q+1},\tilde{\cal H})}{\pa K^{q+1}_{\r_{[p+1]}}}\Big]_{k-j+1}^{\n_{[p+1-j]} }\vert\,_{sym\, of\, C^{*}_j}\vert\,_{{\cal N}=0}
\nonumber\eqn
and setting ${\cal N}=0$ in the last equation  yields, as $\b_{k-j+1}$ is at least linear in ${\cal
N}$, \bqn \frac{\d^L a_k}{\d C^{*}_{j \; \m_{[q]}\vert\,
\n_{[p+1-j]}}} =(-)^j \d(Z_{k+1-j }^{\prime\; \m_{[q]}\vert\,
\n_{[p+1-j]}} )-Z_{k-j}^{\prime\; \m_{[q]}\vert\, [
\nu_{[p-j]},\n_{p+1-j}]} \vert\,_{sym\, of\, C^{*}_j}
\nonumber \\
+(-)^{p+k+1}A_{k-j}[R^{\m_{[q]}}(K^{q+1},\tilde{\cal H})]_{k-j}^{\n_{[p+1-j]}}\vert\,_{sym\, of\, C^{*}_j} 
\,.
\label{eqinv}
\eqn
This proves the part of the induction regarding the equations for the Euler-Lagrange derivatives. We now prove that $Z_{k-j+1}$ verifies Eq.(\ref{basis}).

Substracting Eq.(\ref{eqinv}) from
Eq.(\ref{eqtot}), we get
\bqn
0 \,&= &(-)^j \d  \Big(Z_{k+1-j }^{\m_{[q]}\vert\, \n_{[p+1-j]}}
-Z_{k+1-j }^{\prime\; \m_{[q]}\vert\, \n_{[p+1-j]}}-\b_{k+1-j}^{ \m_{[q]} \vert\, [ \nu_{[p-j]},
\nu_{p+1-j}]}\vert\,_{sym\, of\, C^{*}_j}\nonumber \\
& &+(-)^{j+k+p}A_{k-j}\Big[P^{\m_{[q]}}(\tilde{\cal H})+\frac{1}{s}T^q_{\r_{[p+1]}}\frac{\pa^L R^{\m_{[q]}}(K^{q+1},\tilde{\cal H})}{\pa K^{q+1}_{\r_{[p+1]}}}\Big]_{k+1-j}^{\n_{[p+1-j]} }\vert\,_{sym\, of\, C^{*}_j}\Big)\,. \nonumber
\eqn

As $k+1-j>0$, this implies \bqn Z_{k+1-j }^{ \m_{[q]}\vert\,
\n_{[p+1-j]}}&=&Z_{k+1-j }^{\prime\; \m_{[q]}\vert\, \n_{[p+1-j]}}+
(-)^{j-1} \d \b_{k-j}^{ \m_{[q]}\vert\, \n_{[p+1-j]}}+\b_{k-j+1}^{
\m_{[q]}\vert\, [\nu_{[p-j]}, \nu_{p+1-j}]}\vert\,_{sym\, of\,
C^{*}_j}
\nonumber \\
&& +A_{k+1-j}\Big[P^{\m_{[q]}}(\tilde{\cal H})+\frac{1}{s}T^q_{\r_{[p+1]}}\frac{\pa^L R^{\m_{[q]}}(K^{q+1},\tilde{\cal H})}{\pa K^{q+1}_{\r_{[p+1]}}}\Big]_{k+1-j}^{\n_{[p+1-j]} }\vert\,_{sym\, of\, C^{*}_j} \,,\nonumber
\eqn
which is the expression (\ref{basis}) for $Z_{k+1-j }$.

Assuming that $Z_{k-j}$ satisfies  Eq.(\ref{basis}) , we have thus proved
that the equation for $\frac{\d^L a_k}{\d C_j^{*}} $ has the
desired form and that $Z_{k+1-j}$ also satisfies Eq.(\ref{basis}).
Iterating this step, one shows that all $Z$'s satisfy
Eq.(\ref{basis}) and that the equations involving only $Z$'s have the
desired form.

It remains to be proved that the Euler-Lagrange derivative with respect to the field takes the right form. Inserting the expression (\ref{basis}) for $Z_k$ into Eq.(\ref{EL}) and some algebra yield
\bqn
\frac{\d^L a_k}{\d \phi^{\m_{[q]}\vert\, \nu_{[q]}}} &=&\d
(\tilde{Y}_{k+1 \; \m_{[q]}\vert\, \nu_{[q]}} \vert\,_{sym\, of\,
\phi})+ \b D_{\m_{[q]}\vert\,
\nu_{[q]} \vert\,
 \r_{[p]} \vert\, \s_{[q]} }{Z'}_k^{\s_{[q]}
\vert\,\r_{[p]}  }\nonumber\\
&&+A\,\d^{[\s_{[q]} \a \m_{[p]}\xi]}_{[\n_{[q]}\b \r_{[p+1]}]}
\pa_\a \pa^\b (x_{\xi} \,[R_{\s_{[q]}}(K^{q+1},\tilde{\cal H})]_k^{\r_{[p+1]}})\vert\,_{sym\, of\,
\phi}\,,
\nonumber
\eqn
where \bqn\tilde{Y}_{k+1 \; \m_{[q]}\vert\, \nu_{[q]}} &\equiv & Y_{k+1 \; \m_{[q]}\vert\, \nu_{[q]}}
+ \b D_{\m_{[q]}\vert\,\nu_{[q]} \vert\,\r_{[p]} \vert\, \s_{[q]} }\b_{k+1}^{\s_{[q]}\vert\, \r_{[p]}}
\nonumber \\
&&+\,c \,\d^{[\s_{[q]}\a \m_{[p]} ]}_{[\n_{[q]}\b \r_{[p]}]} \pa_{\a
} \Big[P_{\s_{[q]}}(\tilde{\cal H})+ \frac{1}{s} T^q_{\l_{[p+1]}}\frac{\pa^L R^{\s_{[q]}}(K^{q+1},\tilde{\cal H})}{\pa K^{q+1}_{\l_{[p+1]}}}\Big]_{k+1}^{[\r_{[p]}\b]}\nonumber \\
&&+(-)^{k+q+1}A\,\d^{[\s_{[q]} \a \m_{[p]}\xi]}_{[\n_{[q]}\b \r_{[p+1]}]}
\pa_\a (x_{\xi} \, [R_{\s_{[q]}}(K^{q+1},\tilde{\cal H})]_{k+1}^{[\r_{[p+1]} \b]})
\nonumber \eqn
and  $c\equiv\b \frac{1}{(p+1)!q!}A_k (-)^{p+k+1}$. Defining $Y_{k+1\; \m_{[p]} \vert\, \n_{[q]}}^{\prime}\equiv \tilde{Y}_{k+1 \; \m_{[q]}\vert\, \nu_{[q]}} \vert\,_{sym\, of\,
\phi}\vert_{{\cal N}=0}$ and setting ${\cal N}=0$ in the above equation completes the proof of Lemma \ref{yzinv}. \quad\qedsymbol

\subsection{Euler-Lagrange derivative with respect to the field}
\label{rewriting}
\label{sec7.4}

In this section, we manipulate the Euler-Lagrange derivative of
$a_k$ with respect to the field $\phi\,$.

We have proved in the previous section that it can be written in
the form \bqn \frac{\d^L a_k}{\d \phi^{\m_{[p]} \vert\,
\n_{[q]}}} &=&\d (Y^{'}_{k+1 \; \m_{[p]} \vert\, \n_{[q]}} )+ \b
D_{\m_{[p]} \vert\, \n_{[q]} \vert\,  \r_{[p]} \vert\,  \s_{[q]}
}Z_k^{'\,\s_{[q]}\vert\,\r_{[p]} }\nonumber\\
&&+A\,\d^{[\s_{[q]} \a \m_{[p]}\xi]}_{[\n_{[q]}\b \r_{[p+1]}]}
\pa_\a \pa^\b (x_{\xi} \,[R_{\s_{[q]}}(K^{q+1},\tilde{\cal H})]_k^{\r_{[p+1]}})\vert\,_{sym\, of\,
\phi}\,.\nonumber \eqn 
As $a_k$
is invariant, it can depend on $\phi_{\m_{[p]} \vert\, \n_{[q]}}$
only through $K_{\m_{[p]} \a\vert\, \n_{[q]} \b}$, which implies
that $ \frac{\d^L a_k}{\d \phi^{\m_{[p]} \vert\, \n_{[q]}}}
=\pa^{\a \b}X_{[\m_{[p]} \a]\vert\, [\n_{[q]} \b]}\,,$ where $X$ has the symmetry of the curvature. This
in turn implies that $\d(Y^{'}_{k+1 \; \m_{[p]} \vert\, \n_{[q]}}
)=\pa^{\a \b}W_{\m_{[p]}  \a\vert\, \n_{[q]} \b}$ for some $W$
with the Young symmetry $  [p+1,q+1]\,$.
Let us consider the indices $ \m_{[p]} $ as form indices. As
$H_{k+1}^{n-p} (\d \vert\, d) \cong H_{p+1+k}^{n}(\d \vert\,
d)\cong 0 $ for $k>0$, the last equation implies \bqn Y^{'}_{k+1
\; \m_{[p]} \vert\, \n_{[q]}} = \d A_{k+2 \; \m_{[p]} \vert\,
\n_{[q]}}+ \pa^{\l}T_{k+1 \; [\l \m_{[p]} ] \vert\, \n_{[q]}}\,.
\label{yprime} \eqn 
By the induction hypothesis  for $p+1+k$ , we
can take $A_{k+2}$ and $T_{k+1}$ invariant. Antisymmetrizing
Eq.(\ref{yprime}) over the indices $\m_q \ldots \m_p \nu_1 \ldots
\nu_{q}$ yields \bqn 0=\d A_{k+2 \; \m_1 \ldots \m_{q-1}[\m_q
\ldots \m_p \vert\, \nu_1 \ldots \nu_{q}]}+ \pa^{\l}T_{k+1 \; \l
\m_1 \ldots \m_{q-1}[\m_q \ldots \m_p \vert\, \nu_1 \ldots
\nu_{q}]}\,. \nonumber \eqn The solution of this equation for
$T_{k+1}$ 
is 
\bqn T_{k+1\, \m_0 \ldots
\m_{q-1}[\m_q \ldots \m_p \vert\, \nu_1 \ldots \nu_{q}]}=\!\Big[U_{[\m_q \ldots \m_p \nu_1 \ldots \nu_{q}]}^{(u)}(\tilde{\cal H})\Big]_{k+1}^{\r_{[n-q]}} \ve_{\m_0 \ldots \m_{q-1}\r_{[n-q]}}\hspace{2cm}
\nonumber \\
\hspace{2cm}+
 \d
Q_{k+2 \, \m_0 \ldots \m_{q-1}\vert\, [\m_q \ldots \m_p \nu_1
\ldots \nu_{q}]}+ \pa^{\a}S_{k+1 \, \a\m_0 \ldots
\m_{q-1}\vert\, [\m_q \ldots \m_p \nu_1 \ldots
\nu_{q}]}
\,,\nonumber\eqn
where $U^{(u)}$ is a polynomial of degree $u$ in $\tilde{\cal H}$, present when $k+q+1=n-u(n-p-1)$ for some strictly positive integer $u$.
As $T$ and $U^{(u)}(\tilde{\cal H})$ are invariant, we can use the induction hypothesis for $k^{\prime}=k+1+q$. This implies
\bqn
&T_{k+1\, \m_0 \ldots
\m_{q-1}[\m_q \ldots \m_p \vert\, \nu_1 \ldots \nu_{q}]}=&
 \d
Q^{\prime}_{k+2 \,  \m_0 \ldots \m_{q-1}\vert\, [\m_q \ldots \m_p \nu_1
\ldots \nu_{q}]}\label{S}\\
&&+ \pa^{\a}S^{\prime}_{k+1 \, \a\m_0 \ldots
\m_{q-1}\vert\, [\m_q \ldots \m_p \nu_1 \ldots
\nu_{q}]}\nn \\
&\hspace{2.2cm}+\Big[U_{[\m_q \ldots \m_p \nu_1 \ldots \nu_{q}]}^{(u)}&\!\!\!(\tilde{\cal H})+V_{[\m_q \ldots \m_p \nu_1 \ldots \nu_{q}]}^{(v,w)}(K^{q+1},\tilde{\cal H})\Big]_{k+1}^{\r_{[n-q]}} \ve_{\m_0 \ldots \m_{q-1}\r_{[n-q]}}\,,
\nonumber 
\eqn
where $Q_{k+2 }^{\prime}$ and $S_{k+1 }^{ \prime}$ are invariants and 
$V^{(v,w)}$ is a polynomial of order $v$ and $w$ in $K^{q+1}$ and $\tilde{\cal H}$ respectively, present when $n-q=v(q+1)+w(n-p-1)+k+1$ for some strictly positive integers $v,w$. 
\vspace*{.2cm}

We define the invariant tensor $E_{\a \m_{[p]} \vert\, \b  \nu_{[q]} }$
with Young symmetry $  [p+1,q+1]$ by \bqn E_{\a \m_{[p]} \vert\,
\b \nu_{[q]} }= \sum_{i=0}^{q+1} \a_i S^{\prime}_{k+1\;\r_0 \ldots \r_{i-1}
[ \n_i \ldots \n_q \vert\, \b \n_1 \ldots \n_{i-1}] \r_i \ldots
\r_p} \d^{[\r_0 \ldots \r_p]}_{[\a \m_{[p]}] } \nonumber\eqn
where $\a_i =  \a_0 \frac{(q+1)!}{(q+1-i)! \, i! }$ and $\a_0 = (-)^{pq}
\frac{((p+1)!)^2}{ (p-q)!\, (q!)^2 \, (p-q+1) \,(p+2)\, \sum_{j=0}^{q}\frac{(p-j)!}{(q-j)!}}\,$.
\vspace*{.2cm}

Writing $\pa^{\a \b}E_{k+1\; \a \m_{[p]} \vert\, \b \nu_{[q]} }$
in terms of $S^{ \prime}_{k+1}$ and using Eqs.(\ref{S}) and (\ref{yprime})
yields \bqn Y^{'}_{k+1 \;  \m_{[p]} \vert\, \nu_{[q]}}&=&\pa^{\a
\b}E_{k+1\; \a \m_{[p]} \vert\, \b \nu_{[q]} }+\d F_{k+2 \;
\m_{[p]} \vert\, \nu_{[q]}} \nonumber \\
&+&\pa^{\a} \sum_{i=0}^{q} \b_i \Big[V_{[\a \n_{[i]}\m_{i+1} \ldots \m_p]}^{(v,w)}(K^{q+1},\tilde{\cal H})\Big]_{k+1}^{\r_{[n-q]}} \ve_{\m_{[i]}\n_{i+1} \ldots \n_q\r_{[n-q]}}\,,\qquad\label{yprime2}
\eqn where $F_{k+2}$ is invariant, $\b_i \equiv \a_0 \frac{(p+2) q!}{(p+1)\, i! \,
(q-i)!}$ and $v$ is allowed to take the value $v=0$ to cover also the case of the polynomial $U^{(w)}(\tilde{\cal H})$.

\subsection{Homotopy formula}
\label{sec7.5}

We will now use the homotopy formula to reconstruct $a_k$ from its
Euler-Lagrange derivatives: \bqn a^n_k
&=&\int_0^1dt \Big[ \phi_{
\m_{[p]} \vert\,  \nu_{[q]}}\frac{\d^L a_k}{\d \phi_{\m_{[p]}
\vert\, \nu_{[q]}}} + \sum_{j=1}^{p+1} C^{*}_{j \;  \m_{[q]}
\vert\, \nu_{[p+1-j]}}\frac{\d^L a_k}{\d C^{*}_{j \; \m_{[q]}
\vert\, \nu_{[p+1-j]}}} \Big]\, d^nx\,.
\nonumber \eqn
Inserting the expressions for the Euler-Lagrange derivatives given by Lemma \ref{yzinv} yields
\bqn
a^n_k&=&\int_0^1dt \Big[\d(\phi_{\m_{[p]}
\vert\,  \nu_{[q]}}\,Y_{k+1} ^{\prime\; \m_{[p]} \vert\,  \nu_{[q]}})
+\sum_{j=1}^{p+1} \d (C^{*}_{j \;  \m_{[q]} \vert\, \nu_{[p+1-j]}}Z_{k+1-j}^{\prime\;\m_{[q]}
\vert\,  \nu_{[p+1-j]}}) 
\nonumber \\
&&\hspace{1cm}+\sum_{j=1}^{k} C^{*}_{j \;  \m_{[q]} \vert\, \nu_{[p+1-j]}}(-)^{k+p+1}A_{k-j}[R^{\m_{[q]} }(K^{q+1},\tilde{\cal H})]_{k-j}^{\nu_{[p+1-j]}}
\nonumber \\
&&\hspace{1cm}+\phi_{\m_{[p]}
\vert\,  \nu_{[q]}}A\,\d^{[\s_{[q]} \a \m_{[p]}\xi]}_{[\n_{[q]}\b \r_{[p+1]}]}
\pa_\a \pa^\b (x_{\xi} \,[R_{\s_{[q]}}(K^{q+1},\tilde{\cal H})]_k^{\r_{[p+1]}})
\nonumber \\
&&\hspace{1cm}+\, C^{*}_{k\;  \m_{[q]} \vert\, \nu_{[p+1-k]}} [Q^{(m)\,\m_{[q]}}(K^{q+1})]^{\nu_{[p+1-k]}}\Big]  d^nx 
+ d n^{n-1}_{k}\,.
\nonumber\eqn
Using the result  (\ref{yprime2}) for $Y_{k+1} ^{\prime}$ and some algebra, one finds
\bqn
\!a^n_k \!&\!\!=\!\!&\!\!\!\int_0^1 \!\!dt \Big[\d (K_{\m_{[p+1]}\vert \n_{[q+1]}}E_{k+1}^{\m_{[p+1]}\vert \n_{[q+1]}}d^nx)+ a_v \, K_{\m_{[p+1]}}^{q+1}[V^{(v,w)\,\m_{[p+1]}}(K^{q+1},\tilde{\cal H})]_k^{n-q-1}  
\nonumber \\
&&\hspace{.8cm}+\sum_{j=1}^{p+1} \d (C^{*}_{j \;  \m_{[q]} \vert\, \nu_{[p+1-j]}}Z_{k+1-j}^{\prime\;\m_{[q]}
\vert\,  \nu_{[p+1-j]}}d^nx) 
+a_r [\tilde{\cal H}^{\s_{[q]}}\,R_{\s_{[q]}}(K^{q+1},\tilde{\cal H})]_{k}^{n}
\nonumber \\
&&\hspace{.8cm}+ a_q \, [\tilde{\cal H}^{\s_{[q]}}]_k^{n-m(q+1)} \, Q^{(m)}_{\s_{[q]}}(K^{q+1})\Big]+ d \bar{n}^{n-1}_{k}
\,,\nonumber
\eqn
where $a_v=(-)^{k(q+1)} \sum_{i=0}^{q} \b_i \frac{i!(p-i)!}{p!}\,$, 
$a_r= (-)^{n(p+k+1)+\frac{p(p+1)+k(k+1)}{2}}$
 and 

\noindent $a_q=(-)^k a_r\,$.
In short,
\bqn
a^n_k =\, [P(K^{q+1},\tilde{\cal H})]_k^n  +\d \m^n_{k+1} +d \bar{n}^{n-1}_{k}\nonumber
\eqn
for some invariant $\m^n_{k+1}$, and some polynomial $P$ of strictly positive order in $K^{q+1}$ and $\tilde{\cal H}$.
\vspace{.2cm}

We still have to prove that $\bar{n}^{n-1}_{k}$ can be taken
invariant. 

\noindent Acting with $\g$ on the last equation yields $d (\g
\bar{n}^{n-1}_{k}) =0$. By the Poincar\'e lemma, $\g
\bar{n}^{n-1}_{k}= d (r_k^{n-2})$. Furthermore,  a well-known
result on $H(\g\vert\, d)$ for positive antifield number $k$ (see e.g. Appendix A.1 of \cite{Boulanger:2000rq}) 
states that one can redefine $\bar{n}^{n-1}_{k}$ in such a way that $\g \bar{n}^{n-1}_{k}=0$. As the pureghost number of
$\bar{n}^{n-1}_{k}$ vanishes, the last equation implies that $\bar{n}^{n-1}_{k}$ is an
invariant polynomial.
\vspace{.2cm}

This completes the proof of Theorem \ref{cohoinv} for $k\geq  q$.\quad\qedsymbol


%% file: Schouten.tex
\section{Schouten identities}
\label{schouten}
%
The Schouten identities are identities due to the fact that in $n$ dimensions the antisymmetrization over any $n+1$ indices vanishes. These identities obviously depend on the dimension and relate functions of the fields.

The solving of equations in the sections \ref{interactions} and \ref{sec:def} requires the knowledge of bases for several kinds of functions. When Schouten identities come into play, these bases are not obvious. This appendix is thus devoted to finding these bases, which depend on the structure of the functions at hand and the number of dimensions. 
 
Note that we write the internal indices only when it is necessary.
%
\subsection{Functions of the structure $\ve\, C^*\widehat{T}\widehat{T}$ in $n=4$}
\label{ectt}
%
In order to achieve the four-dimensional study of the algebra deformation in $D$-degree 2, a list of the Schouten identities is needed for the functions of the structure $\ve C^*\widehat{T}\widehat{T}$. 
The space of these functions is spanned by 
\bqn
T_1^{bc}=\ve^{\m\n\r\s}\
C^{*\a}_{\m}\ \widehat{T}^{b~\ \b}_{\n\r\vert}\ \widehat{T}_{\s\a\vert \b}^c\;,\
T_2^{bc}=\ve^{\m\n\r\s}\
C^{*\a}_{\m}\ \widehat{T}^{b~\ \b}_{\n\r\vert}\ \widehat{T}_{\s\b\vert \a}^c\;,\nnn
T_3^{[bc]}=\ve^{\m\n\r\s}\ C^{*\a\b}\
\widehat{T}^{b}_{\m\n\vert \a}\ \widehat{T}_{\r\s\vert \b}^c\,.\nn\eqn
There are two Schouten identities. Indeed, one should first notice that all Schouten identities are linear combinations of identities with the structure
$$\d^{[\a\b\g\d\e]}_{[\m\n\r\s\t]}\ve^{\m\n\r\s} C^*\widehat{T}\widehat{T}=0\,,$$ where the indices $\a\b\g\d\e\t$ are contracted with the indices of the ghosts and where $\d^{[\a\b\g\d\e]}_{[\m\n\r\s\t]}=\d^{[\a}_{[\m}\d^\b_\n\d^\g_\r\d^\d_\s\d^{\e]}_{\t]}\,.$ Furthermore, there are only two independent identities of this type:
$$\d^{[\a\b\g\d\e]}_{[\m\n\r\s\t]}\ve^{\m\n\r\s} C^{*\t}_\a\widehat{T}^b_{\b\g\vert \l}\widehat{T}^{c~\l}_{\d\e\vert}=0\;,\
\d^{[\a\b\g\d\e]}_{[\m\n\r\s\t]}\ve^{\m\n\r\s} C^{*\l}_\a\widehat{T}^{b~\,\t}_{\b\g\vert }\widehat{T}^{c}_{\d\e\vert\l}=0\,.
$$
Expanding the product of $\d$'s, one finds that the first identity implies that $T_1^{bc}$ is symmetric: $T_1^{bc}=T_1^{(bc)}\,,$
while the second one relates $T_2^{bc}$ and $T_3^{[bc]}\,$: $T_2^{bc}=T_3^{[bc]}\,.$

So, in four dimensions, a basis of the functions with the structure $\ve C^*\widehat{T}\widehat{T}$ is given by $T_1^{(bc)}$ and $T_3^{[bc]}\,.$

\subsection{Functions of the structure $\ve\, h^*\widehat{T}\widehat{U}$ in $n=4$}
\label{ehtu}
%
These functions appear in the study of the algebra deformation in $D$-degree 3, $n=4$ . 
They are completely generated by the following terms: 
\bqn
&
T_1=\ve^{\m\n\r\s}h_{ \m}^{*\a\b}\widehat{T}_{\n\g\vert \b}\widehat{U}_{\r\s|\a}{}^{ \g} \,,\, 
T_2=\ve^{\m\n\r\s}h_{\m}^{*\a\b}\widehat{T}_{\n\b\vert \g}\widehat{U}_{\r\s|\a}{}^{\g}\,,\,&\nonumber \\ &
T_3\!=\ve^{\m\n\r\s}h^{*\a}\widehat{T}_{\m\n\vert }{}^{\b}\widehat{U}_{\r\s|\a\b}\,,\, 
T_4\!=\ve^{\m\n\r\s}h^{*}_{\m}\widehat{T}^{\a\b\vert }{}_{ \n}\widehat{U}_{\r\s|\a\b}\,,\,
T_5\!=\ve^{\m\n\r\s}h_{ \m}^{*\a\b}\widehat{T}_{\n\r\vert \g}\widehat{U}_{\s\a|\b}{}^{ \g}.&
\nonumber 
\eqn

There are three Schouten identities:
\begin{eqnarray}
\d^{[\a\b\g\d\e]}_{[\m\n\r\s\t]}\ve^{\m\n\r\s} h^{*~\,\t}_{\a\l}  \widehat{T}_{\b\g\vert \h}\widehat{U}_{\d\e\vert}{}^{\l\h}=0\;,\
\d^{[\a\b\g\d\e]}_{[\m\n\r\s\t]}\ve^{\m\n\r\s} h^{*}_{\a}  \widehat{T}_{\b\g\vert \l}\widehat{U}_{\d\e\vert}{}^{\t\l}=0\;,\nonumber \\
\d^{[\a\b\g\d\e]}_{[\m\n\r\s\t]}\ve^{\m\n\r\s} h^{*~\,\l}_{\a\h}  \widehat{T}_{\b\g\vert \l}\widehat{U}_{\d\e\vert}{}^{\t\h}=0
\;.\nonumber 
\end{eqnarray}
An explicit expansion of these identities yields the  relations
$$T_3+2T_2+2T_5=0\;,\ T_3-T_4=0\;,\ T_1=0\;.$$
%
\subsection{Functions of the structure $\ve \, C^*\widehat{U}\widehat{U}$ in $n=4$}
\label{ecuu}
%
The Schouten identities for the functions of the structure $\ve \, C^*\widehat{U}\widehat{U}$ in $n=4$ are needed for the analysis of the algebra deformation in $D$-degree four.
The functions at hand are generated by $T_1^{[bc]}=\ve^{\m\n\r\s}\
C^{*}_{\a\b}\ \widehat{U}^{b\ \ \a\g}_{\m\n|}\ \widehat{U}^{c\ \ \b}_{\r\s|\ \ \g}\,$ and $T_2^{bc}=\ve^{\m\n\r\s}\ C^{*}_{\m\b}\ \widehat{U}^{b}_{\n\r|\a\g}\ \widehat{U}^{c\ \b |\a\g}_{ \s}\;.$ However, these vanish because of the Schouten identities
$$\d^{[\a\b\g\d\e]}_{[\m\n\r\s\t]}\ve^{\m\n\r\s}C^{*\l}_{\a}\widehat{U}^{b~~\t\h}_{\b\g\vert}\widehat{U}^{c}_{\g\d\vert\l\h}=0\;,\ 
\d^{[\a\b\g\d\e]}_{[\m\n\r\s\t]}\ve^{\m\n\r\s}C^{*\t}_{\a}\widehat{U}^{b~~\l\h}_{\b\g\vert}\widehat{U}^{c}_{\g\d\vert\l\h}=0\;.
$$
Indeed, they imply that
$T_1^{[bc]}+T_2^{bc}=0$ and $T_2^{bc}=T_2^{(bc)}$, which can be satisfied only if $T_1^{[bc]}=T_2^{(bc)}=0\,.$

%
\subsection{Functions of the structure $\ve C \pa^3 h h$ and $\ve C \pa^2 h\pa  h$  in $n=3$}
\label{cpahh}
%
These functions appear when solving $\d a_1+\g a_0=db_0$ in Section \ref{dim3peu}. In generic dimension ($n>4$), there are respectively 45 and 130 independent functions in the sets $\ve C \pa^3 h h$ and $\ve C \pa^2 h\pa  h\,.$  In three dimensions, there are 108 Schouten identities relating them, which leave only 67 independent functions. One can compute all these identities and the relations between the 108 dependent functions and the 67 independent ones. However, given their numbers, they will not be reproduced here. 
